\documentclass[3p]{elsarticle}
\usepackage{amsmath}
\usepackage{amssymb}
\usepackage{epstopdf}
\usepackage[font={small}]{caption}
\usepackage{subcaption}
\usepackage[breaklinks=true]{hyperref}
\usepackage{overpic}
\usepackage{placeins}

\hypersetup{colorlinks, citecolor = {blue}, linkcolor= {blue},
 pdftitle = {Cooperative Surmounting of Bottlenecks}}

\journal{Physics Report}

\begin{document}

\begin{frontmatter}

\title{Cooperative surmounting of bottlenecks}

\author{D. Hennig$^{1}$, C. Mulhern$^{2}$, L.
Schimansky-Geier$^{3}$, G.P. Tsironis$^{4,5}$, and P. H\"{a}nggi$^{2,6,7}$}

\address{
1 Dept. Math., University of Portsmouth,
  Portsmouth, PO1 3HF, UK\\
2 Max Planck Institute for the Physics of Complex
  Systems, N\"othnitzer Str.~38, 01187 Dresden, Germany \\
3 Institut f\"ur Physik, Humboldt-Universit\"at zu Berlin, Newtonstra{\ss}e
  15, 12489 Berlin, Germany\\
4 Dept. Phys., University of Crete, and IESL, FORTH, P. O. Box 2208,
  71003 Heraklion, Greece\\
5 Dept. Phys., Nazarbayev
  University, 53 Kabanbay Batyr Ave., Astana 010000, Kazakhstan\\
6 Institut f\"ur Physik, Universit\"at  Augsburg,
  Universit\"atsstr. 1, D-86135 Augsburg, Germany\\
7 Nanosystems Initiative Munich, Schellingstr. 4, D-80799 M\"{u}nchen, Germany
}

\begin{abstract}
The physics of activated escape of objects out of a metastable state
plays a key role in diverse scientific areas involving
chemical kinetics, diffusion and dislocation motion in solids, nucleation, electrical transport,
motion of flux lines  superconductors, charge density waves, and transport processes
of macromolecules and astrophysics, to name but a few. The underlying activated processes present
the multidimensional extension of the Kramers problem of a single Brownian particle.
In comparison to the  latter case, however,
the dynamics ensuing from the  interactions of many
coupled units can lead to intriguing novel phenomena
that are not present when only a single degree of freedom is involved.
In this review we report on a variety of
such phenomena that are exhibited by systems consisting of chains of interacting units in the presence of
potential barriers.

In the first part we consider recent developments in the case of
a deterministic dynamics driving cooperative escape processes of coupled nonlinear units
out of metastable states. The ability of chains of coupled units
to undergo spontaneous conformational transitions
can lead to a self-organised escape. The mechanism at work is that the energies of the units become re-arranged, while keeping the total 
energy conserved, in forming localised energy modes that in turn trigger the cooperative escape. We present scenarios of significantly enhanced noise-free escape rates if compared to the noise-assisted case.

The second part of the review deals with the collective directed transport of systems of interacting particles
overcoming energetic barriers in periodic potential landscapes.
Escape processes in both time-homogeneous and time-dependent driven systems  are considered for the emergence of
directed motion. It is shown that
ballistic channels immersed in the associated mixed high-dimensional
phase space are at the source for the directed long-range transport.
Open problems and future directions are discussed in order to invigorate readers to engage in their own research. 
\end{abstract}


\end{frontmatter}
\tableofcontents

\section{Introduction}
\label{section:Intro}

Ever since the seminal work by Kramers (for a comprehensive review
see Ref.~\cite{hanggi.1990.romp}) there has been significant interest in the dynamics
of escape processes of single particles, of coupled degrees of
freedom or of chains of coupled objects out of metastable states. To
accomplish the escape the considered objects must cross an energetic
barrier, separating the local potential minimum from a neighbouring
attracting domain. From the perspective of  statistical physics
it is predominantly the thermally activated escape, based on the permanent
interaction of the considered system with a heat bath, which is being
investigated \cite{hanggi.1990.romp,pollak.2005.chaos}. The coupling to the heat bath causes dissipation
and local energy fluctuations and consequently the escape process is conditioned
on the creation of a rare, optimal fluctuation which in turn
triggers an escape process. To put it differently, an optimal fluctuation
transfers sufficient energy to the system so that the system is able
to statistically surmount the energetic bottleneck associated with
the transition state. Characteristic time-scales of such a process
are determined by the inverse of the corresponding rates of escape out
of the domain of attraction. Within this topic, numerous extensions
of Kramers escape theory and of first passage time problems  have
been widely investigated \cite{hanggi.1990.romp,pollak.2005.chaos,hanggi.1986.jstatphys_a,hanggi.1986.jstatphys_b}. Early 
generalisations to
multi-dimensional systems date back to the late $1960$'s
\cite{Langer.1969.AnnPhys}. This method is by now well established and is
commonly put to use in biophysical contexts, and for a great many other
applications occurring in physics and chemistry and related areas
\cite{petukhov.1972.zetf,buttiker.1981.pra,Hanggi.1988.PRL, Hanggi.1996.PRL, Marchesoni.1996.PRL, Sung.1996.PRL, Park.1998.PRE, 
Sebastian.2000.PREb, Lee.2001.PRE_b,
Lee.2001.PRE_a, Kraikivski.2004.EPL, Downton.2006.PRE, Hanggi.2005.AnnDerPhys, astumian.2002.phystoday, reimann.2002.apa,
Cattuto.2003.EPL,hanggi.2009.romp}. More modern developments of rate theory address escape under time-dependent driving, typically
being of periodically varying nature
\cite{reimann.1999.pre,lehmann.2000.prl,maier.2001prl,lehmann.2003.physstatsol,talkner.2005.njp} so that a time periodic asymptotic state emerges as the most stable state.
Applications of such time-dependent driven escape problems involve the objective of escape across sloshing potential barriers or wells or also the stochastic neural dynamics involving a time-periodically varying threshold to ``firing events''.

Distinct from the numerous studies of  thermally induced escape processes driven  by  the coupling
of the system to an external  heat bath
is the {\it microcanonical situation}; i.e., when only the internal total fixed energy of a many-particle system is available to
perform structural transitions -- a situation which been studied to a much lesser extent.
The underpinning key question is: how is it possible that many degrees of freedom subject to a deterministic and conservative
dynamics can overcome a
potential barrier given that  initially its  energy is
almost equally shared among the units so that the system is far
away from the critical equilibrium configuration related with a
saddle point in configuration space. The latter configuration is commonly referred to as a localised transition
state \cite{forst.1971.AcadPress}.

Concerning the attainment of localised structures it is
by now an   established fact that nonlinearity and diffusion in extended systems may cooperate to
give rise to the formation of coherent
structures that emerge even from an initial, almost homogeneous state
\cite{remoissenet.1999.SprVerB}. The concentration of an originally  distributed
physical quantity to a few degrees of freedom in
confined regions of a spatially extended and homogeneous
systems proceeds typically in a self-organised manner. In recent years
the concept of intrinsic localised modes (LMs), or discrete breathers, as
time-periodic and spatially localised solutions of nonlinear
lattice systems has proved  to present the archetype of
excitations describing localisation phenomena in numerous physical
situations \cite{mackay.1994.nonlin,aubry.1997.PhysD,flach.1998.PhyRep,marquie.1995.pre,eisenberg.1998.prl,binder.2000.prl}. For the
creation of localised structures, modulational instability, leading
to a self-induced modulation of an initial linear wave with a
subsequent generation of localised pulses, has proven to be an
effective mechanism. In this way energy localisation in a
homogeneous system is achievable. For example,  breathers have been
successfully applied to describe localised excitations  which
reproduce typical features of the thermally induced  opening dynamics of DNA duplex molecules such
as the magnitude of the amplitude and the time-scale of the
oscillating bubble preceding full strand separation (denaturation)
\cite{peyrard.1989.prl,dauxois.1993.pre,barbi.1999.PLA,cocco.1999.prl,barbi.1999.JBioPhys}.
The appearance of LMs in damped,
driven   deterministic nonlinear lattice systems
has been the subject of the investigations in \cite{marin.1996.nonlin,hennig.1999.pre,vanossi.2000.pre,maniadis.2006.epl}.
Furthermore, the spontaneous formation
of LMs (breathers) from thermal fluctuations in lattice systems,
when thermalised  with the Nos\'e-method \cite{nose.1984.chemphys} has been
demonstrated in \cite{Peyrard.1998.PhysicaD,dauxois.1993.pre,tsironis.1996.prl}.

The objective of  this review is to elaborate on the
scenario of  possible escape from a metastable domain of
attraction; the main mechanism being based on the cooperativity between the interacting units of
{\it purely deterministic} systems.  Thus, no
additional coupling to a thermal bath assists the escape. We consider  macroscopic
discrete, coupled nonlinear oscillator chains with typically $100$
links, as these may appear as realistic models in neuroscience, in
various biophysical contexts or also in networks of coupled
superconductors, e.g. see in
Refs.~\cite{flach.1998.PhyRep, rabinovich.2006.romp, sato.2006.romp, Takeno.2005.PLA}. An
efficient deterministic escape that is driven in absence of noise is
particularly important when dealing with low temperatures for which
the activated escape becomes far too slow, or also for situations
with many coupled nonlinear units in presence of non-thermal
intrinsic noise that scales inversely with the square root of the
system size.

We demonstrate how {\it cooperative behaviour} in
systems of interacting units facilitates the task performance of surmounting potential barriers,
which individual units, due to sheer limiting energetic reasons,
would never be able to accomplish.
In fact, it is the self-organised directed energy redistribution among the units of the coupled chain,
leading to the formation
of the critical LM, that
makes the collective barrier crossing of the  chain  in a
  microcanonical situation possible. Moreover, in the deterministic context
the barrier crossing of a chain turns out to be (by far) more efficient compared with a
  thermally activated chain for small ratios between
the total energy of the chain and the barrier energy.
 Also directed
 particle transport in metastable (multiple-well) periodic potential landscapes necessitates
 appropriately coordinated energy transfer processes among the interacting units enabling
 consecutive barrier crossings.
 While there exists a vast literature on directed transport of single
 particles moving mainly in
 periodic potentials
  based on the so-called ratchet effect (for a reviews see \cite{astumian.2002.phystoday,hanggi.2009.romp,denisov.2014.pr}
  and cited references therein),
  by far less is known regarding the transport of systems coupled particles through
  periodic landscapes.  Especially, the effects stemming from the coupling
  between the particles on
  the emergence of a current require a reinforced study.
  Moreover, in contrast to the driven one-dimensional
case, with its three-dimensional phase space, the dynamics of coupled units evolve in a higher-dimensional phase space, of which the
details of its structure remain more elusive. Thus, an additional goal of
  the current review is to present also the state of the art relating to such collective transport.

The review is organised as follows: In Sec.~\ref{ra_ch1} the cooperative escape dynamics of a
one-dimensional chain from  the
metastable state of cubic external potential is considered.
As an extension to the one-dimensional chain system, in Sec.~\ref{ra_ch2} follows
the investigation of a two-dimensional chain model with 
pairwise nonlinear Morse interaction. The special focus will be on the influence
of the additional degree of freedom on the self-organised
escape process.
Escape from metastable states in a system of higher (geometrical) complexity
in the form of  a ring of interacting units evolving in a Mexican hat
two-dimensional potential landscape is studied in the subsequent section.

In view of controlling the process of barrier crossing we consider in the next section
the collective escape dynamics
of a chain of coupled units in the presence of thermal noise and friction. In addition, the application of
a weak
external ac-field leads to
periodically oscillating landscape of bottlenecks. We show
that the formation of the critical LM can be distinctly
accelerated via the application of a weak external ac-driving.
By use of optimally oscillating barrier configurations it is
feasible that
a far faster escape can be promoted, leading to a drastic
enhancement of the escape dynamics. Particularly at low
temperatures, where the rate of thermal barrier crossing is
exponentially suppressed, such a scenario can be very beneficial.

That even a very weak driving force suffices to trigger
fast escape for a chain situated initially extremely
close to the bottom of the potential well and thus containing
a vanishingly small amount of energy is demonstrated in Sec.~\ref{subsec:noisefreeescape}.

The chaos-assisted mechanism of dissociation of interacting molecules related to an organised barrier crossing is studied in
Sec.~\ref{subsection:dissociation}.

Regarding collective transport we distinguish between the {\it time-independent} (autonomous) and the {\it time-dependent} 
(non-autonomous) Hamiltonian systems.
Beginning with autonomous Hamiltonian systems modeling two coupled particles moving in a periodic washboard
potential
the guiding aim is to understand the conditions under which directed
transport in phase-space is supported.  Novel mechanisms pertaining to directed transport
that are a direct consequence of the interaction between the particles are discussed.

Subsequent augmentation of the system
by the inclusion of time-dependent driving and damping terms  adds further complexity
to the coupled dynamics.
For directed particle transport mediated by periodically driving forces we derive exact results
regarding the collective character of the running solutions.

Interaction-induced absolute negative mobility \cite{eichhorn.2002.prl,eichhorn.2002.pre,eichhorn.2005.chaos,machura.2007.prl,kostur.2008.prb} of two coupled particles evolving in a symmetric and periodic substrate potential which is subjected to a static bias force is the subject of the next section. The bias force of magnitude $F_0$ serves
to tilt the potential landscape such that particle motion in the direction of the bias is favoured. It is shown that within a range of coupling strengths,
the coupled particles can become self-organised and go, as periodic running states, frequency
locked with the driving, against the direction of the bias force.

Regarding features of transport exhibited by systems
with many degrees of freedom we consider the Hamiltonian dynamics of a one-dimensional
chain of linearly coupled particles in a spatially
periodic potential which is subjected to a time-periodic mono-frequency external field.
For a chain escaping collectively from a potential, the
possibility of the successive generation of a directed flow based on
large accelerations is illustrated.

We conclude the review with a discussion of open questions and point
out promising directions which may invigorate readers to undertake their own future studies in this research area.

\section{Deterministic dynamics and self-organised cooperative escape
dynamics}\label{section:escape}

\subsection{Crossing one-dimensional
 anharmonic potential barriers}\label{ra_ch1}

\subsubsection{Escape of a one-dimensional chain}\label{subsubsection:1d-chain_model}
Consider first a one-dimensional chain of $N$ linearly coupled units
of mass $m$ \cite{Gross.2014.WorldSci,Hennig.2007.PRE,hennig.2007.epl}. The chain is positioned in a cubic external potential where its
equally spaced units perform one-dimensional motion with
elongations $q_n(t), ~n=1,\ldots,N$, in parallel directions. That is, the $q_n$
describe transversal elongations that are perpendicular
to the direction of the potential valley (see Fig.~\ref{fig:ts:configs}).
Each unit of mass unity experiences
a nonlinear force caused by the potential
\begin{equation}
V(q_n)=\frac{\omega_0^2}{2}q^2_n-\frac{a}{3}q^3_n.
\end{equation}
The parameter $a>0$ controls the strength of the nonlinearity in the system. This
potential possesses a metastable equilibrium at $q_{min} = 0$
corresponding
to the rest energy $E_{min} = 0$ and the maximum is located at
$q_{max} = \omega_0^2/a$ with
barrier energy $E_{max} \equiv \Delta E =\omega_0^6/6a^2$.
Furthermore, there are  coupling forces acting between neighbouring
units with coupling parameter
$\kappa$ regulating the strength of the interactions. Moreover, each unit possesses a 
corresponding momenta $p_n(t)$ that are canonically
conjugate to $q_n(t)$. Unless otherwise stated, periodic boundary conditions
are imposed throughout this report.
First, we consider the deterministic dynamics in a micro-canonical
situation with a Hamiltonian
set-up.  This implies that the {\it total} energy of
the whole chain, $E$, is conserved.

Passing to dimensionless quantities by the following rescalings,
$\widetilde{q}_n=a/(\omega_0^2)\,q_n$,
${\widetilde{p}_n}\,^2=a^2/(\omega_0^6)p_n$, and
${\widetilde{t}}=\omega_0\,t$, we obtain
a Hamiltonian with only a single remaining parameter, the effective
coupling strength $\widetilde{\kappa}=\kappa/(\omega_0^2)$.
In what follows we omit the tilde notation.

The Hamiltonian of the considered chain thus reads:
\begin{align}\label{eq:hamiltonian1}
  \mathcal{H}=\sum_{n=1}^{N}\left[\frac{p_n^2}{2}
  +\frac{\kappa}{2}\left(q_n-q_{n+1}\right)^2+V\left(q_n\right)
  \right], \qquad V(q_n)=\frac{q_n^2}{2}-\frac{q_n^3}{3}.
\end{align}
The resulting equations of motion become
\begin{align}
  \label{equ_motion_1d}
  \ddot{q}_n+q_n-{q_n}^2-\kappa \left(q_{n+1}+q_{n-1}-2\,q_n\right)=0,
  \qquad q_{N+1}=q_1.
\end{align}

For the study of  escape we initially
place the units of the chain close to the bottom of
the external potential, i.e. nearby  $q_{min}=0$, providing each unit with
very little energy compared to the barrier energy of the external potential.
Due the system dynamics the chain is able to generate a critical
state -- a critical
elongation -- in a self-organised manner, thereby
surpassing the potential local maximum. 
This initiates a
transition of the chain into the unbounded regime $q_n > q_{\mathrm{max}}=1$,
 for all $n=1,\ldots,N$, 
 which we refer to as an escape event. Fig.~\ref{fig:ts:configs}  illustrates  
 various transition states which
will be described in more detail below.
For a single particle to overcome the potential barrier
it needs to be supplied with an energy in excess of $\Delta E=V(1)-V(0)=1/6$.

In what follows we make evident that the generation of these
critical states proceeds within a  highly efficient manner,
even in cases where the total chain energy $E$
is small compared to the cumulative barrier energy of all units; i.e., $E \ll N\cdot \Delta E$. This
low-energy regime
is provided by the following initial preparation of the system:
(i) First, the whole chain is placed at a
fixed position $q_n(0)=q_0$, $n=1,\ldots,N$, near the bottom of the potential
well. (ii)
Then, the position and momenta of all units are iso-energetically randomised
while keeping
the total energy a constant -- i.e., $E = const.$. The random position and
momentum values are
uniformly distributed in intervals
\begin{align*}
|q_n(0)-q_0| \le \Delta q \qquad
|p_n(0)|\le \Delta p.
\end{align*}
The value of $q_0$ is chosen in such a way that the average excitation energy
of a single unit is small compared to the depth, $\Delta E$, of the
potential well. For our numerical simulations we consider chains
comprising $N=100$ units.

\subsubsection{Transition states} For conservative Hamiltonian
systems local minima of the energy surface, $H=E=const.$,
are Lyapunov stable. That is, orbits in the vicinity of a
local elliptic equilibrium point never leave it as their associated energy is
conserved. Only those orbits with energies exceeding the energy associated with
a neighbouring saddle point  are no longer bound to the
basin. Thus, the saddle point is referred to as a transition state, as it
separates bounded from unbounded orbits. Concerning our objective, the
system's energy has to exceed the transition state energy to make an
escape event possible. To determine this transition state, we have
to solve $\nabla U(q_1,q_2,\ldots)=0$, where $U$ denotes the potential
energy (thus the transition state is a fixed-point) and
the solution is such that  all eigenvalues of the Hessian matrix of $U$ are
positive,
except for a single negative one.

In the one-dimensional chain model the transition state configurations
solve the stationary equation\footnote{An alternative approach for the
  one-dimensional chain model is presented in
  \cite{hennig.2007.epl}. It casts the stationary
  equation into a two-dimensional map and links the localised lattice
  solutions to its homoclinic orbits.}
\begin{align}
\label{ts1d:equ_sad_point_1d}
q^*_n-{q^*_n}^2-\kappa \left(q^*_{n+1}+q^*_{n-1}-2\,q^*_n\right)=0,
\end{align}
and satisfy the above mentioned condition on the eigenvalues, $\lambda^H$, of
the
Hessian matrix $H$
\[H_{i,j}=\delta_{i,j}(2\,\kappa+1-2\,q^*_i)-\kappa\,\left(\delta_{i,j+1}
+\delta_{i,j-1}\right).  \]

\begin{figure}
\centering
\includegraphics[width=0.3\linewidth]{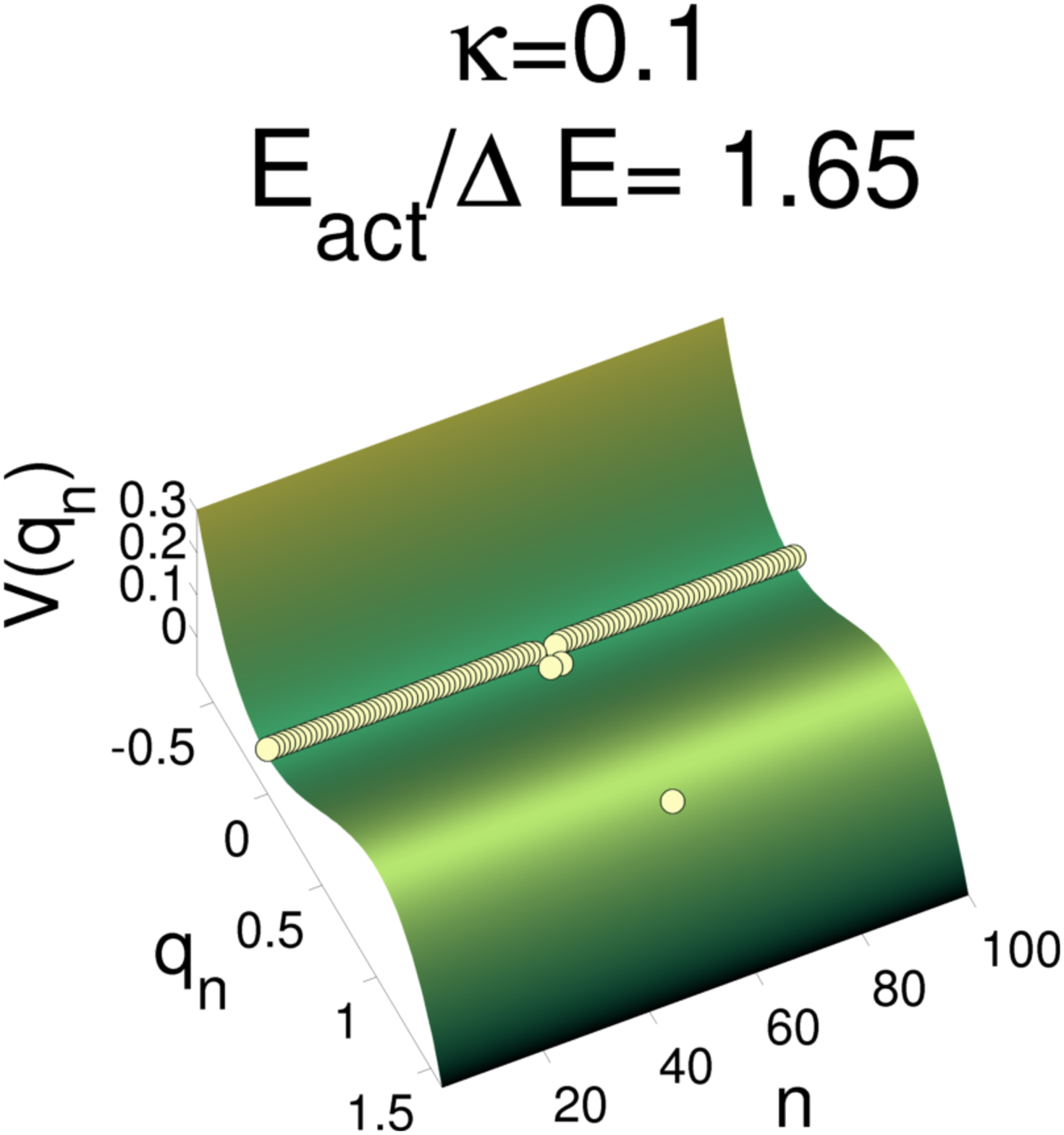}\hfill
\includegraphics[width=0.3\linewidth]{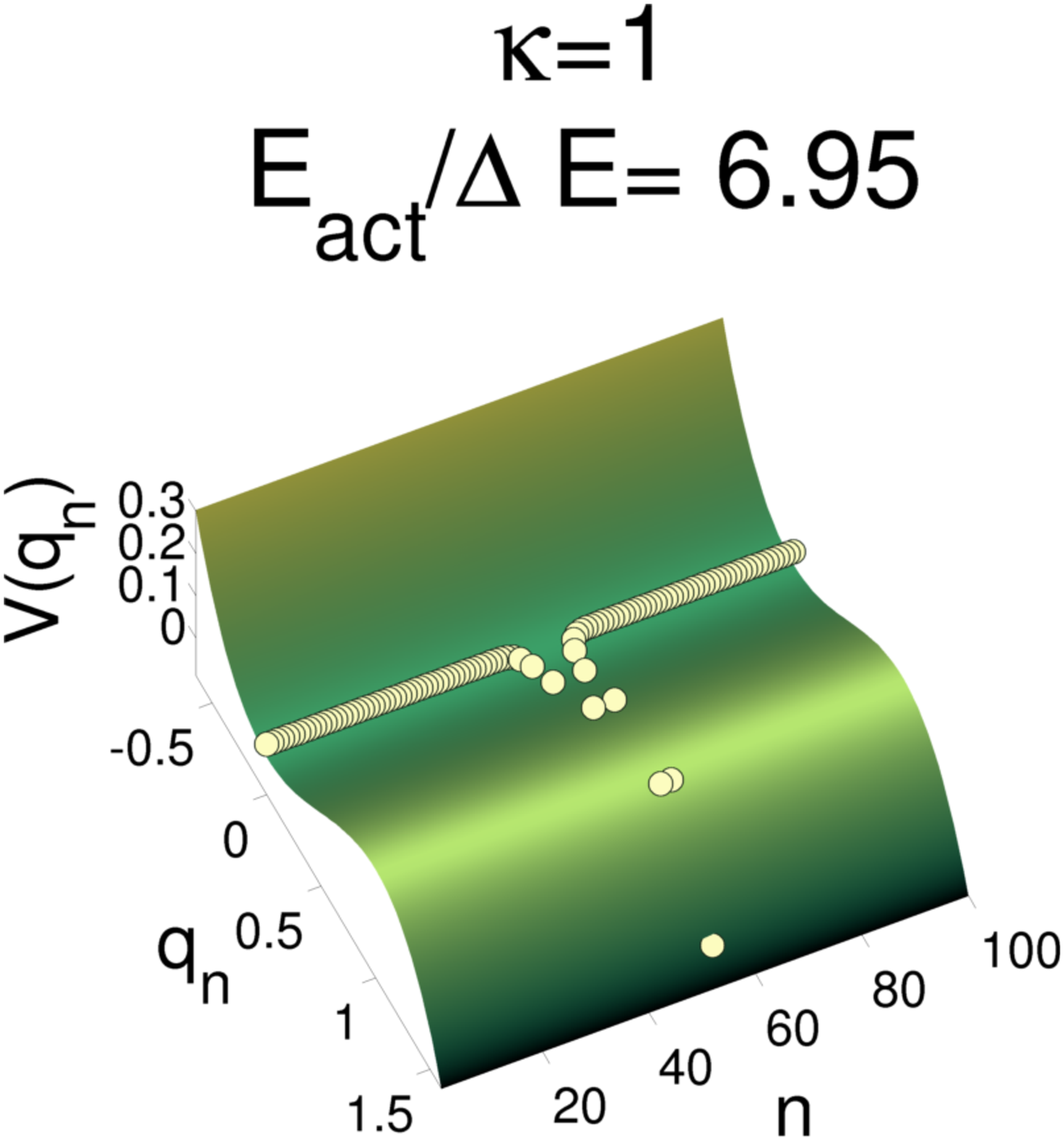}\hfill
\includegraphics[width=0.3\linewidth]{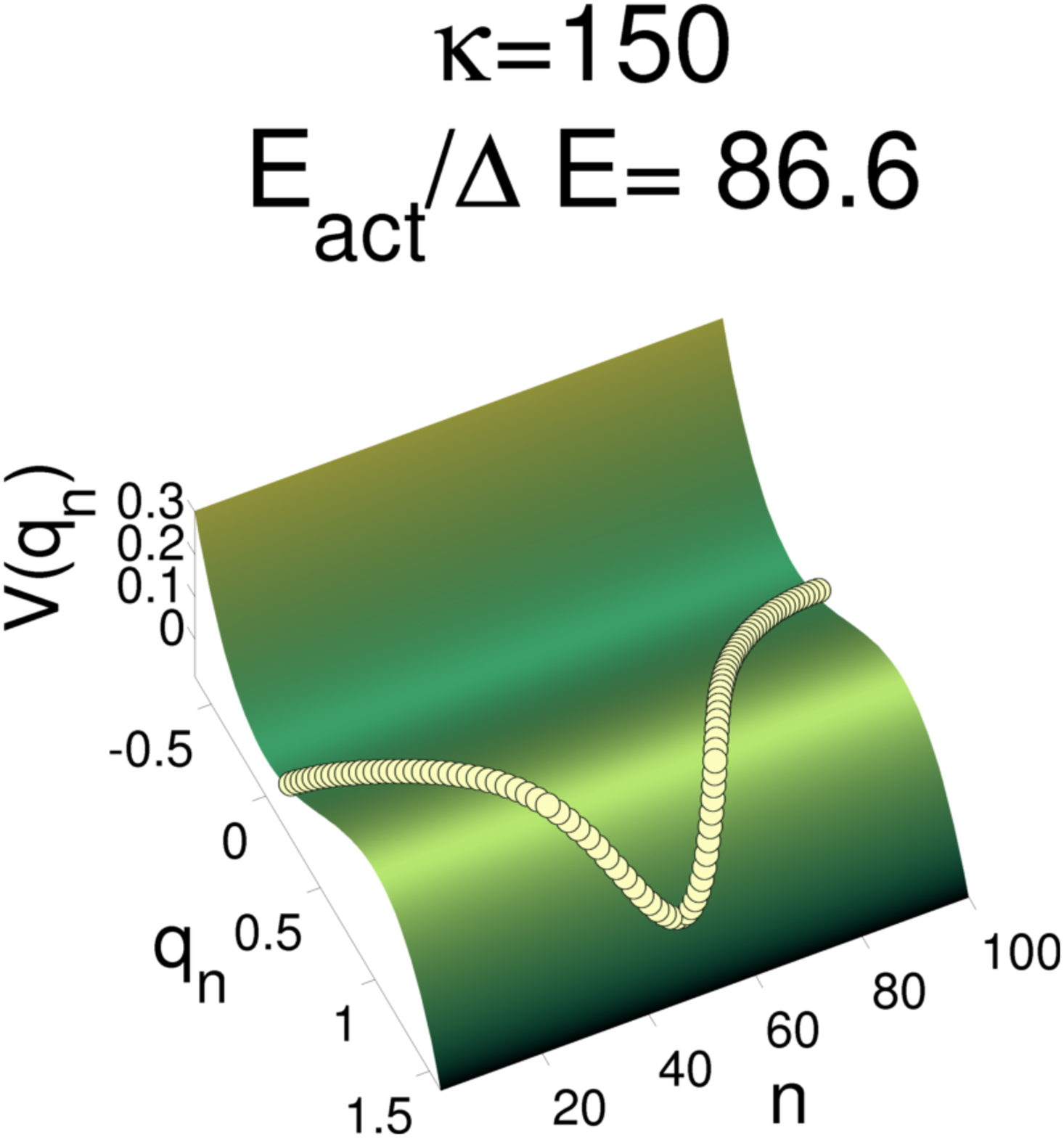}
  \caption{(Colour online) Transition state configurations for a chain placed in a cubic potential for three different values
    of coupling strength $\kappa$. The chain consists of $N=100$
units, whose dynamics obey Eq.~(\ref{equ_motion_1d}). With increasing coupling strength, more units are required for the chain to
adopt
the transition state. Moreover, the ratio $E_{act}/\Delta E$, (see text) of the activation energy required for the chain to adopt the transition
state and the barrier height, increases with $\kappa$ -- that is, more energy is required in the transition state configuration,
as the interactions between the units become stronger. 
Source: Figure taken from Ref.~\cite{Gross.2014.WorldSci}.}
  \label{fig:ts:configs}
\end{figure}

In the case of a vanishing coupling strength the units decouple.
Therefore, the fixed points of the system consist of all the
configurations where each units is placed either on the maximum
of the potential barrier or the potential valley, that is $q_n^*=\left\lbrace
  0,1 \right\rbrace$. In this case also the Hessian matrix, $H$,
becomes diagonal and we can directly read off its eigenvalues,
$\lambda^H_n=1-2\,q^*_n$. Demanding that all but one of the eigenvalues be
negative, the transition states are found to be all the
configurations where all unit are positioned in the potential
valley except for one that is placed on top of the potential barrier. The
according energy is $E_{\mathrm{act}}(\kappa=0)=\Delta E=1/6$.

In contrast, a very large coupling strength corresponds to a situation
where the chain becomes effectively homogeneous and can be considered as a
single unit. Therefore, the
transition state corresponds to a chain configuration where all units
are placed at the maximum of the potential. This can be seen by
taking the limit $\kappa \rightarrow \infty $ in
Eq.\,(\ref{ts1d:equ_sad_point_1d}). If we want $q_n^*$ to take on values
within an energetically bounded regime we must have
$q_{n+1}^*+q_{n-1}^*-2\,q_n^*=0$
in order to satisfy Eq.\,(\ref{ts1d:equ_sad_point_1d}) in this limit.
In the case of periodic boundary conditions this becomes equivalent to
$q_n^*=q^*$ so that Eq.\,(\ref{ts1d:equ_sad_point_1d}) becomes
$q^*(1-q^*)=0$. Which of its two roots corresponds to the transition
state becomes clear from the linear stability analysis of
Eq.\,(\ref{equ_motion_1d}) for this effective one unit problem

\begin{equation}\ddot{q}=-q+{q}^2\approx
-q^*+{q^*}^2+\left(-1+2\,q^*\right)q=\left(-1+2\,q^*\right)q.
\end{equation}

\noindent
Only the case $q_n=q^*=1$ is associated with the inherent instability of a
transition
state. Accordingly, the transition state energy is found to be
$E_{\mathrm{act}}(\kappa\rightarrow\infty)=N\,\Delta E=N/6$.

\begin{figure}
  \centering
  \includegraphics[width=0.5\linewidth]{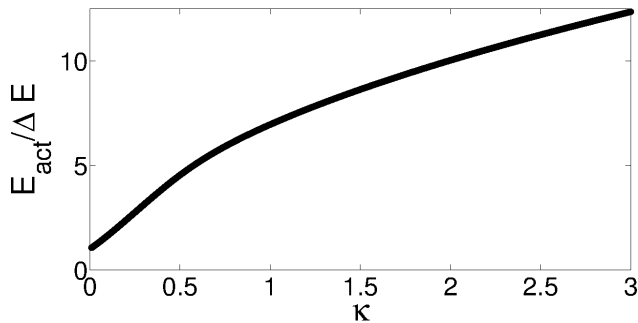}
  \caption{Activation energy (energy of transition state
    configurations) as a function of the coupling strength $\kappa$, for chains consisting of $N=100$ units. Each unit evolves in a
cubic potential and is linearly coupled to its nearest neighbours. There is a clear tendency for the activation energy of the chain to increase with
increasing coupling strength $\kappa$. Source: Figure taken from Ref.~\cite{Gross.2014.WorldSci}.}
  \label{fig:ts:E_act}
\end{figure}

To determine the transition state in the intermediate coupling strength regime 
we used the dimer method, first
introduced in \cite{Henkelman.1999.ChemPhys}, which is a minimum-mode following
method that
solely makes use of gradients of the potential surface. 
The obtained transition states together with their energy are represented in Fig.~\ref{fig:ts:configs}
and Fig.~\ref{fig:ts:E_act}, respectively. 
The latter is computed by substituting the solutions $q^*_n$ to Eq.~(\ref{ts1d:equ_sad_point_1d}) 
into the energy function given in Eq.~(\ref{eq:hamiltonian1}) together with $p^*_n=0$.
 The maximal amplitude of the
hair pin-like transition state configuration grows with increasing
$\kappa$ until it reaches a critical elongation from which it
decreases until the entire chain approaches the maximum of the
potential barrier as described above. Further,
the stronger the coupling is the higher the activation energy,
which is presented in Fig.~\ref{fig:ts:E_act}
in terms of the
barrier height $\Delta E$.

\subsubsection{Modulational instability and spontaneous
localisation}
The energy that is initially almost homogeneously distributed along the
entire chain quickly concentrates into local excitations of single
oscillators. This process is governed by the formation of regularly
shaped wave patterns, so-called breathers which are spatially localised and
time-periodically varying solutions.
Their emergence is due to
a mechanism known as \emph{modulational instability}. In the following we elaborate on the onset of this modulational
instability by applying it to the above situation. Further details are contained in \cite{Kivshar.1992.pra} and
\cite{Daumont.1997.nonlin}.

As an approximation for small oscillation amplitudes we can neglect
the nonlinear term in Eq.~(\ref{equ_motion_1d}). The  resulting equation in
linear
approximation exhibits phonon solutions with frequency $\omega$ and
wave number $k=2\pi\,k_0/N$ (with $k_0\in\mathbb{Z}$ and $-N/2\leq k_0
\leq N/2$) related by the dispersion relation
\begin{align*}
  \omega ^2=1+4\,\kappa\,\sin^2\left(\frac{k}{2}\right).
\end{align*}
We make an Ansatz that only takes into account the first harmonics
(rotating wave approximation)
\begin{align*}
  q_i=F_{1,i}(t)\,e^{-i t}+ F_{0,i}(t)+F_{2,i}(t)\,e^{-2i t}+c.c.
\end{align*}
The amplitudes of the harmonics are expected to be of a lower order of
magnitude ($| F_{0,i}|\ll| F_{1,i}|$, $| F_{2,i}|\ll| F_{1,i}|$).
Furthermore, we assume our envelope functions to vary slowly
($|\dot{F}_{m,i}|\ll |{F}_{m,i}|$) as well as the phonon band to be
small ($1> 4\kappa$). Within the limits of these assumptions we obtain
a discrete nonlinear Schr\"odinger equation (DNLS) for the amplitudes
of the first harmonic.
\begin{align}\label{MI1d:DNLS}
  2i\dot{F}_{1,i}=\kappa
\left(\left(F_{1,i-1}+F_{1,i+1}\right)+2F_{1,i}\right)-\frac{10}{3}\,\left|F_{1,
i}\right|^2F_{1,i}.
\end{align}
We want to study the stability of this equation's plane wave solutions
in the presence of small perturbations $\left|\delta B_i(t)\right|\ll
1$ and $\left|\delta \Psi_i(t)\right|\ll 1$, leading to a new Ansatz
for the envelope function
\begin{align}\label{MI1d:envelope_Ansatz}
  {{F}_{1,n}}^{pert.}=\left(A+\delta B_n(t)\right)
  e^{i\left((k\,n-\Delta\omega\,t)+\delta\Psi_n(t)\right)}.
\end{align}
The perturbations are sufficiently small so that we can expand the
envelope function up to the first order in $\delta$ and neglect all
terms of higher order. Using the Ansatz Eq.~(\ref{MI1d:envelope_Ansatz})
in Eq.~(\ref{MI1d:DNLS}) leads to a complex differential equation for
the perturbation functions $B(t)$ and $\Psi(t)$.  The real and
imaginary part of this equation are independent. Hence, collecting all
terms of first order in $\delta$ results in two linear relations.
\begin{align}
  -A\,\dot{\Psi}_i&=-\frac{\kappa}{2} \left\lbrace A\,\sin k
\left(\Psi_{i-1}-\Psi_{i+1}\right)+\cos k
\left(B_{i+1}+B_{i-1}\right)\right\rbrace -\frac{10}{3} \,A^2\, B_i\\
  2 \dot{B}_i&=-\kappa\left\lbrace A\,\cos k \left(\Psi_{i+1}+
      \Psi_{i-1}-2\Psi_i \right)+\sin k
    \left(B_{i+1}-B_{i-1}\right)\right\rbrace.
\end{align}
Again, the solution to those coupled equations are plane waves
\begin{align}
  \Psi_n=\Psi ^0 e^{i(Q\,n-\Omega\,t)} \qquad
  B_n=B^0e^{i(Q\,n-\Omega\,t)}
\end{align}
with the dispersion relation
\begin{align}
  \label{MI1d:pert.-disp_rel}
  \left(\Omega-\kappa\,\sin k\,\sin Q\right)^2 =
    \kappa\,\cos k\, \sin ^2
    \left(\frac{Q}{2}\right)\left(4\,\kappa\,\cos k \, \sin ^ 2
      \left(\frac{Q}{2}\right)-\frac{20}{3} A^2 \right)
\end{align}

\noindent
which describes the stability of the $Q$-mode perturbation on the
$k$-mode carrier wave. $Q$ and $k$ have a $2\,\pi$ periodicity and can
therefore be chosen to be in the first Brillouin zone. Furthermore, we
can restrict the range of $k$ and $Q$: $k,Q\in\lbrace 0 ,\pi \rbrace$,
because negative values correspond to waves with the opposite
direction of propagation.

\begin{figure}
\centering
\begin{minipage}{0.46\textwidth}
 \includegraphics[scale=0.2]{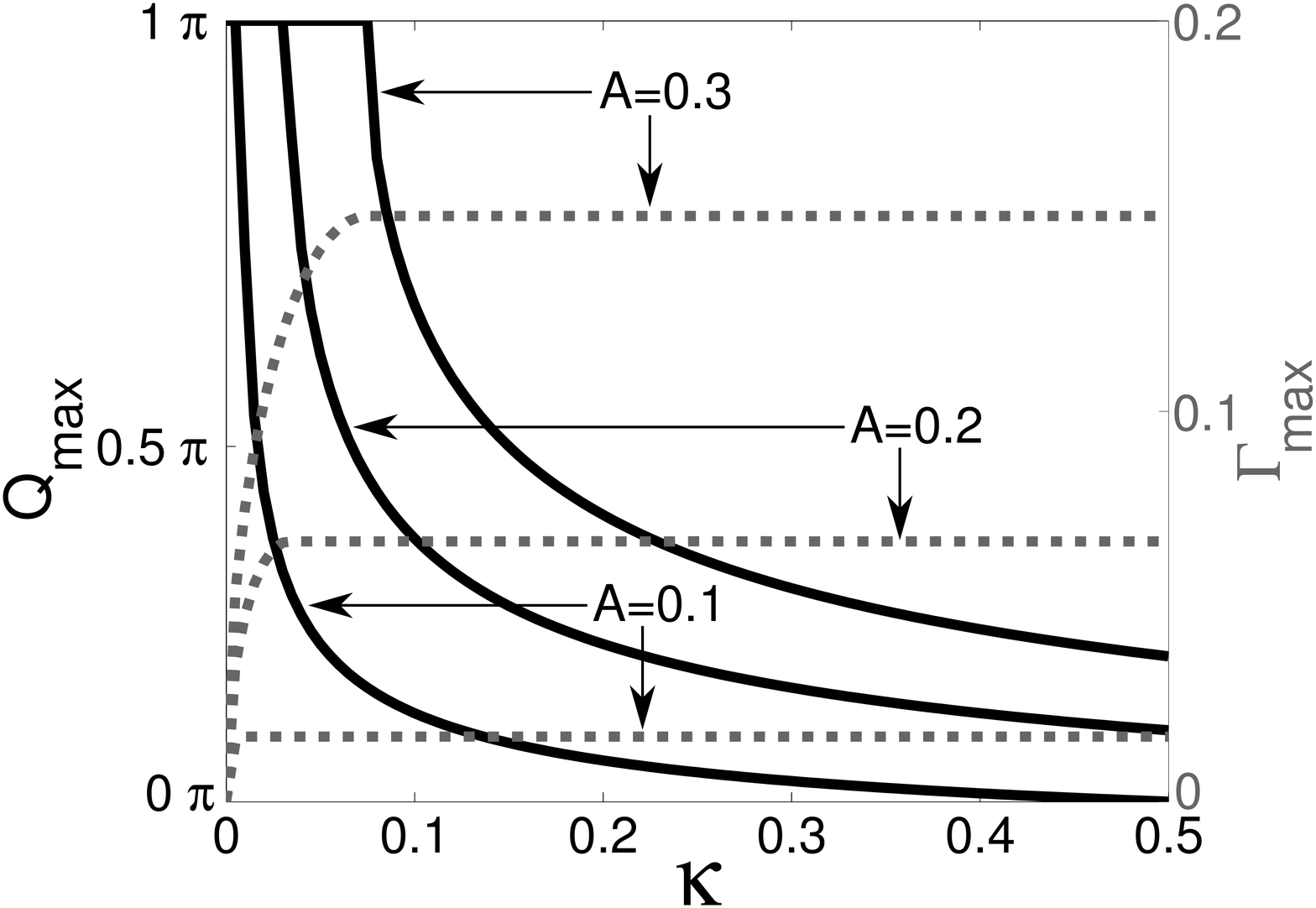}
 \caption{Fastest growing plane wave modes (solid line, left axis) - Eq.~(\ref{MI1d:Q_max}), and their growth rates (dashed line,
right axis) - Eq.~(\ref{MI1d:max_growth_rate}), for a wave number $k=0.0$, both as a function of the coupling strength $\kappa$. The
parameter $A$ controls the amplitude of the envelop function Eq.~(\ref{MI1d:envelope_Ansatz}). Source: Figure taken from
Ref.~\cite{Gross.2014.WorldSci}.}\label{fig:MI:MI_multi}
\end{minipage}
\hspace{0.04\textwidth}
\begin{minipage}{0.46\textwidth}
  \includegraphics[scale=0.2]{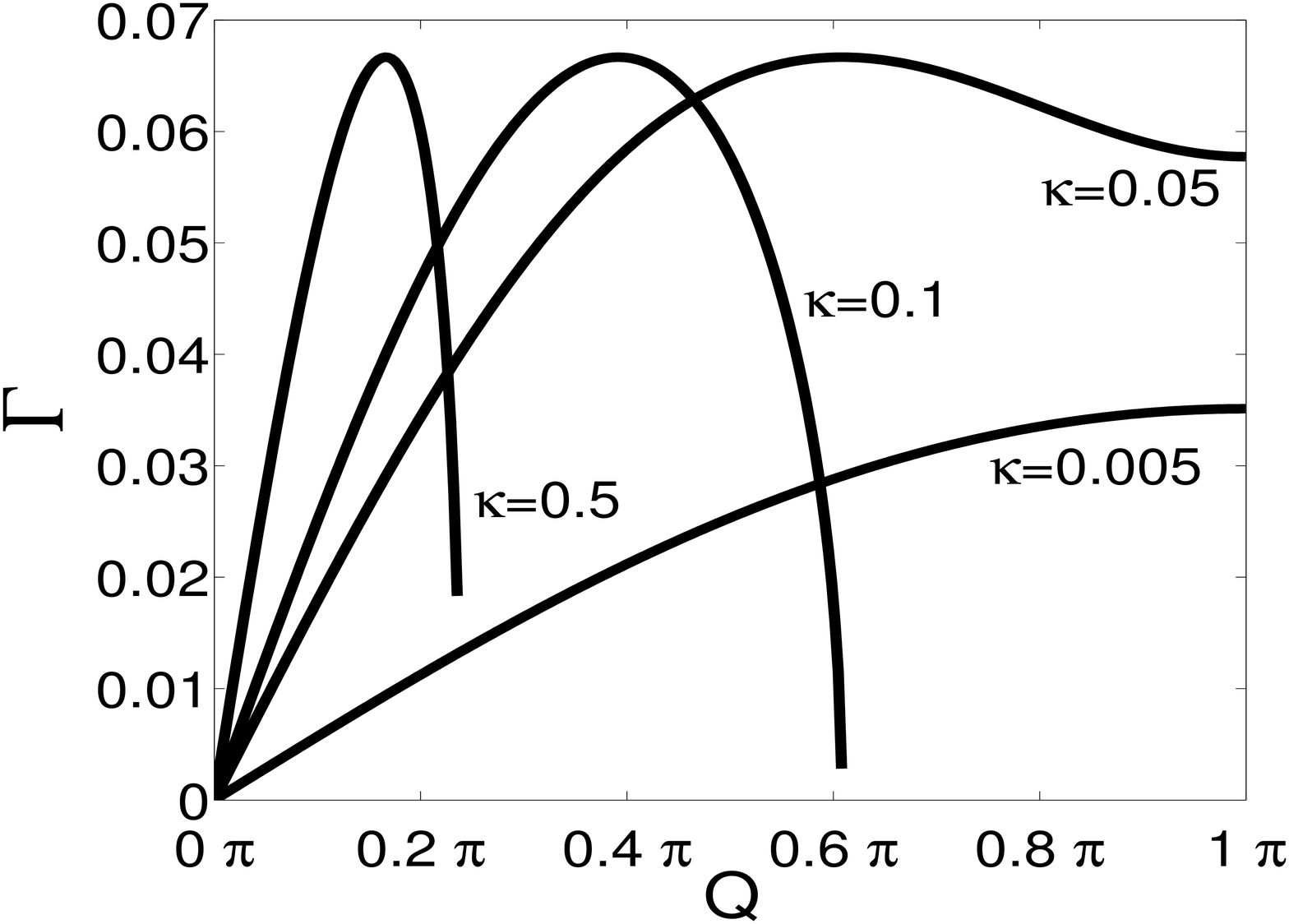}
  \caption{Growth rates $\Gamma$ of  perturbations to the derived Discrete Nonlinear Schr\"odinger equation - Eq.~(\ref{MI1d:DNLS}), as
a function of the unstable plane wave Q-modes according
    to Eq.~(\ref{MI1d:growth_rate}) from a $k=0$ carrier mode with
    $A=0.2$. Each curve is obtained for a different value of the coupling strength $\kappa$, as indicated in the figure. Source: Figure
taken from Ref.~\cite{Gross.2014.WorldSci}.}\label{fig:MI:MI_tau}
\end{minipage}
\end{figure}

The perturbations are stable for $\Omega\in\mathbb{R}$ which is the
case when the right hand side of Eq.~(\ref{MI1d:pert.-disp_rel}) is
positive. Therefore, all carrier waves with $k\in\lbrace\pi/2, \pi
\rbrace$ are stable with respect to all perturbation modes. For
$k\in\lbrace 0 , \pi/2\rbrace$ perturbations will grow, provided that
\begin{align}
  \label{MI1d:unstable_Q}
  \cos k \, \sin ^2\left(\frac{Q}{2}\right)&\leq
  \frac{5\,A^2}{3\,\kappa}.
\end{align}
We can then find an according growth rate
\begin{equation}
  \label{MI1d:growth_rate}
  \Gamma (Q)=\left |\operatorname{Im} (\Omega) \right
|=\sin\left(\frac{Q}{2}\right)\sqrt{\frac{20}{3}\kappa\,\cos
k\,\left({A}^2-\frac{3}{5}\kappa\sin^2 \left(\frac{Q}{2}\right)\cos k\right)}
\end{equation}
which, for the case that $ A^2\leq \frac{6}{5}\kappa \cos k, $ has its
maximum at
\begin{equation}
  \label{MI1d:Q_max}
  Q_\mathrm{max}=2\,\arcsin\sqrt{\frac{5\,A^2}{6\,\kappa\, \cos k}}.
\end{equation}
Otherwise the maximum growth rate is found at $Q=\pi$. The
corresponding growth rates become
\begin{align}\label{MI1d:max_growth_rate}
  \Gamma_\mathrm{max}=
  \begin{cases}\Gamma(Q_\mathrm{max})=\frac{5}{3}A^2&\text{if }A^2\leq
\frac{6}{5}\kappa\cos k\\
    \Gamma(\pi)=\sqrt{\frac{20}{3}\kappa\cos k
      (A^2-\frac{3}{5}\kappa\cos k)}<\Gamma(Q_\mathrm{max})&\text{if
    }A^2> \frac{6}{5}\kappa\cos k\\
  \end{cases}
\end{align}

We recall that our system is
initially prepared in a slightly perturbed $k=0$ mode. This is thus
the only possible carrier wave mode; a result which follows because of the amplitudes, $A$, required for the
other modes (which scale with the amplitude of the perturbation). The reason being that either the amplitudes are
likely too small to generate growing modes -- note the inequality
given by Eq.~(\ref{MI1d:unstable_Q}) -- or the arising maximal growth rates are
suppressed. Evaluating the growth rate of instabilities of the $k=0$
carrier mode for different values of $\kappa$ (Fig.
\ref{fig:MI:MI_tau}), we find that the modulational instability
becomes more mode selective upon increasing $\kappa$. Hence, for large
values of $\kappa$ the only relevant unstable modes are near the
fastest growing mode depicted in Fig. \ref{fig:MI:MI_multi}. In such a
situation we expect the emergence of a regular wave pattern (an array of
breathers) that
efficiently does localise  energy and thereby enhances the escape of the chain.

\subsubsection{The formation of breathers} The mean values of (postive amplitudes) $q_0$
are taken in such a way that the average excitation
energy
of a single unit, $E_0$, is small compared to the depth, $\Delta E$, of the
potential well. Due to the choice of sufficiently small detunings $\Delta q$
the initial lattice state, $q_n(0)=q_0 + \Delta q$,
is close to an almost homogeneous state
and yet such disturbed that there result very small -- but non-vanishing --
initial interaction terms. More precisely, Eq.~(\ref{eq:esingle})
determines the energy of a unit
\begin{equation}
E_n=\dfrac{p_n^2}{2}+V(q_n)+\frac{\kappa}{4}\left( (q_{n+1} - q_n)^2 + (q_{n-1}
- q_n)^2 \right)
.\label{eq:esingle}
\end{equation}

The last term in Eq.~(\ref{eq:esingle}) represents the interaction
energy of an individual unit. The initial set-up discussed above allows for
weak, non-vanishing, interactions between neighbouring units.
Thus an energy exchange
between the coupled units is entailed. The initial energy per unit obeys
$E_n \ll \Delta E$,
but is still sufficiently large to initiate the excitation of nonlinear
modes.
In this realm the formation of localised excitations can be explained by the
above discussed modulational instability.
This mechanism initiates an instability of
a plane wave when small perturbations of non-vanishing wavenumbers are
imposed on the almost homogeneous state close to $q_n(0) = q^* = 0$ for all $n
\in [1,N]$. The instability -- giving rise to an
exponential growth of the perturbation -- destroys the initial configuration at
a critical
wavenumber. Eventually,
a pattern of localised humps gets formed, virtually with equal distance
between them distributed on the chain [15--17]. A detailed study of the
parameter's influences on the creation of the localised humps and in consequence
on the escape process can be found in [17].

For a chain situated near the bottom of the potential the system is
initially prepared in a slightly perturbed $k=0$ mode.
In such a situation analytical considerations establish
the emergence of a regular wave pattern
(an array of breathers) that
efficiently localises energy and thereby
enhances the escape of the chain \cite{Hennig.2007.PRE,hennig.2007.epl}.

Regarding the strength of the interaction between the
units we note that in the limit $\kappa \rightarrow 0$,
leading to uncoupled units,  as well as the
opposite limit of very strong couplings leading to a rigid rod-like chain,
the chain behaves
like a single particle, and so the escape is prevented
on the grounds of too little energy available in each unit.
In between these two limiting cases there exists a value of
the coupling strength $\kappa$ that optimises the escape rate. The latter
can be approximated analytically \cite{Hennig.2007.PRE}.

\begin{figure}
  \centering
   \includegraphics[width=0.6\linewidth,height=6cm]{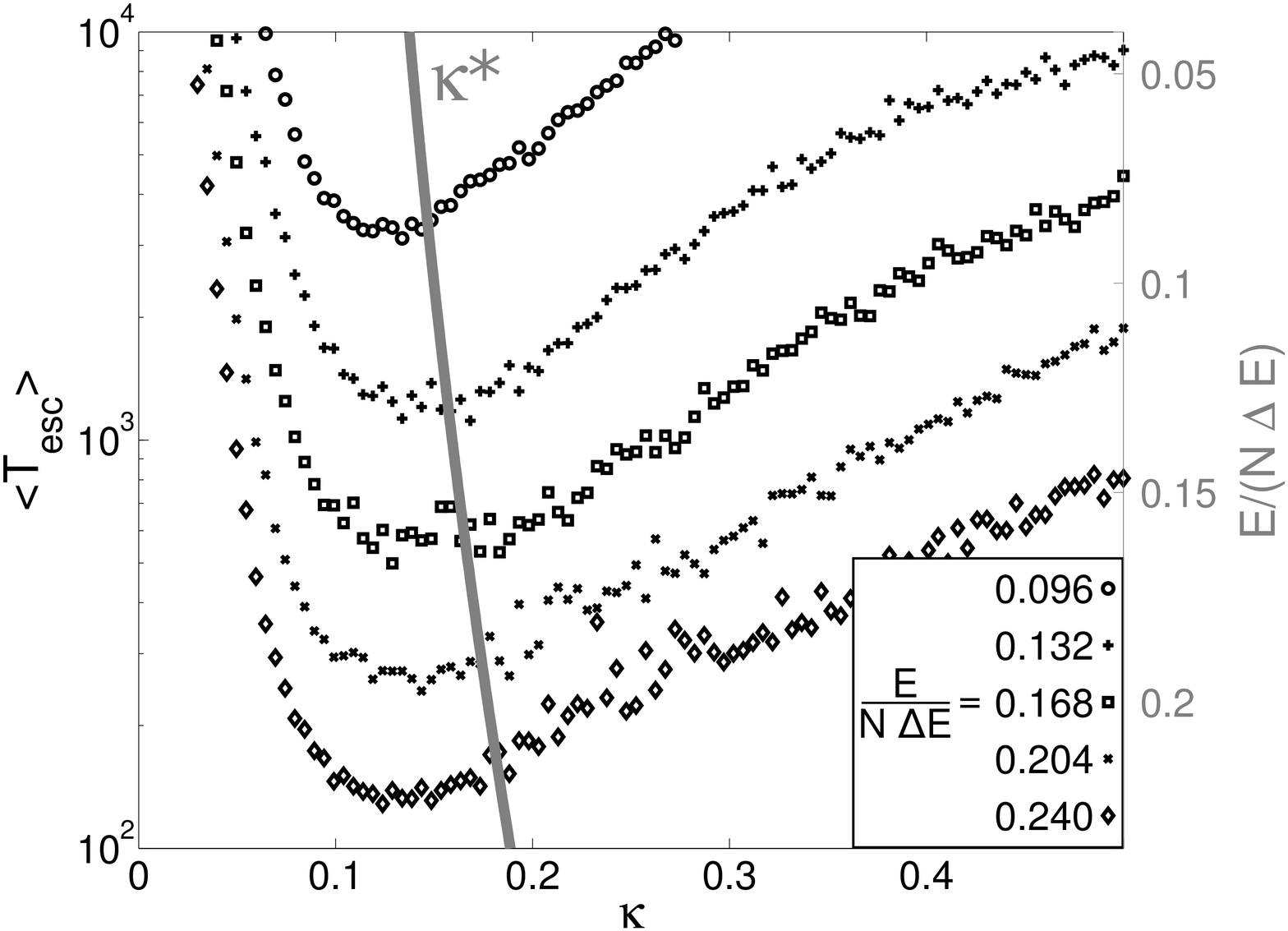}
  \caption{Semi-logscale plot showing the mean escape times (marker symbols) for chains, consisting of $N=100$ units, initially
confined to the metastable minimum of a cubic potential. The averaging was done over 500 realisations of randomised initial conditions,
each of which obey the relations $|q_n(0)-q_0| \le \Delta q$ and $|p_n(0)|\le \Delta p$ with $\Delta q =0.05,
    \Delta p=0.05$. The different symbols represent different ratios of the total energy $E$ and $N$ times the barrier height, as
indicated in the key. The solid grey line represents
    the analytical approximation for the optimal coupling strength
    for energy values given by the
    right-hand axis. Source: Figure taken from
Ref.~\cite{Gross.2014.WorldSci}.}
  \label{fig:ts:T_esc_optimal_kappa}
\end{figure}

The escape time of a unit is defined as the time instant
when the unit passes across a coordinate value  far beyond
the potential barrier. Setting this value to  $q = 20q_{max}$ 
no likely recrossing back into
the potential valley  occurs. The mean escape time
of the chain is then determined by the average of the
escape times of all of its units. For the numerical evaluation of
mean escape times in dependence of $\kappa$ (see Fig.~\ref{fig:ts:T_esc_optimal_kappa}) the system - Eq.~(\ref{equ_motion_1d}) - was
integrated using a fourth order Runge-Kutta scheme. Numerical accuracy was
obtained by ensuring the energy deviation to remain smaller than the
order of $10^{-12}$.
The mean escape times were determined from $500$
realisations of randomised initial conditions at a given energy.
The maximal integration time is $5\cdot 10^5$ time units.

\begin{figure}
\centering
\begin{subfigure}[t]{0.48\linewidth}
\includegraphics[width=\linewidth]{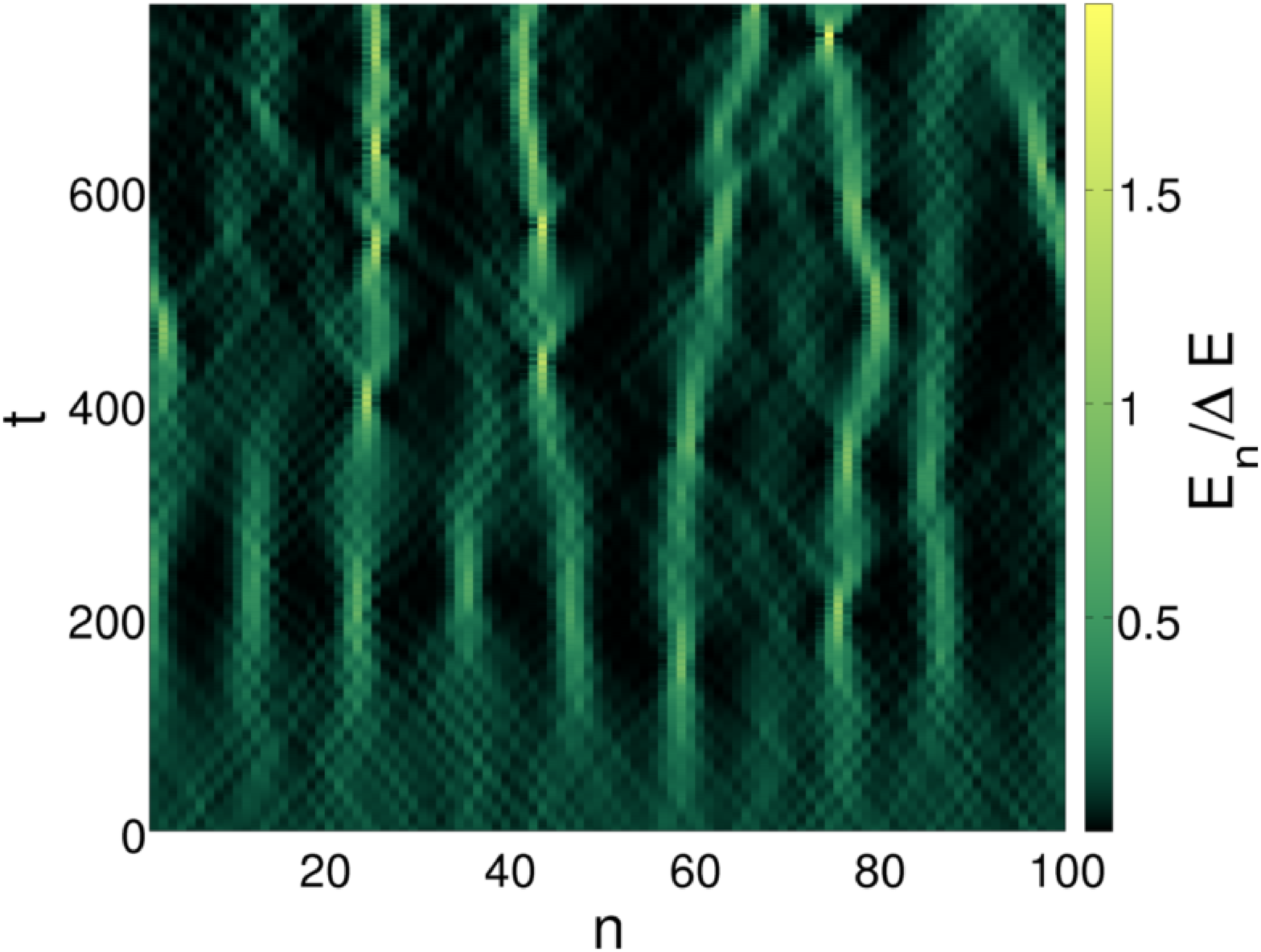}
\caption{$\frac{E}{N\,\Delta E}=0.12$}
\label{fig:ts:energy_evolution_a}
\end{subfigure}
\hfill
\begin{subfigure}[t]{0.48\linewidth}
\includegraphics[width=\linewidth]{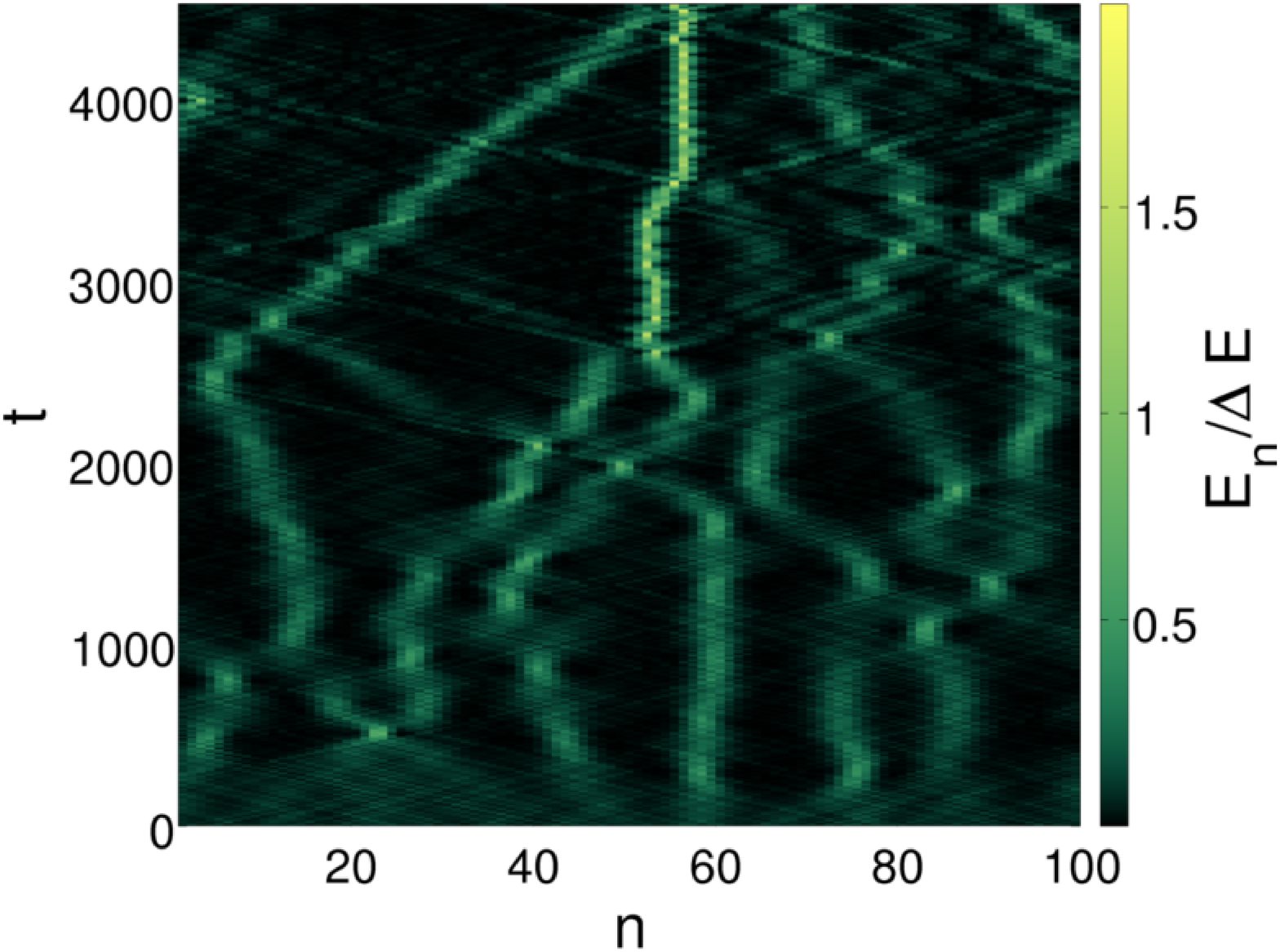}
\caption{$\frac{E}{N\,\Delta E}=0.07$}
\label{fig:ts:energy_evolution_b}
\end{subfigure}
 \caption{(Colour Online) Heat maps showing the temporal evolution of the energy distribution $E_n(t)$ for linear chains evolving in a
cubic potential
- Eq.~(\ref{equ_motion_1d}). Panels a) and b) represent different ratios of the total energy $E$ and $N$ times the barrier height. The
    localisation of energy from an initially homogeneous state causes
    in both cases an escape at the end of the depicted time frame.
    While the chain's energy in Fig. \ref{fig:ts:energy_evolution_a} is
    sufficiently high to let an individual breather from the early
    breather array to surpass the potential barrier, the lower
    energy in Fig. \ref{fig:ts:energy_evolution_b} necessitates a
    merging of breathers to cause the critical chain elongation.
    The parameter values are: $\kappa=0.15, \Delta q =0.05, \Delta
    p =0.05,N=100$. Note the different time scales. Source: Figure taken from Ref.~\cite{Gross.2014.WorldSci}.}
  \label{fig:ts:energy_evolution}
\end{figure}

Fig.~\ref{fig:ts:T_esc_optimal_kappa} demonstrates the
resonance behaviour of the mean escape time as a function of $\kappa$ and also
reveals
a fairly good accordance of the
analytical approximation of the optimal $\kappa=\kappa^*$ with the simulation
results (for details see \cite{Hennig.2007.PRE},\cite{Gross.2014.WorldSci}). Regarding the pronounced variation of the mean
escape times (ranging over several orders of magnitude) for different
energies we remark that a low system energy does
not supply single breathers with an energy sufficient to trigger an
escape event. For example, for $\kappa=0.15$ and $E/(N\,\Delta E)=0.1$ typically
an array with ten or more breathers forms so that each one can 
hold only an energy $E/N_B<\Delta E$, with $N_B$ being the number of breathers.
Nevertheless, an escape takes place,
eventually. This implies a further concentration of energy due to breather
coalescence taking place at later times after the
initial creation of the breather array. The time evolution of the energy
distribution
defined in Eq.~(\ref{eq:esingle}), shown
in Fig.~\ref{fig:ts:energy_evolution}, illustrates this feature.
$E_n$  was monitored in time
(upwards) for two exemplary cases. Energy is localised in both cases
starting from an initially almost homogeneous state. In Fig.~\ref{fig:ts:energy_evolution_a}, corresponding to a relatively high
energy,
we see the appearance of a regular
breather pattern. Every breather concentrates enough energy to certain
units in order to trigger an escape, which in this case happens at $t\approx
700$. In contrast, the lower system
energy in Fig.~\ref{fig:ts:energy_evolution_b} does not allow for a direct
escape of the initially formed breathers. Instead, breathers
start an erratic movement. After an inelastic interaction they merge
and can thereby eventually result in a configuration exceeding the critical
chain elongation, see also \cite{Peyrard.1998.PhysicaD}. However,
this secondary process is slow compared to the (direct) breather formation
which explains the different orders of magnitude of the escape
times scale, in dependence of the energy content, seen in Fig.~\ref{fig:ts:T_esc_optimal_kappa}.

\subsubsection{Thermally activated escape} The previous
investigations  dealt with a deterministic
chain dynamics governing an escape event. In this part we study
how a thermal bath with a temperature $T$ assists the transition over the
barrier, cf. in Ref. \cite{hennig.2007.epl}. For
this purpose we consider the associated Langevin equation driven by additive thermal noise,
\begin{align}
  \label{1d_Langevin_eq}
  \ddot{q}_n+q_n-{q_n}^2-\kappa
  \left(q_{n+1}+q_{n-1}-2\,q_n\right)+\gamma \dot{q}_n+\xi_n(t)=0,
\end{align}
with a common damping parameter $\gamma$ and uncorrelated Gaussian white noise terms
$\xi_n(t)$. In order to be able to compare the deterministic situation
to the thermally activated setting, the associated conserved energy
$E$ in the Hamiltonian case and the average energy $\overline{E}$
transferred from the bath need to be equal. The latter is governed by
the correlation function of the noisy force $\xi(t)$ given by
\begin{align}
\label{fdt}
\left\langle \xi_n(t) \xi_{n^\prime}(t^\prime) \right\rangle=2\,\gamma
\,\overline{E}/N \,\delta_{n,n^\prime}\,\delta(t-t^\prime)\,,
\end{align}
expressing that the mean energy of all particles is given by $\overline{E}$. If
expressed by the bath temperature, every particle assumes on average
$k_B\,T$, {\it i.e.} $\overline{E} = N\, k_B\, T$ with $k_B$ being the Boltzmann
constant.

\begin{figure}
\centering
\begin{subfigure}[t]{0.49\linewidth}
\includegraphics[width=\linewidth]{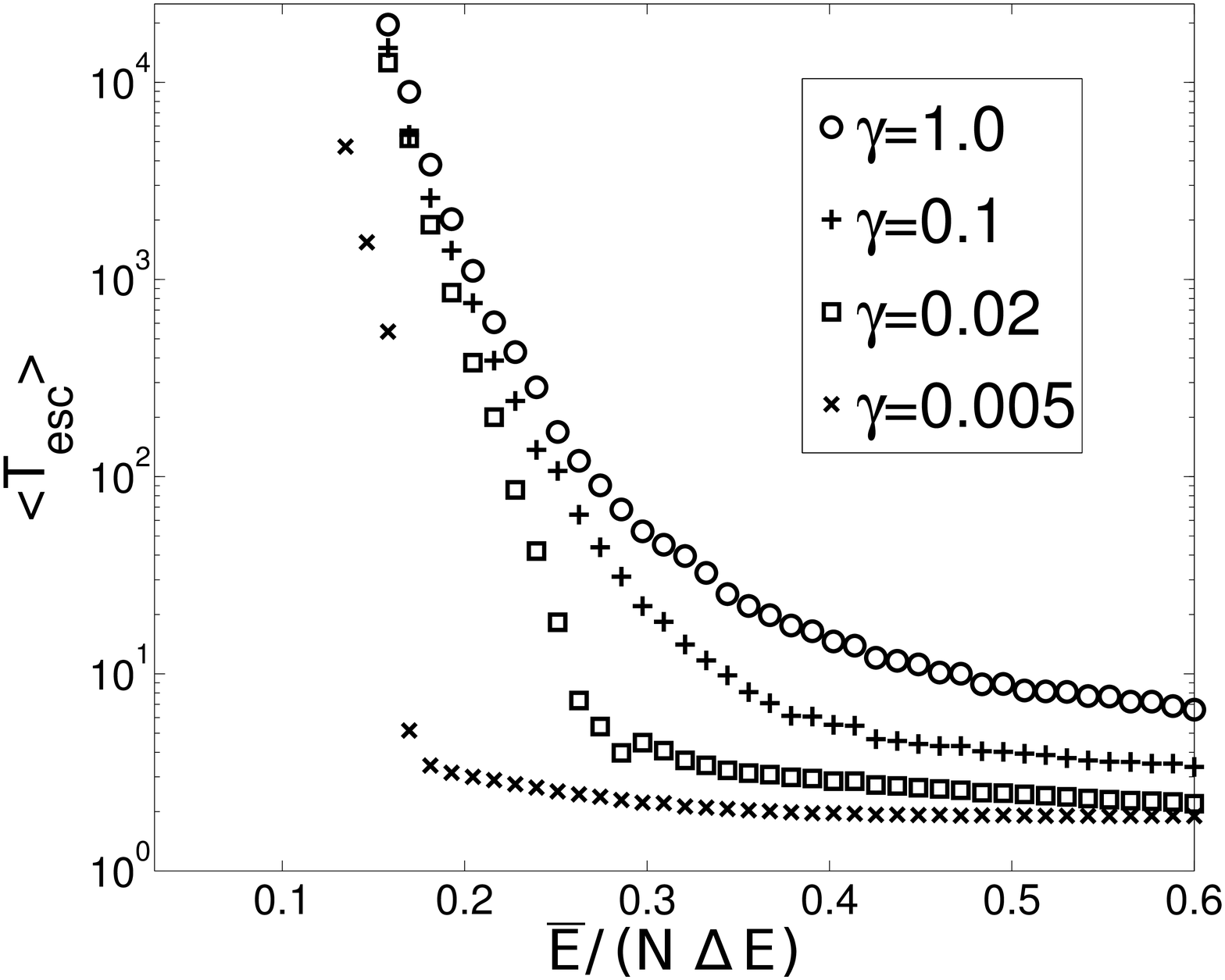}
\caption{Thermally activated setting -- Eq.\,(\ref{1d_Langevin_eq})}
\label{fig:comp_plot1}
\end{subfigure}
\hfill
\begin{subfigure}[t]{0.49\linewidth}
\includegraphics[width=\linewidth]{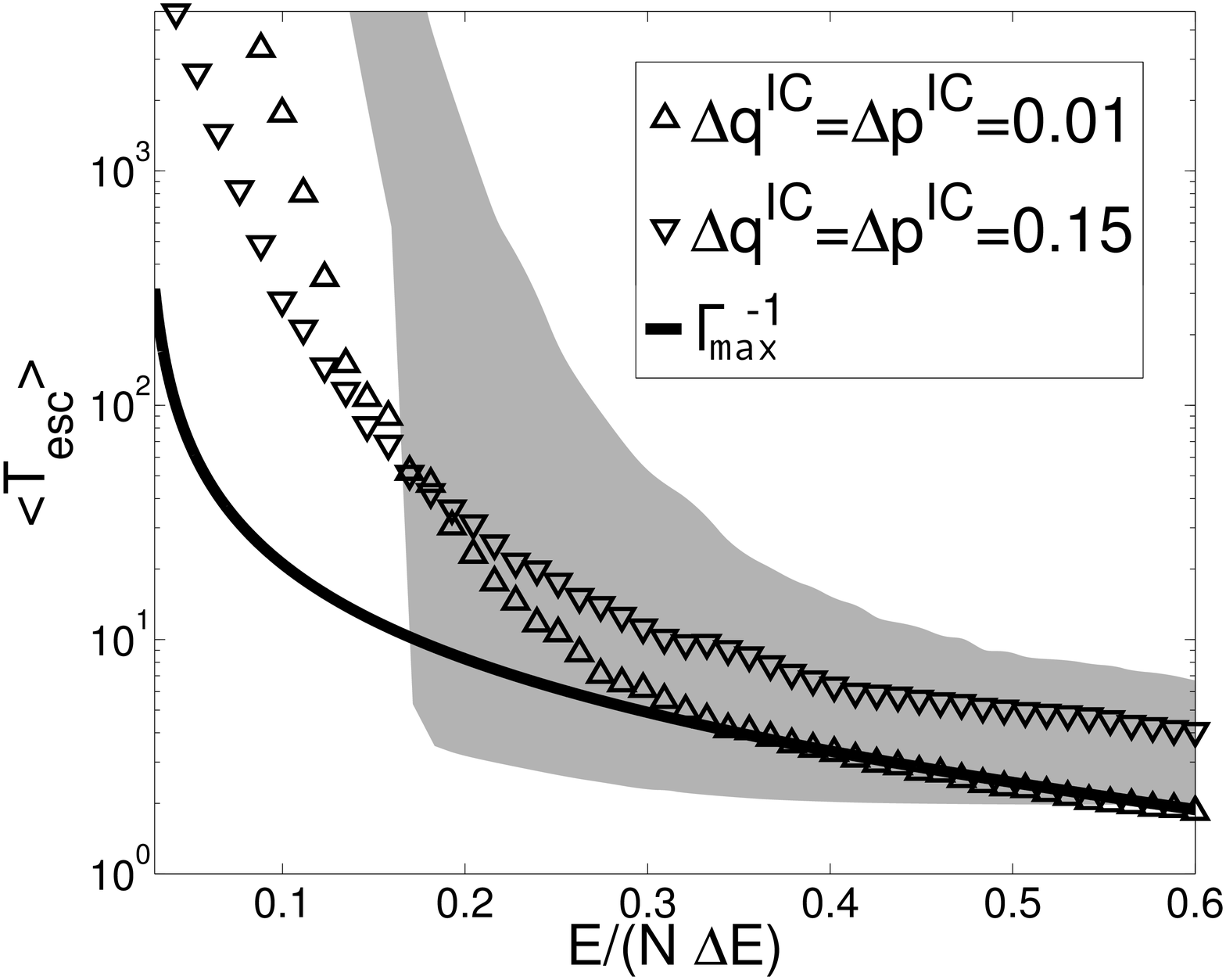}
\caption{Deterministic setting -- Eq.\,(\ref{equ_motion_1d})}
\label{fig:comp_plot2}
\end{subfigure}
\caption{Lin-log plot showing the mean escape times in the thermally activated case -- Eq.~(\ref{1d_Langevin_eq}) -- for
    different values of the damping constant, Fig.~\ref{fig:comp_plot1}, and their comparison to
    the deterministic setting, Fig.~\ref{fig:comp_plot2}, for 500 realisations each. The grey area
    in Fig.~\ref{fig:comp_plot2} sketches the mean escape times for
    the thermally activated case for $0.005 < \gamma < 1.0$ as shown
    in Fig.~\ref{fig:comp_plot1}, whereas the symbols show results
    from the deterministic set-up with initial conditions given in the
    inset. The solid black line in Fig.~\ref{fig:comp_plot2} is the inverse of
the maximal growth rate, $\Gamma_{max}$, obtained from
Eq.~(\ref{MI1d:max_growth_rate}). The parameter values are: $\kappa=0.15, N=100$. In both panels $\overline{E}$ and $E$ are expressed in
units of $N$ times the barrier energy, with $\overline{E}=Nk_bT$ and $E$ being the total system energy in the deterministic case.
Source: Figure taken from Ref.~\cite{Gross.2014.WorldSci}.}
  \label{fig:comp_plot}
\end{figure}

We measure the mean escape time for the system described by
Eq.~(\ref{1d_Langevin_eq}).
The latter is numerically integrated using an
second-order Heun stochastic solver scheme, again with a maximal integration
time of $5\cdot 10^5$ time units.
The system is initialised with
all units set to the minimum of the potential and zero momenta.
The system thermalises and the moment when its energy reaches
$\overline{E}$ is taken as the initial time for our studies of escape.

We study the system for the optimal coupling constant, $\kappa=0.15$.
Fig.~\ref{fig:comp_plot1} shows the mean escape times of 500
realisations for different values of the damping constant in
dependence of $\overline{E}$.
Fig.~\ref{fig:comp_plot2} compares these times (depicted as the grey
surface) to the corresponding mean escape times of the deterministic
system. It additionally shows the characteristic time constant for the
formation of breathers due to modulational instability, given by Eq.~(\ref{MI1d:growth_rate}),
 where  the $k=0$
phonon amplitude, $A$, is related to the system energy via $E(A)=N\,V(A)$, with $V(\cdot)$ is defined in Eq.~(\ref{eq:hamiltonian1}).
$Q$ is the wave number of the perturbation-mode acting on the $k$-mode carrier
wave.

Especially for smaller energies the deterministic escape is considerably
faster than the thermally activated one. Notably for $\overline{E}/(N\,\Delta
E)<0.1$
and quite contrary to the deterministic setting, escape events are practically
absent during our simulations time in the thermal case. For larger
mean energy values this picture can change to a higher efficiency
of the thermal escape process when damping is weak.
Also two deterministic settings with different magnitudes of
the random initial perturbations
swap  their features of escape upon passing from low to high energies.

\begin{figure}
\centering
\begin{subfigure}[t]{0.48\linewidth}
\includegraphics[width=\linewidth]{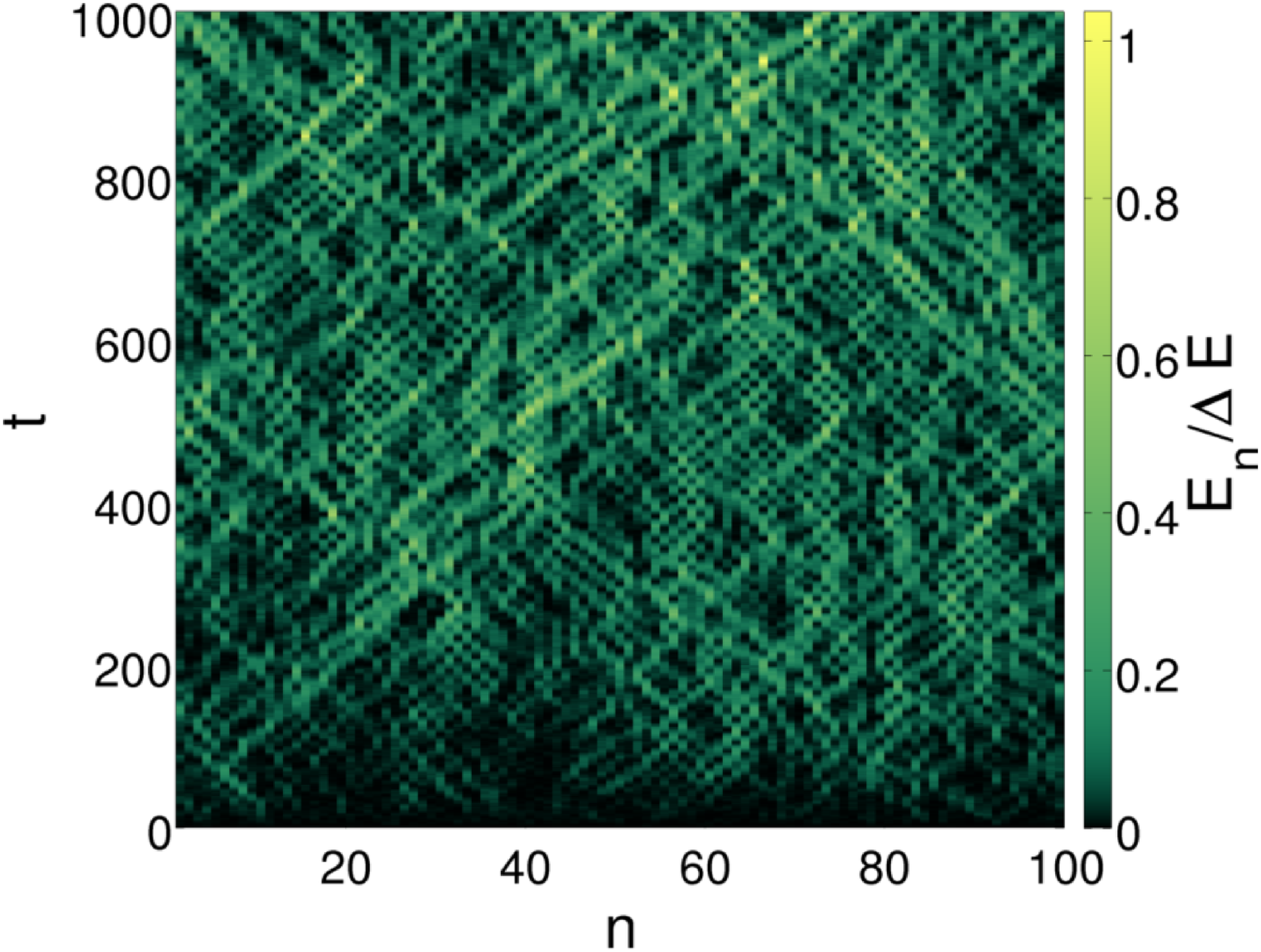}
\caption{$\gamma=0.005$}
\label{fig:1d:E_evo_thermal_1}
\end{subfigure}
\hfill
\begin{subfigure}[t]{0.48\linewidth}
\includegraphics[width=\linewidth]{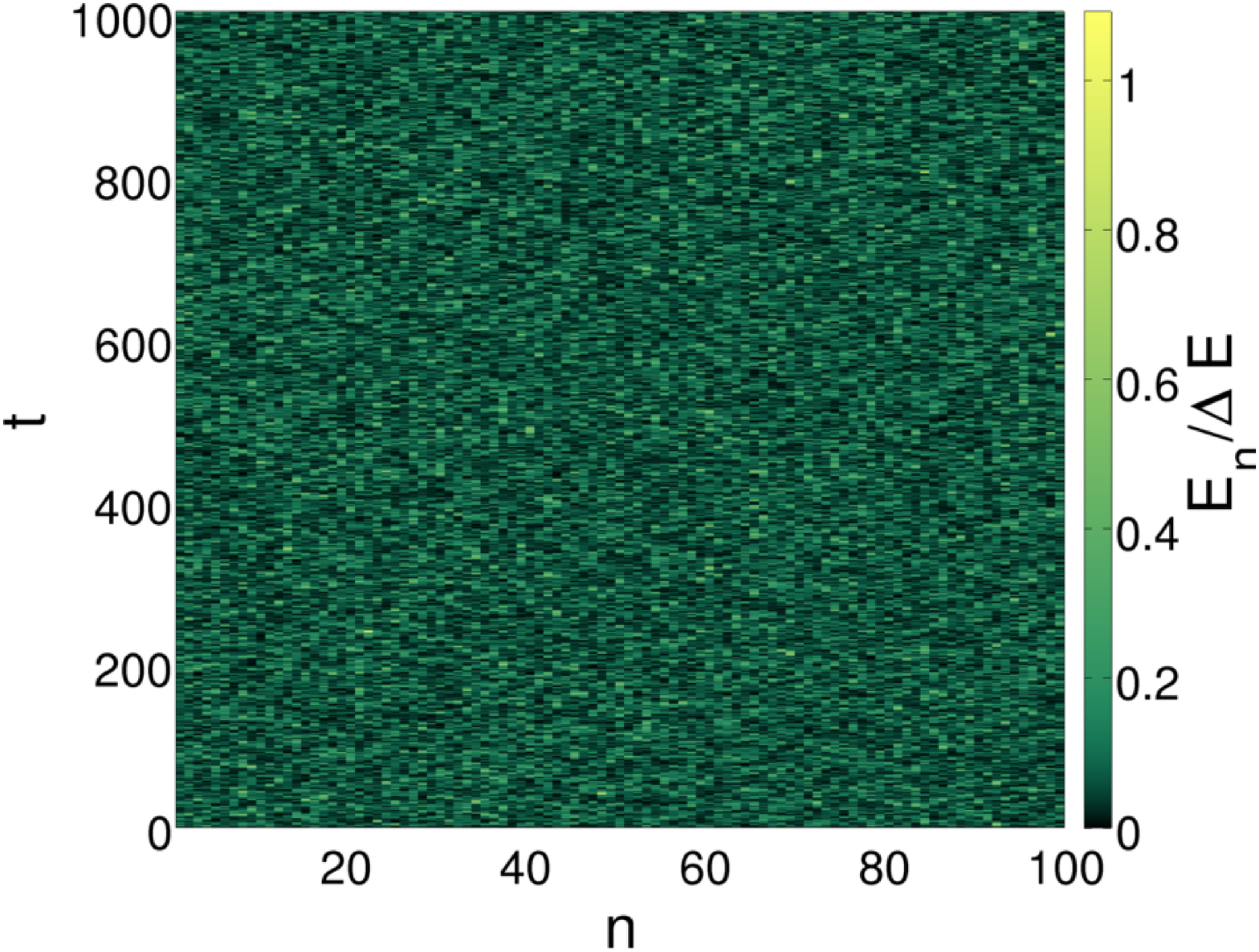}
\caption{$\gamma=1.0$}
\label{fig:1d:E_evo_thermal_2}
\end{subfigure}
  \caption{(Colour Online) Heat maps showing the temporal evolution of the energy distribution $E_n(t)$ in
    the thermally activated case - Eq.~(\ref{1d_Langevin_eq}) - for chains consisting of $N=100$ units. The two panels were obtained
using different values of the damping parameter $\gamma$: A smaller damping constant leads to
    a higher degree of energy localisation. The remaining parameter values are:
    $\overline{E}/(N\,\Delta E)=0.15$, $\kappa=0.15$ with $\overline{E}=Nk_bT$ and $\Delta E$ being the barrier height. Source: Figure
taken from Ref.~\cite{Gross.2014.WorldSci}.}
  \label{fig:1d:E_evo_thermal}
\end{figure}

How can the speedy escape in the deterministic system, in comparison to the
thermally activated
escape,
be explained?  For small values of $\gamma$ the system approaches the
deterministic setting.
This entails a tendency towards a
more efficient and pronounced localised energy distribution as seen when comparing
Fig. \ref{fig:1d:E_evo_thermal_1} (low damping strength $\gamma=0.005$) and
Fig. \ref{fig:1d:E_evo_thermal_2} (high damping strength $\gamma=1.0$).
The relaxation time of the chain scales with the inverse of the damping
constant.
Correspondingly,
the life times of local excitations grow with decreasing $\gamma$.
The outcome is a more heterogeneous energetic structure where thermal
fluctuations are more likely to cause critical chain elongations.
This explains the faster escape comparing small with large damping constants.
But even in the case of very small $\gamma$ the relaxation time is still much
shorter
than the time needed for the coalescence of multiple breathers (which is of the
order of
several hundred time units, see Fig. \ref{fig:ts:energy_evolution_b}).
Therefore,
the long term  cumulative concentration of energy is generally inhibited in the
thermal case.
This explains the virtual impossibility of a thermal escape for small energies.

For higher energies the deterministic escape is mostly assisted
by  arrays of breathers that are formed as the result of modulational
instability
(referred to as initial breathers). The formation time of the initial breather
array can be
estimated by the inverse of the maximal growth rate, $\Gamma_{\mathrm{max}}$
from
Eq.~(\ref{MI1d:max_growth_rate}). Fig.~\ref{fig:comp_plot2} shows the expected
convergence of $\Gamma_{\mathrm{max}}^{-1}$ to the deterministic escape times
for
large energies.
We recall that $\Gamma_{\mathrm{max}}$ was determined
on the basis of a linear expansion in the perturbations.
This explains the better match of $\Gamma_{\mathrm{max}}^{-1}$ and $\langle
T_{esc} \rangle$
in the case of smaller initial perturbations.
The divergence of $\Gamma_{\mathrm{max}}^{-1}$ with the mean
escape time for small energies again gives evidence to the fact that
the initial breather array
does not induce an escape and other mechanisms enhancing the degree of
localisation
are required.

In the thermally activated case with large energy
the mobility of the breathers becomes amplified if noise acts.
If the breathers possess longer life times, {\it i.e.}
for smaller damping, a few (usually not more than two)
breathers can temporarily merge and approach
a critical chain elongation. Starting with a heterogeneous energy structure
after its thermalisation then leads to shorter escape times in the presence of
noise compared
to the deterministic setting that first has to re-allocate energy from an
initially almost
homogeneous state. Conversely, for
stronger damping, the life times of breathers is too short and
the energy distribution remains mostly homogeneous (see Fig.~\ref{fig:1d:E_evo_thermal_2}).
The escape of the chains over the potential barrier then rely entirely on rare, strong enough, spontaneous
fluctuations of the noise term and so the mean escape times generally
become comparatively large.

Regarding the impact of the initial perturbations, $\Delta q,\Delta p$,
of a  homogeneous state on the
escape time (cf. right panel, Fig. \ref{fig:comp_plot}) in the deterministic
case we note that,
as the initial
conditions for smaller perturbations are closer to the $k=0$ phonon mode,
the initial breather array emerges more quickly so that the escape times
are smaller when the energy is high enough for initial breathers to initiate
the escape. Contrarily, when low energy necessitates coalescence,
stronger perturbations lead to a higher breather
mobility which speeds up the breather coalescence so that the escape times
become smaller.

\subsection{Deterministic escape dynamics of two-dimensional coupled nonlinear
chains}\label{ra_ch2}

As an extension to the one-dimensional chain system, as discussed in the previous
section,
we consider next a two-dimensional chain model with
pairwise nonlinear Morse interaction between neighbouring units and study the role 
of the additional degree of freedom on the self-organised
escape process \cite{Fugmann.2008.PRE}. In distinct contrast to the previous study taking
into
account solely motion in the transition direction, in this case dynamical
motion is also allowed {\it transverse} to the barrier along the
well of the external potential. As a  consequence, in addition to
the formation of localised large amplitude breathers, with
amplitudes evolving in transition direction, global
oscillations of the chain transverse to the barrier are observed.
Eventually a few chain links accumulate locally sufficient
energy to cross the barrier. This mechanism is shown to take
place for both linear rod-like and for coil-like configurations
of the chain in two dimensions.

\subsubsection{Two-dimensional coupled unit chain model}
Our study treats a spring mass chain model. The chain
consists of $N$ units which are pairwise connected through nonlinear springs. The
motion of these units takes place in the $x-y$-plane. We denote by
$q_{xn}$ the displacement of the $n$th unit in the $x$-direction,
also referred to as the transition coordinate, while in
the transverse $y$-direction displacements from the rest position are
denoted by $q_{yn}$. The local on-site potential $U$ reads as
\begin{equation}
 U(q_{xn})=\frac{m \omega_0^2}{2} q_{xn}^2-\frac{a}{3}q_{xn}^3,
\end{equation}
with $n=1,\,...\,N$.
We assume a nonlinear interaction potential of Morse type
between adjacent units of the chain
\begin{equation}
 U_M(r_{n+1,n})=D_0\left[1-\exp(-d(r_{n+1,n}-l)^2\right]^2,
\end{equation}
with depth $D_0$, range parameter $d$, $l$ the equilibrium distance
of the units (also referred to as bond length), and $r_{n+1,n}$
the Euclidean distance of two neighbouring units,
\begin{equation}
 r_{n+1,n}=\sqrt{(q_{xn+1}-q_{xn})^2+(q_{yn+1}-q_{yn})^2},
\end{equation}
with $n=1,\,...\,N$.
The corresponding Hamiltonian of the two-dimensional chain model
reads as
\begin{equation}
 H=\sum_{n=1}^N\,\left(\frac{p_{xn}^2}{2}+\frac{p_{yn}^2}{2}+  U(q_{xn}) \right)
 +\sum_{n=1}^{N-1}U_M(r_{n+1,n}).
 \end{equation}
 Passing to dimensionless quantities is achieved with the following rescaling procedure:
$\tilde{q}=qd$,
 $\tilde{p}=pd/(m \omega_0)$, $\tilde{t}=\omega_0t$, $\tilde{a}=a/(d m
\omega_0^2)$. 
Furthermore,  we use
 $\kappa=2D_0d^2/(m \omega_0^2)$. In what follows we omit the tilde notation.
 The intrinsic length scale
 \begin{equation}\label{eq:intrinic_len_scale}
  s=q_x^{max}-q_x^{min}=\frac{1}{a} \;,
 \end{equation}
where $q_x^{max}=m\omega_0/a^2$ and $q_x^{min}=0$ denote the
position of the potential maximum and
minimum respectively,
plays an important role.  To be precise, small ratios $l/s$; i.e.,
the ratio of the bond length $l$ and the intrinsic
length scale $s$ of the system,  cause coil-like
chain configurations, while rodlike states appear for ratios of
the order of one or above. Note that in the limit of vanishing
$a$ the barrier disappears and the intrinsic length scale diverges.

The  equations of motion derived from the Hamiltonian read:
\begin{eqnarray}
 \ddot{q}_{xn}&=&-q_{xn}+aq_{xn}^2\nonumber\\
 &-&\kappa \left[1-\exp\big(-(r_{n+1,n}-l)\big)\right]
 \exp\big(-(r_{n+1,n}-l)\big)\frac{q_{xn}-q_{xn+1}}{r_{n+1,n}}\nonumber\\
 &-&\kappa \left[1-\exp\big(-(r_{n,n-1}-l)\big)\right]
 \exp\big(-(r_{n,n-1}-l)\big)\frac{q_{xn}-q_{xn-1}}{r_{n,n-1}}\label{eq:eomx_2d}\\
 \vspace{1.1cm}\nonumber\\
  \ddot{q}_{yn}&=&-\kappa \left[1-\exp\big(-(r_{n+1,n}-l)\big)\right]
 \exp\big(-(r_{n+1,n}-l)\big)\frac{q_{yn}-q_{yn+1}}{r_{n+1,n}}\nonumber\\
 &-&\kappa \left[1-\exp\big(-(r_{n,n-1}-l)\big)\right]
 \exp\big(-(r_{n,n-1}-l)\big)\frac{q_{yn}-q_{yn-1}}{r_{n,n-1}}\label{eq:eomy_2d} \;,
\end{eqnarray}
for $n=2,\,,,,\,N-1$. Moreover,  open boundary conditions are imposed.

Like in the study of the one-dimensional case, initially the chain is in a  flat
state
of amplitude $q_{x0}$ on which
small perturbations are
exerted by taking random initial amplitudes which are uniformly
distributed in an interval  $|q_{xn}(0)-q_{x0}| \le \Delta q_x$.
The mean values of $q_{x0}$ are taken in such a way that the
average excitation energy of a single unit, $E_0$, is small
compared to the depth, $\Delta E$, of the potential well. Due to the
choice of sufficiently small displacements $\Delta q_x$ the initial
lattice state, $q_{xn}(0) = q_{x0} + \Delta q_{xn}$, is close to an almost
homogeneous state and yet sufficiently disturbed that there result
small but non-vanishing initial interaction terms. Thus an
energy exchange between the coupled units is instigated. The
initial momenta $p_{xn} = 0$ are set zero. The initial
amplitudes in the transverse direction are $q_{yn}(0) = nl$ and the
momenta $p_{yn}(0) = 0$, entailing the conservation of the
centre of mass in the y-direction.

\subsubsection{Energy redistribution process}
In the beginning, the system
energy is essentially equally shared
among all units in the chain, expressed by a homogeneous
elongation of the whole chain in the transition direction,
and  the escape scenario proceeds as
follows: Just like for the one-dimensional study (see previous Section) the
process of
modulational instability governs the
dynamics of the system during an early phase of the evolution.
In particular this triggers the formation of an array of
localised solutions (large-amplitude breathers) in the $x$-direction. Later on,
the influence of the -- compared to a purely one-dimensional unit model -- second,
transverse, degree of freedom
 crucially affects the dynamical processes of the
coupled unit chain.
Interestingly, we observe that in the cases of both very short
bond lengths $l/s \ll 1$ and bond lengths obeying $l/s >1$ the
structure, which is formed by a modulational instability,
persists for very long times, whereas it disappears rather fast for
intermediate values of the bond length.
Such a decrease of the amplitudes of the
localised structures in the $x$-direction comes along with the excitation of
motions in the transverse degree of freedom. As one measure
of the energy content in the $x$- and the $y$-direction the
respective kinetic energy can be taken. Initially the mean of the
$x$-kinetic energy of the $x$-motion is $E_{kin}^x = 0.5E$, whereas
$E_{kin}^y = 0$. Induced by the breather formation in the $x$-direction
an enhanced interaction of neighbouring units is caused
and since the interaction force couples the motion in the $x$-
and in the $y$-direction an energy transfer is initiated.
In fact, one observes that a state of equipartition is reached for which
$E_{kin} = 0.25E$ in both the $x$- and the $y$-direction,
respectively. The time until equipartition is attained decreases with increasing
coupling
strength. We underline, that the energy transfer described
above constitutes a purely nonlinear effect. We could never observe a
complete back transfer of energy from the $y$- to the $x$- motion. Rather
 a breathing-like behaviour, for which the chain contracts and
 relaxes along its axis periodically in time results. This behaviour of
global oscillations of the chain as a whole along the transverse
direction appears in addition to the large-amplitude breathers
evolving in the transition coordinate direction involving fairly strong
energy localisation at certain sites.

\subsubsection{Transition states and collective escape} Whether an
unit involved in a large amplitude
breather state is able to escape from the region of bounded
motion inside the potential well or is held back depends on
the corresponding amplitude pattern as well as on the
coupling strength. The associated critical chain
configuration -- called the transition state -- is determined
as the solution of the corresponding stationary system
which represents a force-free configuration corresponding to a
first-order saddle point in configuration space. There appear two scenarios:
The peak of
the localised amplitude profile of the critical configuration is
situated either at one free end of the chain (referred to as
boundary critical localised mode -- BCLM), or somewhere in
between the free ends (referred simply to as CLM).

Regarding CLMs it is found that, for a given coupling strength $\kappa$,
the transition state is represented by a thin needle shape with the central unit
situated beyond
the barrier if $l<s$ (akin to the one-dimensional situation).
In general, the smaller the bond length $l$ the larger is the extension of the
critical
localised mode along the $x$-direction and the more units
are elongated from the potential minimum. In contrast, for
$l>s$, the central unit is always situated at the barrier,
while its neighbours are arranged in such a way that there
remains no stress arising from the bonds. We underline that
the ends of the chain are free and there act thus no restoring
forces. Hence, in order to reduce the stress arising from the
elongation of the central unit over the barrier, its
neighbours can be displaced force free along the $y$-axis. Furthermore,
due to the strong degree of localisation the obtained structures
remain the same when increasing the number of units in the chain.
It should be stressed that the alignment of the units
along the potential minimum situated at $q_{min}^x=0$ is
completely arbitrary, as long as next neighbours keep their
equilibrium distance $l$. In particular, completely coiled
configurations can be critical transition states, too.

For the BCLMs, the qualitative dependencies on the system's
parameters $\kappa$ and $l$ remain -- compared to the situation
when the critical peak is formed between the ends -- the
same. But -- since the unit beyond the barrier is now
connected to only one neighbour inside the potential well and
thus the acting back-pulling forces are smaller --
the force-free critical state is less elongated.

Concerning the dependence of the activation energy on the
bond length, we observe a decay of $E_{act}$ with enlarging bond
length.  In the limit $l \rightarrow s$ its value approaches $E_{act} = \Delta
E$.
We remark that in the case of $l \ge s$ we always find a critical
stationary solution with activation energy $E_{act} = \Delta E$.
Surprisingly, in the limit of long bonds
the activation energy becomes independent of the coupling
strength and is equal to the net barrier height.
As long as $l<s$, the activation energy grows with
an increasing value of the coupling strength. The growth is
the stronger the smaller is $l$.
In the case of BCLMs the
activation energies are remarkably lower compared to transition
states with a peak somewhere in between the loose ends. A
detailed analysis of the energetic contributions to the
activation energy reveals that the major part of $E_{act}$ is stored as
deformation energy of the springs. Since a BCLM contains
fewer stretched springs, the value of activation energy is
lower. Crucially, the process of barrier crossing of the
chain is not only influenced by the amount of energy provided to the system.
It also depends on the ratio of different
length scales, since the geometry plays a vital role for
motions in a two-dimensional potential landscape. Thus, the
rate of escape will be significantly affected by the choice of the
parameters.

\begin{figure}[!ht]
 \centering
  \includegraphics[width=0.5\linewidth]{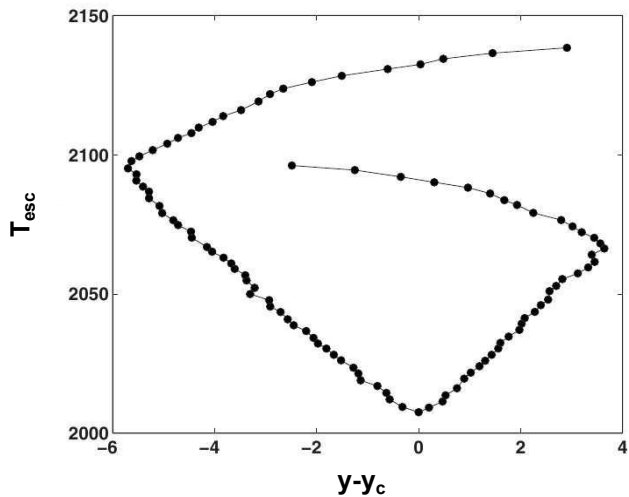}\hfill
  \caption{Escape process for one realisation of
  initial conditions for the two-dimensional chain model - Eqs.~(\ref{eq:eomx_2d})-(\ref{eq:eomy_2d}): Escape times $t_{esc}$ of the
units versus the $y$
   position when the corresponding unit passes the threshold value
$q^{thres}_x$
   far beyond the barrier. The position of the unit which overcomes the
  barrier first is denoted by $y_c$, all other positions are shifted by this
   value. The initial conditions are $q_{x0}=-0.05$ and $\Delta q_x=0.001$,
   yielding $E_0/\Delta E=0.219$.  The parameter values are $a=5$,
  $\kappa = 0.9$, $l/s= 1.25$, and $N=100$. Source: Figure
adapted from Ref.~\cite{Fugmann.2008.PRE}.}
  \label{fig:esc2d}
  \end{figure}

The escape time of a chain is as defined in the one-dimensional case (see
Sec.~\ref{ra_ch1}). In Fig.~\ref{fig:esc2d} a typical escape process is
illustrated by showing
the escape times $t_{esc}$ versus the position of the escaping unit
when the corresponding unit passes $q^{thresh}_x$. First, one unit
moves directly beyond the barrier (since the underlying
dynamics is irregular for different realisations
of initial conditions the incident escape can happen at an arbitrary location
in the chain), and therefore adjacent units are subjected
to pulling forces and a cascade of escapes is initiated in a
relatively short time interval.

Regarding the dependence of the escape time on the coupling strength $\kappa$,
the fraction of escape events, as a function of $\kappa$, has been calculated. Numerically $200$ different
realisations of initial conditions, each with a simulation time
 set to $t_{sim} = 10^4$, corresponding to more than $1500$ periods of linear ground-state oscillations, were considered.
 Since not all simulations lead to
escape events during $t_{sim}$ the mere calculation of the mean
escape time alone is not suitable.

\begin{figure}[!ht]
  \centering
  \includegraphics[width=0.5\linewidth]{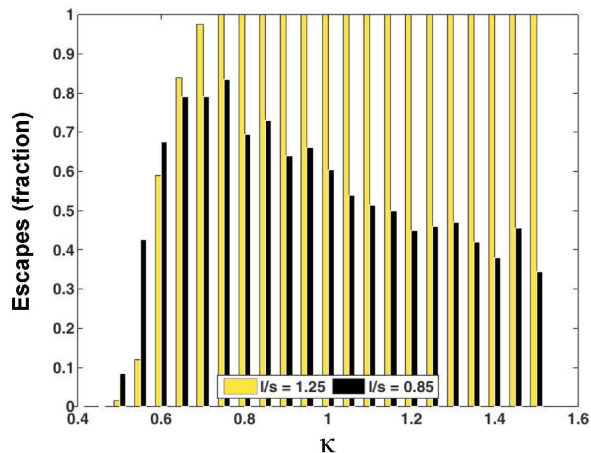}\hfill
\caption{(Colour online) Fraction of completed escapes of the chain as a
  function of the coupling strength $\kappa$ for the two-dimensional chain model - Eqs.~(\ref{eq:eomx_2d})-(\ref{eq:eomy_2d}), with
chains consisting of $N=100$ units. The value of the bond length $l$
  is given in the legend, in units of the intrinsic length scale of the system $s$ - Eq.~(\ref{eq:intrinic_len_scale}).
  The initial conditions are $q_{x0}=-0.05$ and $\Delta q_x=0.001$,
  amounting to $E_0/ \Delta E = 0.219$.
  The remaining dimensionless parameter, which regulates the barrier height of the cubic potential, is $a=5$. Source: Figure
taken from Ref.~\cite{Fugmann.2008.PRE}.}
  \label{fig:esc2dfraction}
  \end{figure}

Fig.~\ref{fig:esc2dfraction} depicts the fraction of successful exits of the
entire chain for two different bond lengths as a function of
the coupling strength $\kappa$. The values of the bond lengths are
chosen in such way that one is smaller and one is larger than
$s$. For $\kappa = 0.45$ one observes the first rare events of escape of
the complete unit chain. The fraction of successful
escape events of the entire chain further increases with
enhanced coupling strength.
Whereas for the larger bond length at $\kappa= 0.75$ the curve
saturates to $1$ -- i.e., all initial preparations lead to an escape
of the whole chain -- the curve for the smaller bond length reaches
there a maximum of $0.835$. With further increasing of the
coupling strength the latter curve descends reaching a value
of $0.345$ at $\kappa = 1.5$.

The different shape of the curves can be explained by the
dependence of the activation energy on the value of the coupling strength for
$l<s$. Here, the effective potential barrier
that must be surmounted during the escape process grows with
$\kappa$. As a result we observe a drastic reduction of the
successful exit events for shorter bond lengths.

\begin{figure}
 \centering
\includegraphics[width=0.5\linewidth]{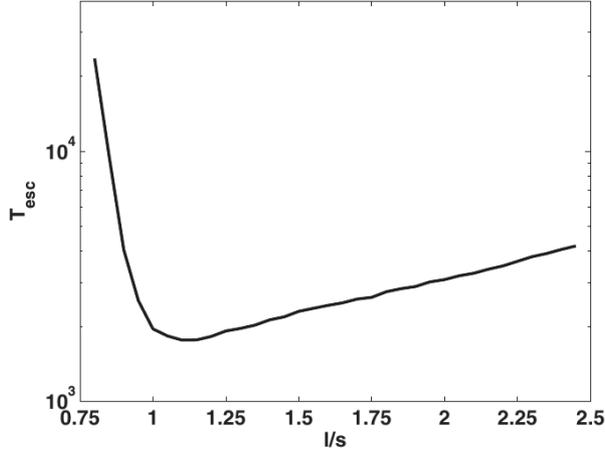}\hfill
   \caption{Mean escape time as a function of the bond length $l$, in units of the intrinsic length scale of the system $s$ -
Eq.~(\ref{eq:intrinic_len_scale}). 500
realisations of the random initial conditions, for chains consisting of $N=100$ units, have been used in the averaging.
  The parameters regulating the initial conditions are $q_{x0} = -0.05$ and
 $\Delta q_x =0.001$, yielding $E_0/\Delta E = 0.219$. The remaining dimensionless parameter, which regulates the barrier height of the
cubic potential, is $a=5$. Source: Figure
taken from Ref.~\cite{Fugmann.2008.PRE}.}
 \label{fig:2descapel}
 \end{figure}

Regarding a study of the influence of the bond length on the
mean escape time of the unit chain the coupling
strength was fixed to $\kappa = 0.9$, guaranteeing an exit of the entire chain
without fragmentation.
The results are shown with Fig.~\ref{fig:2descapel}.
For small bond lengths up
to $l/s = 1$ the curve drops continuously and at $l/s \approx 1.1$ a minimum is
reached. With further increase of the bond length the mean
escape time slightly increases.
The rise of the curve for enlarging bond length can be
explained with the smaller growth rates of the localised
structures created during the process of modulational instability.
The effective interaction is weaker. Thus the process
of energy localisation is slower and thereby all subsequent
exchange processes induced by the modulational instability
are slowed down, too. The reason for the absence of escape for $l/s \lesssim
0.75$
is that  the limit of $l/s \rightarrow 0$, i.e. small bonds
compared to the width of the local potential, and a fairly weak
coupling, the chain tends to fragment.

\subsection{Mexican hat potential}

Extending the previous studies of escape from metastable states
to systems of higher (geometrical) complexity
we investigate a ring of interacting units evolving in a
two-dimensional potential landscape of the form of a Mexican hat \cite{Gross.2014.PRE}.

Besides the conceptual interest in this work,
the results can be applied to the description of
micro bubble surface modes which can be modelled by a closely related system in which
breather modes were verified experimentally \cite{DosSantos.2012.ConfProc}.
These results demonstrate the potential relevance of resonant
wave modes, as well as the escape behaviour in such topological set ups.

\begin{figure}
\centering
\includegraphics[width=0.5\linewidth]{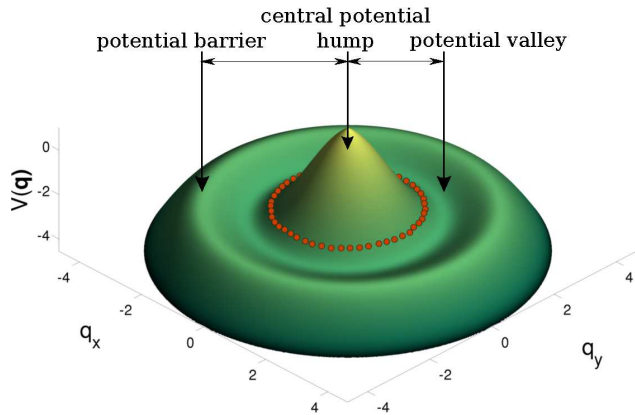}
\caption{(Colour online) A typical initial preparation of a chain in the Mexican hat potential $V\left(\mathbf{q}_i\right)$ -
Eq.~(\ref{eq:mex_hat_pot}). The key features of the potential are indicated in the figure. Source: Figure
taken from Ref.~\cite{Gross.2014.PRE}.}
\label{fig:IC}
\end{figure}

We study a Hamiltonian system consisting of a two dimensional
ring of $N$ linearly coupled units subjected
to an external Mexican-hat-like,  anharmonic potential; i.e.,

\begin{equation}\label{eq:mex_hat_pot}
V\left(\mathbf{q}_i\right)=-\sqrt{\mathbf{q}_i^2}
+\cos{\left( \frac{\sqrt{\mathbf{q}_i^2}}{\lambda}\right)}
\end{equation}
with (conserved) Hamiltonian in dimensionless units given by
\begin{align}\label{scaledHam}\mathcal{H}&=
\sum_{i=0}^{N-1}\left[\frac{\mathbf{p}_i^2}{2}+
\frac{\kappa}{2}\left(\mathbf{q}_i-\mathbf{q}_{i+1}
\right)^2+V\left(\mathbf{q}_i\right)\right]\;,
\end{align}

\noindent
where $\kappa$ is the coupling strength between neighbouring units
and the parameter $\lambda$ regulates the width of the
Mexican hat potential $V\left(\mathbf{q}_i\right)$. Moreover, $\mathbf{q}_i = (q_{xi},q_{yi})$ and $\mathbf{p}_i = (p_{xi},p_{yi})$.
A typical setup is shown in Fig.~\ref{fig:IC}.

The corresponding equations of motion read
\begin{align}
\begin{split}\label{qdyn1}
\mathbf{\ddot{q}}_i=&-\kappa\,\left(2\,\mathbf{q}_i-
\mathbf{q}_{i+1}-\mathbf{q}_{i-1}\right)+
\frac{\mathbf{q}_i}{\sqrt{\mathbf{q}_i^2}}\\
&+\sin{\left(\frac{\sqrt{\mathbf{q}_i^2}}{\lambda}\right)}\,
\frac{\mathbf{q}_i}{\lambda\,\sqrt{\mathbf{q}_i^2}}\qquad i\in0\ldots N-1 \;.
\end{split}
\end{align}

\subsubsection{Wave modes and escape dynamics} We elucidate the influence of different wave modes on the escape
dynamics of the ring from the metastable
state over the potential barrier when
the ring is initially situated in a metastable state in the vicinity of the potential's
bottom corresponding to a local minimum energy configuration.
Given the rotational symmetry of the Mexican hat potential polar coordinates, i.e.
$(r_i,\varphi_i)$,
can be introduced accounting for the motions of
the radial and angular degree of freedom  of each unit. The local minimum
energy configuration is determined by
$r_i=r_0$ and $\varphi_i=i\Delta \Theta$ with $\Delta \Theta =2\pi/N$
complemented with a conditional equation for the radial position, reading

\begin{equation}
\label{r^0-conditional_equation}
-2\,\kappa\,r_0\left(1-\cos\left(\Delta \Theta\right)\right)
+1+\frac{1}{\lambda}\sin\left(\frac{r_0}{\lambda}\right)=0 \;.
\end{equation}

Initially, the units are placed in a slightly perturbed ring-like
structure around the central potential hump, see Fig.\,\ref{fig:IC},

\begin{align*}
\mathbf{q}_i(0)=r_0\cdot\begin{pmatrix}\cos\left(i\,\Delta
\Theta \right)\\ \sin\left(i\,\Delta \Theta \right)\end{pmatrix}+\begin{pmatrix}
\Delta q_i^x\\ \Delta q_i^y
\end{pmatrix},\qquad
\mathbf{p}_i(0)=\begin{pmatrix}
\Delta p_i^x\\ \Delta p_i^y
\end{pmatrix},
\end{align*}
where $\Delta q_i^x$ and $\Delta q_i^y$ as well as $\Delta p_i^x$ and
$\Delta p_i^y$ are random perturbations taken from a uniform distribution
within the intervals
\begin{align*}
\Delta q_i^x,\,\Delta q_i^y\in \left[-0.01,0.01 \right];\quad
\Delta p_i^x,\,\Delta p_i^y\in \left[-0.01,0.01 \right].
\end{align*}

\noindent
In this setting, the angular distance between any pair of neighbouring
units is almost equal (i.e. being close to $\Delta \Theta$) so that the initial
angular acceleration is small. Thus, for short time periods after the system's
preparation, virtually no variations of the angular variables,
{\it viz.} $\varphi_i(t)=\varphi_i^0=i\, \Delta \Theta$,
are expectable. Furthermore each unit's initial
radius is close to $r_0$. Thus, (at least) for short periods of time after the
initialisation of the system, the angular components remain fixed and
the units can only move along equally spaced rays that all
emerge from the origin.

\begin{figure}[!ht]
\centering
\includegraphics[width=0.2\linewidth]{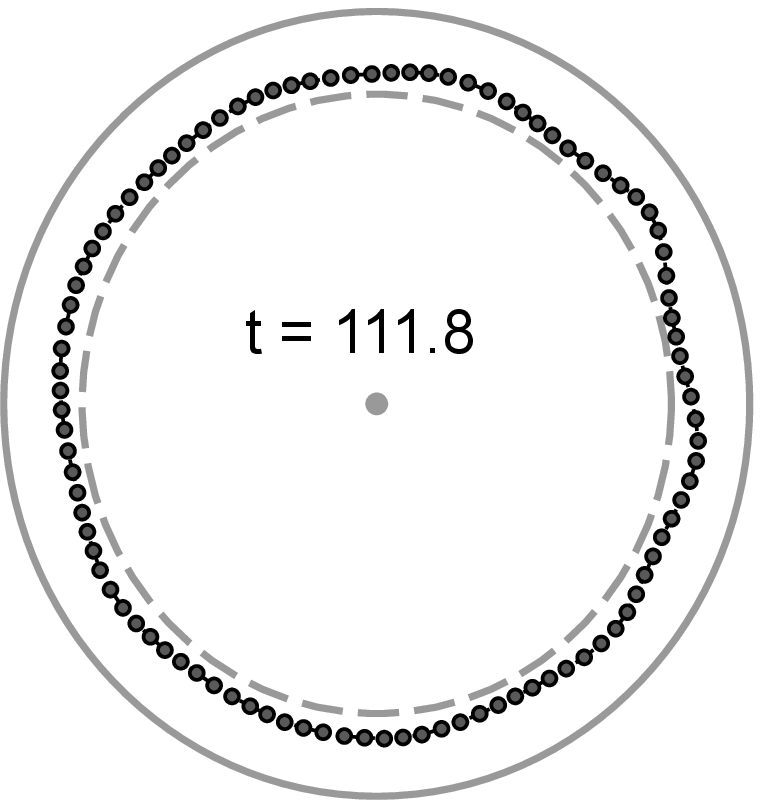}\hfill
\includegraphics[width=0.2\linewidth]{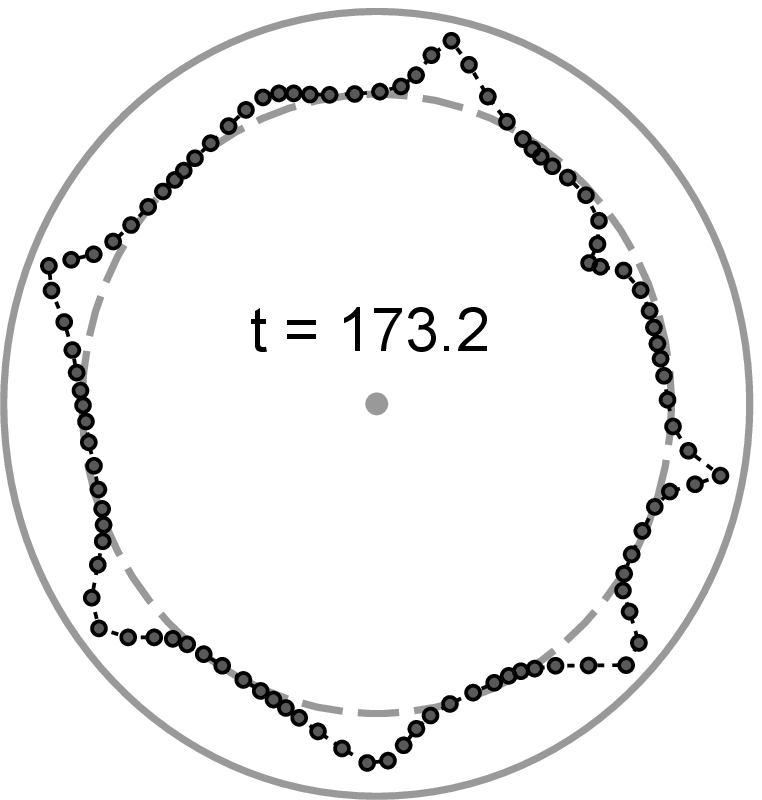}\hfill
\includegraphics[width=0.2\linewidth,angle=90]{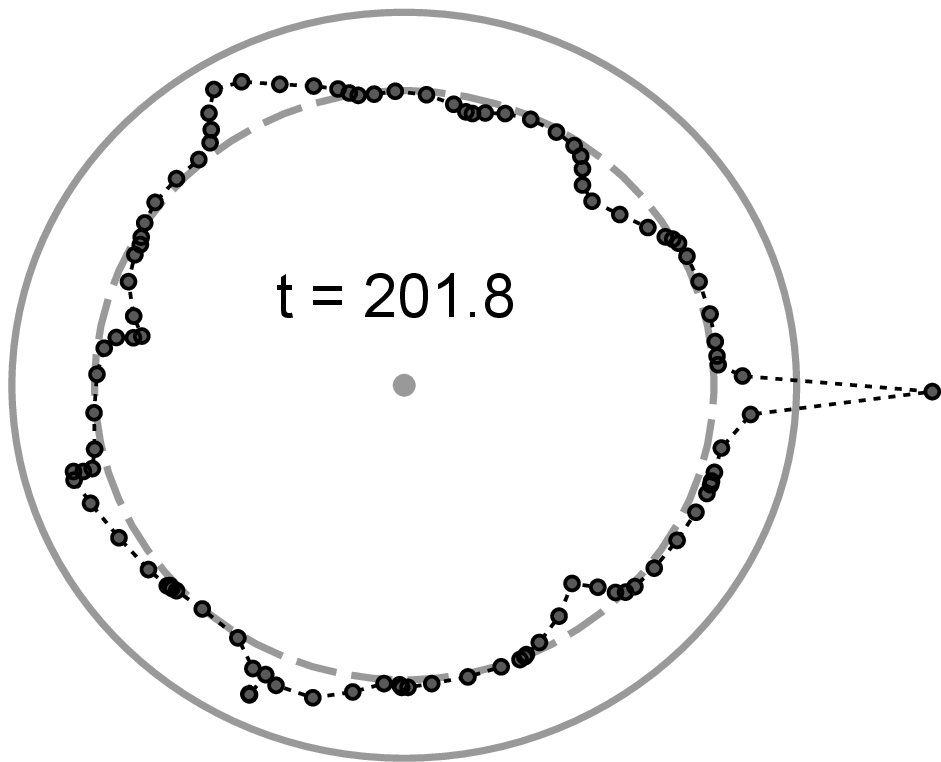}
\caption{Simulation snapshots of a chain in a Mexican Hat potential evolving according to Eq.~(\ref{qdyn1}). The snapshots show the
growth of a radial
breather array from an almost homogeneous initial state due
to modulational instability. The ongoing amplification of
this pattern eventually drives an individual unit over the potential barrier
and thus triggers an escape of type I as described below.
The parameter values are: $\kappa\,\Delta\Theta ^2=0.79\cdot 10^{-4}$, $\lambda=0.85$. Source: Figure
adapted from Ref.~\cite{Gross.2014.PRE}.}
\label{fig:MI_high_wavenumber_escape}
\end{figure}

In more detail, the initial ring-like setup entails that the chain
will first oscillate
in a $k=0$ phonon-like manner. For appropriate parameter values
the mechanism of modulational instability triggers the formation of
 a regularly spaced array of breathers as shown in
Fig.\,\ref{fig:MI_high_wavenumber_escape}. These localised excitations,
forming a transversal wave pattern,
play a crucial role because they concentrate energy in single radial degrees
of freedom and therefore substantially
influence the escape behaviour.

Generally, at an early stage the dynamics is characterised by the transversal wave pattern.
 However, later on the angular components experience changes as well.
 Remarkably, for most of the parameter
choices the angular movement is far from being erratic but instead consists
of regular and pronounced longitudinal wave patterns, as shown in
Fig.\,\ref{fig:longi_reso_example}. As will be seen below, these patterns are fundamental
for the characterisation of the system's dynamics and its  escape behaviour.

\begin{figure}[!ht]
\centering
\includegraphics[width=0.2\linewidth]{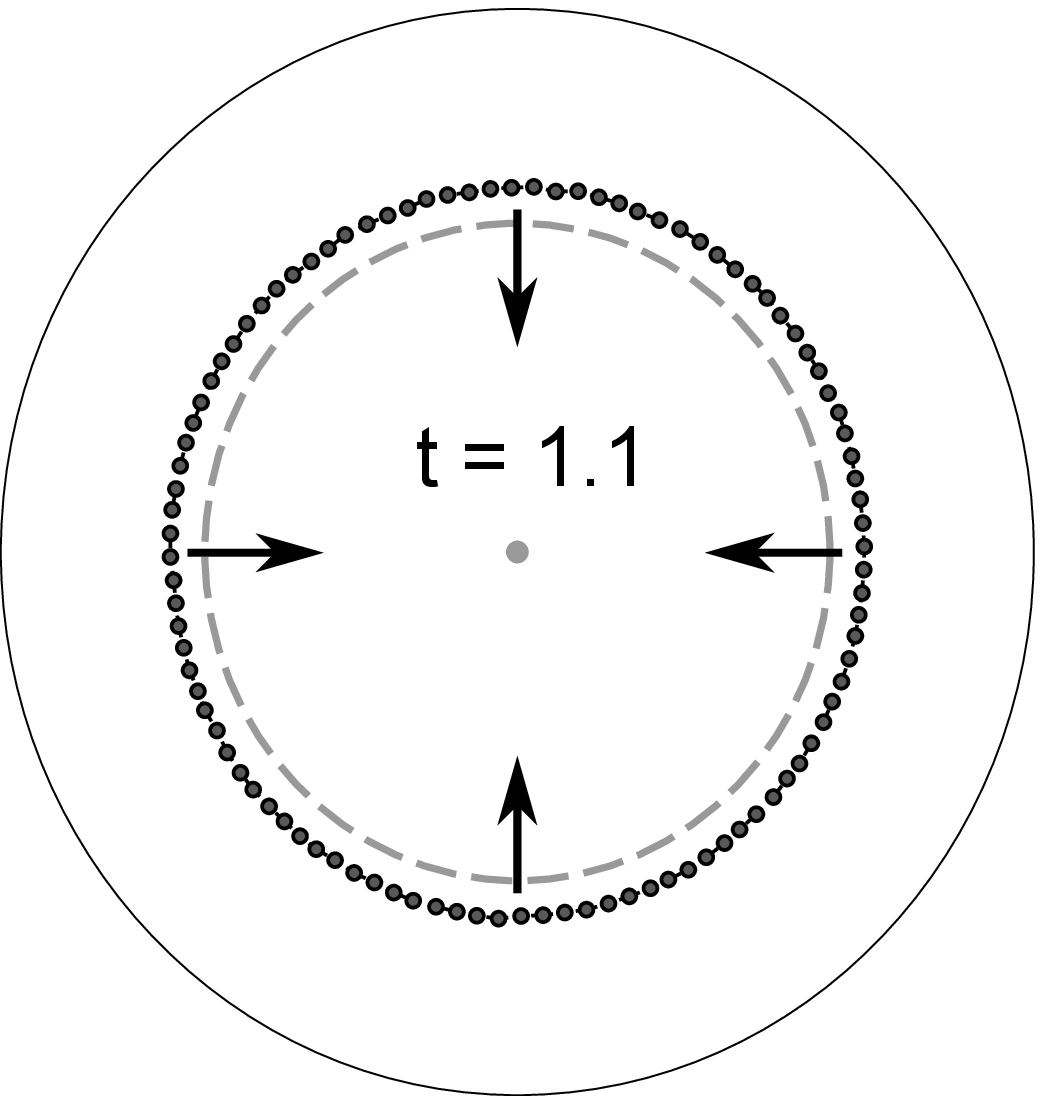}\hfill
\includegraphics[width=0.2\linewidth]{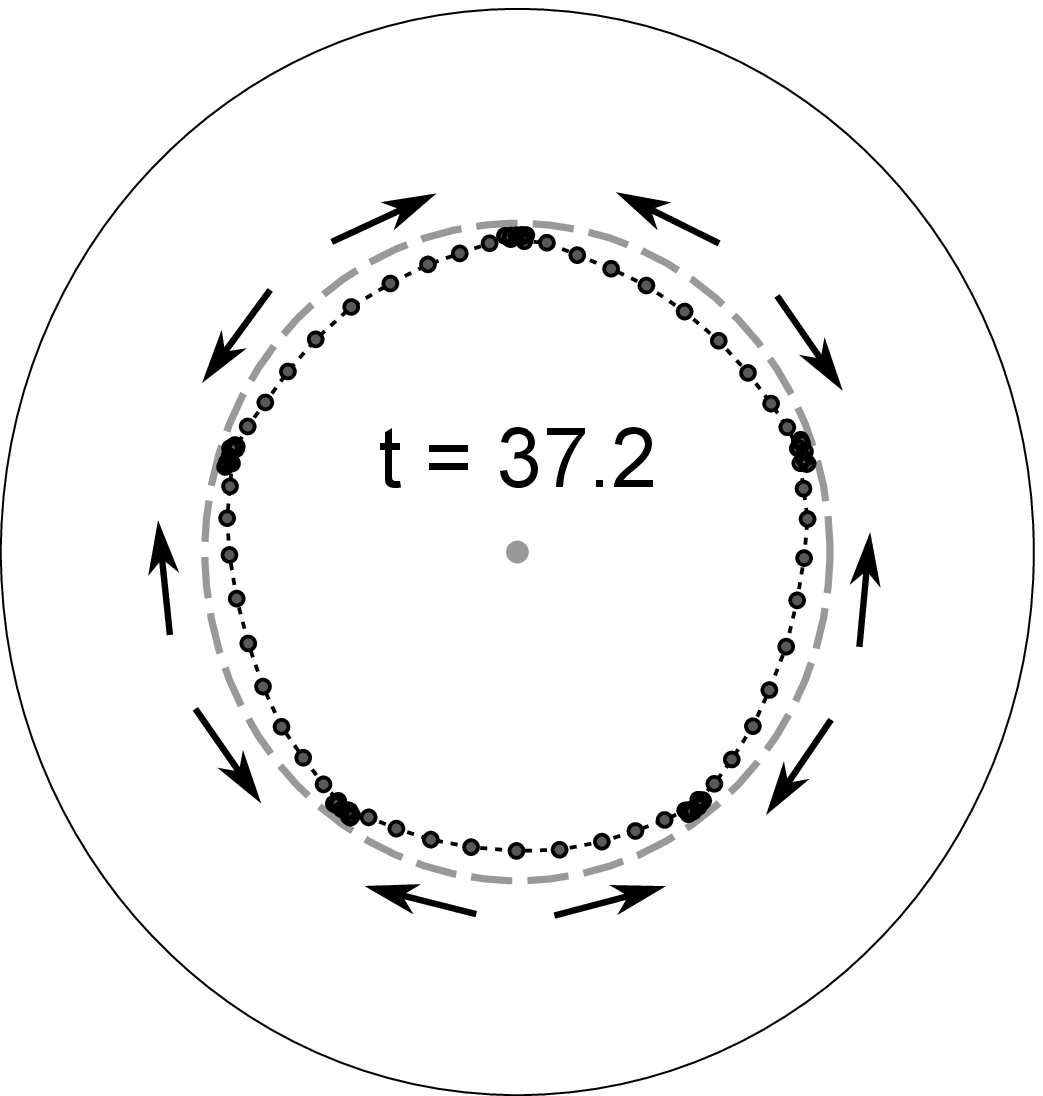}\hfill
\includegraphics[width=0.2\linewidth]{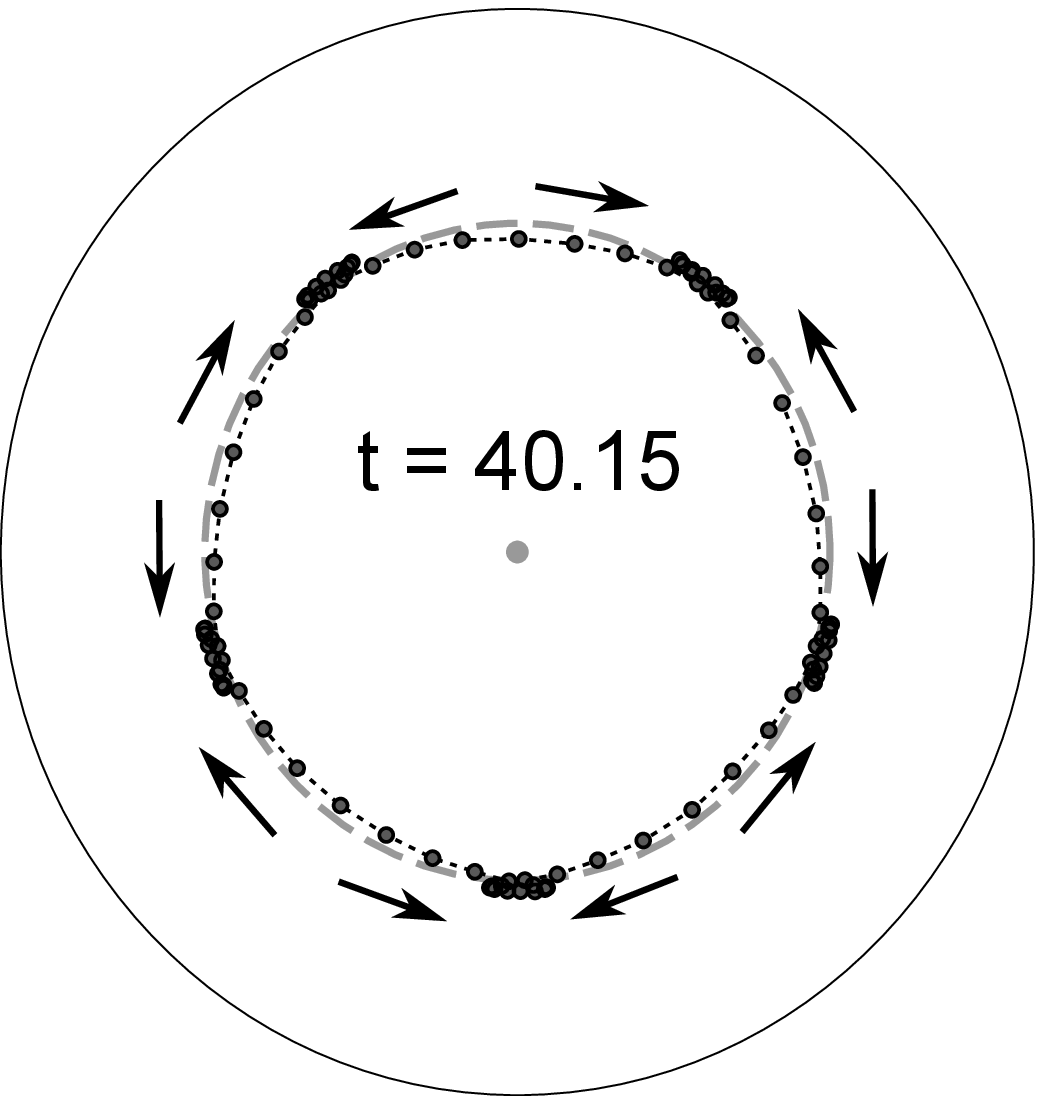}
\caption{Simulation snapshots for a chain in a Mexican hat potential evolving according to Eq.~(\ref{qdyn1}). The snapshots
show the emergence of a
longitudinal wave pattern.
Arrows indicate the chain movement. The parameter values are:
$\kappa\,\Delta\Theta ^2=0.06$, $\lambda=0.4$. Source: Figure
adapted from Ref.~\cite{Gross.2014.PRE}.}
\label{fig:longi_reso_example}
\end{figure}

\begin{figure}
 \centering
 \begin{subfigure}[t]{0.32\linewidth}\centering
 \includegraphics[height=0.125\paperheight]{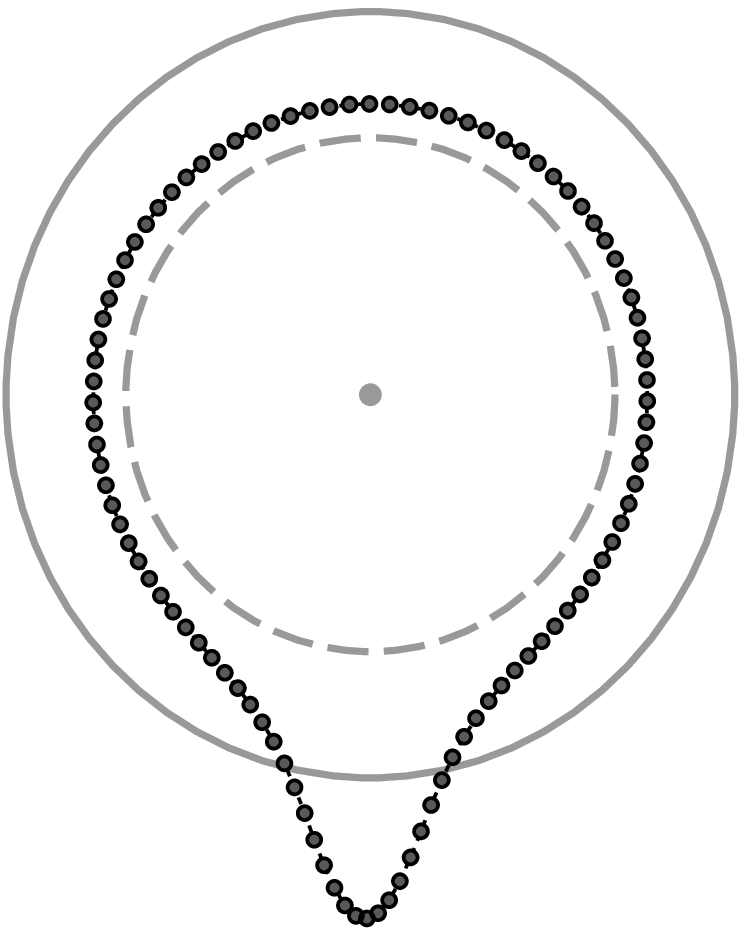}
 \caption*{type I}
 \end{subfigure}
 \hfill
 \begin{subfigure}[t]{0.32\linewidth}\centering
 \includegraphics[height=0.125\paperheight]{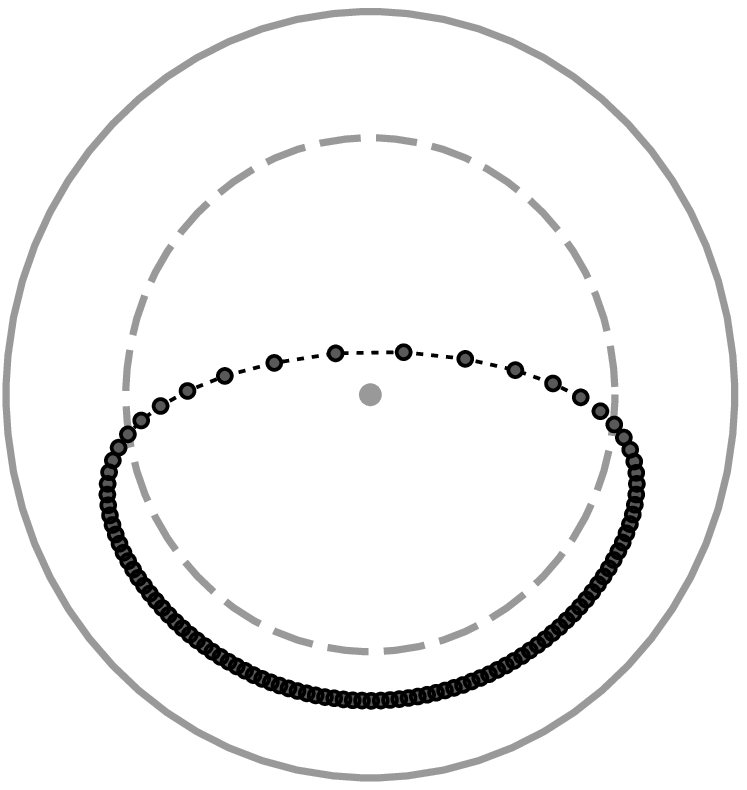}
 \caption*{type IIa}
 \end{subfigure}
 \hfill
 \begin{subfigure}[t]{0.32\linewidth}\centering
 \includegraphics[height=0.125\paperheight]{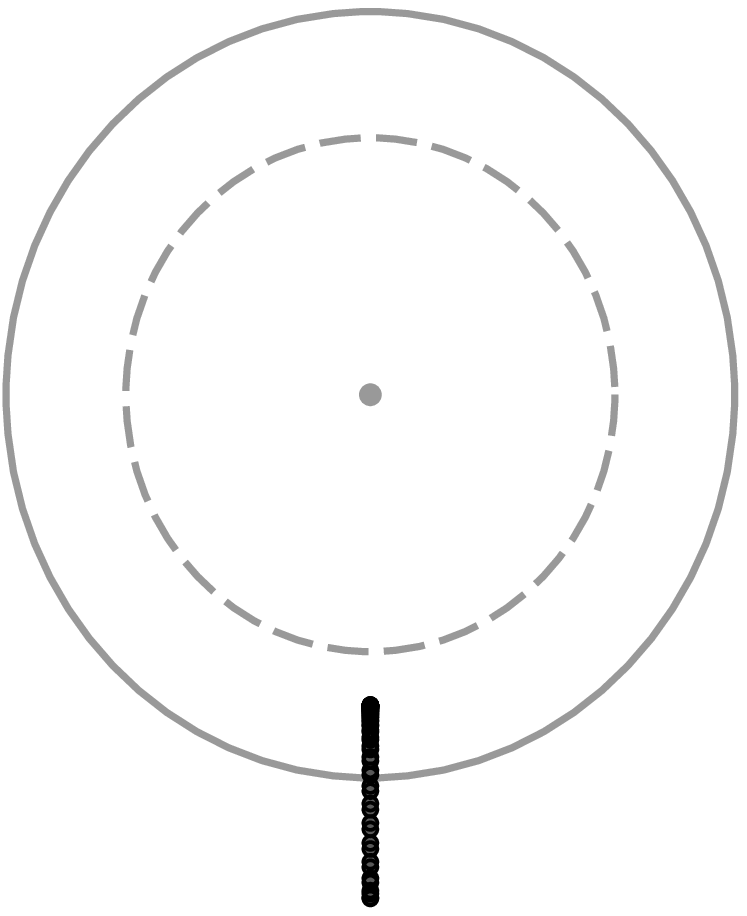}
 \caption*{type IIb}
 \end{subfigure}
 \caption{Relevant transition state types for chains in a Mexican hat potential. The chains, by passing through a transition state, can
escape from the metastable potential minimum. The states differ by the way in which the chain overcomes the central potential
hump, and also the potential barrier (see also Fig.~\ref{fig:IC}). Snapshots of an escape of type I (type II) can be seen in
Fig.~\ref{fig:I_escape} (Fig.~\ref{fig:II_escape}). Source: Figure
taken from Ref.~\cite{Gross.2014.PRE}.}
 \label{fig:overview_ts}
 \end{figure}

The escape can be realised via two escape channels which
are related to different transition states depicted in
Fig.~\ref{fig:overview_ts}.
The escape through each of these channels is driven by breather
modes. They efficiently accumulate energy into single
radial degrees of freedom.

\begin{figure}
\centering
\includegraphics[width=0.2\linewidth]{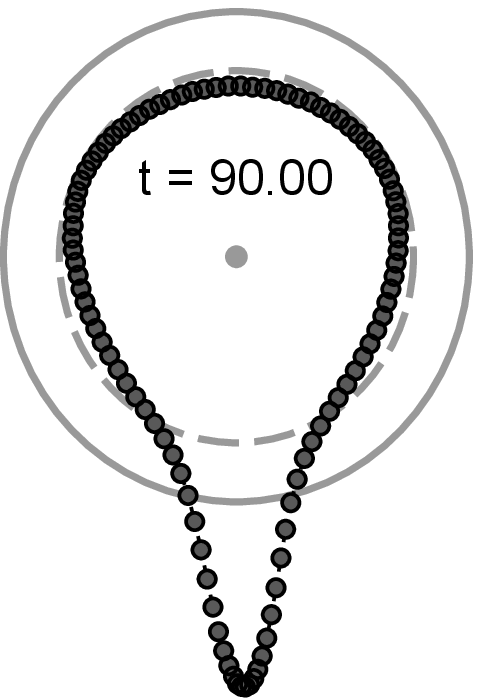}\hfill
\includegraphics[width=0.2\linewidth]{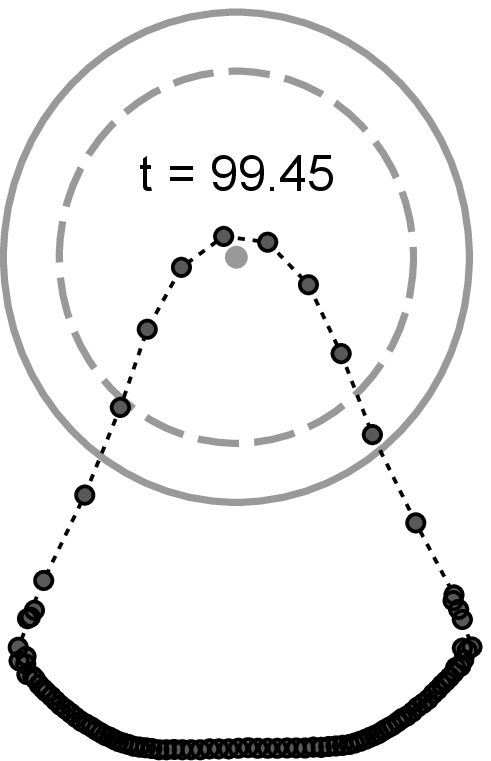}\hfill
\includegraphics[width=0.2\linewidth]{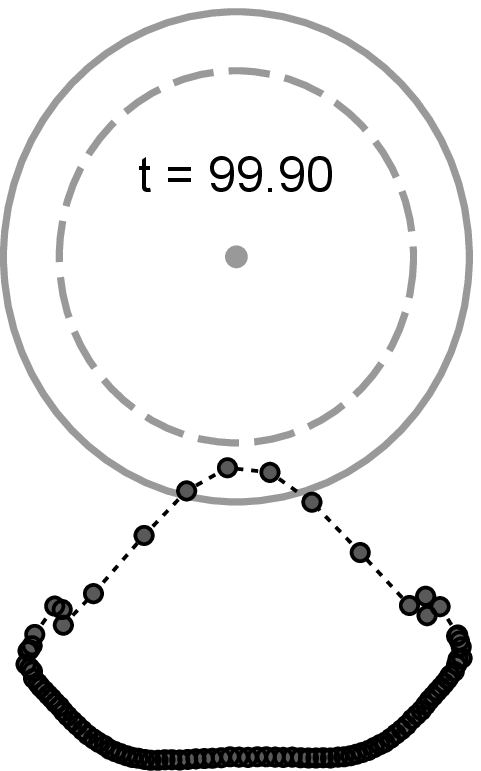}
\caption{Snapshots of an escape process of type I for a chain in a Mexican hat potential evolving according to Eq.~(\ref{qdyn1}).
Notice
that a few units surmount the potential barrier, are driven further down the outer slope of the potential barrier and thereby
pull the entire chain, first over the central potential hump, and subsequently from the potential valley. Parameters: $\lambda=0.8$ and
$\kappa\,\Delta\Theta ^2=0.08$.
Source: Figure
taken from Ref.~\cite{Gross.2014.PRE}.}
\label{fig:I_escape}
\end{figure}

\begin{figure}
\centering
\includegraphics[width=0.2\linewidth]{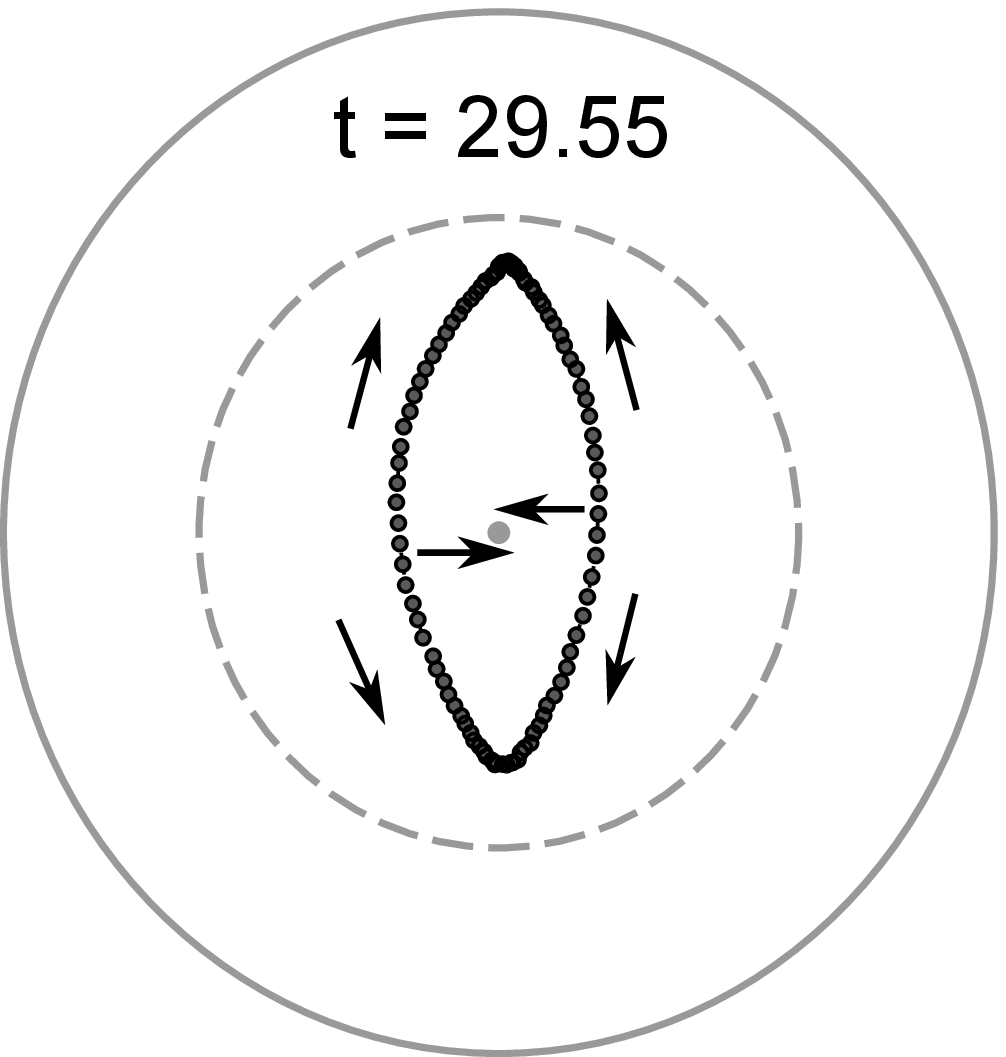}\hfill
\includegraphics[width=0.2\linewidth]{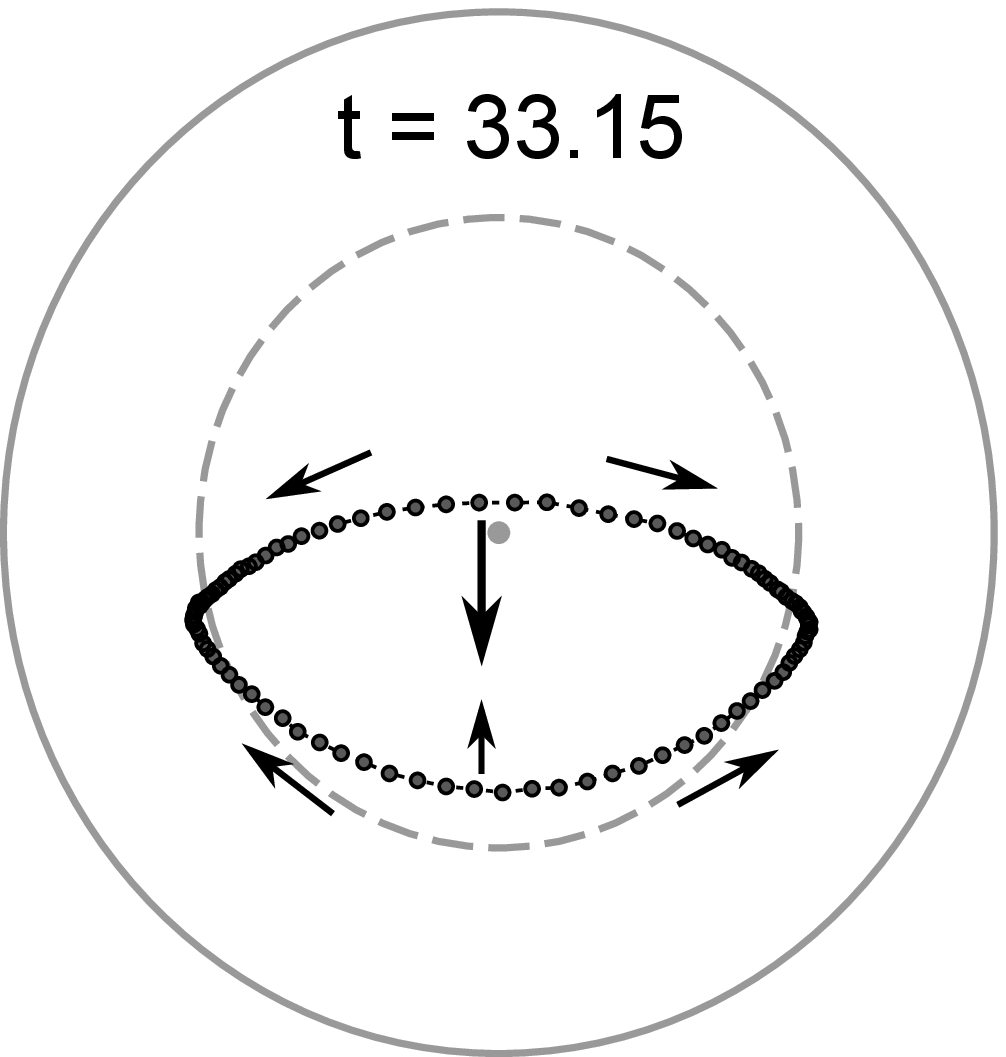}\hfill
\includegraphics[width=0.2\linewidth]{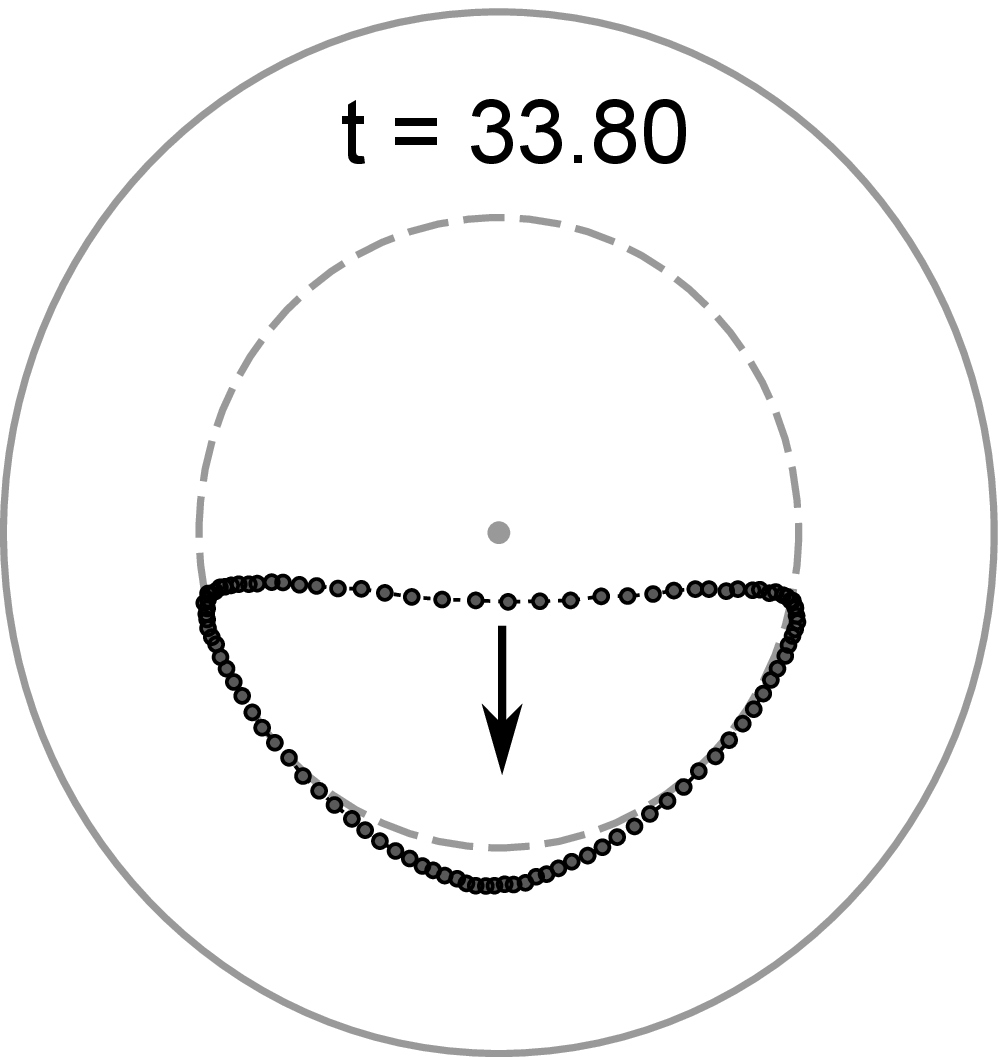}\\
\vspace{1ex}
\includegraphics[width=0.2\linewidth]{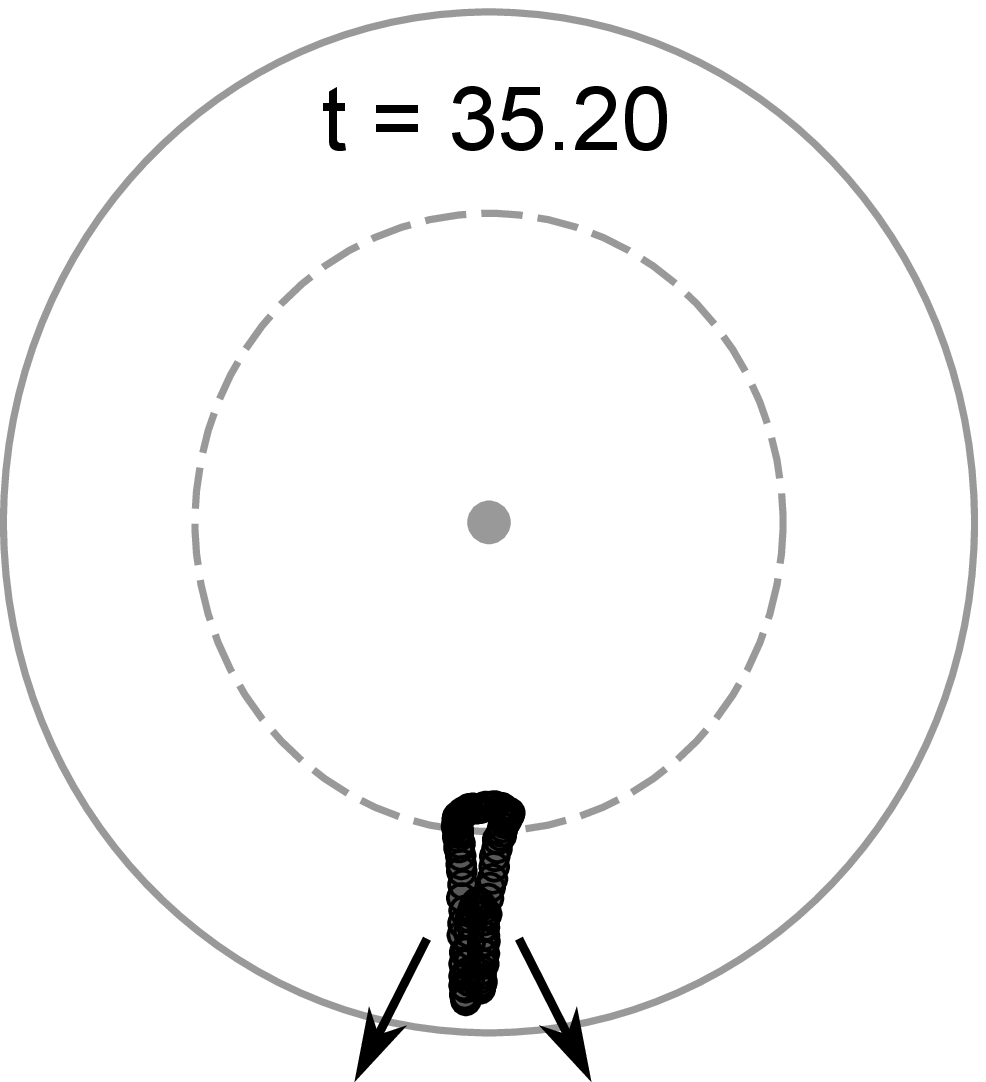}\hfill
\includegraphics[width=0.2\linewidth]{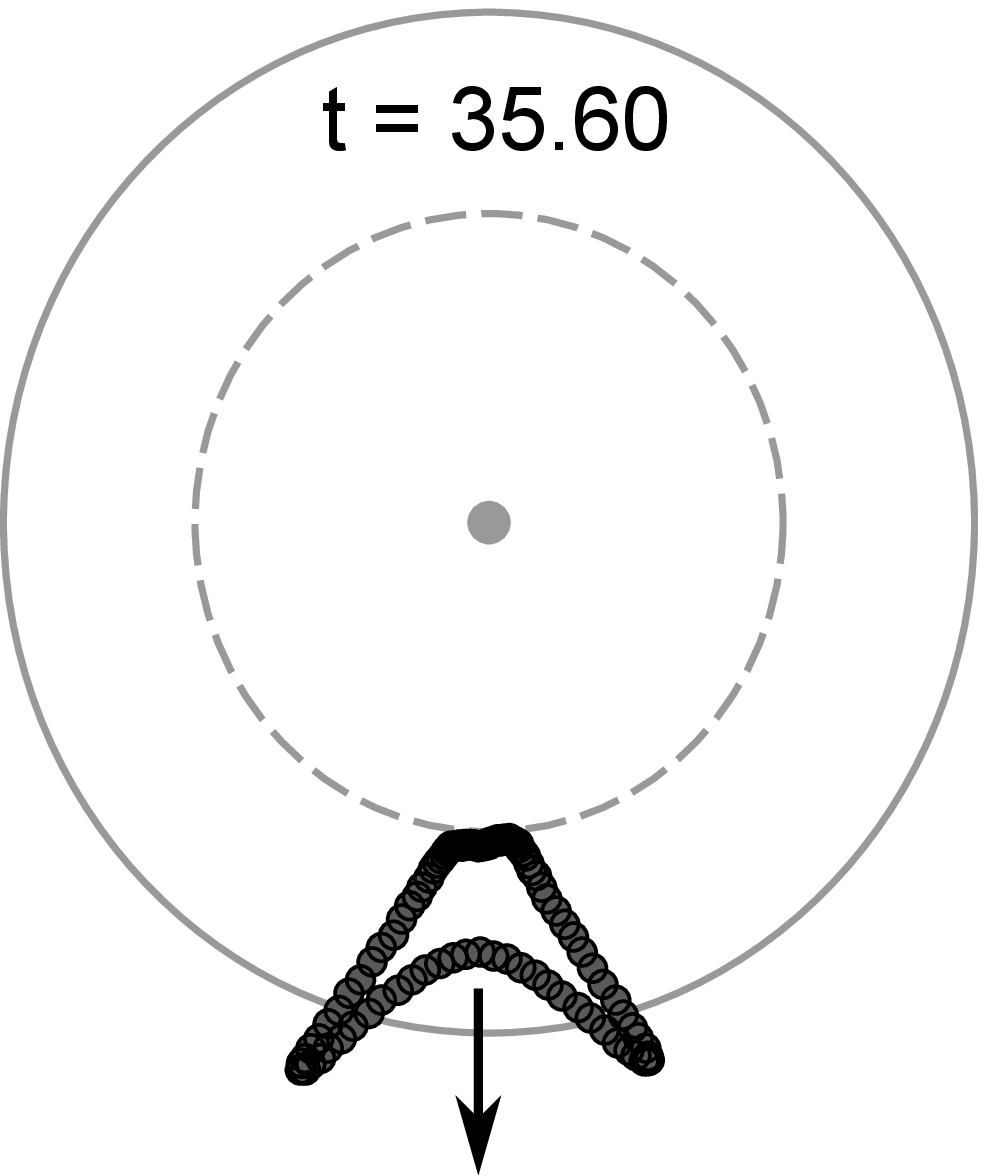}\hfill
\includegraphics[width=0.2\linewidth]{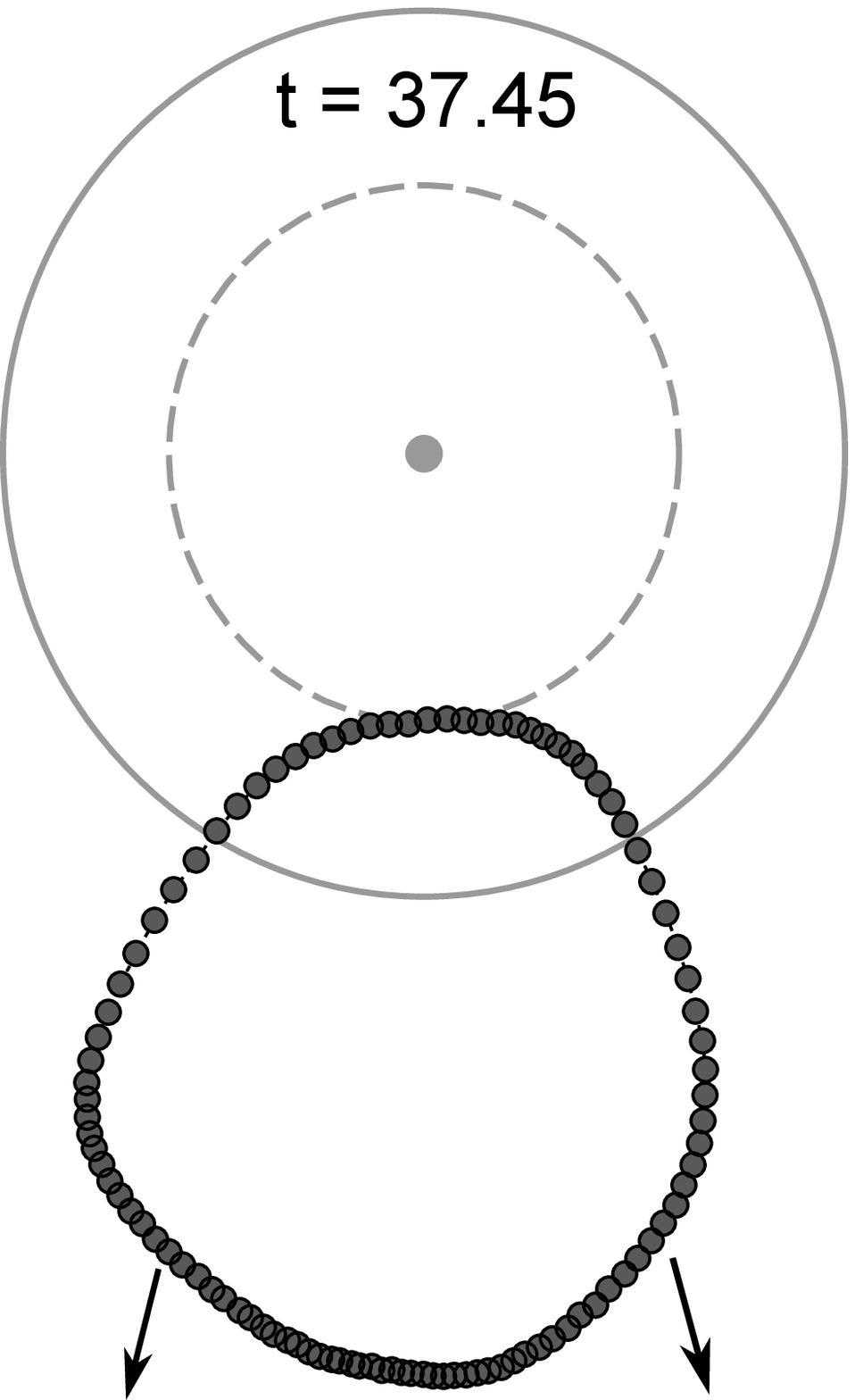}
\caption{Snapshots of an escape process of type II for a chain in a Mexican hat potential evolving according to Eq.~(\ref{qdyn1}).
Arrows indicate the direction of chain movement. In contrast to a type I escape, in a type II escape the chain first overcomes the 
central potential hump, and then the potential barrier. The parameter values are: $\lambda=0.4$ and $\kappa \Delta \Theta =1.00$. 
Source: Figure adapted from Ref.~\cite{Gross.2014.PRE}.}
\label{fig:II_escape}
\end{figure}

The escape related to transition state type I, as depicted in
Fig.\,\ref{fig:I_escape}, indicates a process in which a few units
surmount the potential barrier, are driven further down the outer slope
of the potential barrier and thereby pull out the entire chain from the
meta-stable state. An escape of type II, depicted in Fig.\,\ref{fig:II_escape},
describes the process in which the chain first surmounts the central potential
hump, passing the transition state of type IIa as a bundle, and then overcomes the
potential barrier in the way indicated by transition state type IIb.

In both of these cases  successful  escape events
occur already for energy values in the order of a few times
of the associated activation energy.

Concerning the dynamics on longer time scales,  we can identify three dynamical
regimes characterised by how the energy is transferred into different
degrees of freedom depending on the
 parameter values. This has a crucial impact on typical escape times.

(I) If longitudinal wave modes are absent
 or when the angular displacements of the  units
remain small the system's behaviour is dominated by transversal wave modes.
When the system's dynamics comprises a dominant transversal wave mode,
it forms breathers that promote an escape of type I, as shown in Fig.\,\ref{fig:I_escape}.

(II) In the opposite case, when radial wave modes are absent
(viz. the $k=0$ phonon mode is not affected by
modulational instability),
longitudinal (angular) wave modes arise due to a resonant
excitation from the initial phonon mode. Eventually,
the system attains a state of
periodic energy exchange between the phonon mode and the longitudinal mode.
Both the phonon and
longitudinal mode cause a synchronous oscillation between kinetic
and potential energy for each unit.  In general, an escape
of type I is not expected because of the
 lack of energy concentration into critical radial elongations.
 However, we see an enhancement of an escape of type II.
 In such a case the chain is strongly
stretched in between  two wave nodes. It tends to reduce
 the tension by decreasing the length of the stretched sections,
 rendering them more straight and they thereby surmount the
 central potential hump. The initial perturbations can break the
 symmetry of the longitudinal pattern which can cause one of the
 two stretched segments to overcome the potential hump. Exactly
 this can be observed in the first three snapshots of
 Fig.~\ref{fig:II_escape}.
 The first two snapshots show the two wave nodes
 (first vertically then horizontally aligned) and
 the third  displays how the upper part
 of the stretched
chain segment is carried over the potential hump.

(III) In the simultaneous presence of radial (transversal) and angular (longitudinal) wave
modes the system will evolve into a  chaotic state as the two modes mix in such a way
that the system develops irregular long-term behaviour. In fact, once the initial
wave patterns have ceased, the resulting irregularity leads to an
on average almost homogeneous distribution of the system energy into all
degrees of freedom. Compared to case (I) the  escape events  proceed with a reduced overall
efficiency.

\FloatBarrier
\subsection{Surmounting collectively oscillating bottlenecks}

We next consider the collective escape dynamics of a chain of coupled units in the presence of a weak
external ac-field,  rendering
periodically oscillating barriers \cite{hennig.2008.epl}. To be precise, the escape dynamics for the system of coupled Langevin 
equations
(Eq.~(\ref{1d_Langevin_eq})) augmented by a
 external time-periodic modulation field is considered. The driving field of amplitude strength $f$,
frequency $\omega$ and phase $\theta_0$  globally acts upon the system, with a temporal dynamics obeying:

\begin{equation}
 \ddot{q}_n+\gamma q_n+\omega_0^2 q_n- aq_n^2 +\xi_n(t)-\kappa[q_{n+1}-2q_n+q_{n-1}]
 -f \sin(\omega t+\theta_0)=0,\label{eq:Ldriven}
 \end{equation}

For the deterministic system it is shown in Ref.~\cite{hennig.2008.epl}
that for a  chain  situated initially at the bottom of the potential well,
i.e. there is no net energy contained
in the chain, that for certain values of the frequency $\omega$ of  the  weak external driving field $F(t)=f \sin(\omega t+\theta_0)$,
energy
is now pumped resonantly into the chain, resulting in  a plane wave excitation.
Notably, the energy that
a unit gains on average from the external field, measured by the ratio $E_n/\tilde{E}_{field}$,
can attain a remarkably large ratio
of $\sim 20$ where
we denote by
\begin{equation}
 E_n=\frac{p_n^2}{2}+U(q_n)
\end{equation}
and
\begin{equation}
 \tilde{E}_{field}(t)=-f\sin(\omega t+\theta_0)q_n
\end{equation}
the energy of a unit and the field energy respectively.
Exerting a stationary flat state, represented by a plane wave, additionally to
weak spatially periodic perturbations of certain
critical wave numbers  the deterministic system responses by means
of parametric instability
such that  a pattern of localised states forms.

In the context of the driven Langevin system Eq.~(\ref{eq:Ldriven}) the energy is introduced
in the lattice coherently in the form of a plane-wave
excitation as the response to the external ac-field and
non-coherently through thermal fluctuations.
The stochastic source and the external
 ac-field conspire to produce such an instability mechanism
 of the stationary flat-state (plane-wave) solution yielding
 a spatially localised system state.

\begin{figure}
\includegraphics[scale=0.42]{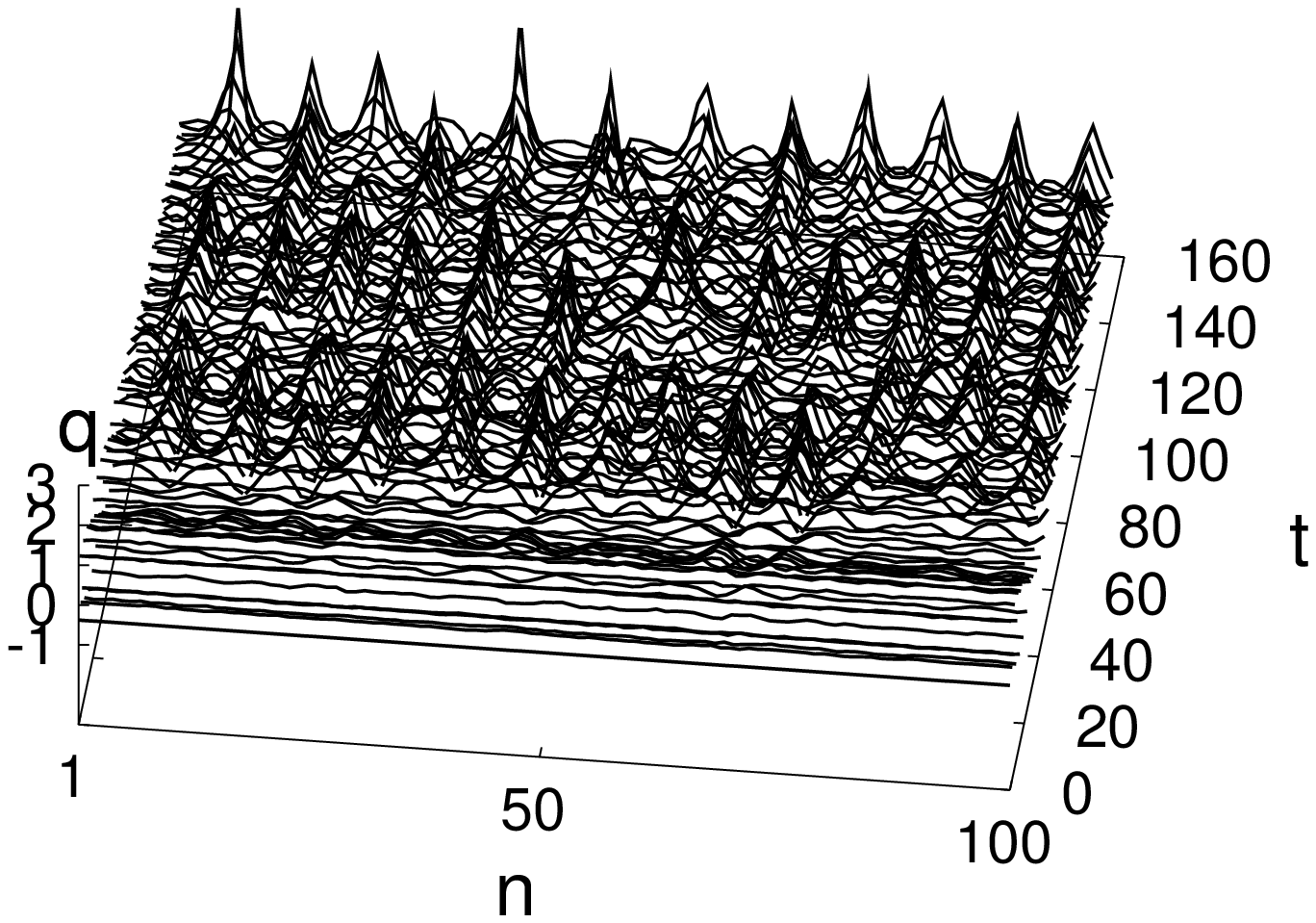}
\includegraphics[scale=0.42]{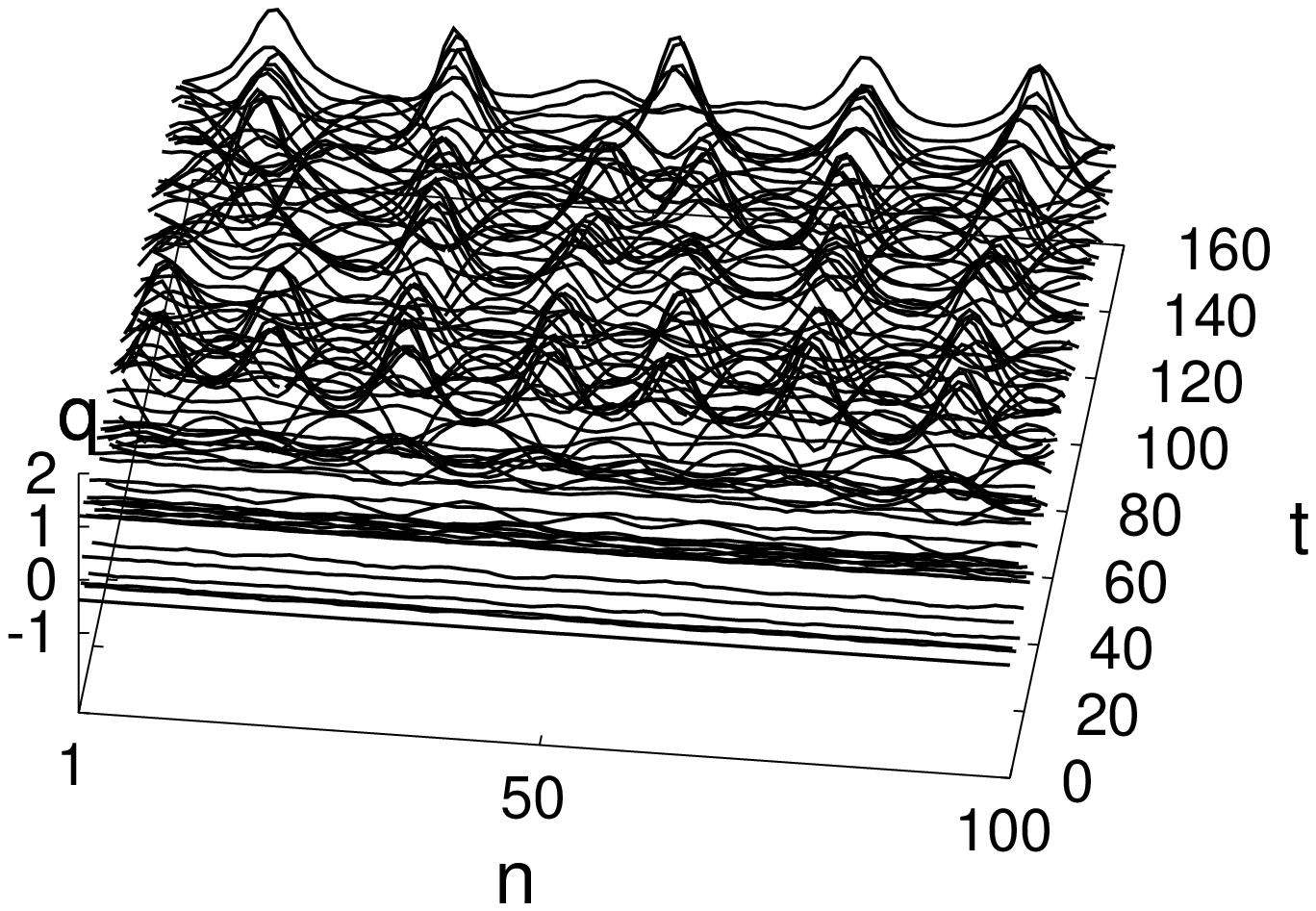}
\caption{Spatio-temporal pattern
of the solutions $q_n(t)$ for chains consisting of $N=100$ units. These chains are driven by an ac-field, damped, and subjected to
random fluctuations - see Eq.~(\ref{eq:Ldriven}). The two panels correspond to coupling strengths $\kappa=0.5$ (left panel) and
$\kappa=2$
(right panel), respectively, for the same realisation of Gaussian white noise
with thermal energy $k_BT=0.001\times \Delta E$. The remaining parameter values
are set at $f=0.15$, $\omega=1.295$, $\theta_0=0$ and friction
$\gamma=0.1$. Source: Figure adapted from Ref.~\cite{hennig.2008.epl}.}
\label{fig:pattern}
\end{figure}

In fact, the additional stochastic term provides perturbations of all wave
numbers and a pattern emerges from the low-energy homogeneous flat
state. (The energy contained in the chain at $t=0$  is vanishingly small
compared to the barrier energy,
i.e. $E/\Delta E \ll 1$.) That is, perturbations provided by the thermal noise grow and
induce a localised mode (LM) consisting of several humps. This is demonstrated in
Fig.~\ref{fig:pattern} which shows the spatio-temporal evolution of the amplitudes $q_n(t)$
for couplings $\kappa=0.5$ and $\kappa=2$. The Langevin equations
were numerically integrated using a second-order Heun stochastic solver scheme.
We note the formation of a LM of certain wave length arising from
the homogeneous state soon after a short time span (after $t\sim 60$)
and we find that the period duration for oscillations near the
bottom of the potential is around {\bf $2\pi/\omega_0\simeq 4.4$}.

The fastest
growing perturbations are those associated with the critical
wave number (see Sec.~\ref{ra_ch1}). Each of these humps
resembles the hairpin shape of the transition state as the critical
escape configuration possessing an energy $E_{act}$ through
which the coupled units have to pass in order to cross the
barrier. The robustness of the LMs is remarkable: a
LM is sustained, despite continuously impacting thermal
noise of strengths up to values $k_B T= 0.2 \times \Delta E$.
Moreover, the formed patterns maintain their
distinct wavelength determined by the critical wave number at which the
parametric resonance occurs.

\begin{figure}
 \centering
  \includegraphics[width=0.4\linewidth, height=5cm]{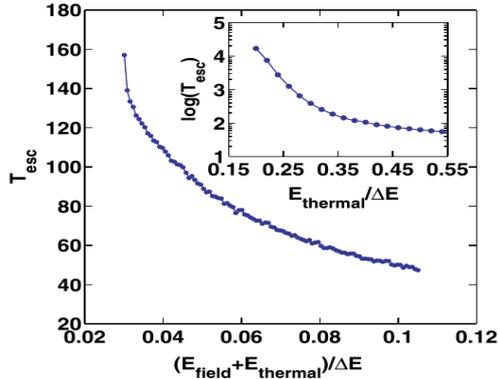}
  \caption{(Colour online) Mean escape time in the one-dimensional Langevin setting - Eq.~(\ref{eq:Ldriven}). Shown is the mean escape 
time vs. the mean
injected energy $E=E_{field}+ k_B T$, for chains consisting of $N=100$ units, averaged over ensembles. The injected energy is measured
in units of $\Delta E$
with fixed field energy $E_{field} = 0.03 \times \Delta E$ provided by an
external modulation field with $\omega =1.295$, $\theta_0=0$ and $f=0.15$.
Here we vary the thermal energy $E_{thermal}=k_B T$. The inset
depicts the unforced case with $f=0$ in semi-logscale. The remaining parameter
values are $N=100$, $\kappa=0.28$ and $\gamma=0.1$. Source: Figure adapted from Ref.~\cite{hennig.2008.epl}.}
  \label{fig:drivingescape}
  \end{figure}

When the noise strength is increased the
growth rate of the humps becomes enhanced, being
reflected in the statistics of the barrier crossing of the
chain in the presence of weak ac-driving. The amplitude
and frequency of the latter are chosen such that the
dynamics exhibits parametric resonance. The dependence
of the mean escape time of the chain on the injected average energy
$E =E_{field} + E_{thermal}$, with $E_{field} = \langle \tilde{E}_{field}(t) \rangle_t$ and $E_{thermal} = k_B T$
(measured in units of the barrier energy $\Delta E$) is displayed
in Fig.~\ref{fig:drivingescape}. The thermal energy $E_{thermal}$, supplied
non-coherently by the heat bath, is varied within the range
$[(10^{-4} - 0.11) \times \Delta E]$.

We observe that the underlying
irregular dynamics serves for self-averaging and thus the
choice of the phase, $\theta_0$, of the coherent, external forcing
does not affect the mean escape time. In the forced as
well as unforced case there occurs a rather rapid decay of
$T_{esc}$ with growing $E_{thermal} = k_B T$ at low temperatures.
This effect weakens gradually upon further increasing
$k_B T$. Most strikingly, for the forced system the escape
times become drastically shortened in comparison with the
unforced case with $f = 0$. Moreover, for the forced system
escape takes place also at very low temperatures for which
in the undriven case not even the escape of a single unit has
been observed during the simulation time (taken here as
$t = 10^5$) implying a giant enhancement of the rate of escape
as compared to the purely thermal-noise-driven rate.
This emphasises the collective mechanism of this
resonance effect which occurs for finite interaction
strength $\kappa \ne  0$ only.

\subsection{Escape assisted by entropic localisation}\label{subsec:noisefreeescape}

In a  deterministic (noise-free) setup the escape
problem of a chain of harmonically coupled units over the
barrier of a metastable potential was studied in \cite{Hennig.2014.PRE}.
Energy is injected into
the system by means of an applied external time-periodic
field.
The corresponding system is given by
\begin{equation}
 \ddot{q}_n+\omega_0^2 q_n- aq_n^2 -\kappa[q_{n+1}-2q_n+q_{n-1}]
 -f \sin(\omega t+\theta_0)=0.\label{eq:entropic}
 \end{equation}

Notably, even for a very weak driving force there results
fast escape for a chain situated initially extremely
  close to the bottom of the potential well and thus containing
  a vanishingly small amount of energy. For a suitably chosen
  driving frequency, almost coinciding with the frequency at the
  lower edge of the phonon band of linear oscillations, as a start,
  an almost uniform oscillating state of the chain is excited.
  The amplitude of the latter rises (slowly) in time, and upon
  entering the weakly nonlinear regime the almost uniform state
  becomes unstable with respect to spatial perturbations. This
  triggers the formation of a few localised humps (standing
  breathers) coexisting with a phonon ``bath'' background in
      between them. Due to the effect of entropic localisation for the
standing breathers, their energy-reduction process is impeded.
In more detail, once a unit has acquired
a sufficiently high energy, it is retained for a fairly long time due to the fact
that in a soft unit the energy dwells relatively more time in the
potential part rather than in its kinetic form. The reason for this is that in soft
potentials the oscillation frequency decreases with increasing
amplitude. In conjunction with the fact that the density of states
increases with increasing amplitude, attaining and preserving
higher amplitudes becomes entropically more favourable.
Thus, during the major part of an oscillation period
of a unit, after it has gained energy from the external field, its
neighbours, or impacting moving breathers, the displacement
of this unit remains large while the velocity is low. Therefore,
this entropic localisation mechanism impedes the energy
exchange of a higher-amplitude unit with the surroundings.
Conclusively, localisation of energy minimises the free energy
as it is favoured, with respect to maximisation of entropy, that
the energy gaining units populate regions in phase space where
the density of states is higher.

In contrast, the process in the other direction is entropically
favoured. That is, due to the fact that the driving frequency lies
just below the phonon band, further resonant energy pumping
by the external field into standing breathers is possible,
 provided a proper phase relation is retained between them.
In fact, the associated growth of the amplitude of the breathers
enhances even entropic localisation. However, as with growing
amplitude, the frequency of a breather diminishes, and there
results a frequency mismatch between the external field and the
 standing large-amplitude breather hampering direct substantial
energy feeding from the external field into it. Therefore, at this
 stage the only way a breather can gain more energy is by
processes of internal energy redistribution along the chain.
Conclusively, choosing the frequency of the external driving
just below the phonon band is advantageous for two reasons:
First, emerging standing breathers can become amplified by
 direct energy gain from the (almost) resonant external field. At
  the same time, externally driving with a frequency almost equal
 to that of harmonic oscillations near the bottom of the potential
well generates permanently a phonon ``bath'' background
between the standing breathers forming the source for the
emergence of mobile chaotic breathers. The emergence of the
itinerant chaotic breathers, and the merging with these standing breathers, contributes
to their growth. Eventually, for overcritical amplitudes a
standing breather adopts the shape and energy content of the
transition state, and by passing across the latter, escape of the
chain over the barrier is instigated.

\begin{figure}
\centering
\includegraphics[height=5cm, width=6.5cm]{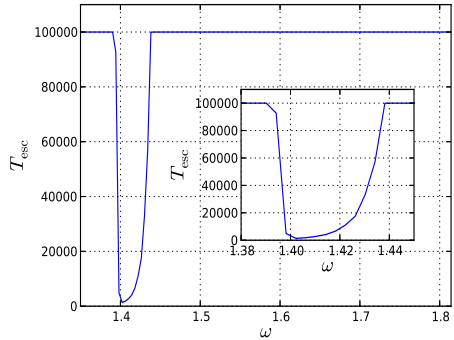}
\caption{(Colour online) Mean escape time for ensembles as a
function of the driving frequency $\omega$. Each chain in the ensemble, consisting of $N=100$ units and evolving in a cubic potential,
has deterministic dynamics and is subjected to a weak periodic driving force with a fixed driving strength
$f = 0.003$. The values of the remaining parameters are $\omega_0 = 2$ and
$a = 1$, regulating the depth and width of the valley in the cubic potential respectively, $\theta_0 = 0$ controlling the phase of the
external driving, and $\kappa = 0.1$ regulating the strength of the nearest-neighbour interactions. Note that $T_{esc}=10^5$ in fact
implies that no transitions for any initial condition in
the ensemble were observed in our simulations. Source: Figure taken from Ref.~\cite{Hennig.2014.PRE}.}
\label{fig:entropic}
\end{figure}

In Fig.~\ref{fig:entropic} the mean escape
time $T_{esc}$ of the chain versus the driving frequency
for a small driving amplitude A = 0.003 is displayed.
The averages were performed over $1000$ realisations
of random initial conditions. There is a window of frequencies
$1.395 \lessapprox \omega \lessapprox 1.437$ for which speedy escape is accomplished,
and outside of this window not a single event of escape
takes place throughout the simulation time $T_s = 10^5$. The
window of the driving frequencies associated with speedy
escape  has an overlap with the phonon
band penetrating the latter from its lower edge, underlining
the fact that permanently impinging phonons (the latter arise as the
result of driving the system within this frequency range) are
paramount for the creation of breathers promoting eventually
the escape process. The resulting  breathers possess frequencies lying
in the window of speedy escape just below the phonon
band.

\subsection[Cooperativity in molecular dissociation]{Cooperativity in molecular
dissociation}\label{subsection:dissociation}
\noindent

The problem of molecular dissociation is usually looked upon in a stochastic
context of Arrhenius or Kramers
theory~\cite{Wilson.1955.McGrawHill,Sage.1981.JWS,hanggi.1986.jstatphys_a,Page.1988.ChemPhys,hanggi.1990.romp}.
Nevertheless, it is well known that the bond breaking mechanism may appear within
the context of nonlinear dynamics
in the form of breaking up of  a separatrix lines and the emerging
chaotic dynamics~\cite{HENNIG.1994.PRE}.  This chaos induced dissociation constitutes a relatively
slow
process where the chaoic dynamics now take on the role of a stochastic dynamics that drives the
dissociation in the Kramers
escape regime.  As an alternative, a ``coherent" mechanism exists involving the organised
dynamics of interacting
molecules that leads to a cooperative barrier crossing~\cite{Takeno.2005.PLA},
similar in nature to the topics  treated in this review above.
This is a non-perturbative approach of anharmonic molecular vibrations
with a particular concern to the exploration of the general
principles for large-amplitude molecular vibrations, leading eventually to
molecular dissociation.  Since generally the dissociation involves a small number of
constituent parts, a reasonable
minimal model involves a three-atom molecule composed of linearly arranged
identical atoms.

\subsubsection[Mechanism of a triatomic dissociation]
{Triatomic dissociation}\label{subsubsection:triatomic}

We consider here the vibrations of a linear molecule composed of three identical
atoms A, B, C each with atomic masses $m$  and respective
displacements from their equilibrium position denoted as ${\displaystyle u_1, u_2}$ and
${\displaystyle u_3}$, assuming  only
nearest-neighbour interactions.  The interatomic potential $U_i$, $i=1$ for the A-B
interaction and $i=2$ for B-C interaction
depend solely on relative distance;  explicitly those are chosen  as  Morse
potentials; i.e.,
\begin{eqnarray}
U_i(|u_i-u_{i+1}|)=D_i{\left(1-\exp[-a_i|u_i-u_{i+1}|]\right)^2},\ \ \ i=1,2\;,
\label{eq:diss1}
\end{eqnarray}
where $D_i$ and $a_i$ denote potential parameters.
The equations of motion for the atoms in the molecule in  relative
displacements
$x_1=u_1-u_2$  and $x_2=u_3-u_2$ read
\begin{eqnarray}
\ddot{x}_1=-\frac{1}{\mu}\frac{dU_1(x_1)}{dx_1}-\frac{1}{m}\frac{dU_2(x_2)}{dx_2
}
\label{eq:diss2}
\end{eqnarray}

\begin{eqnarray}
\ddot{x}_2=-\frac{1}{m}\frac{dU_1(x_1)}{dx_1}-\frac{1}{\mu}\frac{dU_2(x_2)}{dx_2
} \label{eq:diss3}
\end{eqnarray}
where $\mu=m/2$ denotes the reduced mass. Under the nearest neighbour approximation,
the three-body problem reduces to a pair of coupled
nonlinear differential equations, viz. Eqs.~(\ref{eq:diss2})-(\ref{eq:diss3}), that are generally
non-integrable \cite{Takeno.2005.PLA,HENNIG.1994.PRE}.
For the further
analysis of the energetics of the dissociation processes it is useful to introduce
 bond energy variables $h_1 (t)$, $h_2 (t)$  for the A-B, B-C bonds respectively,
via the relations
\begin{eqnarray}
h_i(t) \equiv h_i = \frac{\mu}{2} \dot{x}_{i}^2+U_i(x_i )  \ \ \, i=1,\ 2 \;.
\label{eq:diss4}
\end{eqnarray}

\noindent
By use of the equations of motion one finds that \cite{Takeno.2005.PLA}:
\begin{eqnarray}
\frac{{dh_1}}{dt} = -\frac{\mu}{m}\frac {dU_{2}(x_2)}{dx_{2}} \, \dot{x}_1 \;,
\label{eq:diss5}
\end{eqnarray}

\begin{eqnarray}
\frac{{dh_2}}{dt} = -\frac{\mu}{m}\frac {dU_{1}(x_1)}{dx_{1}} \, \dot{x}_2 \;,
\label{eq:diss5b}
\end{eqnarray}

\noindent
Upon multiplying the latter two relations we obtain that

\begin{eqnarray}
\dot{h}_1\dot{h}_2=\frac{\mu^2}{m^2}\dot{U}_1\dot{U}_2 ~~\label{eq:diss5c}.
\end{eqnarray}
From this relation it is seen that for identical oscillators and identical initial conditions the system of nonlinear oscillators  
decouples in which case the coefficients
of the Morse potential become renormalised.

In the more general case, however,
we may recast
Eqs.~(\ref{eq:diss5})-(\ref{eq:diss5b}) as~\cite{Takeno.2005.PLA}
\begin{eqnarray}
\dot{h}_1-\frac{\mu}{m}f(t)\dot{U}_1=0, \ \ \ \
\dot{h}_2-\frac{\mu}{m}\frac{1}{f(t)}\dot{U}_2=0,  ~~\label{eq:diss6}
\end{eqnarray}
where
\begin{equation}
f(t) = \frac{\mu}{m}\dot{U}_2/\dot{h}_2= \dot{h}_1/\Big(\frac{\mu}{m}\dot{U}_1\Big) \;,
\end{equation}
is some function of time. Eqs.~(\ref{eq:diss6}) show that each
unit experiences an effective  force that
involves both the
intrinsic nonlinearity of the problem as well as the  implied initial conditions.
It is not obvious from the form
of these equations alone whether any coherent processes may affect the dissociation
process.  This analysis may be done through the study of the
potential landscape of the problem and the numerical solution of the
equations of motion.  We note that molecular dissociation of say bond $2$ at a
given time $t_0$ signifies that with force $F\rightarrow 0$ as $t\rightarrow 0$ we have
\begin{eqnarray}
h_2 \gg h_1 \hspace{0.5cm} \text{as} \hspace{0.5cm} t \rightarrow  t_0 \;.
~~\label{eq:diss7}
\end{eqnarray}
This in turn shows  that bond breaking is  generally associated with
dissociation.

\begin{figure}[!t]
\centering
\includegraphics[width=0.7\linewidth, height=10cm]{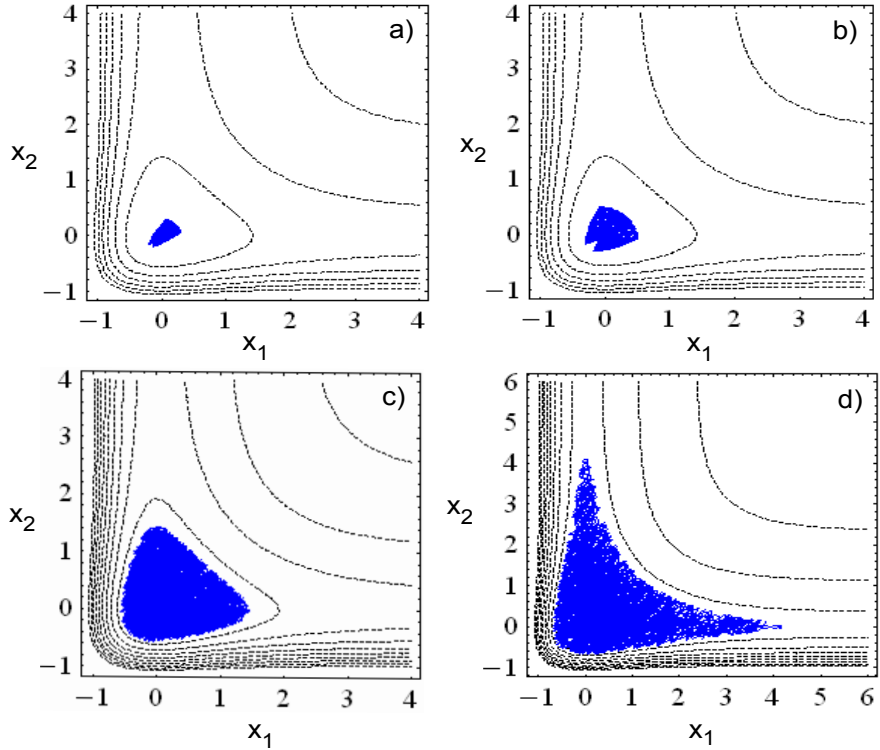}
\caption{(Colour online) Molecular trajectories in the potential energy landscape of the linear three
atom
molecule for various initial relative velocities in bond 2.  In (a) $v=0.4$, (b)
$v=0.8$, (c) $v=1.5$ (Type-I trajectories: fast
energy exchanges between bonds)
 (d) $v=1.9$ (Type-II trajectories: delimit  stable linearised
energy exchange between the units and unbounded motion). Source: Figure adapted from Ref.~\cite{Takeno.2005.PLA}.}\label{fig:diss1}
\end{figure}

\begin{figure}[!t]
\centering
\includegraphics[width=0.7 \linewidth,height=5cm]{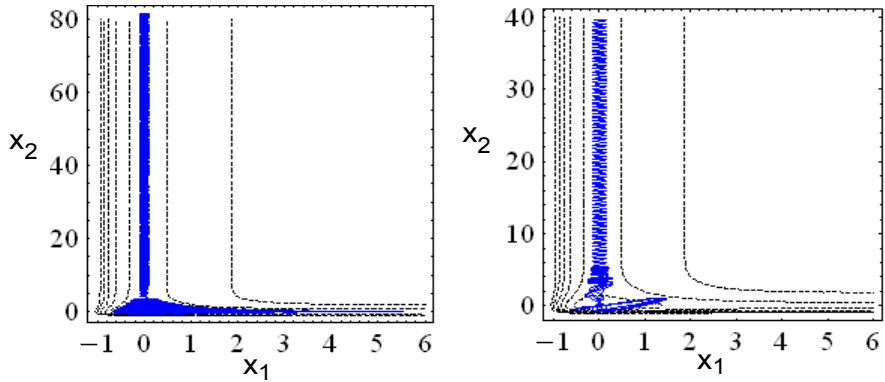}
\caption{(Colour online) Molecular trajectories in the potential energy landscape  for
unbounded
motion leading to
bond breaking and dissociation.
Type-III motion with $v=1.935 $ (left panel) and $v=1.95 $
(right panel). Source: Figure adapted from Ref.~\cite{Takeno.2005.PLA}.
}\label{fig:diss2}
\end{figure}

\subsubsection[Cooperative reaction paths]
{Cooperative reaction paths}\label{subsubsection:reaction_paths}

We introduce the two-dimensional potential function and the potential
energy surface
\begin{figure}[!t]
\centering
\includegraphics[width=0.85 \linewidth,height=8cm]{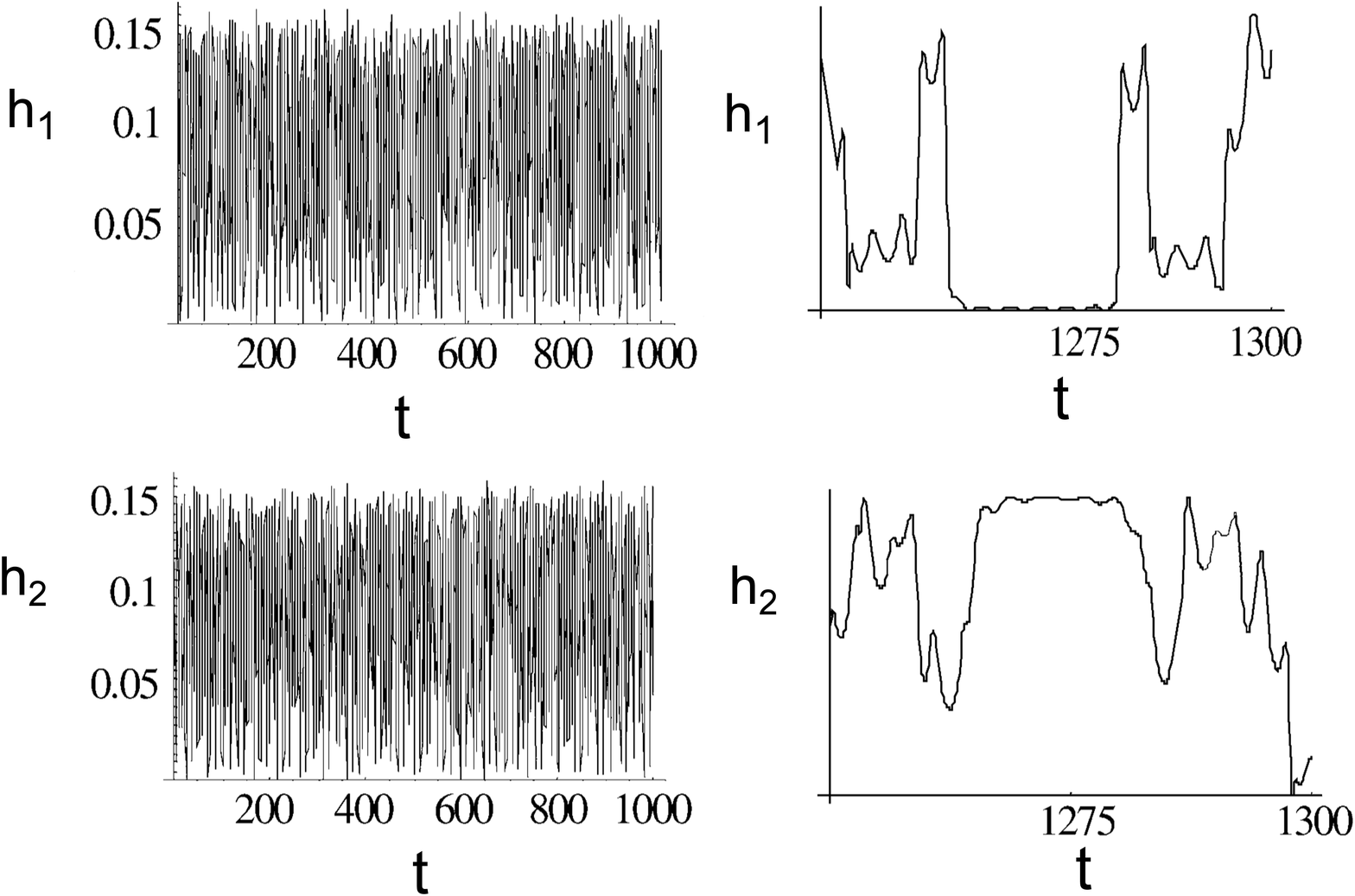}
\caption{Total energies $h_1 (t)$ and $h_2 (t)$ accumulated in each bond as a
function of time.
The left two
plots correspond to type-I motion ($v=0.8 $) while the right two figures
correspond to
type-II, intermittent motion ($v=1.9$). The continuous energy exchange seen in
the
linearised regime is sharply contrasted by the intermittent yet coherent
predissociative exchange
in type-II motion. Source: Figure adapted from Ref.~\cite{Takeno.2005.PLA}.}\label{fig:diss3}
\end{figure}

\begin{figure}[!t]
\centering
\includegraphics[width=0.85 \linewidth,height=8cm]{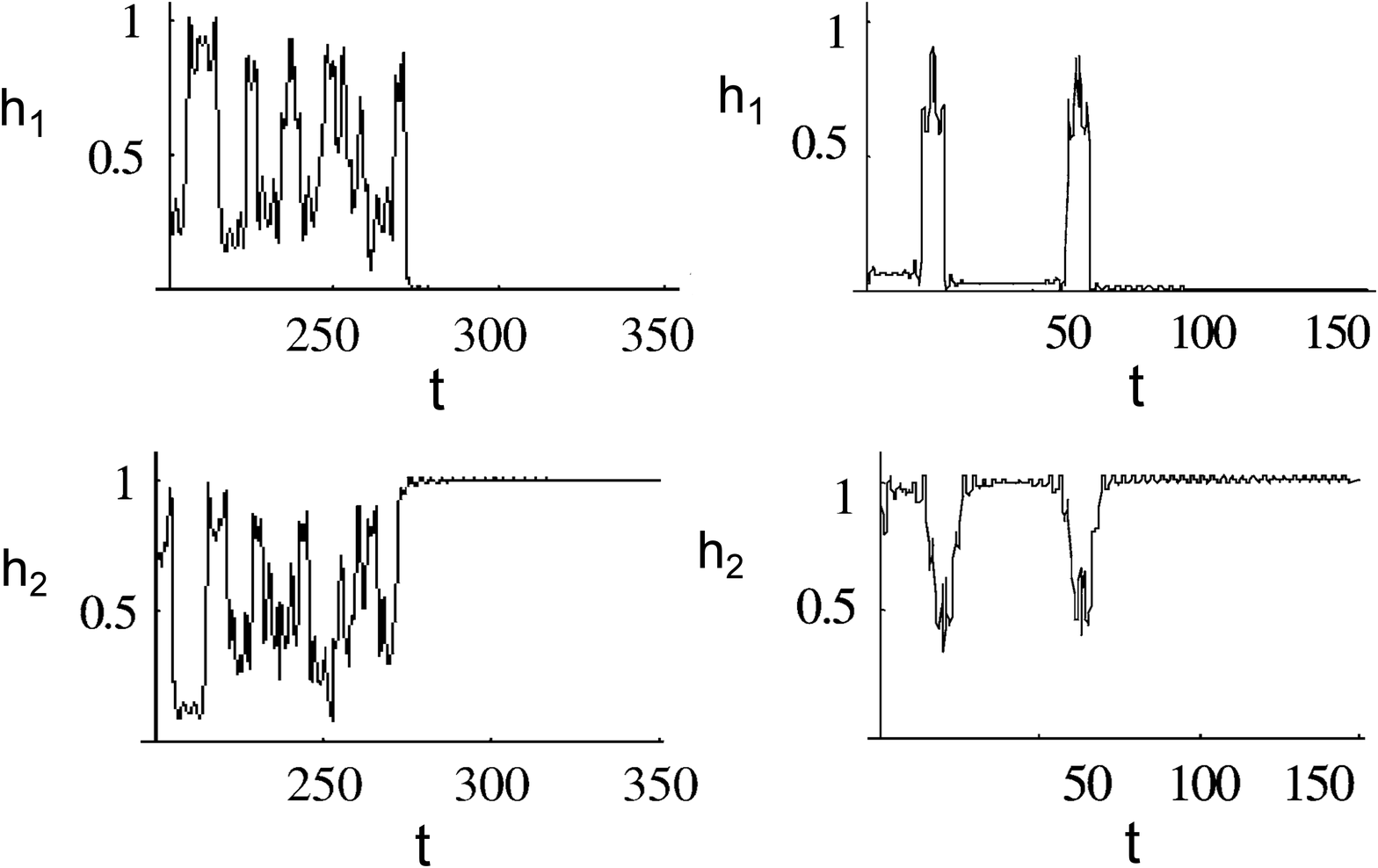}
\caption{Total energies $h_1 (t)$ and $h_2 (t)$ accumulated in each bond as a
function of time
for type-III motion. The left panels correspond to $v=1.935$ while the right
panels to
$v=1.5$.  Dissociation occurs after coherent initial energy localisation and
exchange of the
localised energy between the two bonds. As initial energy increases, bond
breaking occurs faster
and with less or eventually no exchange of localised energy
``packets''. Source: Figure adapted from Ref.~\cite{Takeno.2005.PLA}.}\label{fig:diss4}
\end{figure}
\begin{equation}
U(x_1,x_2)=U_1(x_1)+U_2(x_2)
\label{eq:diss8}
\end{equation}

\noindent
and consider the various trajectories for different total energies that follow from
 the numerical solution of Eqs.~(\ref{eq:diss2})-(\ref{eq:diss3}) for the
parameters ${\displaystyle a_1=a_2=D_1=D_2=1}$, and with
the initial conditions
$x_1(t=0)=x_2(t=0)=0$ and $\dot{x}_1(t=0)=0.5, ~ \dot{x}_2(t=0)=v$.
It can be deduced from Figs.~\ref{fig:diss1} and Fig.~\ref{fig:diss2} that varying the initial velocity $v$ we
may access different regimes of the three unit problem. Three regimes can be identified and are discussed in the following.

In Fig.~\ref{fig:diss1} we show the
potential landscape projections for type-I motion with:
(a) $v=0.4$, (b) $v=0.8$, and (c) $v=1.5$,  as well as a type-II motion for (d)  $v=1.9$.
In contrast, in Fig.~\ref{fig:diss2} we have a type-III motion with (a) $v=1.935$  and  a type III-2 with (b) $v=1.95$.
It is seen that from case I (a) to (c) that the area covered by the trajectory
around the
origin at $(0,0)$ changes from a cross section of a cone lying along the
(1,1)-direction to a regime which covers almost the whole region enclosed by a potential-energy
contour. In the separating type-II case, the domain covered by the orbit tends to extend
almost equally in the
direction of the rays of the potential function. In type-III motion  shown in
Fig.~\ref{fig:diss2}, the trajectory elongates indefinitely along the
${\displaystyle x_2}$ direction,
showing that a break-up of bond $2$ occurs. This line is identified as a
reaction path.  Thus, only for type-I motion does
normal mode theory  apply with bond energy equipartition.

In Fig.~\ref{fig:diss3} we depict  the energy quantities $h_1 (t)$ and
$h_2 (t)$ as a function of time for
type-I (top and bottom figures on lhs) and type-II motion (top and bottom figures on
rhs).  The first pair clearly shows
a rapid energy exchange between the bonds. For
type-II motion we display the quantities
${\displaystyle h_1(t), h_2(t)}$
for ${\displaystyle 1250 \leq t \leq 1300}$  and observe, in distinct contrast
to the type-I regime an energy localisation in each bond that is subsequently
transferred to the neighbouring  one.  The process of energy exchange proceeds  in ``packets'' and not in
a continuous manner as for the type-I linearised motion.  This process of exchange of localised
energy
between the
bonds continues indefinitely and, as a result, we may consider this case as
the pre-bond breaking regime. The value ${\displaystyle v_c \approx 1.92}$ is a critical value because
for $v$ slightly larger than ${\displaystyle v_c}$, the energy-versus-time relation
in the
two bonds becomes irregular.
As $v$ increases further, the system crosses over into type-III motion, as  displayed with
Fig.~\ref{fig:diss4}.

From the numerical
calculation of ${\displaystyle h_1(t)}$ and ${\displaystyle h_2(t)}$
we observe that bond $2$ breaks  although almost complete transfer of
energy from bond $2$ to bond 1 may take place.  These bouncing exchanges become
less pronounced
as $v$, and thus the initial energy, increases.
Physically, breaking of a bond in chemical dissociation
is equivalent to a transition from a bounded motion to an unbounded one.
In the latter case equipartition of the system energy between kinetic energy and
the potential energy
breaks down and
almost all system energy is transformed into potential energy localised in the
breaking bond.  This accumulation of potential energy in bond $2$ at the expense
of
kinetic energies and potential energy in bond $1$ leads to molecular
dissociation.

Molecular dissociation  occurs here as a process that is induced by collective
effects and
nonlinear interactions; spontaneous energy accumulation in the form of discrete
breathers~\cite{SIEVERS.1988.PRL}
appears to play a role. In the simple three atom case one finds, in addition to
a ``trivial'' low energy regime,
a collective, yet intermittent regime that is predissociative and is
characterised by spontaneous energy accumulation
and non-continuous energy transfer. At yet higher energies a spontaneous
dissociation regime also appears where motion becomes
unbounded. Dissociation occurs
through system energy concentration to potential
energy of a given bond at the expense of kinetic energy; this resonant transfer
is cooperative and
possibly related to   targeted energy transfer \cite{Kopidakis.2001.PRL}.
This resonant energy accumulation and exchange may take place for a
substantially long time before leading to molecular dissociation.  This process is
dependent on the
initial bond energy: as the initial kinetic energy becomes larger, localised
energy oscillations
become less frequent, leading to an almost instantaneous bond breaking at sufficiently
large initial energies.

\section[Collective transport]{Collective transport}\label{section:transport}

\subsection[Transport in general]{Transport in general}

The topic of particle transport is typically concerned with the movement
of particles across potential landscapes. Studies of transport in such systems
focus on how forces on the particles resolve themselves to allow for directed transport;
i.e. net motion that is (on average) in one direction or another. Of particular interest
is the \emph{ratchet effect}: the possibility to obtain directed transport by
using zero mean perturbations. The state of the art with respect to transport
in spatially periodic systems out of thermal equilibrium was presented
recently in \cite{astumian.2002.phystoday,hanggi.2009.romp,denisov.2014.pr}.

Some of the earliest studies of particle transport were in the area of
celestial mechanics, in an effort to understand, for example, the motion
of planets orbiting a star. Since the advent of numerical computation
\cite{ZABUSKY.1981.CompPhys}, investigations into particle transport have increased
at an almost exponential rate. This new tool has complemented analytical
and experimental work already being carried out. In addition, the twentieth
century saw the birth of new fields of research, fueled by application
areas such as Josephson junctions, cold atom systems, and Bose-Einstein
condensates, where transport properties shed light on the features of
these systems. Thus the study of transport properties in the presence and absence of random perturbations or noise is a very active
area of research \cite{hanggi.2009.romp,denisov.2014.pr}.

\subsubsection{Individual particles}
A large portion of the literature has focused on the transport properties of
individual particles. In \cite{MACKAY.1984.PhysicaD} the authors carried out relevant work on transport
in (autonomous) Hamiltonian systems (to be precise for area-preserving maps). In particular,
they outlined the structures contained in a largely chaotic phase-space,
most notably cantori and KAM-tori, that play a key role for the occurrence
of transport. They showed that the particle flow through the partial barriers,
created by cantori, was controlled by \emph{turnstiles} that could trap particles
in transporting channels. The phase-space of the
standard map exhibits coexisting chaotic and regular regions (details of the standard map
can be found in \cite{Reichl.1992.SprVerB}). In fact, magnification of any of the boundaries between regular
and chaotic regions will reveal more intricate structures embedded in the chaotic
regions. In particular, the hierarchy of cantori will be revealed through
successive magnifications of particular regions \cite{radons.1989.ACP}.

Analogous features would take prominence in time-dependent driven (i.e. non-autonomous)
Hamiltonian  systems (flows) \cite{Yevtushenko.2000.PRE,Denisov.2002.PREa,Denisov.2006.epl}. In many studies
it has been shown that the emergence of a directed current is triggered by an external
time-dependent field of zero mean. Important in this respect are the spatio-temporal
symmetries (cf. Section \ref{subsubsection:symm_prop}) of the system. With regard to the emergence
of a non-zero current, all symmetries that, to each trajectory, generate a counterpart
moving in the opposite direction need to be broken. Further, it has been shown that
a \emph{mixed phase-space} is required \cite{Schanz.2001.PRL,Schanz.2005.PRE,Denisov.2006.epl,denisov.2014.pr}. The mixed phase-space,
containing regular and chaotic components allows for directed transport in
the chaotic component of the phase-space. The regular and irregular components
are separated by impenetrable KAM-tori which are in turn surrounded by a
hierarchy of cantori. These cantori, although appearing to form closed curves,
are interspersed by an infinite number of gaps, which allow particles
to pass through and become stuck in
\emph{ballistic channels}\protect\footnotemark{}\protect\footnotetext{Ballistic
channels exist inside the chaotic component of a mixed phase space at the boundary
with the regular regions. Motion inside a ballistic channel is characterised by
long periods of non-zero average velocity. In contrast, motion that takes place
inside the chaotic component, but not at the boundary of a regular region,
will usually have vanishingly small average velocity.} (similar to the turnstiles
idea from maps). It is these sticking episodes that allow for directed transport. The general phenomena
is sometimes called \emph{intermittency} \cite{Cvitanovic.2010.NBI}.

The observation of a non-vanishing current, as an average velocity in
coordinate-space, based on the chaotic ratchet effect as discussed
in \cite{Flach.2000.PRL, Denisov.2002.PREa, Denisov.2002.PhysicaD, Denisov.2002.PREb,Denisov.2006.epl} has even been extended
to chaotic ratchet acceleration expressed in terms of an averaged
velocity in momentum space \cite{Gong.2004.PRE}. In both cases the sum rule
derived in \cite{Schanz.2005.PRE} for the chaotic transport velocity in driven
one-dimensional systems assures the existence of a persistent chaotic
ratchet  current.

The case of particles in non-Hamiltonian systems has also attracted
considerable interest. While the Hamiltonian systems remain an active
area of research, some have focused on less idealised systems that are
dissipative (possibly including noise), and driven. Brownian motors extract
work from thermal fluctuations in out-of-equilibrium conditions. With
regard to transport under such thermal fluctuations, the `constructive
role of Brownian motion' is crucial \cite{hanggi.1996.lnp,reimann.2002.apa,astumian.2002.phystoday,hanggi.2009.romp}. A particular
type are Brownian motors which find applications in various fields such as physics, engieneering, chemistry and biological transport \cite{astumian.2002.phystoday,hanggi.2009.romp}, to name but a few. Motion in such
ratchet-like devices is confined to a periodic and asymmetric  landscape.
In the presence of  out-of-equilibrium conditions they are able to rectify thermal fluctuations.
Interestingly, under certain conditions, it is possible to derive analytical
solutions pertaining to transport of Brownian particles via the \emph{Gambler's Ruin}
model \cite{Cheng.2007.PRL}. For Brownian ratchets, the broken
spatial symmetry combined with external forces with time-correlations were
shown to be sufficient ingredients for transport \cite{MAGNASCO.1993.PRL,bartussek.1994.epl}.
The interdependence of the confining potential landscape and of the thermal
fluctuations, for the emergence of transport, has further been studied in
\cite{Malgaretti.2012.PRE}.

In general though, the dynamics of particle motion in periodic potentials
at finite temperatures is an extensively studied field \cite{Risken.1989.SprVer}.
In a similar domain, \cite{Hennig.2009.PRE} investigated the motions of driven,
under-damped Brownian particles, evolving in a \emph{washboard potential}
(a spatially symmetric and periodic potential),
under the influence of a time-delayed feedback term. They found that, at finite
temperatures, the time-delayed feedback term can in fact enhance the transport
features of the system such that there is an increase in the overall net motion
of the system, which is not observed when the feedback term is switched off.
These results were related to a desymmetrisation of the relevant attractors
supporting directed transport.

Transport of particles in potentials with multiple wells,
at its most fundamental level, is characterised by escape
processes over potential barriers as discussed in the first part of the review.

\subsubsection{Transport with two or more degrees-of-freedom}
Increasing the number of degrees-of-freedom to two (two and a half),
and by consequence the dimension of the phase-space to four (five),
by coupling two individual particles together adds further complexity
to the system's dynamics. For those studies which look at the transport
features of systems of coupled units, the objective is to investigate
the conditions which lead to said transport features. In particular,
for directed transport to occur, it is quite often the case that the
two particles will work {\it cooperatively} to achieve this directed transport.
Importantly, under the same conditions, a dimer (a compound made up of two
particles) has distinct transport properties compared with those of the
single particle \cite{Heinsalu.2008.PRE}.

Dimer systems have been shown to exhibit a complicated dependence, with
respect to observables of interest, on the coupling between the subsystems
making up the dimer. For example, the value of the net transport for a dimer
system can change erratically as the coupling parameter is varied \cite{Hennig.2009.PhysicaD}.
Similarly, \cite{Fugmann.2008.PhysicD} considered the conservative and deterministic escape
dynamics of two coupled particles out of a metastable potential. The scenario
considered is such that neither particle can escape independently (on energetic
grounds), and thus cooperation is required. It was shown that the escape times
of the dimer out of the metastable state become severely inhibited for coupling
strengths that are either too large or too small.
In \cite{Hennig.2008.PRE} it was demonstrated that
dynamical detrapping of a dimer from a potential well of a periodic potential followed
by unidirectional motion, i.e. a running solution,
depends critically on the coupling between the two subsystems.

Increasing the number of degrees-of-freedom further  renders the analysis of these
systems more cumbersome. Some of the already illusive phase-space structures
become difficult, if not impossible, to detect. However, in spite of this, it
is still possible to explore the transport properties in these higher dimensional
systems in a fashion similar to the studies of escape in the first part of this review.


\subsubsection{Anomalous transport}
For systems modelling the dynamics of particles evolving in periodic potentials,
the inclusion of driving and damping can produce some interesting and unexpected
behaviours. For such a particle, where the underlying potential is of the ratchet
type and the driving is of zero average, \cite{jung.1996.prl,Mateos.2000.PRL,Mateos.2001.APPB} observed a current
reversal via an increase of the driving amplitude -- that is, the current, going
in one direction, passes through zero and changes direction as the driving amplitude
is increased. A ratchet subjected to an unbiased external force that periodically
modulates the inclination of the potential, is called a \emph{rocking ratchet}.
Current reversals in such a system are unexpected due to the inherent bias contained
in the ratchet potential. In this situation, the current reversal, when it occurs,
has been shown to coincide with a bifurcation from chaotic to regular motion
\cite{jung.1996.prl,Mateos.2000.PRL}. In a similar study \cite{Mateos.2003.PhysicaA} related current reversals
to the basins of attraction for the system's
coexisting attractors that produce counter-propagating motion. This study
differs from the previous two in that a current reversal is obtained
through the appropriate selection of  (completely symmetric)  initial conditions, rather than
through the modification of a control parameter. These studies have been
extended to consider the case of two coupled driven and damped particles
evolving in a ratchet potential \cite{Vincent.2010.PRE}. Not surprisingly, the
addition of the second particle can have important consequences for the current
reversals observed in the case of the single particle. It appears that current
reversals exist for coupling strengths below some critical value. However, beyond
this critical coupling strength, the particles become synchronous -- the single
particle dynamics are restored -- and no further current reversals are observed.
Little explanation is given as to why no further current reversals appear. However,
from the \cite{Mateos.2000.PRL} study it is known that current reversals are restored, in
the case of the single
particle, for stronger driving amplitudes. In a separate study, \cite{Cubero.2010.PRE}
investigated the dynamics of a single particle evolving in a periodic and spatially
symmetric potential landscape. In such a potential, there is no inherent bias with
respect to transport. Thus, for a current to emerge at all, some symmetry of the
system needs to be broken (see  in
Sec.~\ref{subsubsection:symm_prop}). In \cite{machura.2010.chemphys,Cubero.2010.PRE} this was achieved  by using a bi-harmonic
driving term, resulting in a system that is driven out of equilibrium by an asymmetric
external force. The authors provide evidence that current reversals, in this situation,
are induced by the symmetry-breaking effect of the system's damping.

Other types of counter-intuitive transport can be collected under the term
\emph{negative mobility}. To motivate this, let us consider two examples.
The first, coined the \emph{Brazil nut effect}\cite{bug.1987.prl}, occurs when
granular media of different sizes are mixed. Applying a rocking force to the mixture
can cause the unexpected result that the larger granules rise \cite{Mobius.2001.Nature}.
Secondly, an effect known as \emph{induced demand} can help explain,
for example, the counter-intuitive rise in traffic when new roads are created
(explanations include people wanting shorter journey times, access to better roads, etc).
For particle transport,  a negative-valued mobility also plays an interesting role \cite{eichhorn.2002.prl,eichhorn.2002.pre,machura.2007.prl,kostur.2008.prb}. The work
in \cite{eichhorn.2005.chaos} gives an overview of this topic, together with
a list of possible applications. In particular, they describe two types
of motion of this kind; namely, \emph{absolute negative mobility} (ANM)
and \emph{differential negative mobility} (DNM). For illustration,
consider a simple system consisting of a single particle in
equilibrium, which has a spatially static homogeneous force $F$ applied. It
is generally expected that the response of this system to the bias force $F$
is in the direction of, and proportional to, this force. However, for systems
driven out of equilibrium (and possibly with the inclusion of noise), these new
types of motion can be observed. ANM refers to motion whose response, on average,
to the sufficiently {\it small} bias force $F$ around $F=0$ is in the opposite direction of $F$. DNM
on the other hand refers to motion that, although in the same direction as $F$,
slows down as the magnitude of $F$ increases. These ideas have been expanded upon
to further understand the response of these systems to noise \cite{kostur.2006.physicaA,machura.2007.prl,Speer.2007.EPL, Speer.2007.PRE,kostur.2008.prb}.

The case of coupled particles has also been considered. For under-damped particles
evolving in spatially periodic and symmetric potentials, subjected to periodic
driving and an additional static bias force, \cite{Mulhern.2011.PRE} related the occurrence
of negative mobility to a bifurcation from chaotic to regular motion. Further,
a heuristic description of the mechanism that allows for such motion is outlined.
In short, the particles must together work cooperatively in conjunction with the
periodic driving so that `downhill' motion is minimised, while `uphill' motion is
promoted. Thus, `uphill' motion ensues. The corresponding solutions remain
stable under low temperature fluctuations (see the discussion in
Sec.~\ref{subsubsection:negative}). In contrast, outside the observed
windows of negative mobility, it is the chaotic dynamics that emerge resulting
in motion that is in the same direction as the bias. A later investigation by
\cite{Speer.2012.PRE} can be considered as an extension of this study to the case of
over-damped particles. Again, ANM was observed. They were able to prove
that ANM is not possible for an over-damped dimer where the interaction
potential is convex. That is, for a system of two coupled particles in the
over-damped limit, subject to the forces discussed above, and with an
interaction potential $W(x)$ such that

\begin{equation}
W''(x) > 0 \hspace{0.89cm}\forall x,
\end{equation}

\noindent
the possibility of ANM is excluded.

\subsubsection[Transport in irregular domains]{Transport in irregular domains\sectionmark{Billiards}}

Extensions of the above concept address the case of autonomous Hamiltonian systems of
one-dimensional billiard
chains \cite{Acevedo.2003.POTPS, Schanz.2005.JPA}. The necessity of
creating chaos requires at least two degrees-of-freedom. As an example
for such a system, a classical magnetic billiard for particles carrying
an electric charge has been studied in \cite{Acevedo.2003.POTPS}. In order to break
the time-reversal invariance,
an external static magnetic field, penetrating the plane of motion perpendicularly,
has been
applied. In addition, achieving directed transport requires breaking of the remaining
spatial
symmetry which can be achieved, e.g. by properly placed asymmetric obstacles inside
the
billiard \cite{Acevedo.2003.POTPS, Schanz.2005.JPA}. Uni-directional motion in a serpent billiard
chain has been reported in \cite{Horvat.2004.JPA}.

\begin{figure}
\begin{center}
\includegraphics[height=4cm, width=4cm]{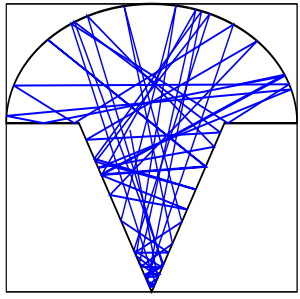}
\caption{(Colour online) Transport in an irregular domain: A \emph{mushroom billiard}, together with an example
trajectory. Source: Figure taken from Ref.~\cite{Mulhern.2012.Thesis} and the corresponding data was obtained from 
\cite{kantz.2007.web}.}
\label{fig:mushroom}
\end{center}
\end{figure}

The work in \cite{Bunimovich.2001.Chaos} introduced a novel class of billiard systems which the author called the
\emph{Mushroom Billiard}. For an example we refer to  Fig.~\ref{fig:mushroom}. Its
novelty comes from the fact that it has the remarkable property of having
a phase-space consisting of a single (regular) KAM-island and a single (chaotic)
ergodic region. Such billiards offer insight into the dynamics of Hamiltonian
system with a more complicated phase-space.
In fact \cite{Altmann.2005.Chaos, Altmann.2006.PRE} looked at the stickiness of chaotic trajectories
to the single KAM island, using recurrence time statistics, in mushroom billiards.
It was shown that the sticking episodes are facilitated by orbits known as
\emph{marginally unstable periodic orbits}. Marginally unstable refers to the
fact that perturbations grow linearly (rather than exponentially) in time.
These orbits, even though being of measure zero, govern the main dynamical
properties of the system. Most notably, they are responsible for a power-law
behaviour
observed in the recurrence time statistics -- something that is often related
to the partial barriers created by cantori in a mixed phase-space.

\subsubsection{Properties that determine transport features}\label{subsubsection:symm_prop}
A symmetry analysis of a system of equations can illuminate important
transport properties.
An important quantity related to transport is the time averaged, ensemble
averaged, momentum. This is typically called the \emph{current}. Let $p(t)$
represent the momentum of a particle at time $t$, then the current is given by

 \begin{align}
 J = \dfrac{1}{T_{s}}\int_{0}^{T_{s}} dt\Bigg(\dfrac{1}{N}\sum_{n=1}^{N}p_{n}(t)\Bigg),
 \label{eq:curr_gen}
 \end{align}

\noindent
with $N$ being the number of initial conditions in the ensemble, and
time $T_s$ taken in the asymptotic limit $T_s \rightarrow \infty$.
The direction and magnitude
of the current is inextricably linked with a system's symmetry properties.

To give an example, \cite{Flach.2000.PRL,denisov.2014.pr} considered the symmetry properties of a system
consisting of a particle evolving in a spatially periodic potential subjected to
driving and damping. The equation of motion is given by

\begin{equation}
\ddot{X} + \gamma \dot{X} + f(X) + E(t) = 0.
\label{eq:ddp}
\end{equation}

\noindent
Here $E(t)=E(t+T)$ is a time-periodic external field of period $T=2\pi/\omega$ and
frequency $\omega$, and $f(X) = f(X+2\pi)$ is a periodic potential function. Both
$E(t)$ and $f(X)$ are assumed to be bounded, and max$(|f(X)|) \sim 1$. The authors
defined system symmetries related to the properties of the underlying potential
and external field. These properties (shown in a modified form which allows for
greater applicability -- due to \cite{Lade.2010.Thesis}, for example) for a given function
$g(a)$ are

\begin{tabular}{ l l l }
  $g_s~:$ \quad $g(a+\tau) = g(-a+\tau)$ & \quad for some $\tau$ & \quad (symmetric) \\
  $g_a~:$ \quad $g(a+\tau) = -g(-a+\tau)$ & \quad for some $\tau$ & \quad (anti-symmetric) \\
  $g_{sh}:$ \quad $g(a) = -g(-a+\tau)$ & \quad for some $\tau$ & \quad (shift-symmetric) \\
\end{tabular}

\noindent
where $g(a)$ can represent either a spatial or temporal function, i.e.
the potential or the time-dependent external field, respectively. If $f(X)$
is anti-symmetric, and $E(X)$ shift-symmetric ($f_a$ and $E_{sh}$),
then Eq.~(\ref{eq:ddp}) is invariant under the symmetry $\hat{S}_a: X \mapsto (-X+2\chi)$,
$t \mapsto t + T/2$ for some appropriate argument shift. In the dissipationless case,
$\gamma = 0$, a second symmetry can be obtained. If $E(t)$ possesses the
shift-symmetry $E_{sh}$ then Eq.~(\ref{eq:ddp}) is invariant under the
symmetry $\hat{S}_b: t \mapsto -t+2\phi$, again for some appropriate argument shift.

It then follows that for a given trajectory $X(t;t_0,X_0,P_0)$, $P(t;t_0,X_0,P_0)$
with initial condition $t_0, X_0, P_0$, it is possible to generate new trajectories
given by

\noindent
\begin{tabular}{l l}
$\hat{S}_a: -X(t+T/2; t_0, X_0, P_0) + 2\chi, -P(t+T/2; t_0, X_0, P_0)$ & \{$\hat{f}_a, \hat{E}_{sh}$\},\\
\vspace{0.2cm}
$\hat{S}_b: X(-t + 2\phi; t_0, X_0, P_0), -P(-t+2\phi; t_0, X_0, P_0)$ & \{$\hat{E}_s, \gamma=0$\}.
\end{tabular}

\noindent
Importantly, these transformations change the sign of $P$. This has the consequence
that the original trajectory, and the corresponding trajectory generated through
the symmetry transformation yield time-averaged values of $P$ that differ only by sign.
Going further, for a system with $\hat{S}_a$ or $\hat{S}_b$ symmetry, the net current will be zero
as each trajectory will have a counterpart that negates the others contribution to the
current. The implication being that in order to generate a non-zero current,
both the symmetries $\hat{S}_a$ and $\hat{S}_b$ need to be broken. Note that $\hat{S}_a$ holds
in both the dissipation and the dissipationless cases, whereas $\hat{S}_b$ holds only
for $\gamma=0$. The work \cite{Denisov.2002.PREa} identified an additional symmetry, $\hat{S}_c$,
of Eq.~(\ref{eq:ddp}), this time in the over-damped case where inertial effects
become negligible, namely:

\begin{tabular}{ l l}
$\hat{S}_c: X(-t; t_0, X_0, P_0) + \chi/2, -P(-t; t_0, X_0, P_0)$ & \{$\hat{f}_{sh}\;,
\hat{E}_{a}, m=0$\}.
\end{tabular}

\noindent
It is worth noting that the symmetries $\hat{S}_a, \hat{S}_b$ and $\hat{S}_c$
require that the time-dependent external field satisfies certain properties.
Thus, an appropriate choice of $E(t)$ can be sufficient to break all three
symmetries.

The authors in Ref. \cite{Yevtushenko.2000.PRE} investigated the symmetry properties of the Hamiltonian
version of Eq.~(\ref{eq:ddp}) ($\gamma=0$), where the underlying potential is
spatially periodic and symmetric. A lowering of the dynamical symmetry, controlled
by the phase of the external field, leads to a directed current.


\subsubsection{Ballistic transport}

Others have focused on the dynamical mechanisms that allow for a directed
current in a mixed phase-space. While the appearance of a dc-output can be
expected using symmetry analysis, its appearance and magnitude are due to
dynamical mechanisms of motion inside the stochastic layer. The paper \cite{Denisov.2001.PRE}
looked at the structures in phase-space and considered how they influence
the magnitude and direction of current. To ensure that the appropriate
symmetries were broken, thus allowing for a directed current, they chose
the external field

\begin{equation}
E(t) = E_1\cos(t) + E_2\cos(2t + \phi),
\end{equation}

\noindent
where $E_2\neq 0$ and $\phi\neq 0,\pi$. The phase-space of this system is
characterised by a stochastic layer which emanates from the separatrix of
the unperturbed system ($E_1=E_2=0$). Inside the stochastic layer there exists
a hierarchical structure of resonance islands that are responsible for the
creation of ballistic channels in phase-space. That is, the resonance islands
form partial barriers such that when a particle enters a ballistic channel
it may be stuck there for large durations, thus contributing to an overall
non-zero net current. The authors relate
the emergence of a directed current to a desymmetrisation of the ballistic
channels bringing the particles in opposite directions. Going further, they
analytically derive an expression for the current from the geometry of the phase-space.
In particular, each resonance island has associated to it a winding number $\omega_i$, a
probability of `sticking' to the resonance island $\rho_i$, and mean sticking time
$\langle t_i\rangle$. In addition the mean time between sticking episodes is
$\langle t_r\rangle$. The authors in \cite{Denisov.2001.PRE} define the current as

\begin{equation}
J = \displaystyle\sum\limits_{i=1}^N \omega_i
\rho_i \langle t_i\rangle \; /\Big(\displaystyle\sum\limits_{i=1}^N \rho_i
\langle t_i\rangle + \langle t_r\rangle \Big)
\end{equation}

\noindent
where $N$ is the number of resonance islands. This definition of the
current suffers two limitations. Firstly, the four unknowns in the equation
will, in general, need to be computed numerically. The second is that there may
be resonances of all orders. To even locate resonances of increasing order becomes
computationally impractical. However, this does not pose much of a problem as
it is only a few resonance islands that are relevant for obtaining the net
current. Higher order resonances have sticking times that are close to zero
and therefore their contribution to the net current is negligible.


\subsubsection{Beneficial role of chaos}
All of the studies discussed above look at systems with at least (effectively)
three variables. Thus the dynamics of these systems then typically display a  chaotic dynamics. Although
chaotic motion seems to be inherently counter-productive with respect to
net directed  motion  it can, for example, allow trajectories
to visit (transporting) ballistic channels associated to resonance islands with
non-zero winding numbers \cite{Schanz.2005.PRE}. However, such ballistic channels will
only exist in non-hyperbolic systems: systems that contain mixed regular and chaotic
regions. These chaotic regions are born out of
(nonintegrable) perturbations to an integrable system, with the strength of perturbations
determining the prevalence of chaos. This is true in general; i.e. nonintegrable
perturbations to an underlying integrable system are the source of chaos. The systems under investigation
in the following exhibit several forms of chaotic motion, including transient chaos,
permanent chaos and motion on strange attractors. In particular, transient chaos can be beneficial for
current rectification.

\subsubsection{Cooperative transport mechanism}

This section explores the deterministic dynamics
of systems of several {\it coupled} units. In particular, the focus
of the review is concerned with the transport features which this complexity generates.
Of interest is how particles work together, cooperatively,
to achieve directed transport. For this reason, the strength of the coupling
between the particles serves as the main control parameter. Further, ensemble
dynamics serve to highlight some of the collective effects of these systems.

The section is split into two parts: The first part  looks at a class of time-independent (autonomous)
Hamiltonian systems, while the second part  considers a class of time-dependent driven (non-autonomous) and damped systems.
A common feature of these systems is that they contain a spatially open component that
facilitates long range transport. More precisely, transport proceeds in a spatially symmetric
and periodic multiple well potential. Thus transport is characterised by particles overcoming
successive energetic barriers created by the potential landscape.

The cooperative effects between the particles becomes apparent in Sec.~\ref{subsection:autonomous}
when the
autonomous Hamiltonian systems are considered. In the uncoupled limit the low dimensional systems decompose
typically into two integrable subsystems and the dynamics are fully understood. However, the dynamics become
more complicated when the particles are coupled. As these systems are conservative a coordinated
energy exchange between the particles is often required for directed transport to ensue. Interestingly enough,
these systems contrast well with the non-autonomous one-and-a-half degree-of-freedom Hamiltonian systems,
where transport occurs through intermittent periods of directed motion in so-called ballistic channels.
The autonomous two degree-of-freedom counterpart considered here relies on a rather different mechanism
for directed transport that is provided solely by regular structures in phase-space.

With the inclusion of external driving and damping (Sec.~\ref{subsection:non-autonomous})
the transport dynamics are
controlled by various coexisting attractors in phase-space. The nature and stability of these
attractors is determined by the system parameters. As before, cooperative effects  play a
key role when it comes to particle transport. Notably, the coupling between
the particles can result in a suppression of chaos that allows, for example, for collective
periodic motion of rotational type.

\subsection{Transient-chaos induced transport in autonomous systems}\label{subsection:autonomous}

\noindent
In this part we review the works on autonomous Hamiltonian systems modelling two coupled units.
The guiding aim is to understand the conditions under which directed transport in phase-space is supported.
In particular, analytical and numerical results are produced illustrating the effects
that the interaction between the two units has on the direction and velocity of transport.

The focus of this section is on the deterministic transport properties of
systems of two coupled particles or units. In particular, much attention
is given to the relationship between various transport
scenarios and the coupling between the units. Notably,
the nature of the interaction between the units that allows
for directed transport (on average) is scrutinised.

The work presented in the first part of this section serves as a bridge between single unit systems
and many unit systems. Importantly, two unit systems are the first
non-trivial step from the lower of these limits, bringing with it new types
of motion, such as hyperchaos \cite{rossler.1979.pla}, that have important implications when it comes
to transport. More than that, the work presented
in this section explores novel mechanisms pertaining to directed transport
that are a direct consequence of the interaction between the units.

In Sec.~\ref{subsubsection:modelone} the first example treated is a spatially symmetric
system containing two open components allowing that either unit can undergo directed
transport. By way of symmetry analysis, it is clear that if a constant energy surface
is entirely populated by initial conditions then no current can emerge. However, this
may not be the case for other, more physically relevant, sets of initial conditions.
For one such set, some qualitatively different transport scenarios are outlined and
their relation to the net current described. As a general point for the class of systems
discussed in Sec.~\ref{subsection:autonomous}, the mechanism promoting directed
transport in these
systems is quite distinct from systems where transport proceeds over finite periods
in so-called \emph{ballistic channels}, each period being separated by an interval
of chaotic motion. The novel mechanism for transport presented here, where chaos is
required only in a transient period of the dynamics (after which transport is provided
solely by regular structures), is illustrated and the implications discussed.

The Hamiltonian systems discussed here are of the form:

\begin{equation}
H(\mathbf{p},\mathbf{q}) = \dfrac{p^2}{2} + \dfrac{P^2}{2} + U(q) + V(Q) + H_{\mathrm{int}}(q,Q)\;,
\label{eq:Ham_Gen}
\end{equation}

\noindent
where $\mathbf{p}~(=(p,P))\in \mathbb{R}^2$, $\mathbf{q}~(=(q, Q))\in \mathbb{R}^2$
are the canonically conjugated positions and momenta of coupled units.
Further, we  assume from now on that these units are of unit mass.
The units evolve in a potential given by
$U_\mathrm{eff}(\mathbf{q}) = U(q)+V(Q)+H_{\mathrm{int}}(q,Q)$,
where $U(q)$ and $V(Q)$ are positive semi-definite functions, and
in addition, are coupled via an interaction potential $H_{\mathrm{int}}(q,Q)$.
It may be the case that $U$ and $V$ describe the same potential
landscape. However, to keep the results as general as possible,
we consider both cases, i.e. when the potential landscapes are the
same, and also when they differ. Crucially, a prerequisite for the
occurrence of transport is that these systems contain an open component.
That is, on surfaces of constant energy the system must be unbounded in
at least one of the spatial coordinates, thus allowing for the
possibility of unbounded and directed transport. Therefore, we
assume that all systems explored here contain an open component.
The equations of motion, corresponding to this class of Hamiltonian
systems, are given by

\begin{equation}
\ddot{q} = -\dfrac{\partial U_\mathrm{eff}(\mathbf{q})}{\partial q}
\qquad
\&
\qquad
\ddot{Q} = -\dfrac{\partial U_\mathrm{eff}(\mathbf{q})}{\partial Q}.
\label{eq:eom_gen}
\end{equation}

Before moving onto the first subsection in this part, it is worthwhile discussing a particular potential
landscape that is used in all of the forthcoming sections.
This potential, often called the \emph{washboard} potential, is periodic
(of period $1$) and spatially symmetric. It is described by the equation

\begin{equation}
U(q) = U(q+1) = \dfrac{1-\cos(2\pi q)}{2\pi}.
\label{eq:washboard}
\end{equation}

\noindent
$U(q)$ has minima at $q_{min}=n$, with $U(q_{min}) = 0$, and maxima at $q_{max}=n+0.5$,
with $U(q_{max}) = 1/\pi$ ($\approx 0.318$), where $n\in \mathbb{Z}$.
As mentioned, the potential is
spatially symmetric, i.e. $U(q) = U(-q)$.

For now let us suppose that a single unit is evolving in a washboard
potential with no external forces present. The occurrence of transport can then be viewed as a
string of consecutive escape
processes where the unit overcomes the potential barriers located at $q_{max}^{n+0.5}$
($n\in \mathbb{Z}$) with increasing $\lvert n\rvert$. The only requirement
for directed transport
is that the system possesses a sufficient amount of energy so that the
unit can overcome these barriers.

An analogous statement regarding transport can be made for the case of two coupled units.
However, directed transport in this case may require not just a sufficient amount of system
energy, but also a coherent energy exchange between the units.
This is elaborated
upon in the next section. To conclude this section an interaction
potential used in the
coming section is presented and its properties briefly discussed.
The interaction
potential is of the form

\begin{equation}
H_{\mathrm{int}}(q,Q) = D\left[1 - \dfrac{1}{\cosh(q - Q)}\right],
\label{eq:H_int}
\end{equation}

\noindent
which is dependent on the distance $d = |q - Q|$. The strength of
this coupling
is regulated by the parameter $D$. Like the washboard potential, the interaction
potential is also spatially symmetric --- $H_{\mathrm{int}}(q,Q) = H_{\mathrm{int}}(-q,-Q)$.
It is important to note that the gradient $dH_{\mathrm{int}}(x)/dx$
goes to zero asymptotically; i.e. as the relative distance $|q - Q|$ increases,
the related interaction forces, $\partial H_{\mathrm{int}}/\partial q$
and $\partial H_{\mathrm{int}}/ \partial Q$, vanish asymptotically, allowing
for transient chaos~\cite{BLEHER.1990.PhysicaD, Contopoulos.1993.PhysicaD, Zaslavsky.1985.HarwoodNY, Zaslavsky.1998.ImpColPress}.
That is, for large distance $|q-Q|\gg 1$, the interaction vanishes with the
result that the two degrees-of-freedom decouple, rendering the dynamics regular.
This is crucial for what is presented in the coming sections.

Transport in autonomous Hamiltonian systems is remarkable due to the fact that
a system requires no additional input of energy, in the form of an external
time-dependent drive for example, for said transport to occur. Rather an
internal energy distribution must take place before  transport can
take place. This effect is even more remarkable in systems of coupled units
when the system's energy does not suffice to allow that both units can undergo rotational
motion\protect\footnotemark{}\protect\footnotetext{In this
review \emph{rotational motion}  refers to motion where
a unit(s) overcomes consecutive potential barriers in a periodic fashion.} at the same time. In this case the subsystems, related
to
each unit, must work cooperatively to achieve transport.
We next show  that the strength of the
coupling between the subsystems is crucial to the resulting dynamics.
More than that, some general features of
systems of coupled units are exposed.


\subsubsection[Symmetric Model]{Case of a spatially symmetric model}
\label{subsubsection:modelone}

Let us  introduce the model that is used in this section.
This model is minimal in two respects. The first is that $U$ and $V$ describe
the same potential landscape -- the so-called \emph{washboard potential}. The
second is that this system contains only one parameter, namely the parameter
that regulates the strength of the bond between the two units. Thus,
there are two units evolving in the washboard potential (Eq.~(\ref{eq:washboard}))
that are coupled via the interaction potential (Eq.~(\ref{eq:H_int})).

The equations of motion are given by \cite{Mulhern.2011.PREb}

\begin{align}
 \ddot{q} = -\sin(2\pi q) - D\left[ \dfrac{\tanh(q - Q)} {\cosh(q - Q)}\right],
\end{align}

\begin{align}
 \ddot{Q} = -\sin(2\pi Q) + D\left[ \dfrac{\tanh(q - Q)} {\cosh(q - Q)}\right].
\end{align}

\noindent
For the numerics in the coming sections the initial set-up is as follows.
The two units are separated by a sufficient distance such that
they are effectively uncoupled, i.e. the energy contained in the
interaction potential saturates: $E_{int}\approx D$. One unit
is situated at the origin, while the other unit is
in a potential well that is sufficiently far from the origin, so
that the effect of one unit on the other is almost negligible.
Further, the unit at the origin is at rest. The additional
unit is given an initial velocity that sees it move towards
the unit at rest (these units are henceforth be called
unit $A$, associated with the variables $(p,q)$, and unit
$B$, associated with the variables $(P,Q)$, respectively).
Of course, the energy supplied to unit $A$ must be greater
than that required to overcome potential barriers of the washboard
potential -- that is $E_A(0)>E_b=1/\pi$, where $E_A(0)$ is the
energy possessed by unit
$A$ at time $t=0$ and $E_b$ is the barrier height of the washboard
potential. Thus, as the relative distance $|q - Q|$ decreases, the
energy exchange becomes more pronounced (depending on the value $D$)
and the system dynamics become more complex.

For $D \neq 0$ the units can interact via the interaction potential
and exchange energy. This exchange excites the additional
(initially resting) unit and, to varying degrees, influences
the motion of the unit that has entered the interaction region.
Again, it is important to note that both components of this system
are open and thus it is feasible that either unit  escapes.
For large $|q - Q| \gg 1$ the interaction between the units
vanishes, and again the dynamics is represented by regular
rotational motion (assuming sufficient system energy). For
the systems considered here, the energy is kept
sufficiently low such that the possibility of both
units escaping independently is excluded,
and the cooperative effects between the units come to the fore.

As mentioned earlier, the initial conditions for unit
$B$ are $Q = P = 0$. The unit $A$ starts as a virtually
free unit in the {\it asymptotic} region, i.e. it approaches
the interaction region from a far distance. The initial amount
of energy $E=0.9$ lies above the highest possible energy of
the saddle-centre points but, for not too low coupling, below
almost all of the saddle-saddle points of the effective potential
(see \cite{Hennig.2011.Chaos}). The initial positions
of the units $A$ are contained within the well whose minimum is
located at $q\simeq -25$ and the corresponding initial momenta
are determined as those points populating, densely and uniformly,
the level curve

\begin{align}
 E = \dfrac{1}{2}p^2 + U(q) + H_{\mathrm{int}}(q, 0),\label{eq:Elevel}
\end{align}

\noindent
in the $(q, p)$-plane. Asymptotically, the interaction potential
attains a value approaching $D$. Therefore, as the units begin
in the asymptotic region and as the initial conditions depend
explicitly on $D$, no two sets of initial conditions are
the same.
The energy is fixed at $E = 0.9$,
which is less than three times the barrier height of the
washboard potential, $E_{b} = 1/\pi \approx 0.3183$. It
should be emphasised that for unit $B$ to escape, it
must gain a sufficient amount of energy from its interaction
with unit $A$. With no interaction this system
contains a strong positive current, as unit $A$ can escape
to infinity feeling no effect from unit $B$. It is worth
adding that for these initial conditions with $D\neq 0$ the
problem becomes a unit scattering problem with the stationary
unit playing the role of the scatter.

We illustrate some of the
qualitatively different transport scenarios that are present in this
system by varying the strength of the coupling parameter $D$.
Before this however, it is useful to present a table of
$D$ values that are used in this section along with
their respective currents. Unit current is assessed quantitatively
by the mean momentum, which is defined by taking the averaged momentum
of an ensemble of units, i.e.

 \begin{align}
 \bar{p} = \dfrac{1}{T_{s}}\int_{0}^{T_{s}} dt\langle p(t)\rangle,
 \label{eq:current1}
 \end{align}

\noindent
where $T_{s}$ is the simulation time, and the ensemble average is given by

 \begin{align}
 \langle p(t)\rangle = \dfrac{1}{N}\sum_{n=1}^{N}\sum_{i=1}^{2}p_{i,n}(t),
 \end{align}

\noindent
 with $N$ being the number of initial conditions. The current, along
 with details of the calculation, is discussed in detail below. Below is a table of representative coupling
 strengths with their respective current values.

\begin{center}
\begin{tabular}{|c|c|}
\hline
\textbf{\textit{D}} & $\bar{\textbf{\emph{p}}}$\\
\hline
0.3 & 0.925\\
0.5613 & -0.239\\
0.5617 & 0.262\\
0.5672 & 0.009\\
0.58169 & -0.0001\\
\hline
\end{tabular}
\end{center}

\begin{figure}[!t]
\includegraphics[height=4.8cm, width=4.8cm]{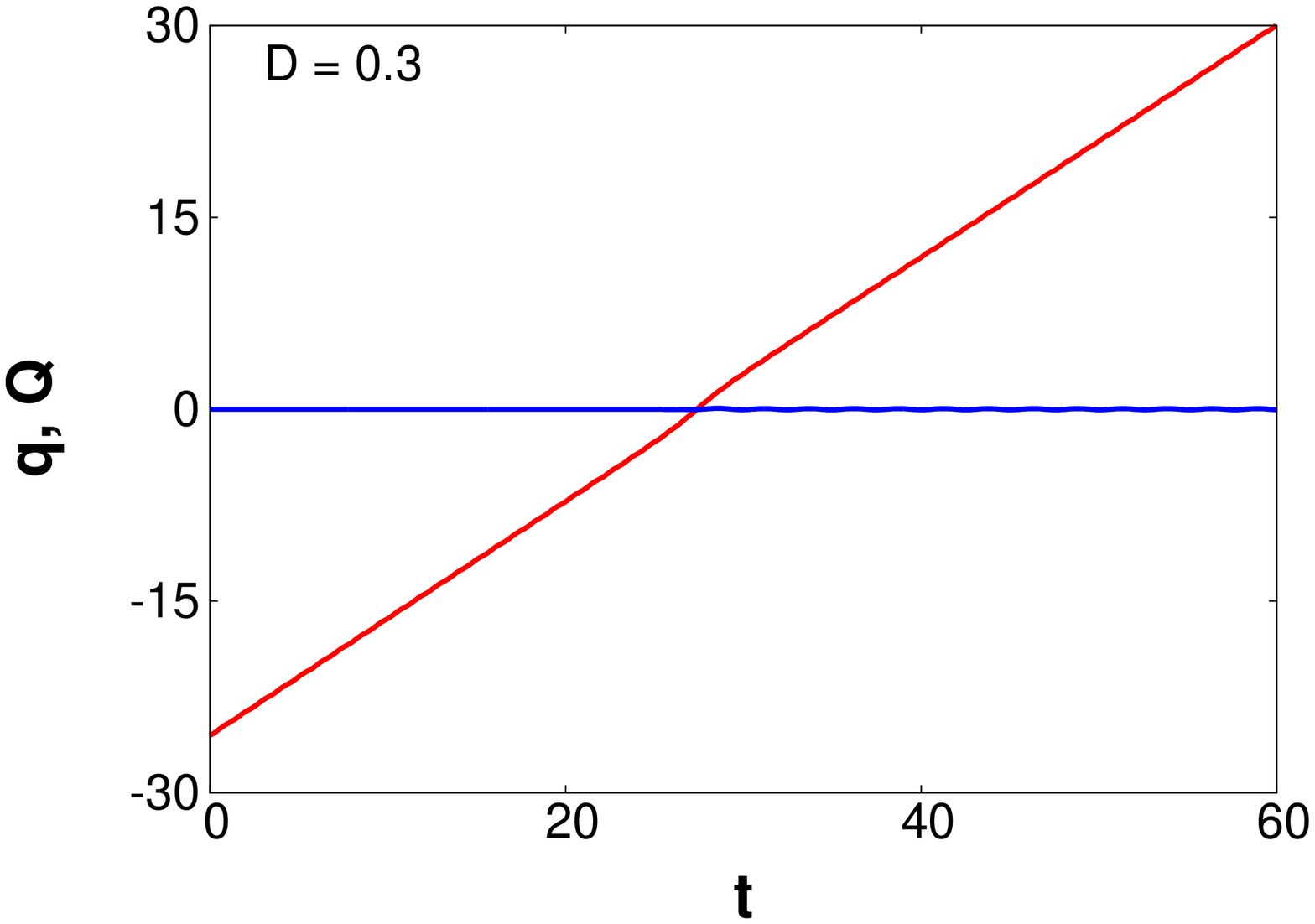} (a)
\includegraphics[height=4.8cm, width=4.8cm]{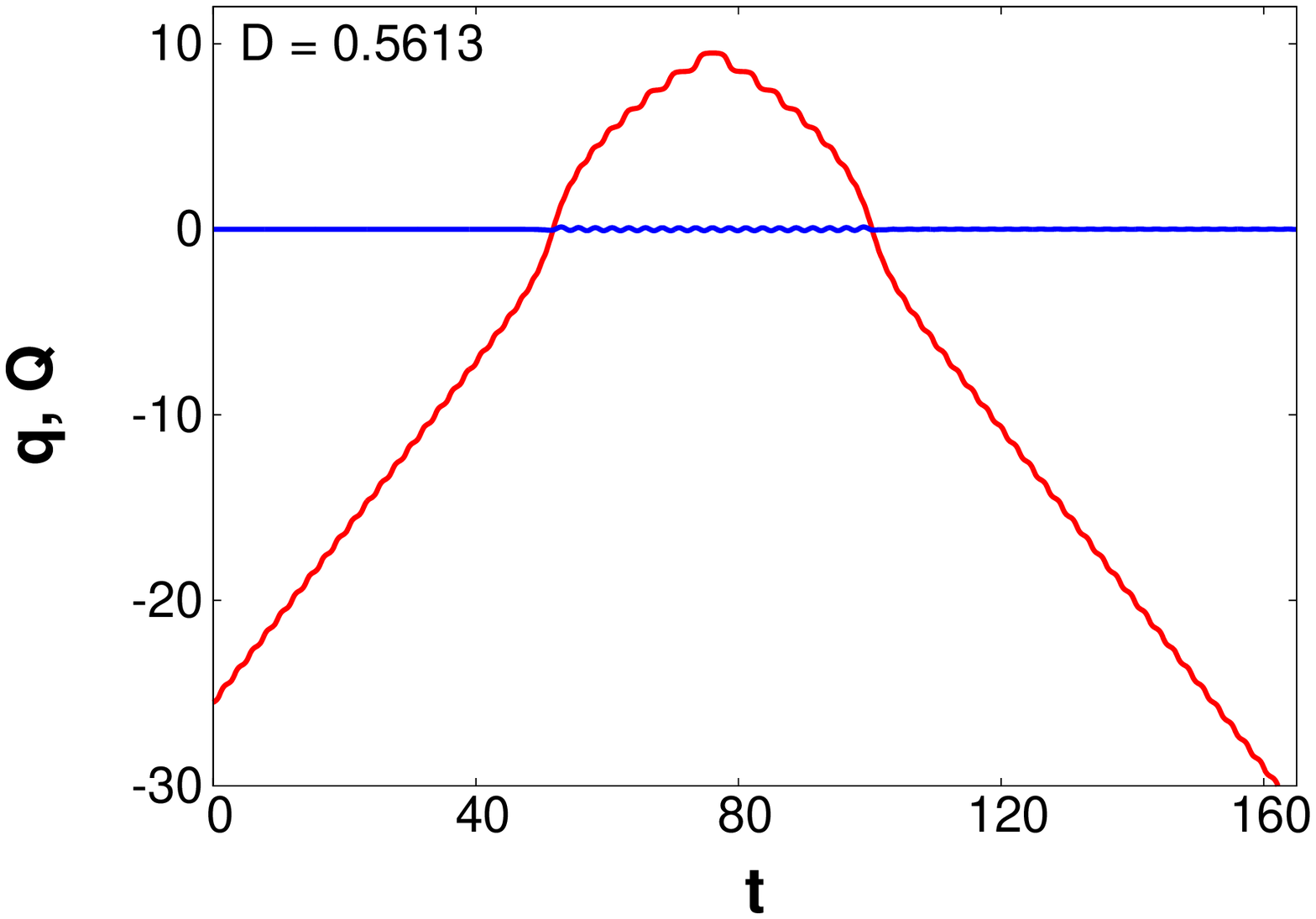} (b)
\includegraphics[height=4.8cm, width=4.8cm]{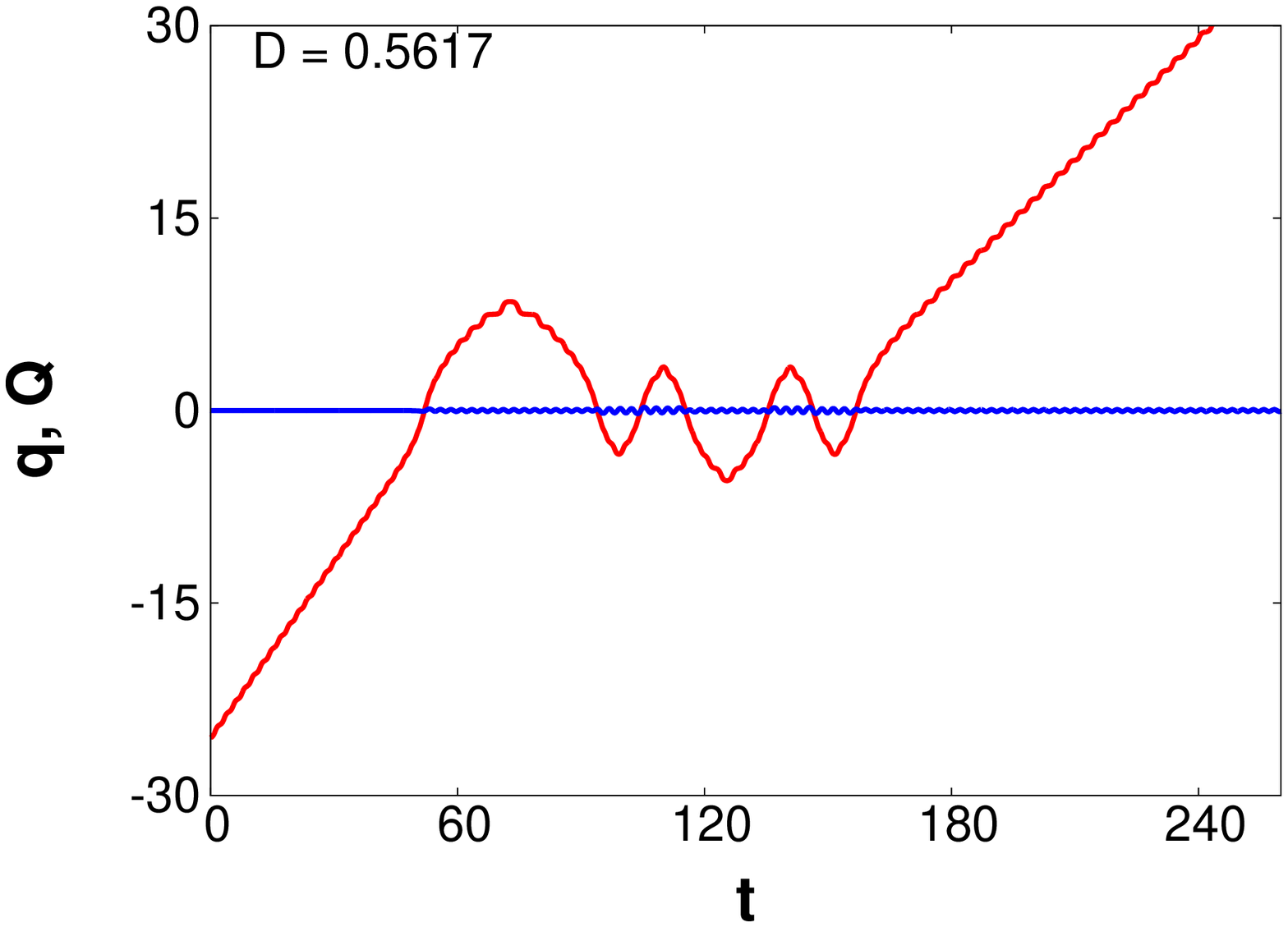} (c)
\includegraphics[height=4.8cm, width=4.8cm]{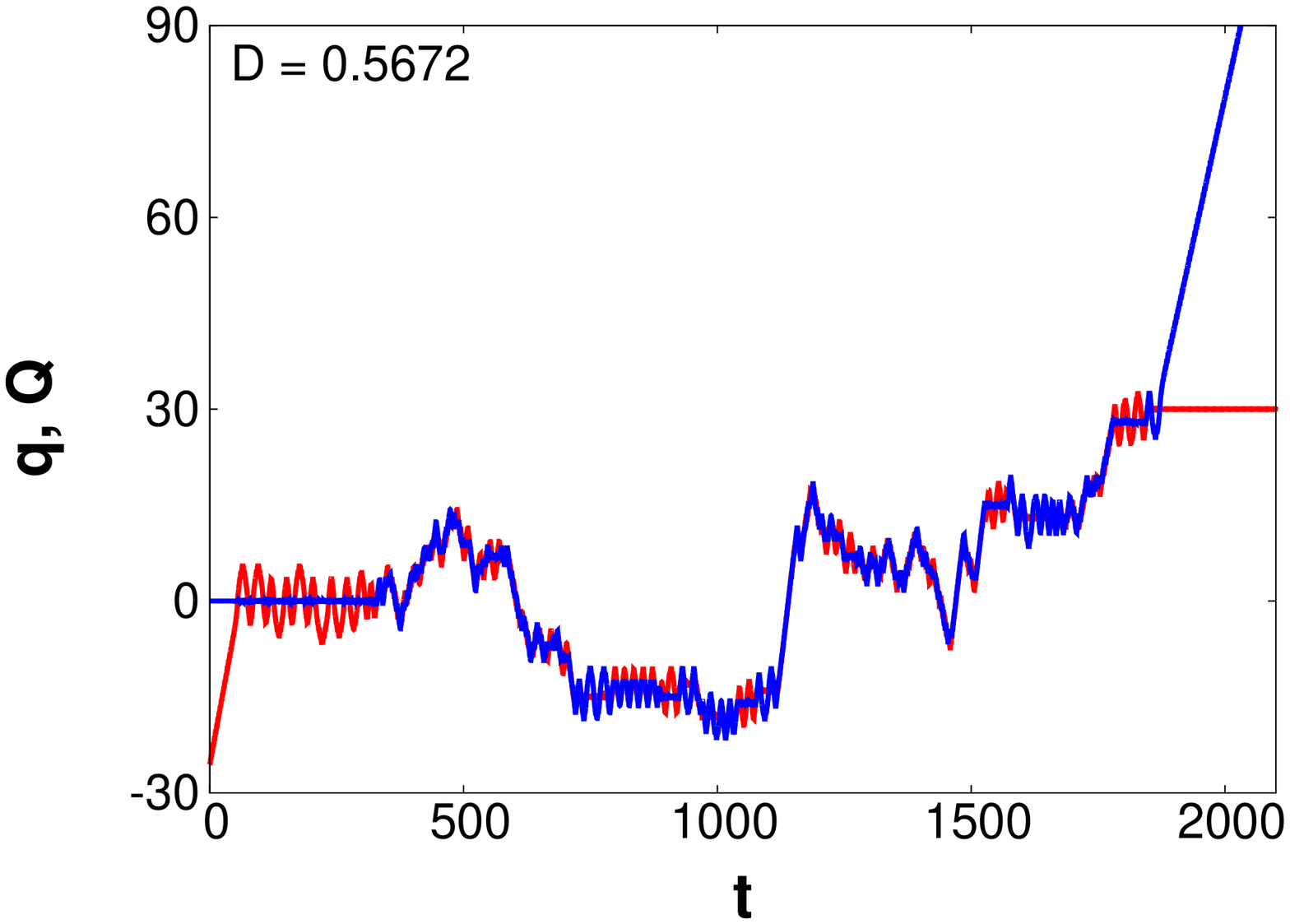} (d)
\includegraphics[height=4.8cm, width=4.8cm]{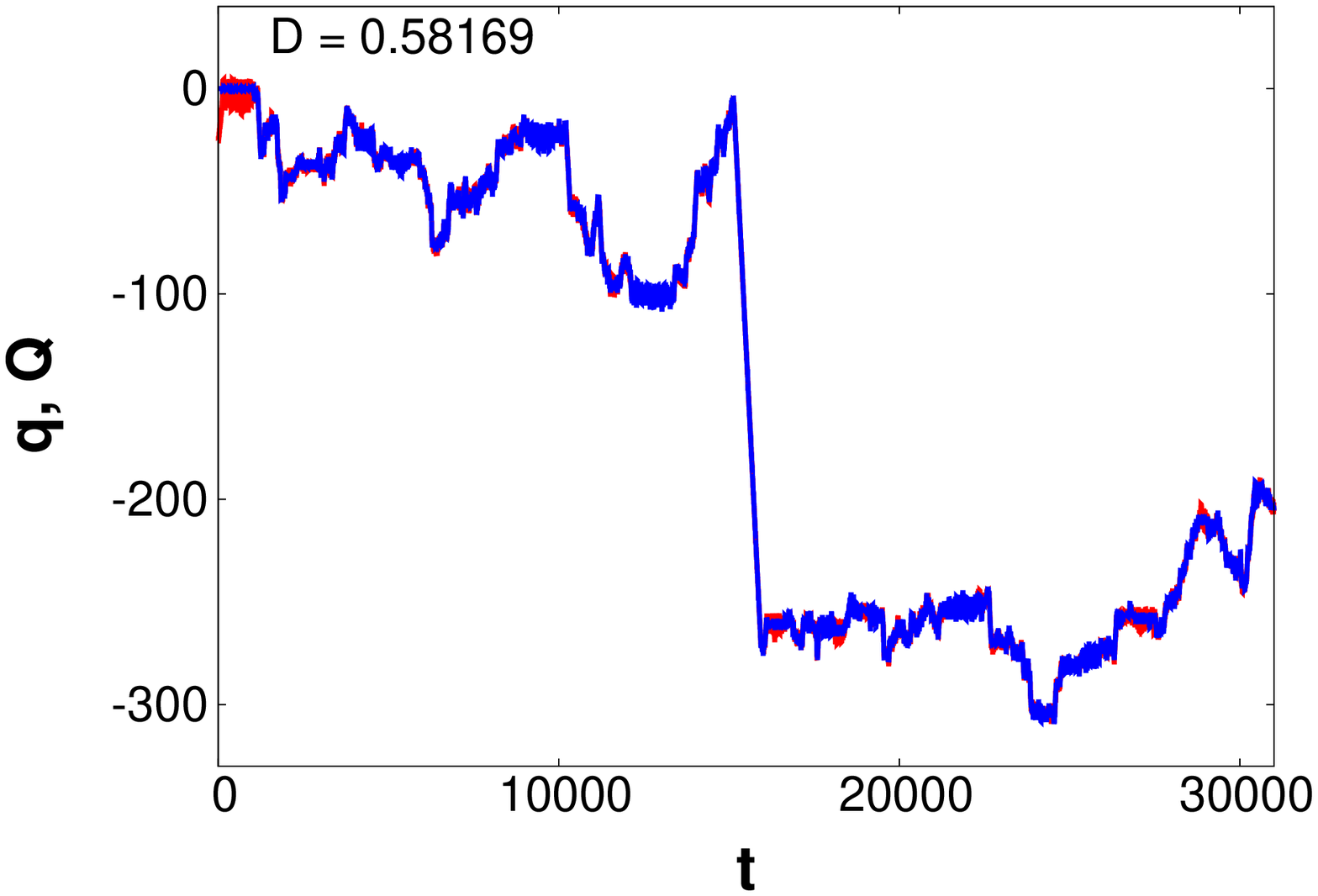} (e)
\caption{(Colour Online) Example trajectories for an autonomous system consisting of two nonlinearly coupled units, both evolving in a
periodic and
symmetric potential. Each panel differs by the strength of the parameter $D$ regulating the strength of the coupling between the
units.
The red line shows the temporal evolution of Unit $A$, which has an initial condition $q(0) = -25.5$ and $p(0)$
as obtained from Eq.~(\ref{eq:Elevel}), while the
blue line shows the time evolution of unit $B$ with initial
condition for each trajectory chosen as $Q(0)=P(0) = 0$. Note the different time-scales. Source: Figure adapted from
Ref.~\cite{Mulhern.2011.PREb}.} \label{fig:traj}
\end{figure}

Fig.~\ref{fig:traj} contains plots showing the temporal evolution of
the coordinates $q$, $Q$ for the five $D$ values contained in the table.
For comparison, for each $D$ value, the initial positions of the pair of
units are the same, i.e. with $q(0) = -25.5$ and $Q(0) = 0$, and
the initial momentum $p(0)$ of unit $A$ follows from the relation
given in Eq.~(\ref{eq:Elevel}), while unit $B$ has zero momentum,
$P(0)=0$. Slightly altering these initial conditions can have a large
impact on the path that the units take, as for a large range
of the coupling strength the dynamics is chaotic.

The $D$ values in the table above have been chosen as they represent,
in addition to typical system dynamics, transport scenarios with
varying contributions to the net current. With $D = 0.3$
(Fig.~\ref{fig:traj}~a) ) we see that unit
$A$ is able to pass straight through the interaction region
almost unscathed. Unit $B$ does receive some energy from
the interaction, but this energy only allows for small oscillations
about its starting position. This set-up favours a strong, positive
current. With regard to unit $B$ leaving its initial potential
well, there appears a blow-up at $D \approx 0.562$, after which
we can expect both units to travel multiple potential wells
together. As can be seen in Fig.s~\ref{fig:traj}~b) \& c),
both with $D < 0.562$, unit $B$ can largely influence the path
of unit $A$ without actually leaving its starting potential well.
Setting $D$ to $0.5613$ (Fig.~\ref{fig:traj}~b) ) we see
that the dynamics of the system is quite
different. The interaction between the units is such that
unit $A$ can pass through the interaction region (to a certain extent)
and subsequently be pulled back, escaping in the negative $q$ direction
and thus contributing to \emph{current reversal}. Again unit $B$
receives little energy from the interaction.
 A similar phenomenon can be seen for
$D = 0.5617$ (Fig.~\ref{fig:traj}~c) ). This time
unit $A$ oscillates around $q = 0$ a number of times before
escaping in the positive $q$ direction maintaining the original
direction of the current. 

Some of the most interesting behaviour
observed in this system can be seen in the remaining two figures.
Figs.~\ref{fig:traj}~d) show a trajectory with $D = 0.5672$.
There are number of striking things that can be noted about this trajectory.
Firstly, the duration of time that the trajectories `stick' together
before one escapes. In this case unit $B$ escapes in the positive
$q$ direction.
This is substantially longer than the escape times presented in the
previous figures. Also, both units take excursions to the left
and right before the escape of unit $B$. However, the most
notable thing about this figure is that \emph{it is unit $B$
that escapes, not unit $A$} as for the previous $D$ values.
Thus, unit $B$ is able to gain enough energy to escape from
its starting potential well, and subsequently from any force that
it feels from unit $A$. Unit $A$ has sacrificed its energy
and has become trapped. This situation describes an
\emph{interchange of the roles} played by the units,
with the initially free unit becoming trapped and
the initially trapped unit becoming free. 

The final
figure (Fig.~\ref{fig:traj}~e) ),
with $D = 0.56169$, show similar behaviour in that the
units seem to `stick' together. However, neither unit
escapes, but instead are, in some sense, stuck to each other
for the duration of the simulation. This is a process known
as \emph{dimerisation}, where the units, each initially acting as
a monomer, form a bound unit. This process is evident in some of the
previous figures, however in this case, the process is permanent.
Both units undergo large excursions, closely following the
line $q = Q$. For
this particular $D$ value, the units are in a continual
and most importantly, a substantial energy exchange. This allows
the units to travel together in an erratic fashion undergoing
multiple changes of direction and visiting multiple potential wells.
Although an independent escape for one of the units remains a
statistical possibility, it requires an optimal energy fluctuation
that sees one unit sacrifice all of its energy to the other.
This is highly unlikely given the fairly strong coupling between
the units.

This symmetric model, which consists
of two washboard potentials that are coupled via an interaction potential,
proves to be a rich source for complex dynamics which produces numerous interesting results.
In fact, a number of transport scenarios
are possible. The units can undergo independent motions: either, a single
unit travels through the potential landscape (the other remaining trapped),
or  both units are transporting. Alternatively, the units can travel
through this landscape in close proximity continuously exchanging energy
(this includes the possibility of complete synchronisation where there is
no energy exchange). The coupling strength determines which of these
scenarios are possible. Furthermore, with fine tuning of the system parameter
which regulates coupling strength,
unit scattering leading to the emergence of a non-zero net current,
or bond formation (dimerisation) yielding a zero net current, are possible
outcomes in this model.

A novel aspect of these autonomous Hamiltonian systems is that directed transport,
when it occurs, is regular and permanent. Chaos is needed only
in an initial phase of the dynamics to guide trajectories beyond
separatrices into the range of unbounded motion. This contrasts with the
transport observed in non-autonomous Hamiltonian systems where there is
a mixed phase-space \cite{Denisov.2001.PRE}. In these non-autonomous systems, finite bursts
of almost regular transport are separated by periods of chaotic motion. Thus the autonomous
systems, where directed transport is provided solely by regular motion, appear
to be favourable with respect to directed transport. However, it should also
be emphasized that the transport observed in this autonomous case is predominantly 
a cooperative effect which uses a favourable energy exchange between subsystems.

An interesting observation is that according to the theory of time-reversal symmetry,
this symmetric model should produce a zero net current for all values of $D$. The explanation of
why this case, and systems alike, can still produce a directed current
is the subject of Sec.~\ref{subsubsection:model2}. This section looks at time-reversal
symmetry for the class of systems whose equations of motion are given by
Eq.~(\ref{eq:eom_gen}) (of which our model is a member). It is explained
in detail how this symmetry is broken in practice, thus allowing for the emergence
of a non-vanishing net current. Further, a second model is introduced below that
is similar to the case studied here but with the exception that one of the washboard potentials
is replaced by a different potential. This new potential serves to break one of
the spatial symmetries of the system. The consequences of this broken symmetry
are also  examined.

In systems that satisfy certain spatial and temporal symmetries it is
possible to find, in phase-space, two trajectories that nullify each
others contribution to the net current. That is to say, for each trajectory
in phase-space there exists another complementary trajectory such that
collectively both trajectories produce zero net current. Therefore, if a
system is to express a non-zero net current, some of these symmetries must
be broken. Quite often this is achieved through the addition of a periodic
(but non-symmetric) time dependent drive to the system \cite{hanggi.2009.romp,Denisov.2002.PREa,denisov.2014.pr}.
Analogously, in the autonomous case the introduction of a static bias
force that penetrates the plane of motion, usually suffices
when breaking the spatial symmetry, thus allowing for the emergence of
a non-zero current \cite{Speer.2007.PRE}.

We search for the mechanism that serves to break the spatio-temporal
symmetries and thus allow for the possible occurrence of a non-zero current in
autonomous systems modelling the interaction of coupled units.
Crucially, in systems (with a mixed phase-space) that rely on regular
interludes between periods of chaotic motion for directed transport, the
chaotic periods are seen as destructive with regard to directed transport,
in that the average velocity of trajectories in the chaotic component of phase-space
will be close to zero. In contrast, the systems looked
at here \emph{require chaos} in an initial stage of the
dynamics so that trajectories can be captured by hyperbolic
structures allowing them to escape. The emergence of a non-zero
net current is still dependent on other factors.
Particular attention is given to the symmetry properties induced
by the inclusion of an interaction potential. The symmetry properties
derived are illustrated via an example model system.
This model is introduced next.

\subsubsection[Asymmetric model]{Case of a spatially asymmetric model }\label{subsubsection:model2}
The above discussed symmetric model was idealised in that both coordinates of the system obeyed a spatial symmetry; namely the system remains invariant under a
change of sign of both coordinates. This is not true of the following asymmetric model  were a new potential is introduced that has the effect of breaking a
spatial symmetry for one of these two subsystems. Rather than having two units each evolve in a washboard potential, as in the symmetric model case, in this model
only one unit evolves in a washboard potential. This unit is coupled to a second unit (which is unable to contribute to the
net current due to energy constraints) that serves as an energy deposit from which the unit evolving in the washboard potential can draw energy. The second potential is
an anharmonic unit and interactions with this unit are local in nature due to the type of interaction potential and the finite
amount of energy injected into the system. It is defined by

\begin{equation}
V(Q) = \exp(-Q) + Q - 1. \label{eq:an_unit}
\end{equation}

\noindent
Unlike the washboard potential which has bounded potential energy $U(q)\le E_b = 1/\pi$,
this anharmonic unit has unbounded potential energy for
$\pm Q$; i.e $V(Q)\rightarrow \infty$ as $Q\rightarrow \pm \infty$.
Note the asymmetry of this potential, i.e. $V(Q) \neq V(-Q)$, as
this is important when the symmetry properties of this model are
discussed. Before moving on to the main focus of this section it is worthwhile exploring some of the coupled system dynamics. The
equations of motion are \cite{Hennig.2010.JPA}

\begin{eqnarray}
 \ddot{q} &=& -\sin(2\pi q) - D\left[ \dfrac{\tanh(q - Q)} {\cosh(q - Q)}\right],\label{eq:eom2a}\\
 \ddot{Q} &=& \exp(-Q) - 1 + D\left[ \dfrac{\tanh(q - Q)} {\cosh(q - Q)}\right].\label{eq:eom2b}
\end{eqnarray}

Let us assume a finite system energy. For $D=0$, the system decouples into two integrable subsystems and the
dynamics is characterised by individual regular motions of the
unit in the washboard potential, and bounded oscillations of the
additional degree-of-freedom (due to the energetic constraints), respectively. For $D\ne 0$, the subsystems
interact, thereby exchanging energy.  While the $Q$-unit performs solely
bounded motion there is the possibility that, for an escaping
unit, the corresponding coordinate, $|q|$, attains large values
and thus the related interaction forces, $\partial H_{\mathrm{int}}/\partial q$ and
$\partial H_{\mathrm{int}}/ \partial Q$, vanish asymptotically, allowing transient
chaos~ \cite{Zaslavsky.1985.HarwoodNY, Zaslavsky.1998.ImpColPress,Ott.1992.CamUniPress}.  That is, for large distance
$|q-Q|\gg 1$, the interaction vanishes with the result that the two
degrees-of-freedom decouple, rendering the dynamics regular.

\begin{figure}
\vspace{1ex}
\parbox{0.32\linewidth}{a)}\hfill
\parbox{0.32\linewidth}{b)}\hfill
\parbox{0.32\linewidth}{c)}\hfill
\vspace{1ex}
\includegraphics[trim=0cm 0.3cm 0.cm 1cm, clip,height=4.8cm, width=4.8cm]{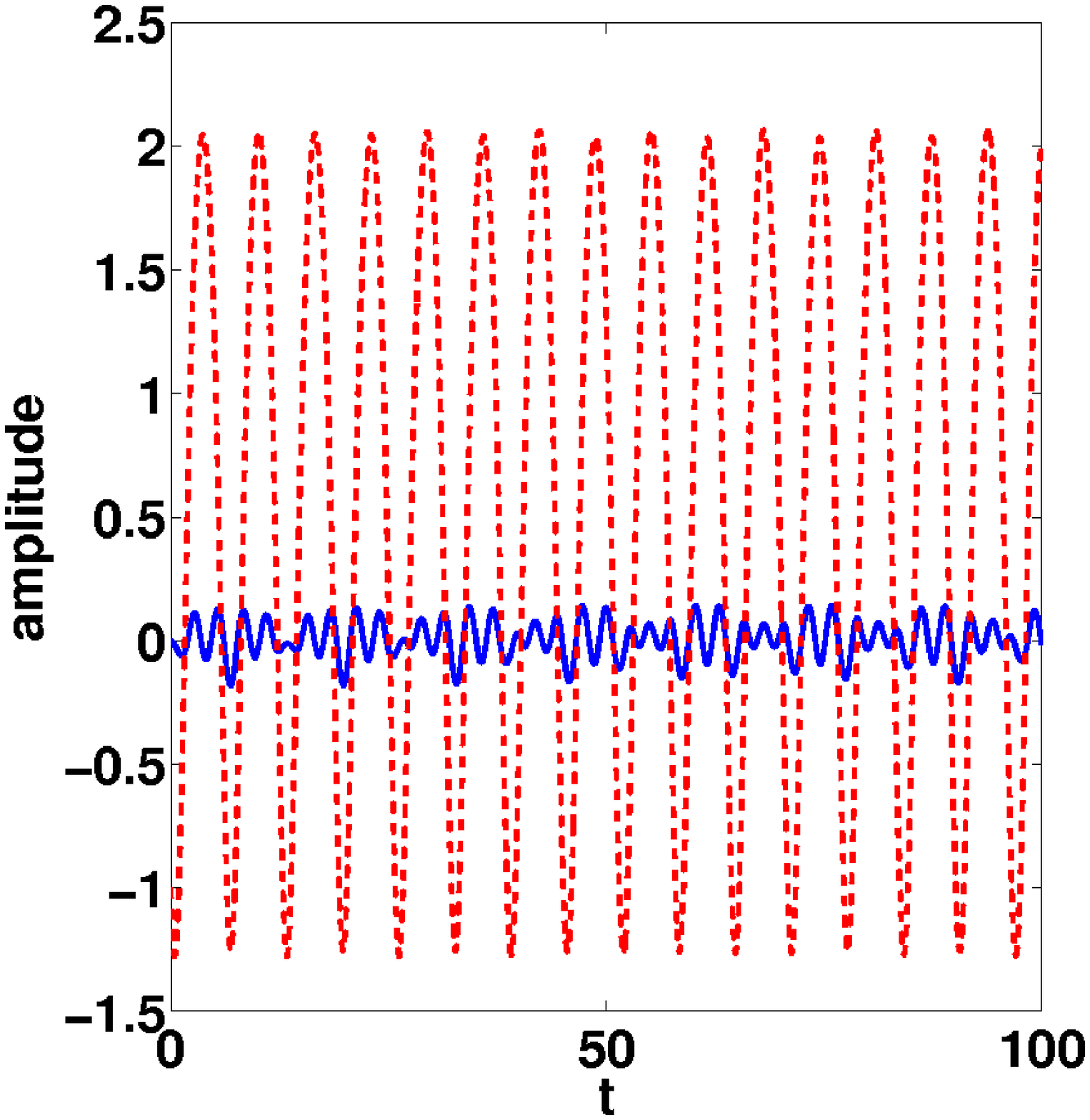}
\includegraphics[trim=0cm 0.3cm 0.cm 1cm, clip,height=4.8cm, width=4.8cm]{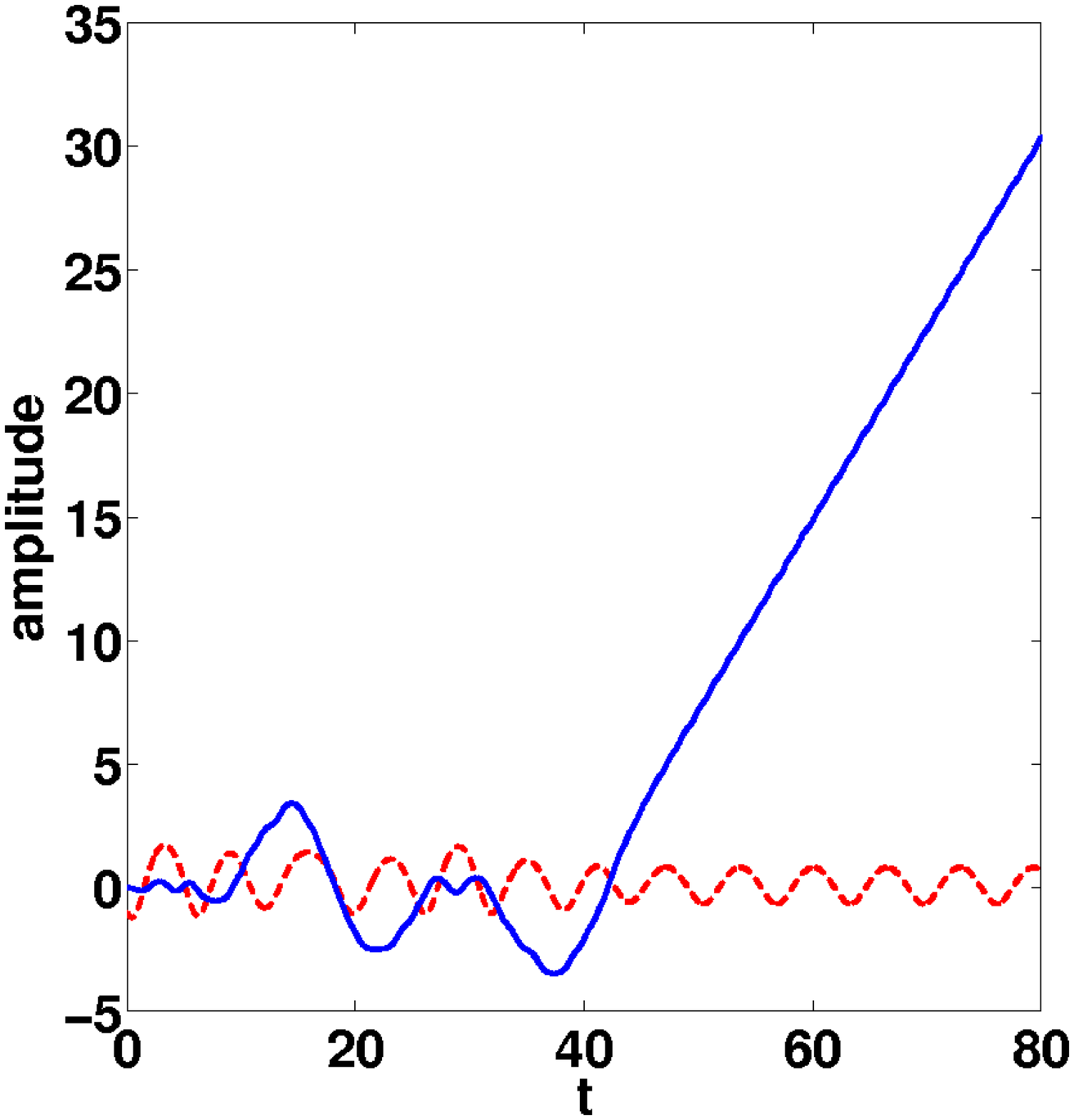}
\includegraphics[trim=0cm 0.3cm 0.cm 1cm, clip,height=4.8cm, width=4.8cm]{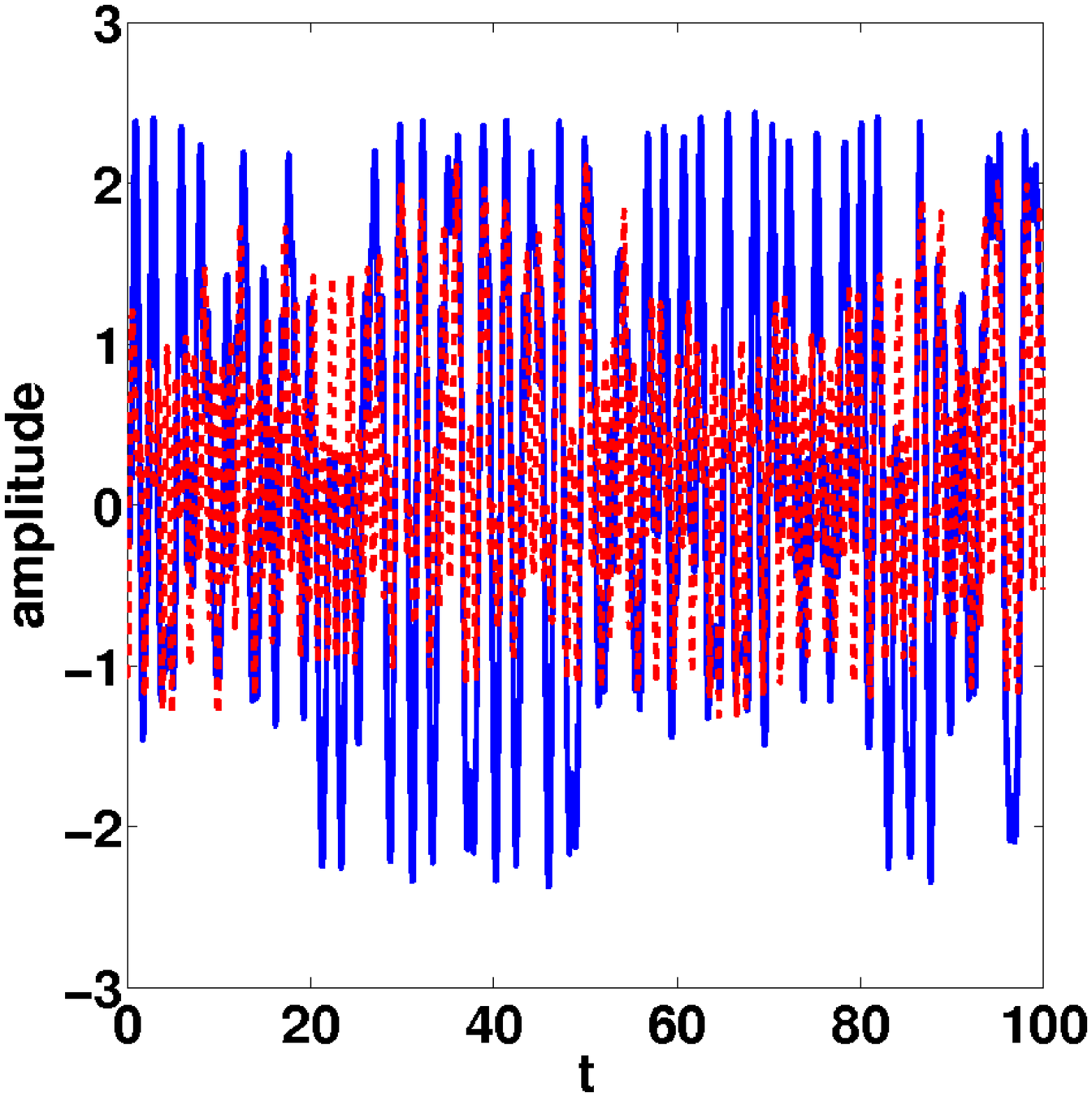}\\
\includegraphics[trim=0cm 0.3cm 0.cm 1cm, clip,height=4.8cm, width=4.8cm]{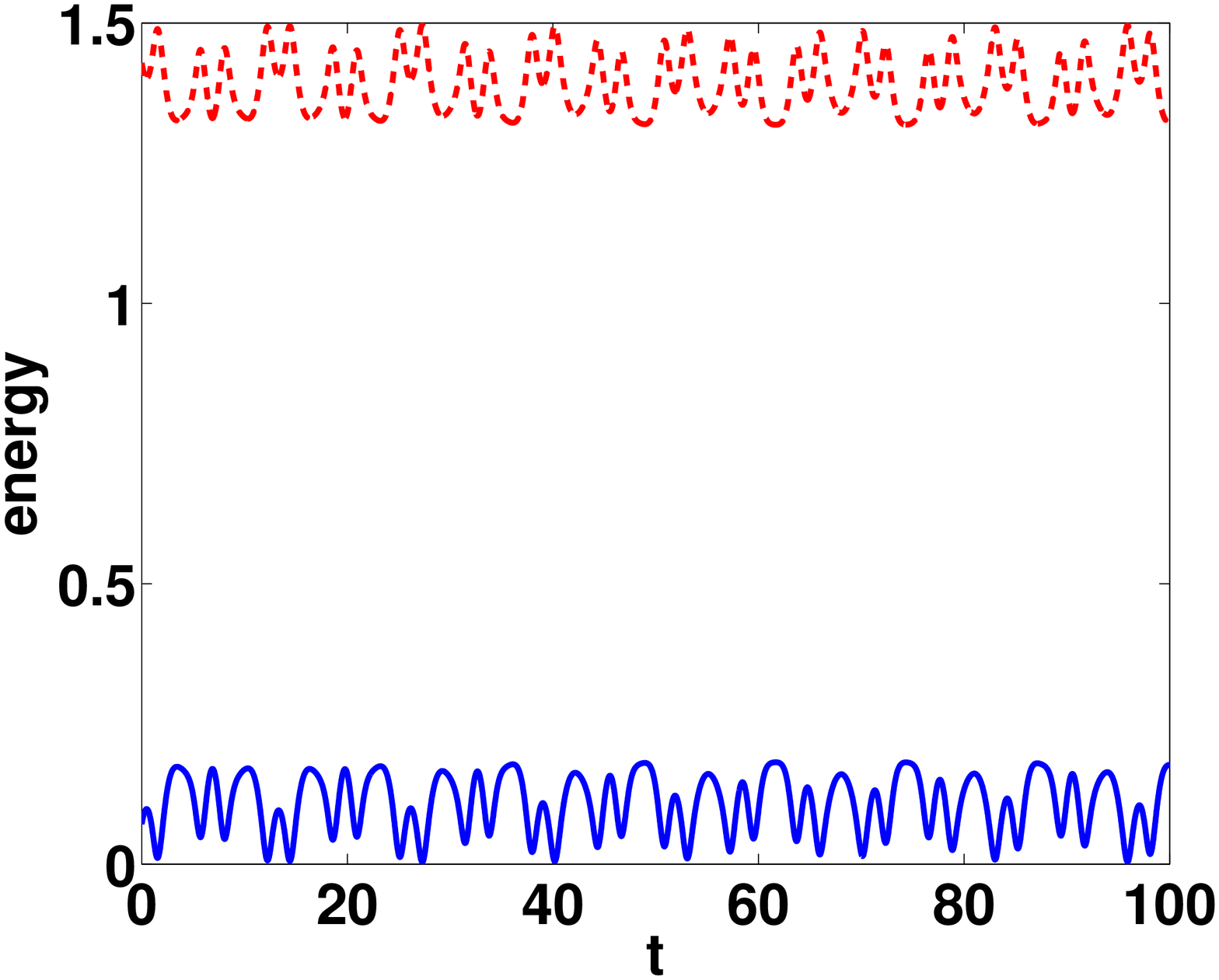}
\includegraphics[trim=0cm 0.3cm 0.cm 1cm, clip,height=4.8cm, width=4.8cm]{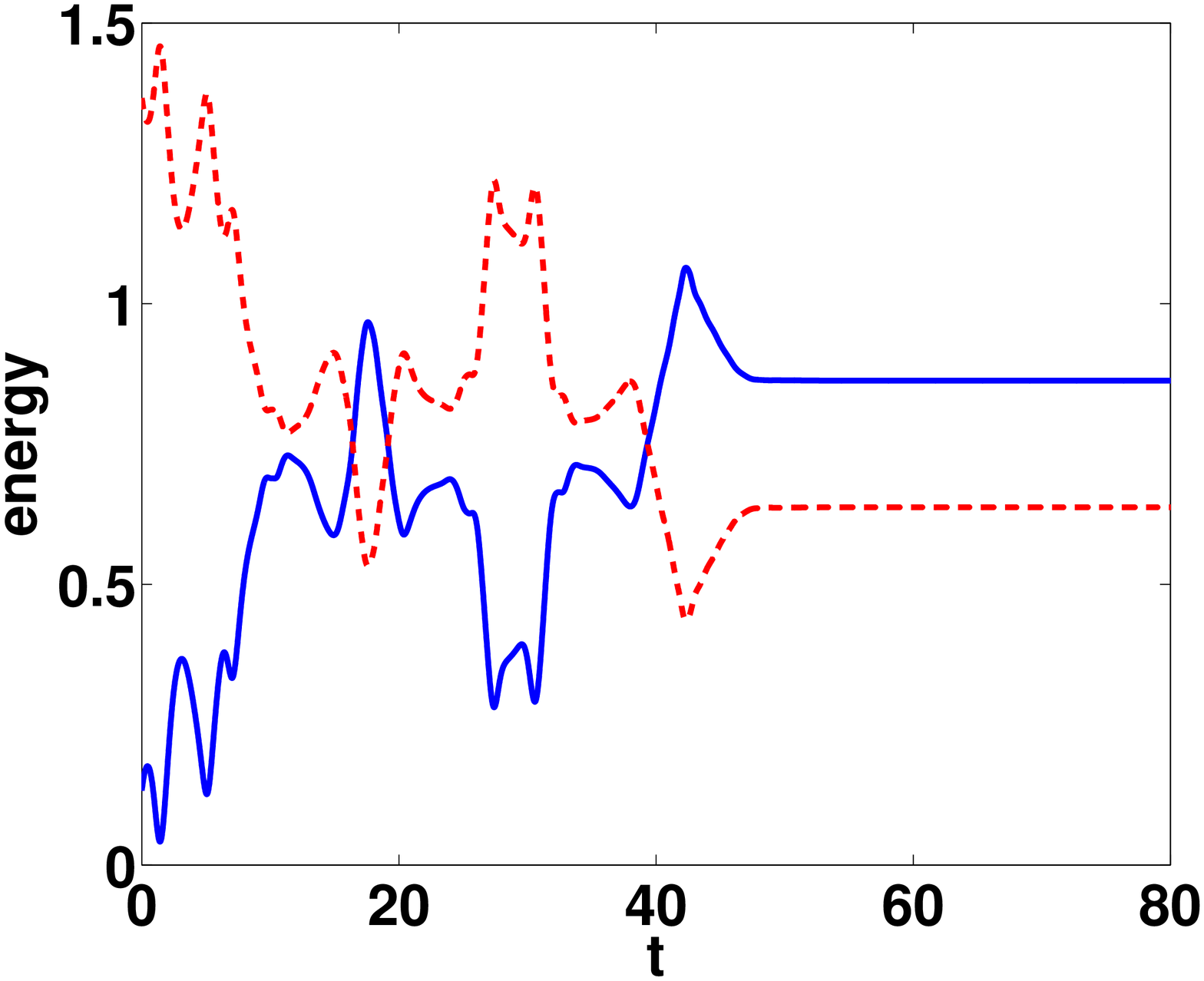}
\includegraphics[trim=0cm 0.3cm 0.cm 1cm, clip,height=4.8cm, width=4.8cm]{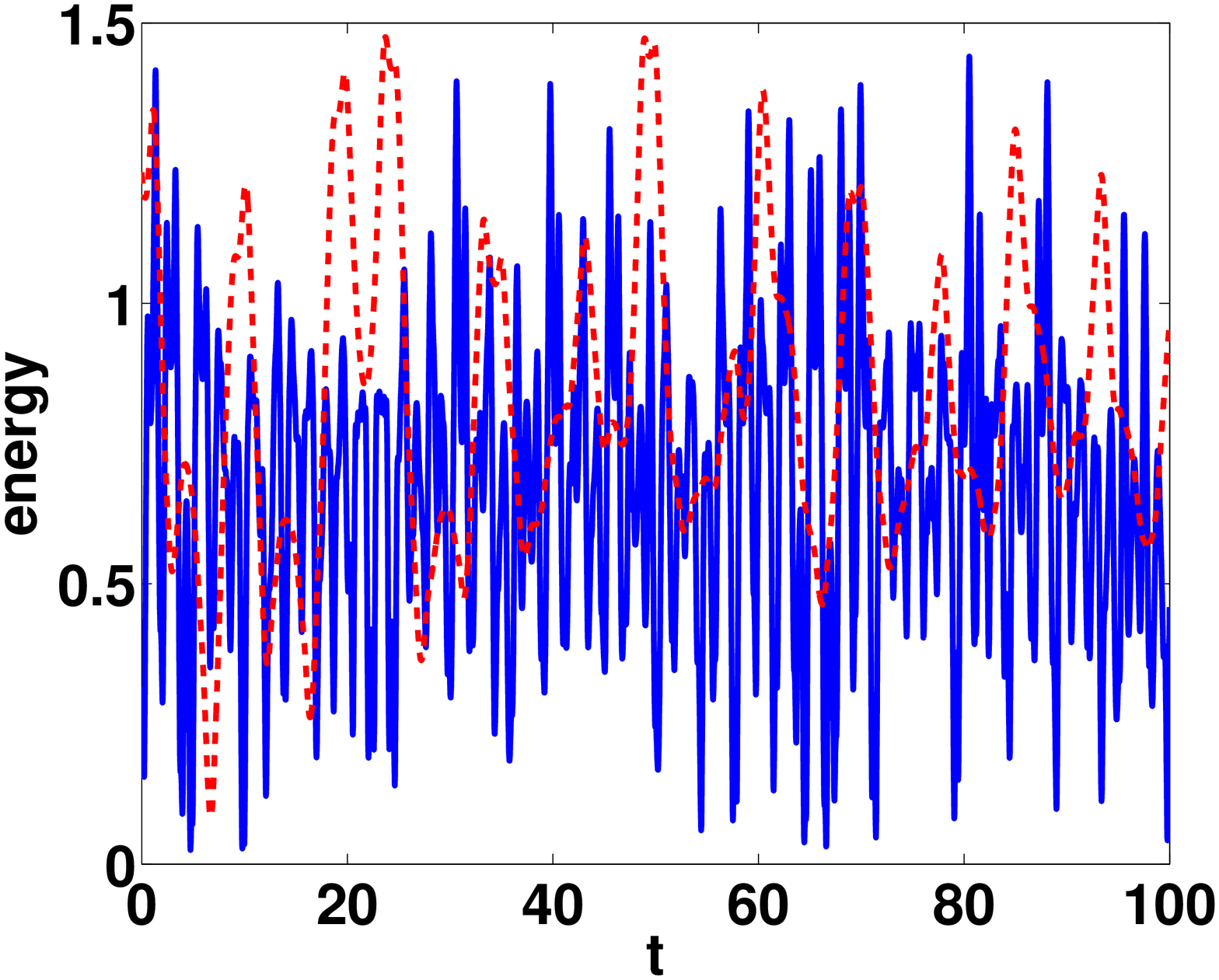}

\caption{(Colour Online) Example trajectories in the spatially asymmetric model described by Eqs.~(\ref{eq:eom2a})-(\ref{eq:eom2b}). The
top panels
show the time evolution of
the coordinates $q$ (solid blue line) and
  $Q$ (dashed red line) for three different values of the coupling
  strength $D$: (a) $D=0.4$, (b) $D=0.75$, and
  (c) $D=1.5$. The bottom panels show the corresponding evolution of the energies. The unit with coordinate $q$ has dynamics
that take place in a periodic and symmetric potential, while the unit with coordinate $Q$ has dynamics that take place in an asymmetric
and anharmonic potential. The units are coupled via an symmetric anharmonic potential. Source: Figure taken from
Ref.~\cite{Hennig.2010.JPA}.}\label{fig:ex_traj2}
\end{figure}

Looking at example trajectories for three representative coupling values reveals some of the system's dynamics. Fig.~\ref{fig:ex_traj2}
presents the time evolution of the coordinates $q$ \& $Q$ for $D=0.4, 0.75, 1.5$. In
addition, the  corresponding partial energies of the unit and the deposit degree-of-freedom are presented. The partial energies for
both units are given by

\begin{equation}
E_q=\frac{1}{2}\dot{q}^2+U(q)+\frac{1}{2}\,H_{\mathrm{int}}(q,Q),\qquad
E_Q=\frac{1}{2}\dot{Q}^2+V(Q)+\frac{1}{2}\,H_{\mathrm{int}}(q,Q)\,,
\end{equation}

\noindent
where the interaction energy has been evenly divided between the two units.

Three qualitatively different transport scenarios are presented. Note that for
the numerics the system's energy is fixed at $E=1.5$. Further, the initial conditions
have been chosen so that all the system's energy is initially in the deposit
degree-of-freedom. For the low value $D=0.4$ (left column) the washboard unit
undergoes small amplitude oscillations about a potential minimum. Crucially,
with regard to transport, these oscillations are much lower than the barrier
height $E_b=1/\pi$ of the washboard potential. In contrast, the deposit
degree-of-freedom sees oscillations of much larger magnitude. Looking at
the partial energies of the washboard unit and the anharmonic unit,
it can be seen that there is an insufficient energy exchange to allow for the
washboard unit to overcome the barrier height $E_b$ -- a prerequisite for
the occurrence of transport. The dynamics changes drastically when the coupling
strength is increased to $D=0.75$ (middle column). The early dynamics ($t<10$) is similar to
the
situation described above. Subsequently, the washboard unit escapes from
its starting potential well and travels to multiple wells in both directions.
At $t\approx40$ the washboard unit gains sufficient energy to allow it to
undergo independent directed transport. That is,
after a chaotic transient, the two subsystems decouple rendering the dynamics
regular. The chaotic exchange of energy preceding the directed transport of
the washboard unit is clearly visible in Fig.~\ref{fig:ex_traj2}. It is
also clear that after the chaotic transient, the energy exchange between the
two units terminates. A further increase in the coupling strength to $D=1.5$
(right column) results in a third qualitatively different transport scenario. It appears
that the washboard unit is free to travel multiple potential wells. However,
for the duration of the simulation it is confined to potential wells in the range
$-2.5<q<2.5$. The reasons for this are two-fold. Firstly, as is proven rigorously
\cite{Mulhern.2012.Thesis}, with $D=1.5$
(and $E=1.5$) the possibility of directed transport for the washboard unit
is excluded. Secondly, the system's (finite) energy means that oscillations of
the anharmonic potential are bounded. Combined, this results in two units
that remain in close contact and under constant chaotic energy exchange.


\vspace*{0.3cm}

\subsubsection{Symmetry breaking and emergence of a current}\label{subsubsection:symmetries}

It was shown that with a suitable choice of parameters the second asymmetric model
can exhibit directed transport, which is a necessary,
but not sufficient, condition for a system to show a non-zero net current. However, the class
of systems that this asymmetric model belongs to possesses time-reversibility symmetry and the implications of this with regard to the net current are
extremely important.
These implications, and a mechanism for destroying this symmetry, was discussed in Ref.~\cite{Hennig.2010.PRE} (and further in
Ref.~\cite{Mulhern.2012.Thesis}) and will be the focus of this section.

The class of systems in question are Hamiltonian and of
the form

\begin{equation}\label{eq:ham_gen3}
H(\mathbf{p}, \mathbf{q}) = \dfrac{1}{2}\mathbf{p}^2 + U_\mathrm{eff}(\mathbf{q})
\end{equation}

\noindent
where $\mathbf{p}, \mathbf{q} \in \mathbb{R}^n$, $\mathbf{p}$
and $\mathbf{q}$ are the canonically conjugated momenta and positions, and $U_\mathrm{eff}(\mathbf{q})$ is the potential function. With
transport and directed current being of interest,
it is assumed that $U_\mathrm{eff}(\mathbf{q})$ provides an open component. To reiterate, this means
that on constant energy surfaces, the system may be
unbounded in one, or more, of its coordinates.

The corresponding Hamiltonian equations $\dot{p}_i = -\partial H$/$\partial q_i$ and $\dot{q}_i = \partial H$/$\partial p_i$, $1 \le i
\le n$, exhibit the time-reversibility symmetry, i.e. there exists a time-reversal operator $\hat{\tau}$ such that if $\mathbf{X}$ is a
solution, then so is $\hat{\tau} \mathbf{X}$. In more detail, suppose that solutions take the form $\mathbf{X}(t) = [\mathbf{p}(t),
\mathbf{q}(t)]$.
Applying the time-reversal operator yields
$\hat{\tau}[\mathbf{p}(t), \mathbf{q}(t)] = [-\mathbf{p}(-t), \mathbf{q}(-t)]$.
This operation is involutory as $\hat{\tau}^2[\mathbf{p}(t), \mathbf{q}(t)] = [\mathbf{p}(t), \mathbf{q}(t)]$. As for the implication
of
time-reversibility with respect to the net current, consider a solution with initial condition (at $t=0$) given as $\mathbf{X}(0)$.
Given a finite observation time $T$ (relevant for numerical and experimental studies), let $\mathbf{X}(t)$ evolve from $\mathbf{X}(0)$
to $\mathbf{X}(T)$.
This trajectory is called the \textit{forward}
trajectory. At this point the time-reversal operator is applied which switches
the sign of the momenta and changes the direction of time. This creates a new initial condition $\hat{\tau} \mathbf{X}(T)$ which can be
evolved in (negative) time.
This trajectory is called the \textit{backward} trajectory. In fact, when the system is evolved from this new initial condition it
traces over the forward trajectory in coordinate-space. Note that the forward and backward trajectories coincide in coordinate-space,
but not in phase-space due to the change in the sign of momenta. Given the general nature of the above initial condition
$\mathbf{X}(0)$, we can conclude that on constant
energy surfaces, for each such initial condition there exists a corresponding initial condition $\hat{\tau} \mathbf{X}(T)$ such that
they cancel each others contribution to the net current. Therefore, for systems with time-reversibility symmetry there is no preferred
direction of the flow thus preventing the emergence of a directed current.

\begin{figure}
\centering
\includegraphics[height=4cm, width=6cm]{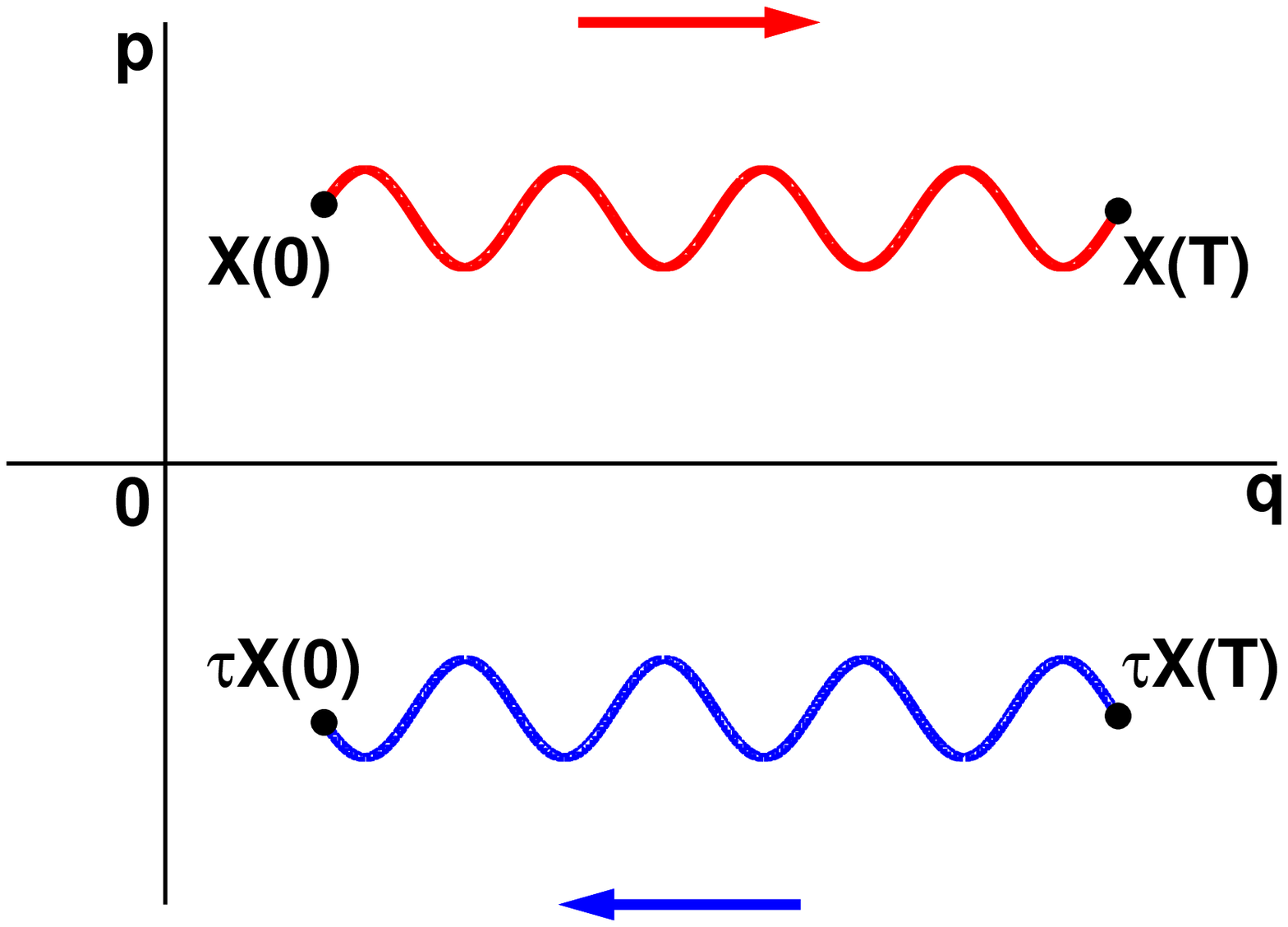}
\begin{overpic}[height=4cm, width=6cm]{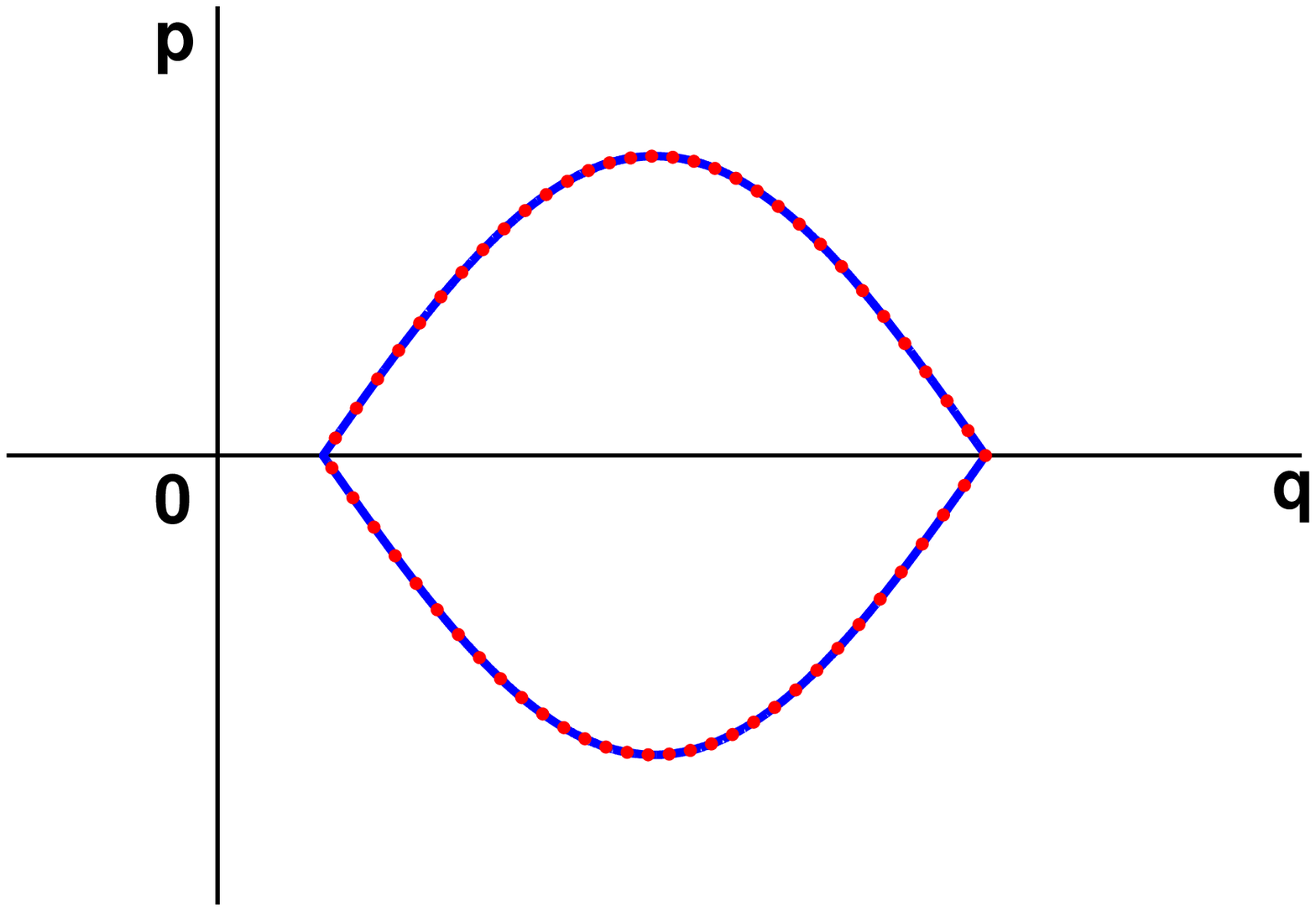}
\put(80,14.5){\includegraphics[height=0.8cm, width=0.8cm]{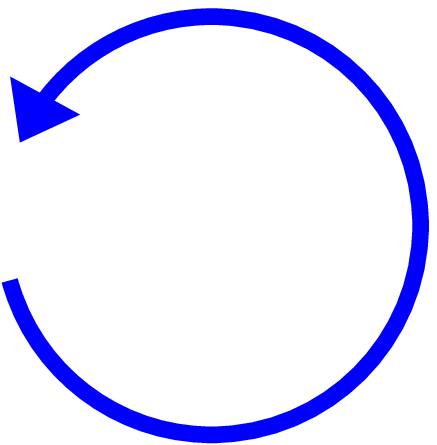}}
\put(80,44.5){\includegraphics[height=0.8cm, width=0.8cm]{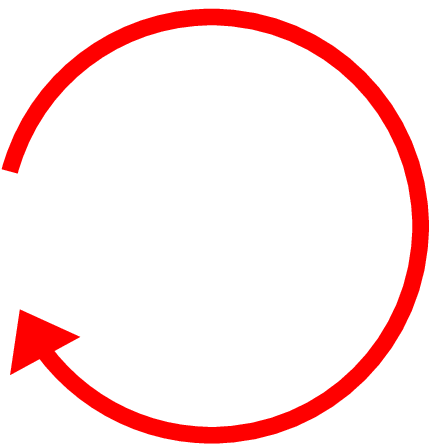}}
\end{overpic}
\caption{(Colour Online) Time reversal symmetry in the case of a simple pendulum. Left: Schematic illustration of the forward (red) and
backward (blue)
trajectories for a unit in the regime of rotational
motion. The forward trajectory has initial condition $X(0)$ and terminal coordinate $X(T)$, whereas the time-reversed trajectory has
initial condition $\tau X(T)$ and terminal coordinate $\tau X(0)$. Right: Schematic illustration of the forward (red) and backward
(blue) trajectories for a unit in the regime of librational motion. Source: Figure taken from
Ref.~\cite{Mulhern.2012.Thesis}.}\label{fig:tr_ex}
\end{figure}

The content of the above discussion is shown schematically in Fig.~\ref{fig:tr_ex}. Imagine $q=\theta$ is the angle of rotation of a
pendulum, and $p=\dot{\theta}$ is the corresponding angular velocity. Then the top half of the figure (in red) shows the phase portrait
of a pendulum, with initial condition $X(0)=(p(0),q(0))$, undergoing rotational motion. The trajectory terminates at
$X(T)=(p(T),q(T))$.
The bottom half of the figure (in blue) is the time-reversed counterpart of this trajectory, $X(T-t)$, with the initial condition
$\tau X(T) = (-p(T),q(T))$. With a view to the present work, we can imagine a single unit, with position $q$ and momentum $p$,
undergoing rotational motion in a washboard potential.

Applying the time-reversal operator to
the system and the original initial condition $X(0)$ produces
another possible motion of the system \cite{Lamb.1998.PhysicaD}. However, it is not always
the case that these two trajectories produce average velocities that are equal
in magnitude but opposite in sign. For example, its possible to envisage a potential
such that a unit moving to the
right will fall into a `trap', while the unit moving to the left will experience
unbounded motion. Clearly, the sum of the two average velocities will not equal zero.
The point emerges that to produce (and guarantee that) two trajectories with zero
average velocity, the first needs to be evolved to some terminal time $t=T$ at
which point the time reversal operator is applied.

It should also be mentioned that the act of selecting an initial condition can be sufficient in itself to violate time-reversibility
symmetry. That is, even though the equations of motion are time-reversal symmetric, not all solutions need necessarily have this
symmetry. This is the case for trajectories where the initial condition and its time reversed counterpart follow distinct paths in
phase-space. As an example of such a trajectory consider a continuously rotating pendulum (cf. Fig.~\ref{fig:tr_ex}). So by creating a
trajectory with time-averaged velocity $\nu\neq 0$, over an observation time of duration $t=T$, and initial condition $X_0$,
time-reversibility symmetry has been broken, unless a second initial condition $\hat{\tau} X_T$ is chosen that produces a trajectory
with time-averaged velocity $-\nu$. This is not true for self-reversed trajectories where the initial conditions $X_0$ and $\hat{\tau}
X_T$ produce trajectories that coincide in phase-space. The librating trajectories of a pendulum are
self-reversed. See Fig.~\ref{fig:tr_ex} for a schematic illustration of this. The same principles apply to trajectories that wander
chaotically in phase-space.

Given what has just been discussed, it seems a rather hopeless situation to a find Hamiltonian of the form given in
Eq.~(\ref{eq:ham_gen3}) that expresses a non-zero net current, because every initial condition is related to another that negates its
contribution to the current. This statement is true as long the entire energy surface is populated with initial conditions. However,
for
systems with an open component, i.e. unbounded in at least one of its coordinates, it is not feasible to populate an entire energy
surface
with initial conditions. Therefore, it is more natural to define a finite set of initial conditions which, given the infinite extent of
(at least one of) the coordinates can be regarded as \emph{localised in space}. Such sets of initial conditions are frequently used in
applications (see \cite{Hennig.2011.PRE} for an example), and are chosen to be physically relevant, such as for the problem of a
unit flow
emerging when the units are initially trapped in a single well of a spatially infinitely
extended multiple well potential.
Indeed this is done when the second model is further examined in a later section.

Looking more closely at the implication of choosing localised initial conditions, it is supposed that the coordinates are localised in
the domain $q_{j,l} \leq q_j(0) \leq q_{j,r}$ with $1 \leq j \leq n$ representing the index of each degree-of-freedom (which for the
present discussion is not restricted to two), and the subscripts $l$ and $r$ denote left and right respectively. Let a trajectory (with
regard to a finite observation time $T$) be \textit{transporting} if (i) at least one of the coordinates $q_j(t)$ escapes from the
domain of the localised initial conditions in some time $0 < t_{escape} \ll T$ and (ii) it subsequently undergoes directed motion, that
is, $\langle p_j(t) \rangle \neq 0$ for $t_{escape} \leq t \leq T$ where $\langle \cdot \rangle$ denotes the average with respect to
time. This gives a trajectory moving away from the set of localised initial conditions such that at the end of the observation time one
of the terminal coordinates obeys $q_j(T)<q_{j,l}$ or $q_j(T)>q_{j,r}$ for some $j$.
 Thus, the situation has arisen where the initial condition of the corresponding backward trajectory, which
would compensate the contribution of the forward trajectory to the current,
is not contained in the set of localised initial conditions.
This seems to suggests that in systems where time-reversibility has been broken, via the use of localised initial conditions,
there will
be a non-zero directed current.
This is not necessarily the case. In fact other symmetries need first to be violated, i.e. spatial symmetries.
This is seen more clearly when the symmetry properties of the second, asymmetric model are considered. First, let us examine the conditions that allow for
the occurrence of a non-zero current in the symmetric model case.

\subsubsection{The emergence of a non-zero current in the symmetric case} 
Before moving on, let us return to the previously discussed symmetric model and  examine the
initial conditions used there. It is worth noting that as the system is spatially
symmetric and yet a non-zero current can emerge in the ensemble dynamics,
the choice of initial conditions must consequently be  important. First, let
us recall what these initial conditions are. Let us denote some initial
condition by $X(0)$ where $X(0) = (p(0), q(0), P(0), Q(0))$ with $Q(0)=P(0)=0$, $p(0)>0$
and $q(0)\in (-25.5,-24.5)$.
Thus, the initial dynamics will see the coordinate $q(0)$ approach $Q(0)$.

Applying the time-reversal operator to $X(0)$ yields
$\tau X(0) = \hat{X}(0) = (-p(0), q(0), -P(0), Q(0))$. The flow,
generated by the equations of motion with initial conditions $X(0)$ and $\hat{X}(0)$, do not necessarily produce
zero-averaging counter-propagating trajectories. This is clear when one considers that with $X(0)$ the coordinate
$q(0)$ will
approach $Q(0)$. However, under the flow with initial condition $\hat{X}(0)$ the distance between these coordinates is
monotonically increasing. This holds true for all such initial conditions defined above. Thus, the presence of the
interaction potential breaks time-reversibility for this set of initial conditions. This echoes Loschmidt's paradox
in that the underlying system obeys time-reversibility, yet some ensembles do not obey the symmetry. Thus, this
system has helped to illuminate an important point, from the point of view of current generation. Namely, a
system (with initial condition $X(0)$) under time-reversal does not necessarily produce
counter-propagating trajectories, $X(0)$ \& $\tau X(0)$, that combined have zero averaged current.

In fact, the appropriate initial condition $\hat{X}(0)$ that would produce the counter-propagating trajectory
negating the current contribution of the trajectory with initial condition $X(0)$ is given by
$\hat{X}(0) = (-p(0), -q(0), 0, 0)$. Crucially, the set of initial conditions described above does not
contain $\hat{X}(0)$, and thus this explains the emergence of a current. It should be noted that $\hat{X}(0)$
is not created through any time-reversal operation. Rather, this initial condition is generated from the system's
spatial symmetries. Another point to note is that if the original initial conditions $X(0)$ are evolved for any
time $T>10$ (which the simulations exceed by far) then the time-reversed initial condition $\tau X(T)$ is outside
the set of localised initial conditions (as described above). Thus the temporal \emph{and} spatial symmetries have
been violated through the choice of initial conditions.

\subsubsection{Time-reversal symmetry manifolds} The system   Eq.~(\ref{eq:ham_gen3}) possesses an energy integral
\begin{equation}
E = \dfrac{1}{2}\mathbf{p}^2 + U_\mathrm{eff}(\mathbf{q}).
\end{equation}
There exists a closed maximum equipotential surface
$U_\mathrm{eff}(\mathbf{q})=E$ bounding all motions and one has
$p=P=0$, on this surface.
Moreover, for Hamiltonian systems of the form Eq.~(\ref{eq:ham_gen3}) time-reversibility
is manifested in coordinate-space
in the symmetry features induced by reflections on the time-reversibility symmetry manifolds.

In the case of
the  two models discussed, or any two-degree-of-freedom system for that matter, the symmetry manifolds are
represented by symmetry lines. In general, for n-degree-of-freedom systems,
these symmetry manifolds are obtained by setting the velocities (momenta) equal to zero.
This produces an
$n$-dimensional `mirror' plane,

 \begin{equation}
 M = \{\mathbf{q}~|~\mathbf{p}=\mathbf{0}\} \,\,\,\,\,\,\,\,\,\,\,\,
 \mathbf{q}, \mathbf{p} \in \mathbb{R}^n
 \end{equation}

 \noindent
 where the trajectories starting on this plane will follow the same path in
coordinate-space in forward and
 backward time. This is clear when one considers the form of the Hamiltonian being even in the
momenta $\mathbf{p}$ and the fact that the time-reversal
operator changes the sign of the
momenta $\mathbf{p} \rightarrow -\mathbf{p}$. More accurately,
the time reversibility manifolds are given by

\begin{equation}
S_k:\quad -\dfrac{\partial U_\mathrm{eff}}{\partial q_k} = F_k(\mathbf{q}) = 0,
\quad\quad\quad 1 \leq k \leq n.
\end{equation}

\noindent
Let us consider reflections of a trajectory, projected onto coordinate
space, on the symmetry manifolds $S_k$ induced by the corresponding operators
$\hat{R}_k$. First note that the reflections spoken of here are
not spatial reflections.
Rather, these reflections map each point of a trajectory onto another,
$\mathbf{q} \rightarrow \hat{R}_k(\mathbf{q})$,
on equipotentials,
$U_\mathrm{eff}(\mathbf{q}) = U_\mathrm{eff}(\prod_{k=1}^m\,\hat{R}_k(\mathbf{q}))$,
such that the sign on the right hand side of the
equations of motion for $\dot{p}_k$ are reversed,
${\rm sign}(F_k(\hat{R}_k(\mathbf{q})))\ne {\rm sign}(F_k(\mathbf{q}))$, with $1 \leq k \leq n$.

Observe that upon reflecting
on all symmetry manifolds the relation
\begin{equation}
 U_\mathrm{eff}(\mathbf{q})=U_\mathrm{eff}(\prod_{k=1}^m
\,\hat{R}_{k}(\mathbf{q})),\,\,\,1\le m \le n
\end{equation}
is left invariant under
permutations of the reflection operators.  In fact, time-reversing
symmetry is in coordinate-space tantamount to invariance with respect
to reflections on the symmetry manifolds. In more detail, any
self-reversed trajectory, projected onto coordinate-space, repeatedly
crosses every symmetry manifold $S_k$ upon which each time the sign of
the corresponding force $-\infty <F_k(\mathbf{q}) < \infty$ changes. Moreover, each
crossing subsequent to the previous one occurs from the opposite
direction. Thus there must be turning points for the trajectory, contained in the maximum
equipotential surface $U_\mathrm{eff}(\mathbf{q})=E$,
implying bounded motion and no directed flow can arise.  Notice that
no assumptions with regard to the spatial symmetries of the trajectory
are needed.  In contrast, as transporting (unbounded) trajectories are
not invariant with respect to reflections on the symmetry manifolds,
preservation of time-reversing symmetry is not possible. A
transporting trajectory may escape without having crossed a symmetry
manifold at all. However, if it does cross then after all such
crossings of a symmetry manifold, the escaping trajectory promotes
directed transport. Nevertheless, reflections on the symmetry
manifolds, mapping a transporting trajectory onto another transporting
one, can induce spatial symmetries such that these two trajectories
mutually compensate each others contribution to the net flow. Let the
point $\mathbf{q}_O$ in coordinate-space be an initial condition associated
with a transporting trajectory.  Reflecting in coordinate-space on the
symmetry manifolds $S_{k}$ transforms an original point, $\mathbf{q}_O$, into
its image point, $\mathbf{q}_I$, according to $\hat{R}_{k}\mathbf{q}_O=\mathbf{q}_{I,k}$
reversing the sign on the r.h.s. in the equations of motion for $\dot{p}_k$
according to $\mathrm{sign}(F_k(\mathbf{q}_O)) \ne \mathrm{sign}(F_k(\mathbf{q}_{I,k}))$.
However, the magnitude of the gradients $F_k=-\partial U_\mathrm{eff}/\partial q_k$ is not
necessarily maintained.  Reflection on all of the symmetry manifolds
yields
\begin{equation}
 \prod_{k=1}^n \,\hat{R}_{k}{\mathbf{q}}_O=\mathbf{q}_I\,\,\,\mathtt{and}\,\,
 \,U_\mathrm{eff}(\mathbf{q}_0)=U_\mathrm{eff}(\mathbf{q}_I)
\end{equation}

\noindent
reversing the sign in
each of the r.h.s. of the equations of motion for the evolution of the
momenta $\mathrm{sign}(F_k(\mathbf{q}_I)) \ne \mathrm{sign}(F_k(\mathbf{q}_O))$,\,$1\le k \le n$.  With the
time evolution of a coordinate expressed as
\begin{equation}
 q_k(t)=q_k(0)+\int_0^t
dt^{\prime}\{p_k(0)+\int_0^{t^{\prime}} dt^{\prime \prime} [-F_k(\mathbf{q}(t
  ^{\prime \prime}))]\},
\end{equation}
we conclude that, for the pair of
trajectories emanating from $\mathbf{q}_O$ and $\mathbf{q}_I$, symmetry (zero net flow)
results if $(\mathbf{p}_I,\mathbf{q}_I)=(-\mathbf{p}_O,-\mathbf{q}_O)$ so that
$F_k(\mathbf{q}_I)=
-F_k(\mathbf{q}_O)$, \,$1\le k \le n$. This is the case when the potential is
even in the coordinates, that is $U_\mathrm{eff}(\mathbf{q})=U_\mathrm{eff}(-\mathbf{q})$
(the symmetric model  for example).  Then there exist pairs
of current-annihilating {\it counterpropagating} trajectories,
${X}(t)$, starting from ${X}(0)$, and $-{X}(t)$, starting from
$-{X}(0)$. In other words, {\it reversion symmetry} under reflections
on the symmetry manifolds is needed for zero net flow which, together
with invariance with respect to changes of the sign of the momenta,
amounts to parity-symmetry of the system $H(\mathbf{p},\mathbf{q})=H(-\mathbf{p},-\mathbf{q})$.
Conversely, violation of reversion symmetry with respect to at least
one of the coordinates $q_k$ establishes a prerequisite for the
occurrence of directed flow.

\subsubsection{The effect of broken symmetries} It is useful to see how the above theory on spatio-temporal symmetries can be
applied
in practice, and in particular to observe the effects of breaking these symmetries. For this reason, the spatio-temporal
symmetry properties of the asymmetric model are next examined. Particular attention is given to the phase-space dynamics
and to current generation
where the effects of broken
symmetries is most clearly visible.

The asymmetric model has an effective potential given (in short) by

\begin{equation}
U_{\mathrm{eff}}(\mathbf{q}) = U(q) + V(Q) + H_{\mathrm{int}}(q,Q)
\end{equation}

\noindent
where $U(q)$ is the washboard potential, $V(Q)$ is the anharmonic (deposit) potential,
and $H_{\mathrm{int}}(q,Q)$ is the interaction potential. Some properties of the washboard and
interaction potentials were discussed in the introduction to Sec.~\ref{section:transport}. However,
it is worth reiterating those that are relevant in this section. The washboard potential is of
period one and observes the coordinate symmetry $U(q) = U(-q)$. This amounts to symmetry with
respect to $q_n=n/2$ for every integer $n$. Likewise, the interaction potential is invariant under reflections of its argument,
namely $(q-Q)\leftrightarrow -(q-Q)$. In contrast, the deposit
degree-of-freedom $V(Q)$ does not obey such a reflection symmetry. That is, $V(Q)\neq V(-Q)$
resulting in equations of motion (cf. Eq.~(\ref{eq:eom2a}) \& Eq.~(\ref{eq:eom2b}))
that do not remain invariant under reflections in $Q$.

Even with the anharmonic part $V(Q)$, which breaks a reflection symmetry of the system, the asymmetric model case
still possesses time-reversal invariance. Thus, if the phase-space is entirely populated with initial
conditions, then the system will produce a zero net current. This raises the question of what effect
localised initial conditions imply for the system's current output.

At this point it is worthwhile describing exactly the set of initial conditions that were
used in the numerical analysis of the asymmetric model. These initial conditions were chosen such that
the washboard unit was at rest at the origin, with the system's energy initially residing
in the deposit degree-of-freedom and the interaction potential. In more detail, at time $t=0$,
the washboard unit's position and velocity were given by $q(0)=\dot{q}(0)=0$. Thus the washboard
unit begins its time evolution with zero energy. Assuming system energy $E$, the set of initial
conditions for the remaining degrees-of-freedom are chosen to populate uniformly and densely the level curve

\begin{equation}
E = \dfrac{1}{2}\dot{Q}^2 + V(Q) + H_{\mathrm{int}}(0,Q)
\end{equation}

\noindent
in the $(Q,\dot{Q})$-plane. This set is topologically a circle. Importantly, these initial
conditions are unbiased in the velocity, i.e. $\dot{Q} \leftrightarrow -\dot{Q}$.

Returning to the symmetry analysis we now turn our attention to the time-reversal symmetry
manifolds. As this is a two-degree-of-freedom system, the symmetry manifolds will exclusively
be termed symmetry lines. Setting the velocities in Eq.~(\ref{eq:eom2a}) \& Eq.~(\ref{eq:eom2b}) equal to zero obtains
Ref.~\cite{Hennig.2010.JPA}

\begin{eqnarray}
S_1:&\quad \sin(2\pi q) + D\dfrac{\tanh(q-Q)}{\cosh(q-Q)} = 0,\label{eq:tr_s1}\\
S_2:&\quad \exp(-Q) - 1 + D\dfrac{\tanh(q-Q)}{\cosh(q-Q)} = 0.\label{eq:tr_s2}
\end{eqnarray}

\noindent
The symmetry line $S_1$ exhibits the following symmetry:

\begin{equation}
Q \rightarrow -Q , \quad \dfrac{n}{2}+q \rightarrow -\dfrac{n}{2}-q: \quad S_{1,n} \rightarrow -S_{1,-n}
\end{equation}

\noindent
with $n$ labelling the branches of the symmetry line as $S_{1,n}$. The occurrence of the
multiple branches of the $S_1$ symmetry line can be understood by considering the symmetries
of the washboard potential. In contrast, $S_2$ yields a single branch containing no apparent symmetries.

\begin{figure}
\centering
\includegraphics[height=5.5cm, width=7.5cm]{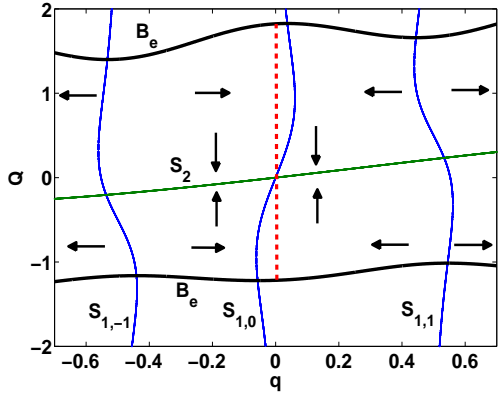}
\caption{(Colour Online) Time-reversal symmetry lines in the coordinate plane $(q,Q)$ - see Eqs.~(\ref{eq:tr_s1})-(\ref{eq:tr_s2}). The
coupled particle (conservative) dynamics take place in a spatially asymmetric potential. The system's energy
is $E=1.5$ and the energetically accessible region is bounded by the two curves indicated by $B_e$.
The dashed (red) line corresponds to the location of the initial conditions projected onto coordinate-space.
Arrows indicate the direction of flow in different regions, as determined by the sign of the forces
$-\partial{U_{\mathrm{eff}}}/\partial{q}$ and $-\partial{U_{\mathrm{eff}}}/\partial{Q}$. Source: Figure taken from
Ref.~\cite{Hennig.2010.JPA}.} \label{fig:trs1}
\end{figure}

The two cases, coupled and uncoupled, result in markedly different dynamics. Notably, uncoupling the two
subsystems produces integrable dynamics. The dynamics becomes non-integrable (chaotic) when
the two subsystems are coupled. This complexity is also manifested in the symmetry lines $S_1$ and $S_2$.
To see this, consider first the uncoupled case.
With the coupling parameter $D=0$, the equations representing the symmetry lines
are simplified and solutions take the form $q=n/2$ for all $n \in \mathbb{Z}$ and $Q=0$.
For $D\neq 0$ the solutions to the equations for $S_1$ and $S_2$ become more complicated.
This point is illustrated in Fig.~\ref{fig:trs1} which shows the time-reversal symmetry lines when $D=0.75$.
For illustration, only the branches of the symmetry line $S_{1,n}$, with $n=-1,0,1$, related to the starting potential well are shown.
The direction of flow, as determined by the sign of the forces $-\partial{U_{\mathrm{eff}}}/\partial{q}$ and
$-\partial{U_{\mathrm{eff}}}/\partial{Q}$, is indicated by arrows in the
different regions in the coordinate plane. Boundaries of the energetically-allowed region
in coordinate-space are represented by the two lines labelled $B_e$. Reflections of a trajectory, projected onto coordinate
space, on the symmetry lines $S_k$ are induced by the corresponding operators $\hat{R}_{k}$
mapping each point on the trajectory to another one
on equipotentials
\begin{equation}
U_\mathrm{eff}(q,Q)=U_\mathrm{eff}(\prod_{k=1}^2 \,\hat{R}_k(q,Q)),
\end{equation}
such that the sign on the r.h.s. in the equations of motion is reversed, i.e.,
\begin{equation}
\mathrm{sign}(\partial U_\mathrm{eff}(q,Q)/\partial q) \ne
\mathrm{sign}(\partial U_\mathrm{eff} (\hat{R}_1(q,Q))/ \partial q)
\end{equation}
upon reflection on
$S_1$, and
\begin{equation}
\mathrm{sign}(\partial U_\mathrm{eff}(q,Q)/\partial Q) \ne
\mathrm{sign}(\partial U_\mathrm{eff} (\hat{R}_2(q,Q))/ \partial Q)
\end{equation}
upon reflection on $S_2$. Note that the magnitude of the gradients
$\partial U_\mathrm{eff} (q,Q)/ \partial q$ and $\partial U_\mathrm{eff} (q,Q)/ \partial Q$
is not necessarily maintained.

It is clear that the symmetry lines for the uncoupled and coupled system are significantly different.
Coupling the two subsystems has the effect of contorting the symmetry lines. It should be stressed again
that even though the time-reversal symmetry lines differ for each value of the coupling parameter, the
result for the net current is the same when the energy surface is entirely populated by initial conditions.
That is, when the energy surface is entirely populated by initial conditions, the resulting current is zero
regardless of the value $D$ which regulates the coupling strength.

Now the issue of localised initial conditions can be addressed. The initial conditions described above are shown,
projected onto the $(q,Q)$-plane, as the red dashed line in Fig.~\ref{fig:trs1} which shows the time-reversal
symmetry lines for $D=0.75$. Notice that one branch of the symmetry lines $S_{1,0}$ divides the initial conditions
into two segments, each promoting transport in different directions. For the segment lying to the left of $S_{1,0}$ the
flow is in the direction of positive $q$, while in the segment to the right of $S_{1,0}$ the flow moves in the direction
of negative $q$. Crucially, there is an imbalance in the size of the segments, and thus initially there is an unequal
number of trajectories moving towards the two chaotic saddles located at the intersections of $S_{1,1}$ and
$S_{1,-1}$ with $S_2$. This imbalance and the fact that the system contains an open component allows for the
emergence of a non-zero directed current.

Previously, it has been stated that broken spatial-symmetries are of no consequence (with respect to the current)
for systems with an energy surface entirely populated with initial conditions. However, the above discussion shows that for
localised initial conditions broken symmetries play an important role.
Namely, the asymmetric $V(Q)$ breaks time-reversibility for the localised initial conditions described above.
Moreover, using a different potential $V(Q)$, which is invariant under
reflections in $Q$, would restore the symmetry between counterpropagating trajectories, leading to a zero current.

It is worth re-emphasising that the mechanism responsible for the
non-zero net current presented in this section is novel in that it
doesn't require a mixed phase-space, induced by time-periodic driving,
consisting of regular and chaotic components \cite{Yevtushenko.2000.PRE,Flach.2000.PRL,denisov.2014.pr}.
Rather, for the autonomous systems discussed here chaos is only needed in
an initial stage of the dynamics to guide trajectories onto regular paths
\cite{Hennig.2009.PhysicaD},\cite{Hennig.2010.JPA},\cite{Hennig.2010.PRE}.
After the finite period of transient chaos, the units subsequently
undergo regular rotational motion. This is in contrast to the sticking
episodes close to tori, of finite duration, that provide a non-zero
net current in the non-autonomous Hamiltonian case.

A step in this direction, i.e. of directed transport in autonomous systems,
was taken by \cite{Dittrich.2010.ChemPhys}. In a three degree-of-freedom system, modelling
a molecular motor, the external time periodic driving was replaced by an
autonomous degree-of-freedom that acts as an energy store. They observed
that energy was able to propagate through the system resulting in directed
transport. However, their results differ from the results presented in
this section in two respects. Firstly, transport in the molecular motor
model was aided by thermal fluctuations. Secondly, although on the one
hand noise aided transport, on the other it also had the effect of
destroying transport. Thus, the intervals transport (which are of
finite duration) are separated by periods of bounded motion.

\subsection[Non-autonomous systems]{Collective transport in time-dependent driven systems}\label{subsection:non-autonomous}
\noindent
In Sec.~\ref{subsection:autonomous} the Hamiltonian dynamics of autonomous systems of two coupled units
were explored. This part  continues in a similar vein. However, the equations of motion
are augmented
by the inclusion of time-dependent driving and dissipative terms, thus adding further complexity to the coupled dynamics. Further,
the short range interaction potential Eq.~(\ref{eq:H_int})
(which allows for the occurrence of certain phenomena,
in particular transient chaos) is replaced by a potential that allows for long range interactions between
the particles.

In Sec.~\ref{subsubsection:running} the driven and damped dynamics
of two interacting
particles evolving in a symmetric
and spatially periodic potential is considered. The latter is exerted to a
time-periodic modulation of its inclination.
For directed particle transport mediated by rotating periodic motion, exact results
regarding the collective character of the running solutions are derived. Sec.~\ref{subsubsection:negative} reports on the
cooperativity-induced negative
mobility in the dynamics of
two coupled particles climbing in unison against the direction of the bias force. Sec.~\ref{subsubsection:adiabatic} deals with the
transport dynamics of systems with many degrees of freedom
in form of an extended linear
chains evolving in a washboard potential. By means of
adiabatically slow modulations of the potential
landscape the extended chain of coupled particles escapes from the confinement of a
potential well, and subsequently enters a regime of long lasting transients where the entire
chain transports in a ballistic fashion.

\subsubsection[Collective periodic running states in coupled particle dynamics]
{Collective periodic running states in coupled particle dynamics}\label{subsubsection:running}

The considered systems consist of two coupled particles each evolving in a washboard potential. The particles
are driven by a time periodic force and damped. These systems are nonautonomous and without conservation of energy.
They posses coupled inertial dynamics  of the form \cite{Mulhern.2013.EPJB}

\begin{eqnarray}
\ddot{q}_1&=&-\sin(2\pi q_1)-\gamma \dot{q}_1+\,F_1\sin(\Omega\, t+\theta_0)-\kappa(q_1-q_2)\,\,,\label{eq:q1}\\
\ddot{q}_2&=&-\sin(2\pi q_2)-\gamma \dot{q}_2+F_2\sin(\Omega\, t+\theta_0)+\kappa (q_1-q_2)\,\,,\label{eq:q2}
\end{eqnarray}

\noindent
where $\gamma$ is the strength of the damping, $F_{1,2}$ are the driving amplitudes,
and $\Omega$, $\theta_0$ are
the driving frequency and phase respectively. $\kappa$ represents the strength of the
linear coupling between the
two particles.

\begin{figure*}[ht!]
\centering
\includegraphics[scale=0.35]{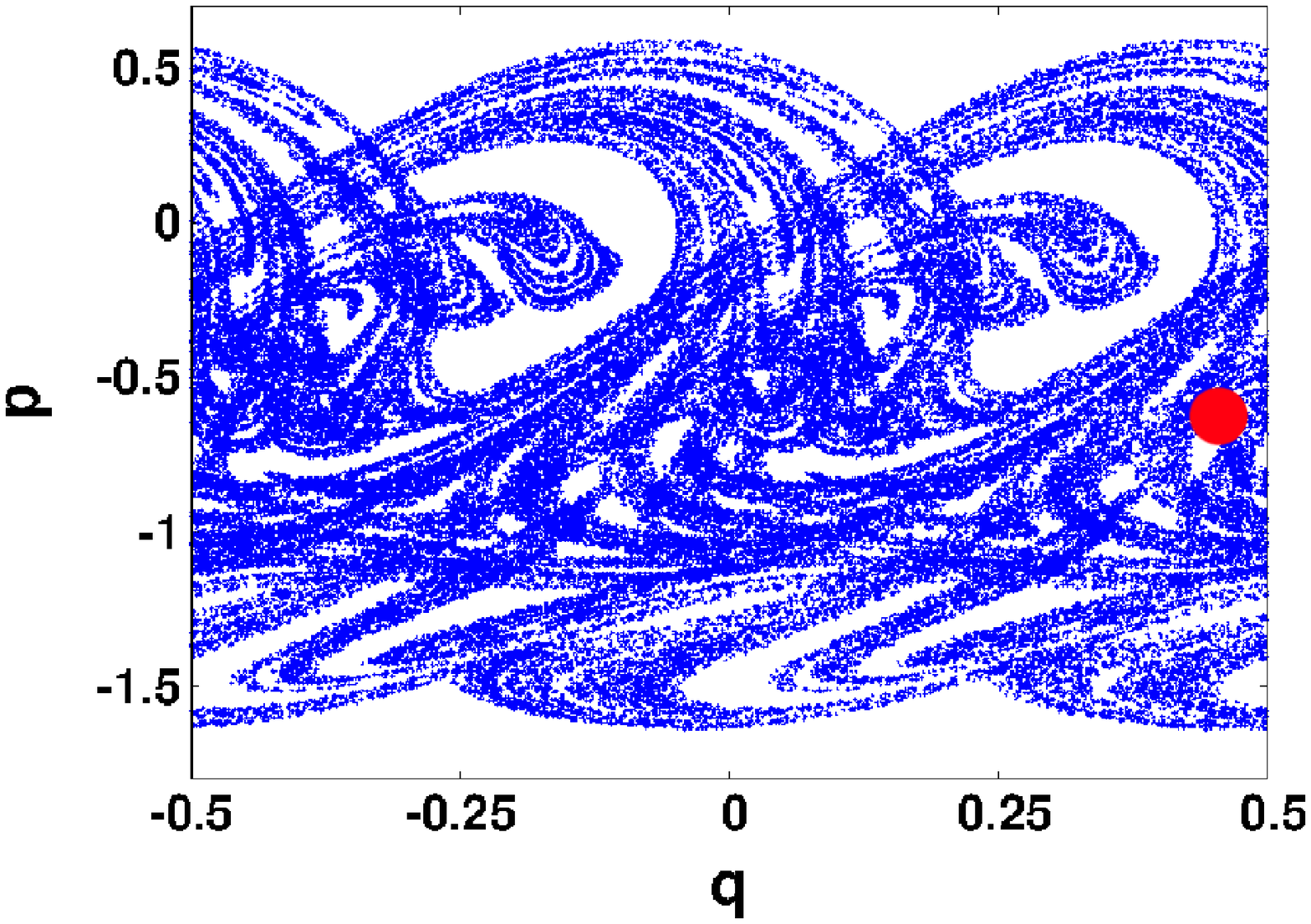}
\includegraphics[scale=0.35]{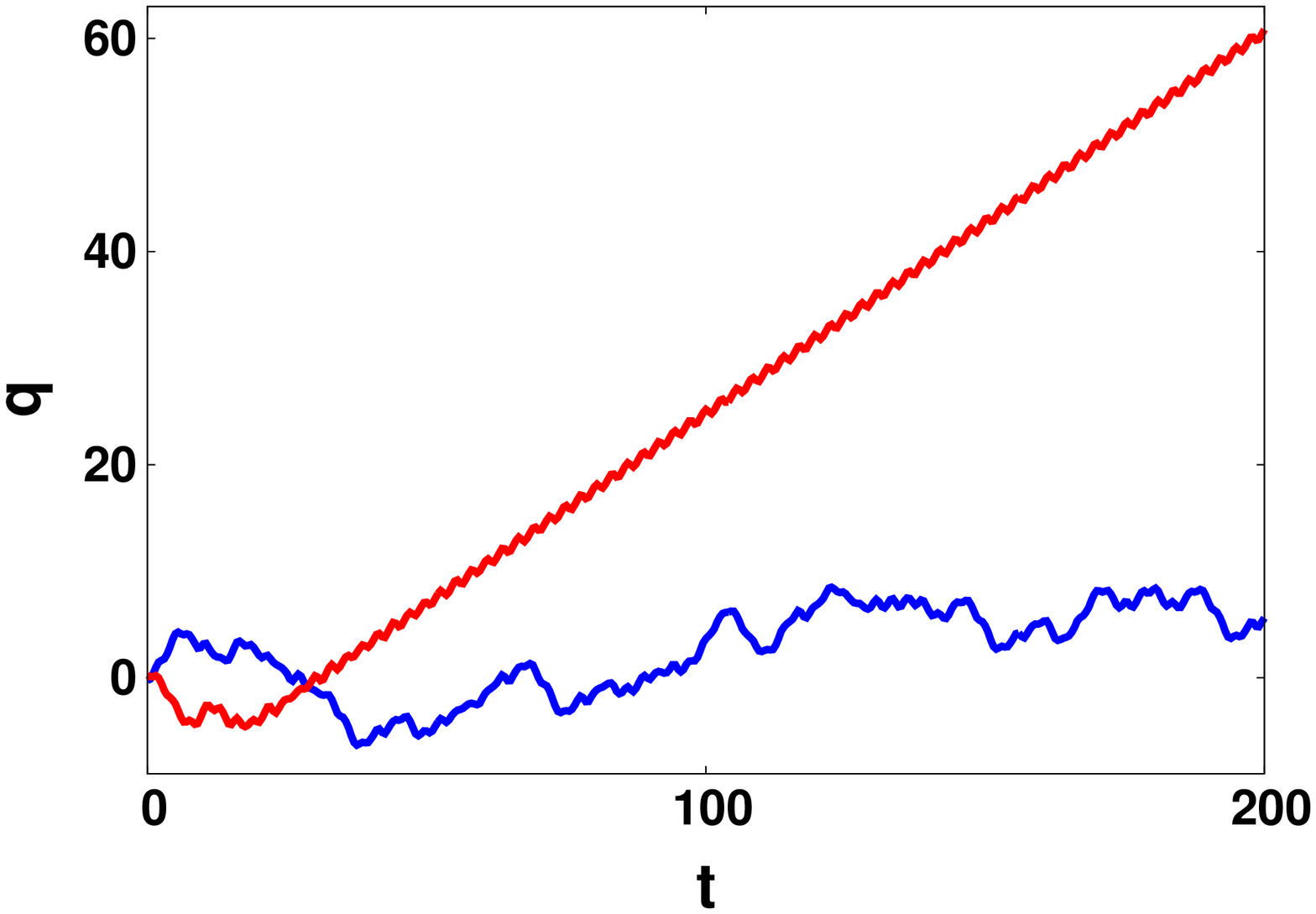}
\caption{(Colour Online) Deterministic driven and damped dynamics of two coupled units, as defined by
Eqs.~(\ref{eq:q1})-(\ref{eq:q2}). Left panel: Stroboscopic map with sampling rate $2\pi/\Omega$, with $\Omega$
being the frequency of the two external periodic driving forces. Shown
are the strange (blue)
and periodic (red) attractors corresponding to driving strengths $F_1=1.3$ and
$F_2=1.5$ respectively, in
the uncoupled $\kappa = 0.0$ regime. The remaining system parameters are
the damping strength $\gamma=0.1$, the driving frequency $\Omega=2.25$, and the driving phase $\theta_0 = 0$. The coordinates $q$ are
given $\mod(1)$. The dot (red) has
been enlarged for emphasis. Right
panel: A corresponding example trajectory for motion on the
strange (blue line) and periodic (red line) attractor. Source: Figure taken from Ref.~\cite{Mulhern.2013.EPJB}.}
\label{fig:attractors}
\end{figure*}

The focus of this section is on how the coupling strength influences the dynamics
of the system above
(Eq.~(\ref{eq:q1}) and Eq.~(\ref{eq:q2})). However, it is useful for what is to come
to understand the uncoupled
dynamics ($\kappa=0$) of this system \cite{Hennig.2009.PRE}. That is, the dynamics of a
single driven and damped particle
evolving in a washboard potential. The dissipative nature of this system means
that all orbits will eventually evolve =
to one of the systems, possibly coexisting, attractors. The type of attractors
present will depend on the parameters
used, while the particular attractor that an orbit evolves to is dependent on the
initial condition. This is
illustrated in Fig.~\ref{fig:attractors} where two qualitatively different attractors
are shown. The parameters
used are $\gamma=0.1$, $\Omega=2.25$, $\theta_0 = 0$. The strange chaotic attractor
(blue) results when the driving
amplitude is $F=1.3$. Increasing this driving amplitude to $F=1.5$ results in the
periodic attractor (red). With
regard to a net current the transporting orbits evolving on the periodic attractor
will yield a non-zero net current,
whereas the trajectories landing on the strange chaotic attractor are, on long time
scales, typically expected to produce
a vanishingly small contribution to the net current. However, as is seen later,
it is possible for trajectories
evolving on a chaotic attractor in a higher dimensional phase-space to produce a
non-zero current.

Thus, the single particle system exhibits a rich source of interesting dynamics.
The full system, Eq.~(\ref{eq:q1})
and Eq.~(\ref{eq:q2}), presents a further opportunity for new and interesting
behaviours. For now consider the
uncoupled case $\kappa=0$. Thus, depending on the strength of the driving
amplitudes $F_1$ and $F_2$ (with the
remaining parameters as given above), there are a number of possible combinations
of attractor for the underlying
subsystems. For driving amplitudes $F=1.3$ (strange chaotic attractor) and $F=1.5$
(regular attractor) there are
three combinations of attractors for the underlying subsystems - regular/regular,
regular/strange, and strange/strange.
The complexity of this system arises when the two subsystems are coupled, i.e $\kappa \neq 0$.

With a view to the discussion in Sec.~\ref{subsection:non-autonomous} regarding symmetries, it
is worth briefly exploring the
symmetry properties of this system which now includes driving and damping terms.
To begin, let us assume equal driving amplitudes $F_1=F_2$. Firstly note that the
particle exchange symmetry $q_1 \rightarrow q_2$, $q_2 \rightarrow q_1$ is preserved.
Thus synchronous solutions, for example, are permitted. However, time-reversibility
is broken here due to damping, meaning that these systems have a clear direction of time
(which is easily verified by applying the time-reversal operator to the equations of motion).
This is a generic feature of dissipative systems. The implication is that applying the
time-reversal operator to a \emph{forward} trajectory does not necessarily produce
another trajectory that is permitted under the equations of motion. Going further,
the corresponding \emph{backward} trajectory that would cancel the forward trajectories
contribution to the net current is not obtained through
the time-reversal operation. In this sense, all sets of initial condition are
be biased. This is only a necessary condition for the generation of a non-zero
net current. The actual current is determined by the basins of attraction
in which the initial conditions lie, and by extension their corresponding attractors.
Finally, in the case of unequal driving amplitudes $F_1 \neq F_2$ the above discussion
on time-reversibility is still valid. The difference now is that the particle exchange
symmetry is broken, and thus synchronous motion becomes hindered (except in the
range of very strong coupling between the units).

We  now look at the the transport
features present in the system.

\subsubsection{Features of collective driven transport} In this section we  examine a very particular type of solution for the full
system.
This solution, which is named the \emph{periodic running solution},
is desirable for achieving directed particle transport.

The periodic running solutions are characterised by

\begin{equation}\label{eq:prs}
q_i(t+T) = q_i(t) + m_i, \quad \dot{q}_i(t+T) = \dot{q}_i(t), \quad i = 1,2,
\end{equation}

\noindent
where $T$ is the duration of the period, and the $m_i$ are constants representing
the distance travelled over the course of a period.
This is a periodic running solution in that over one period the particles
each travel a uniform distance, and yet the momentum variables are
periodic in the standard sense. Notice that such a solution has
a non-zero average velocity. That is

\begin{equation}
 \langle \dot{q}_i\rangle =\frac{1}{T}\int_{0}^{T}dt\, \dot{q}_i(t)=\dfrac{m_i}{T}\ne 0\,.\label{eq:average}
\end{equation}

\noindent
Thus particle $i$ runs to the right (left) when $m_i>0$ ($m_i<0$), for $i=1,2$.
For the dimer the net transport may be zero if $m_1 + m_2 = 0$, that is $m_1 = - m_2$ and
the particles run in a counterpropagating fashion with the same average velocity.
However, this solution is not possible.

The main results regarding the character of periodic running solutions are contained in a theorem in
\cite{Hennig.2011.Chaos} and can be summarised as follows:
By assuming such a solution, described by Eq.~(\ref{eq:prs}),  it has been possible to deduce
important features of transport when the particles are in a periodic
running state. Firstly, in such a state the particles will travel an
equal distance and counterpropagating trajectories are excluded,
i.e. the particles must travel in the same direction, viz. $m_1=m_2$. Further,
non-trivial periodic solutions are impossible without the
time-periodic external modulations. That is, all periodic running
solutions must be frequency locked to certain multiples of the external time-periodic modulations.
In more detail, let $F(t) = F\sin(\Omega t+\Theta_0)$, with period $T_0 = 2\pi/\Omega$,
represent the external time-periodic driving of the system.
For solutions that are frequency-locked to $F(t)$, it holds that the distance between
the particles performs periodic oscillations, i.e.

\begin{equation}\label{eq:per_diff}
q_1(t+T) - q_2(t+T) = q_1(t) - q_2(t)
\end{equation}

Moreover, the period $T$ is determined by

\begin{equation}
T = 2lT_0
\end{equation}
\noindent
for some $l \in \mathbb{Z}$. The coordinates obey
\begin{equation}
q_1\left(t+\frac{1}{2}T\right)=q_2(t)+k\,,\,\,\,q_2\left(t+\frac{1}{2}T\right)=q_1(t)+k\,,\label{eq:distance1}
\end{equation}
with $k\in \mathbb{Z}\setminus \{0\}$ and hence
\begin{equation}
q_i(t+T)=q_i(t)+2k\,,\,\,\,i=1,2\,.\label{eq:distance2}
\end{equation}

The existence of periodic running solutions is remarkable as with the help of a multidimensional Melnikov method it can
be proven that the system exhibits Smale horseshoe chaos in its dynamics and hence, is nonintegrable
(for details see \cite{Hennig.2011.Chaos}).

\begin{figure}
\centering
\includegraphics[scale=0.3]{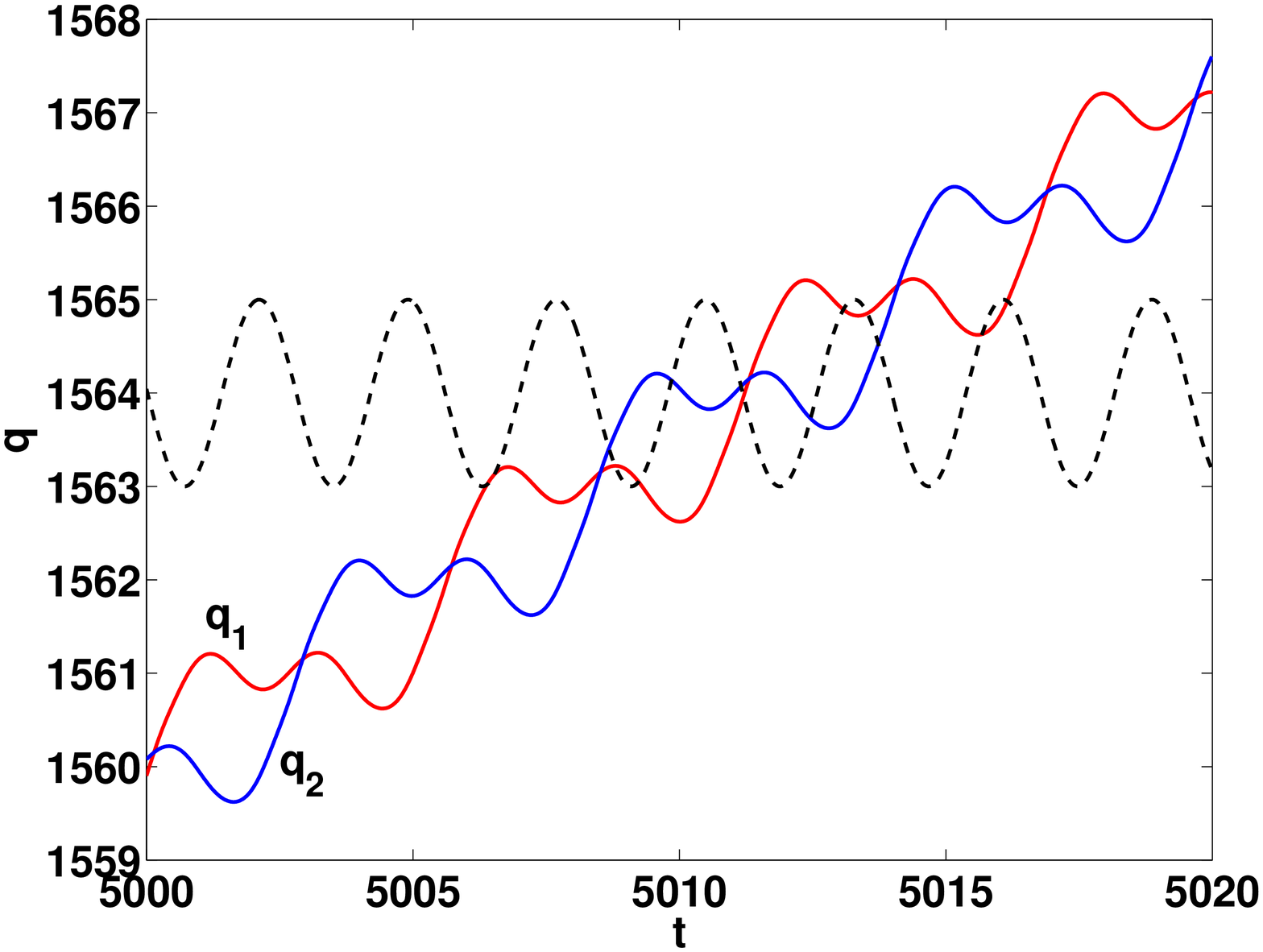}
\caption{(Colour Online) Time evolution of the coordinates $q_{1,2}(t)$ of the two interacting driven and damped units in a periodic
running
state that is frequency-locked to the external
time-periodic modulation. The system dynamics are defined by Eqs.~(\ref{eq:q1})-(\ref{eq:q2}). The parameter values are given by
$\Omega=2.25$, $F_1=F_2=1.3$, $\theta_0=0$, $\gamma=0.1$, and $\kappa=0.46$.
For comparison, $\sin(\Omega t)$ oscillating around $q=1564$
with unit amplitude and frequency $\Omega=2.25$ is shown (dashed line). Source: Figure taken from Ref.~\cite{Hennig.2011.Chaos}.}
\label{fig:running}
\end{figure}

Let us look closer at the periodic running solutions described by Eq.~(\ref{eq:prs}).
It is clear that in the uncoupled regime, $\kappa=0$, a periodic running solution
is not possible given that both particles have trajectories that evolve on strange attractors.
Turning to the coupled regime, notice then that in the case of identical
particle motion, or in-phase motion, the particles effectively decouple.
Thus the dynamics is determined by two independent units which evolve
on strange chaotic attractors. This holds for all values of $\kappa$ showing
that transporting in-phase (synchronous) motion supported by a regular periodic
attractor is excluded. However, this does not exclude the possibility of a periodic
running solution where the two particles have motions that are out of phase. Such
a solution can be seen in Fig.~\ref{fig:running} where the time evolution of the
coordinates $q_{1,2}$ is shown, and the coupling strength is chosen as $\kappa=0.46$.
It is apparent that both are frequency locked
to the external driving. This is also illustrated in Fig.~\ref{fig:running} by the
inclusion of the function $F(t)=\sin(2.25\, t)$ oscillating around $q=1564$.
This figure highlights the {\it cooperation} between the particles which allows directed
transport to take place. One unit will move backward in order for both units to move forward.
In an alternating manner, one unit will sacrifice for the benefit of the dimer.

Importantly, from Fig.~\ref{fig:running} it can be deduced
that the temporal behaviour of the coordinates follows the
relations given in Eqs.~(\ref{eq:per_diff})-(\ref{eq:distance2}).

Concerning the influence of the coupling,
let us take as example coupling strengths $\kappa=0$, where there exist strange
attractors for both particles, and $\kappa=0.46$, where the motion can
be periodic, it seems that
the coupling strength plays a key role for the dynamics of the system.
The fact that the system exhibits rich and complex dynamics is expected
given that the phase-space is five dimensional. Conversely, the fact that
the two chaotic subsystems can combine to produce regular periodic motion
is less expected.
This is quite important as it confirms that there is a phenomenon in which
\emph{chaos combines with chaos to form regular periodic motion}. By this, it
is meant that in the uncoupled regime both subsystems support chaotic motion,
but in the coupled regime there exist periodic solutions.

With regard to the emergence of a non-zero current, the favourable coupling strengths
are those that are
associated with motion on periodic attractors in phase space.
It may be the case though that there are multiple coexisting attractors, each contributing to overall
dynamics with a different weight, and it is these weights that determine the strength of the
resulting net current. These weights can of course be related to the size of the corresponding
basis of attraction \cite{Hennig.2011.Chaos}. Conversely, there exist values of the coupling strength
yielding symmetry between coexisting attractors which take trajectories in opposite directions.
Thus an equal number of trajectories travel in the range of positive and negative coordinates, i.e.
the basins of attraction, promoting motion in different directions, are of the same size. This results
in a zero current.

\FloatBarrier
\subsubsection{Interaction-induced negative mobility}\label{subsubsection:negative}

Here we review work on the driven and damped dynamics of two coupled units evolving in a symmetric and
periodic substrate potential which is subjected to a static bias force of magnitude $F_0$ serving
to tilt the potential landscape such that unit motion to the right is favoured \cite{Mulhern.2011.PRE}.
In addition, each unit is driven by an external time-dependent modulation of amplitude $F$, frequency $\Omega$,
 and phase $\theta_0$ with the same magnitude,
but out-of-phase, to its counterpart. It is shown that, within a range of coupling strengths,
the coupled units can become self-organised and go, as periodic running states, frequency
locked with the driving, against the direction of the bias force.

The equations of motion for this system are given by \cite{Mulhern.2011.PRE}

\begin{equation}
 \ddot{q_{1}} = -\sin(2\pi q_{1}) - \gamma\dot{q_{1}} - F\sin(\Omega t + \theta_{0}) - \kappa(q_1 - q_2) + F_0,
\label{eq:e.o.m1a}
\end{equation}
\begin{equation}
 \ddot{q_{2}} = -\sin(2\pi q_{2}) - \gamma\dot{q_{2}} + F\sin(\Omega t + \theta_{0}) + \kappa(q_1 - q_2) + F_0.
\label{eq:e.o.m1b}
\end{equation}

\noindent Notice the out-of-phase character of the periodic modulation of the two units expressed by
 the different sign of the modulation amplitude $F$. The additional parameters $\gamma$ and $\kappa$ regulate the
 strength of the damping and coupling respectively.

We stress that in our system the two units, forming a dimer, are supposed to perform one-dimensional motion in parallel
 directions each of them in a  washboard potential. That is, for equal coordinates $q_1=q_2=q$ the axis of the
 dimer (virtual line connecting the two units) is perpendicular to the $q-$direction.

In general, the system will exhibit a rich and varied behaviour as a function of its parameters
$\gamma$, $F$, $\kappa$, $\omega$, $\theta_0$, and $F_0$. However, one of our main objectives,
as was previously mentioned, is to explore coupling phenomena within the system, and therefore
 in much of this study we will fix the remaining parameters
while varying the coupling parameter $\kappa$.

\begin{figure}
\begin{center}
\includegraphics[height=5cm, width=6.5cm]{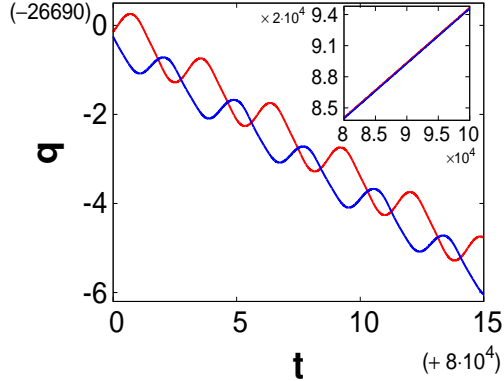}
\caption{(Colour Online) Example trajectories of two driven and damped units in a biased periodic potential with dynamics given by
Eqs.~(\ref{eq:e.o.m1a})-(\ref{eq:e.o.m1b}) (red and blue curves respectively). Inset: Time evolution of the two uncoupled units 
($\kappa = 0$),
travelling with the bias force. Although not visible on the scale of this figure, the two curves do not overlap due to the 
individual units experiencing different initial transient dynamics before they enter the periodic regime with motion in the 
direction of the bias force. Main 
panel: Time evolution of the two coupled units ($\kappa = 0.327$)
 which are travelling in the opposite direction to the bias force. The two lines represent the trajectories
 of the individual units. The remaining parameters are as follows: the constant external bias force $F_0 = 0.1$, damping $\gamma =
0.11$, periodic driving strength $F = 1.3$, driving frequency $\Omega = 2.22$, and driving phase $\theta_0 = 0$. Note the different
time scales. Source: Figure adapted from Ref.~\cite{Mulhern.2011.PRE}.} \label{fig:climbing}
\end{center}
\end{figure}

Fig.~\ref{fig:climbing} displays the existence of negative mobility,
i.e. solutions in which the motion goes against the direction of bias. Below we demonstrate that the mechanism
allowing for such motion is cooperation between the units where they pull each other over consecutive
potential barriers. In more detail, a coordinated energy exchange between the units allows them to
collectively climb against the direction of the tilt.

\begin{figure}
\centering
\includegraphics[height=2.6cm, width=3.9cm]{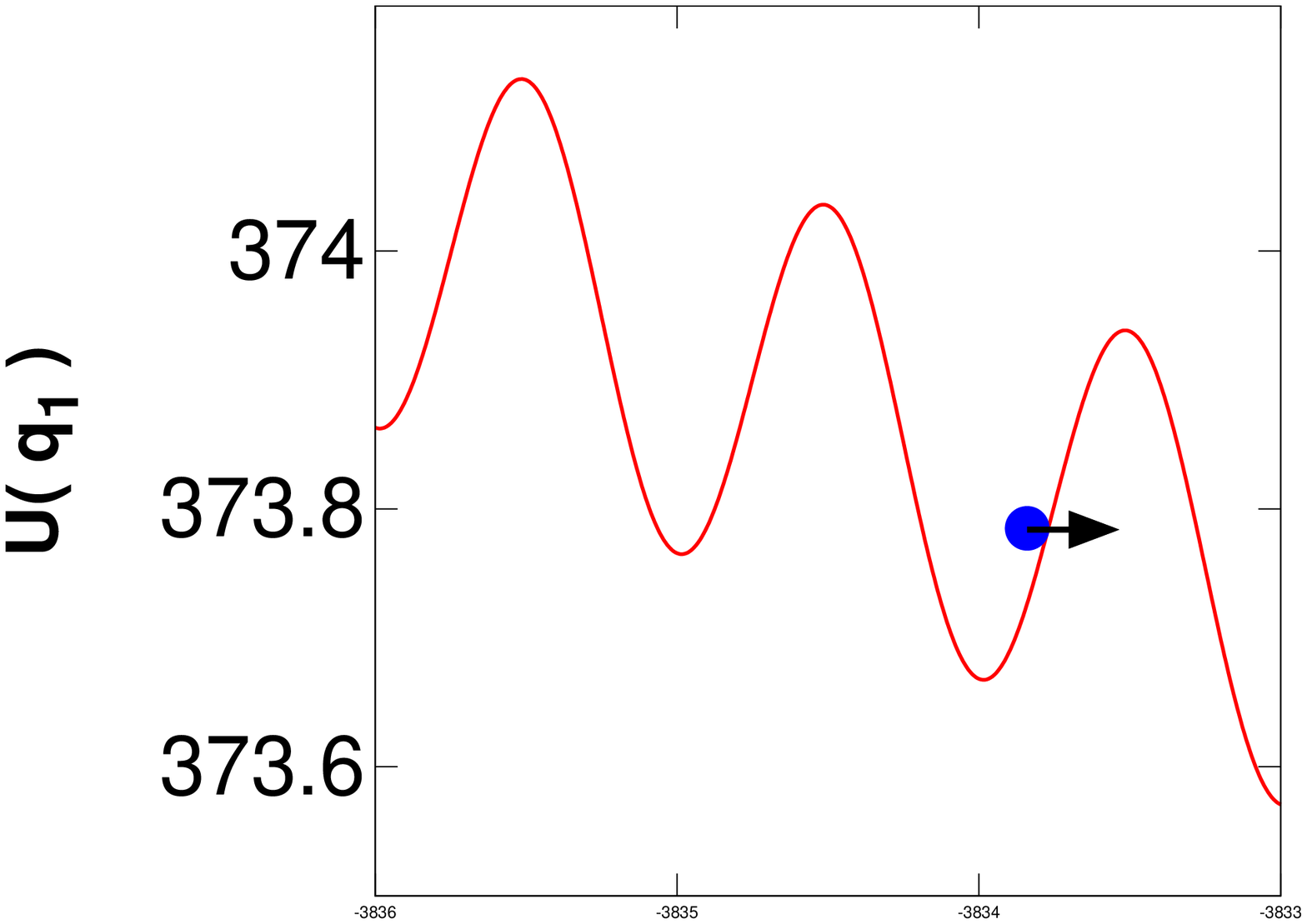}
\includegraphics[height=2.6cm, width=3.9cm]{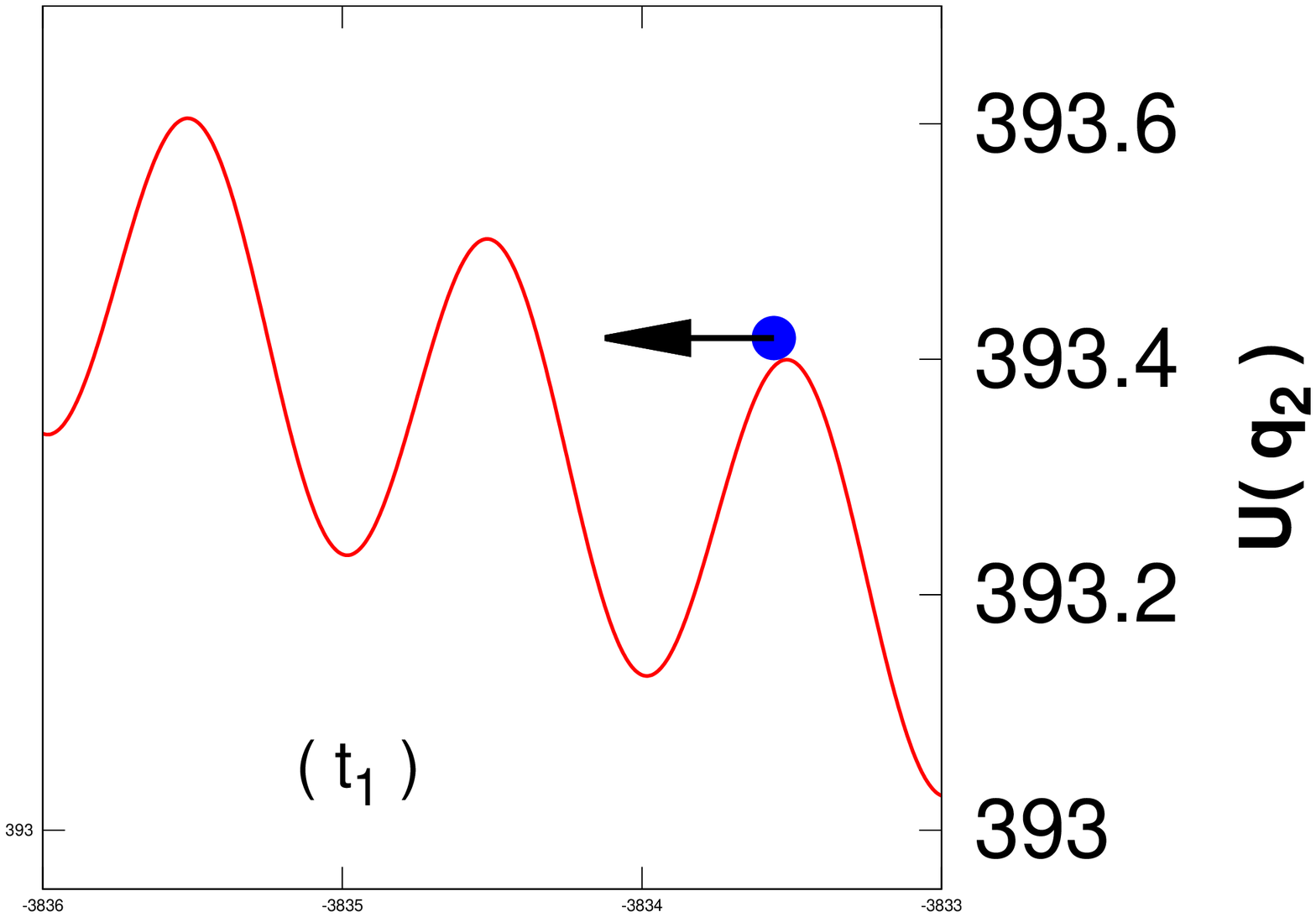}\\
\includegraphics[height=2.6cm, width=3.9cm]{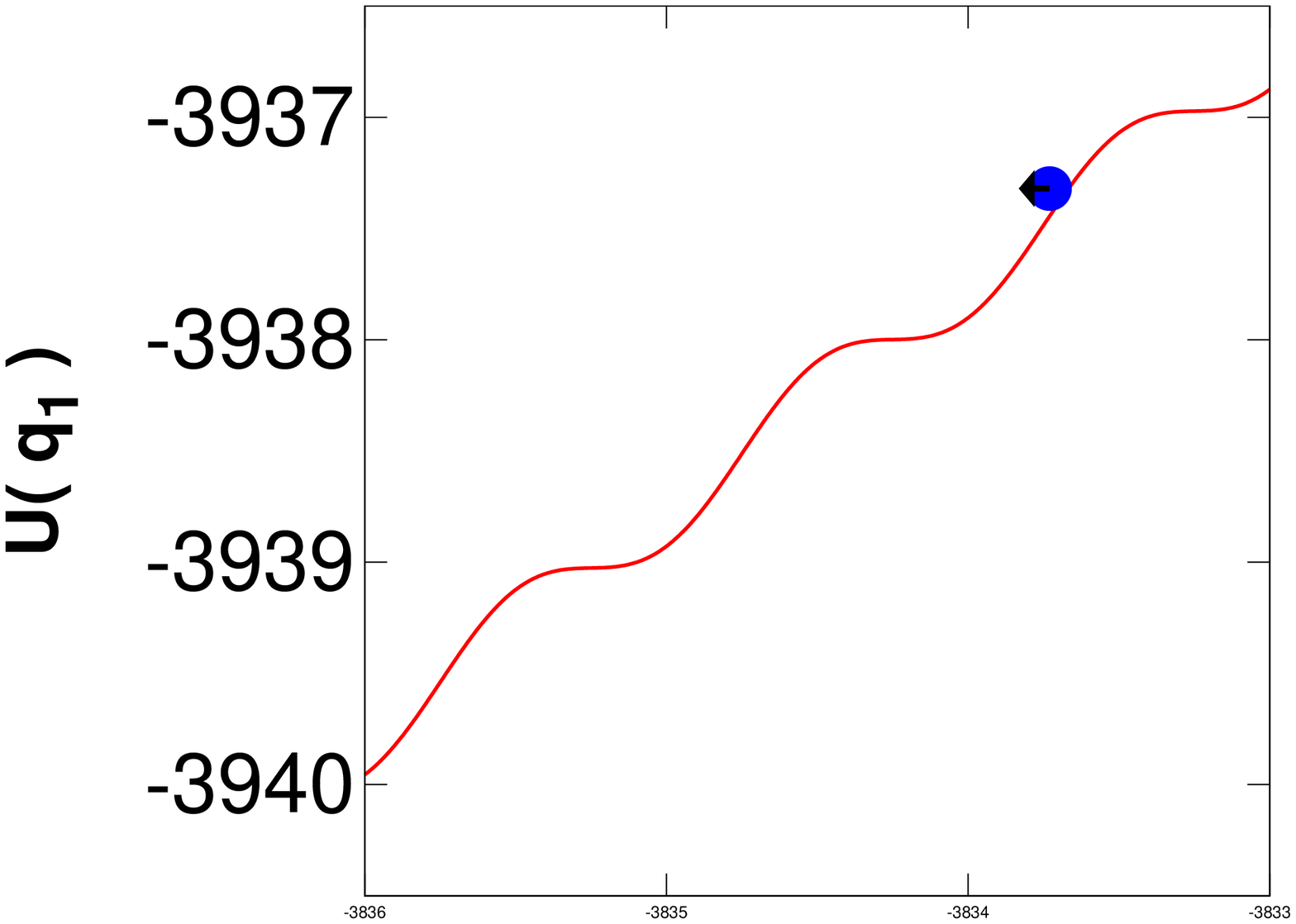}
\includegraphics[height=2.6cm, width=3.9cm]{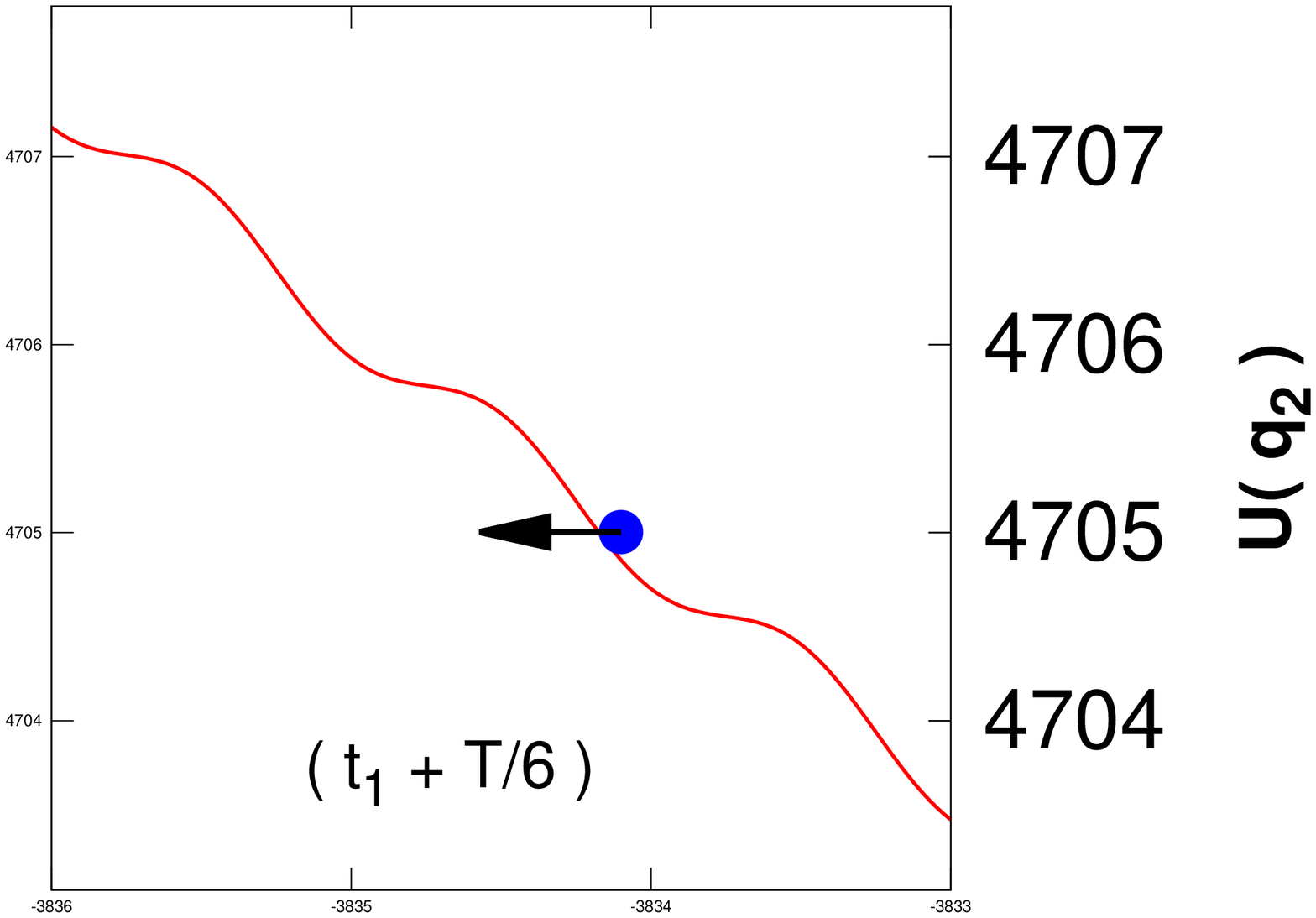}\\
\includegraphics[height=2.6cm, width=3.9cm]{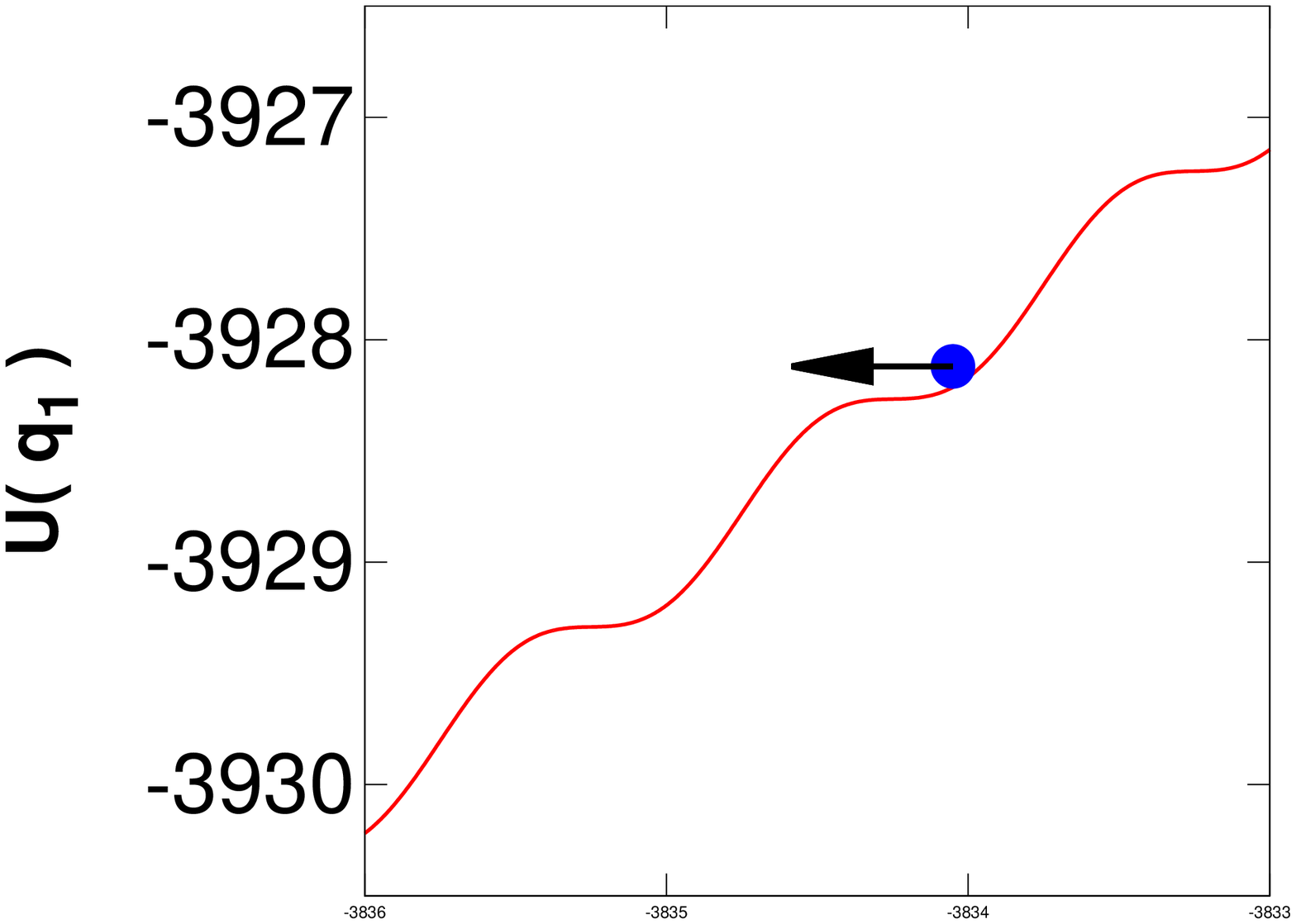}
\includegraphics[height=2.6cm, width=3.9cm]{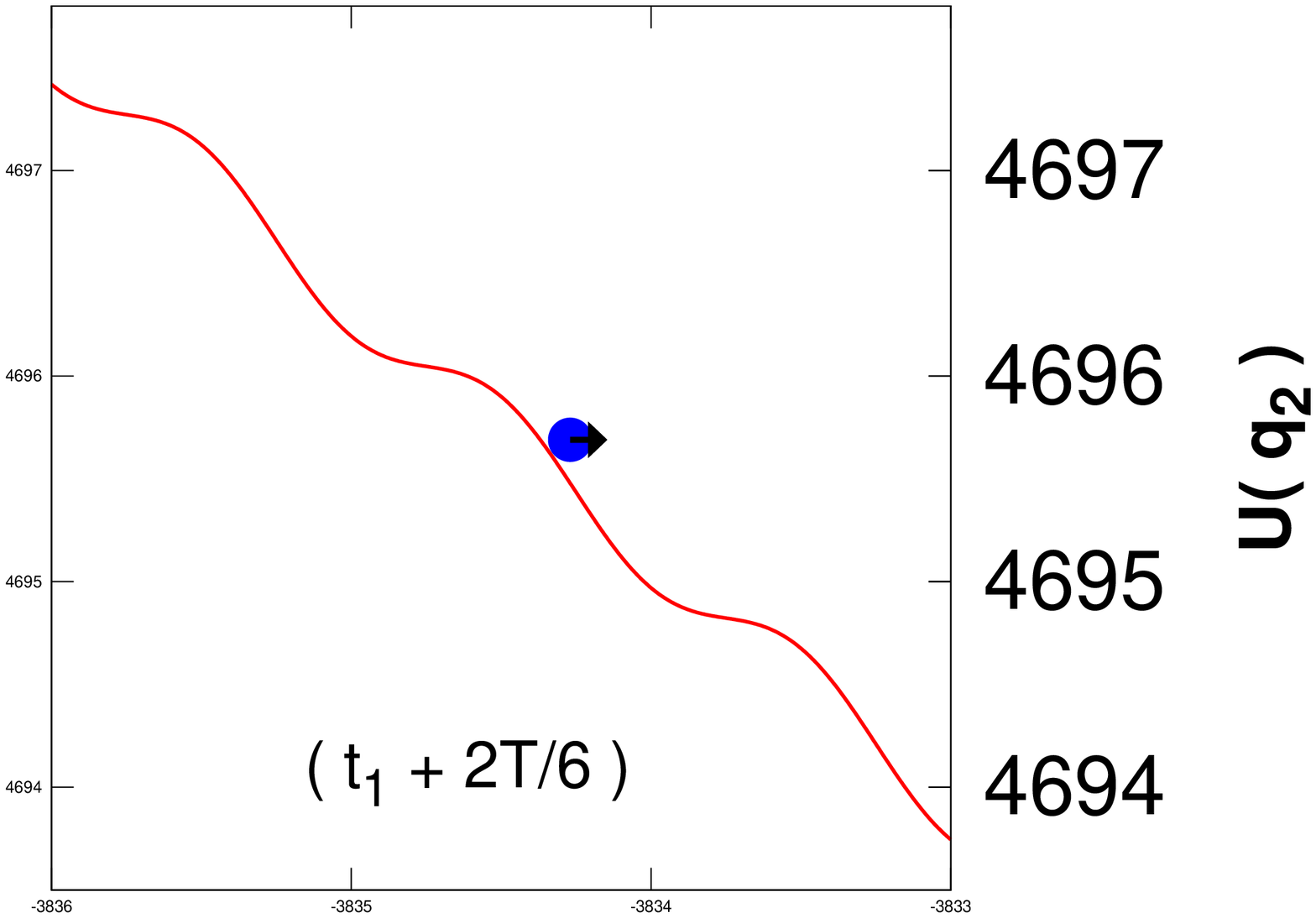}\\
\includegraphics[height=2.6cm, width=3.9cm]{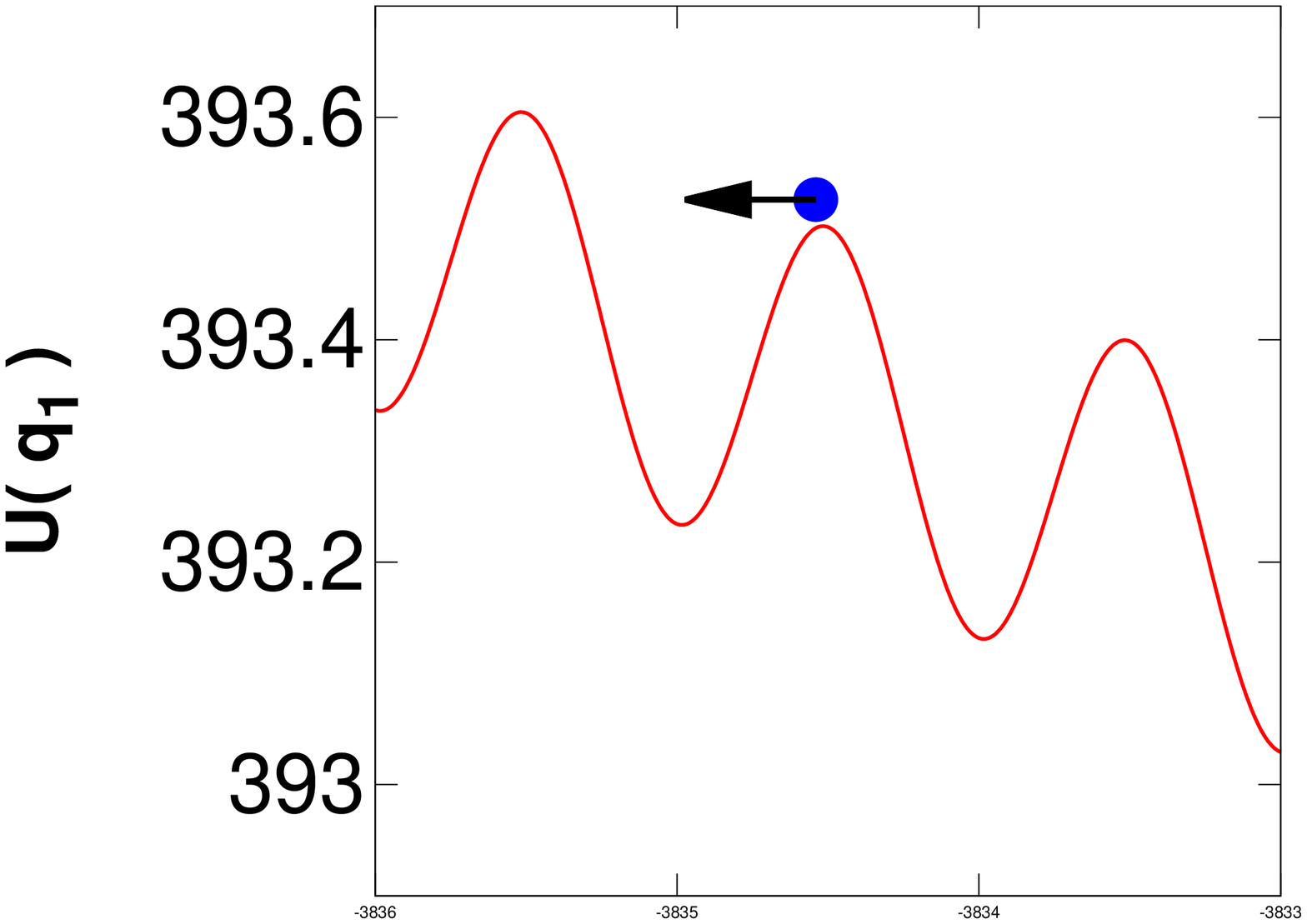}
\includegraphics[height=2.6cm, width=3.9cm]{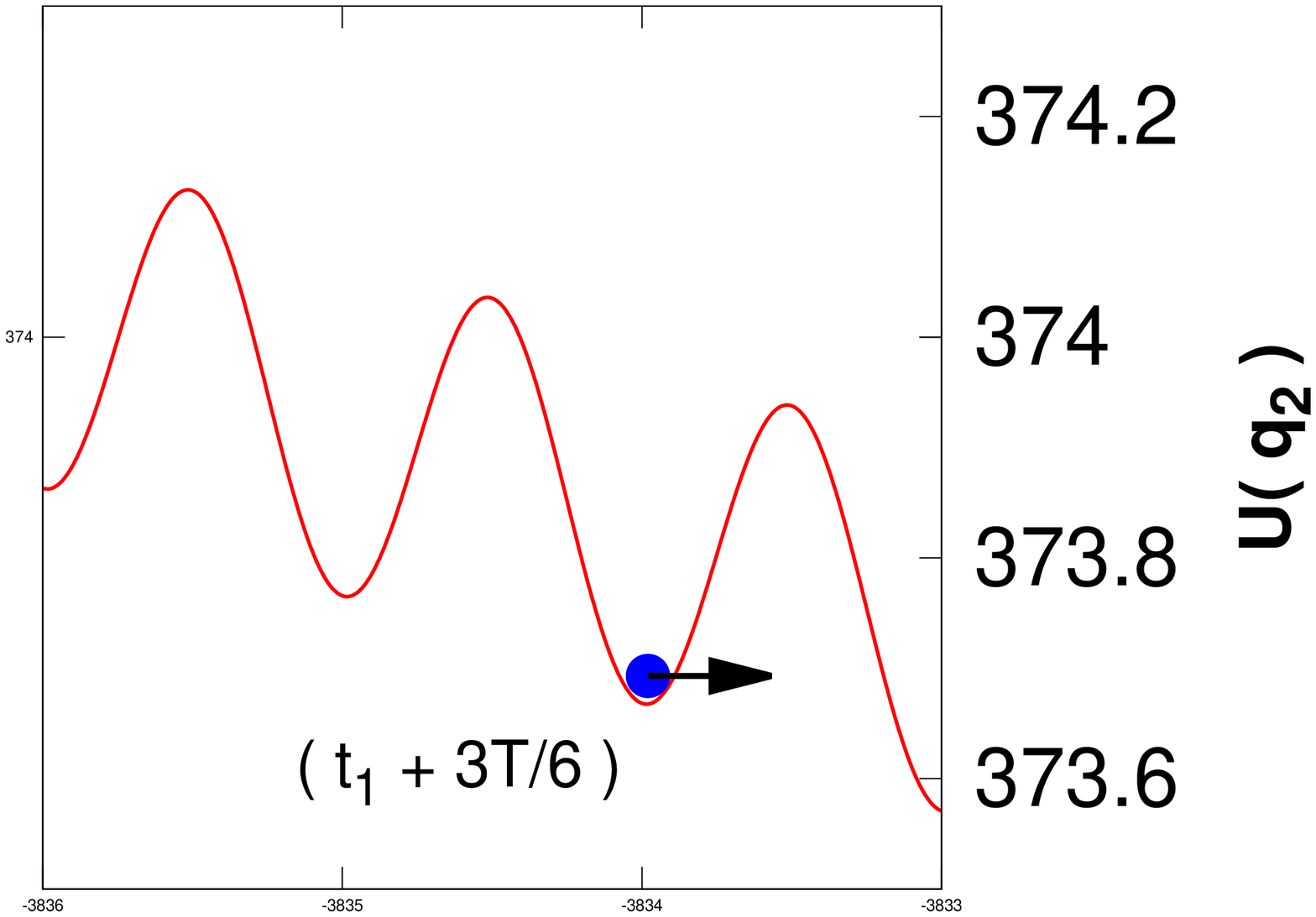}\\
\includegraphics[height=2.6cm, width=3.9cm]{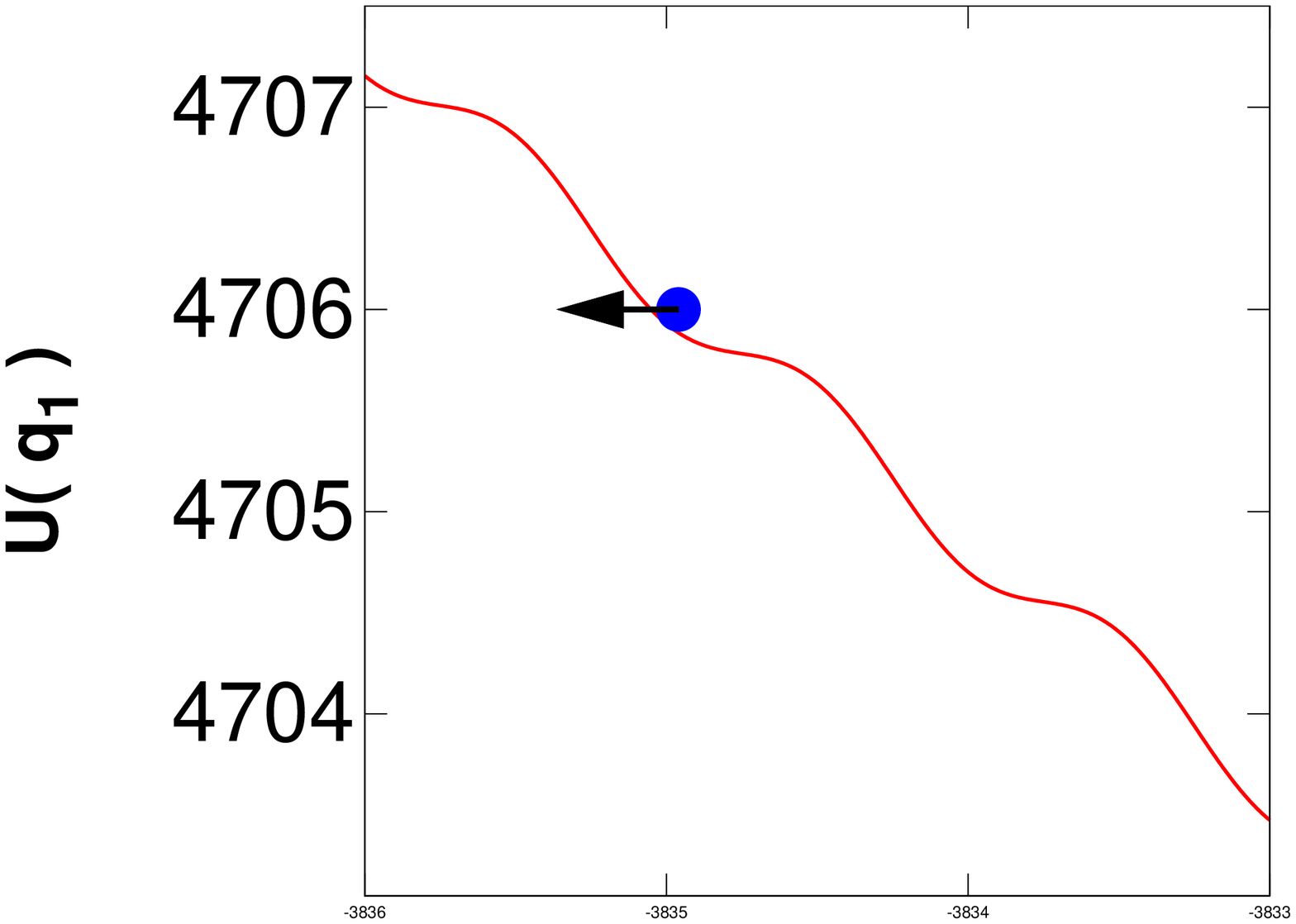}
\includegraphics[height=2.6cm, width=3.9cm]{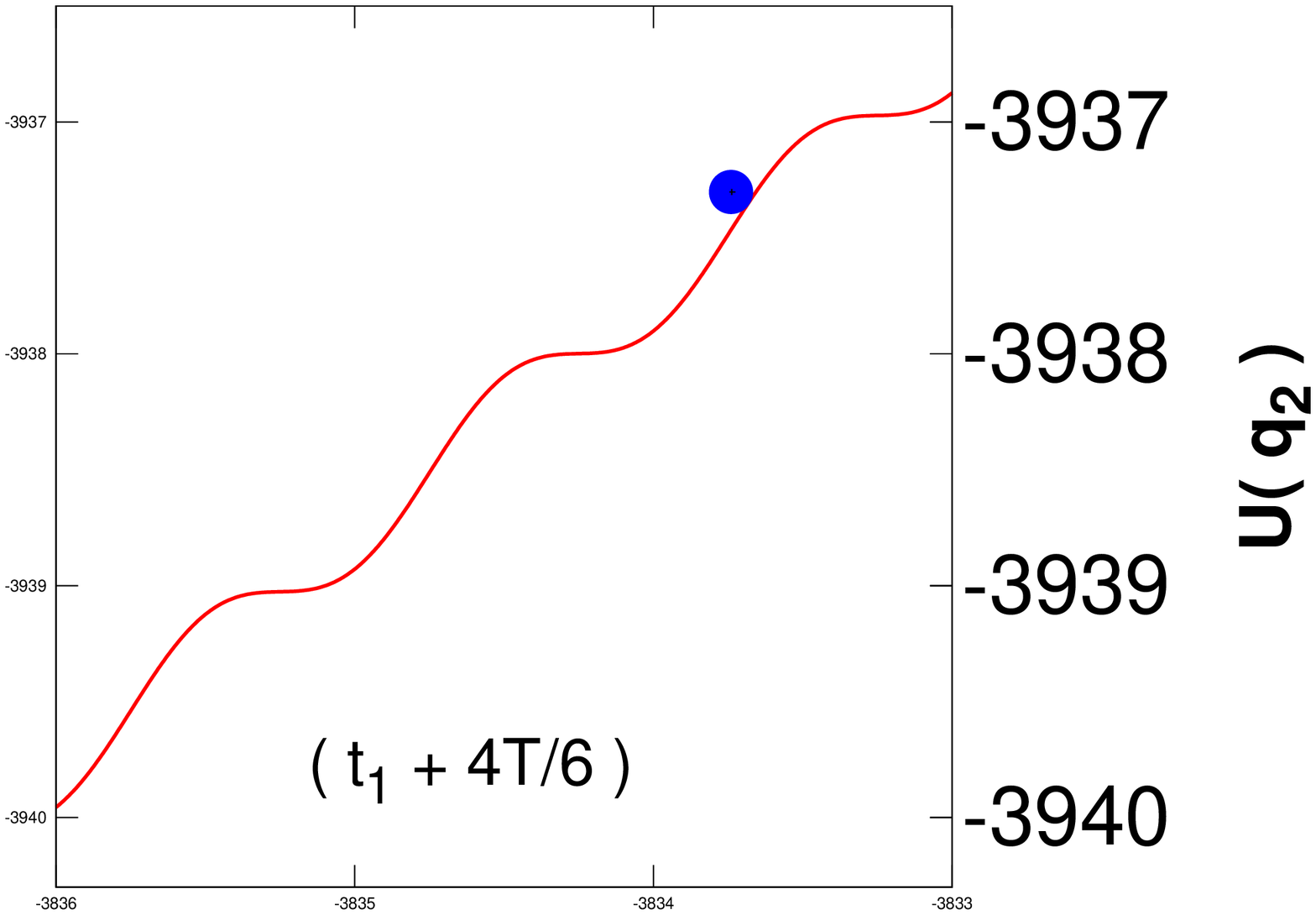}\\
\includegraphics[height=2.6cm, width=3.9cm]{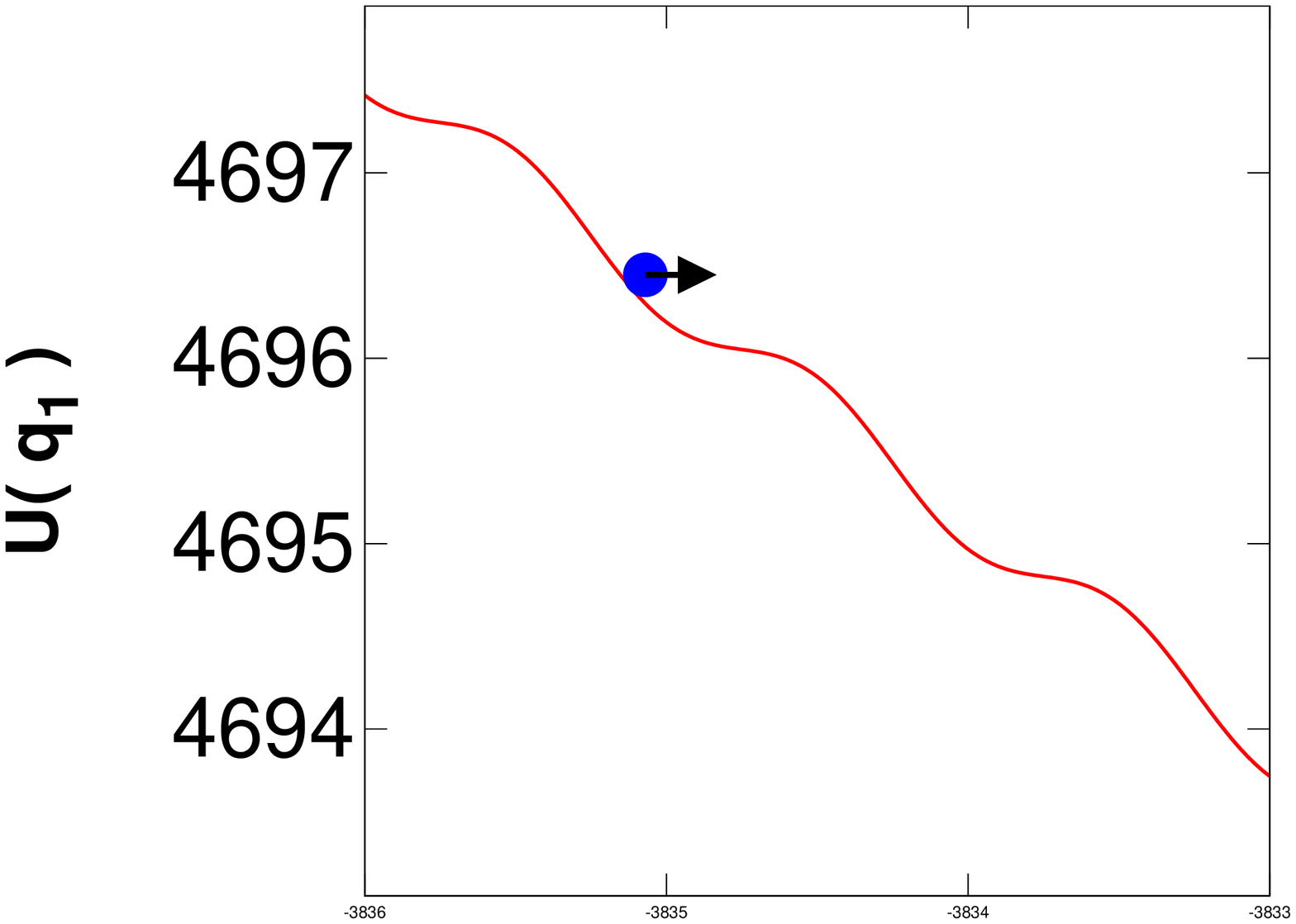}
\includegraphics[height=2.6cm, width=3.9cm]{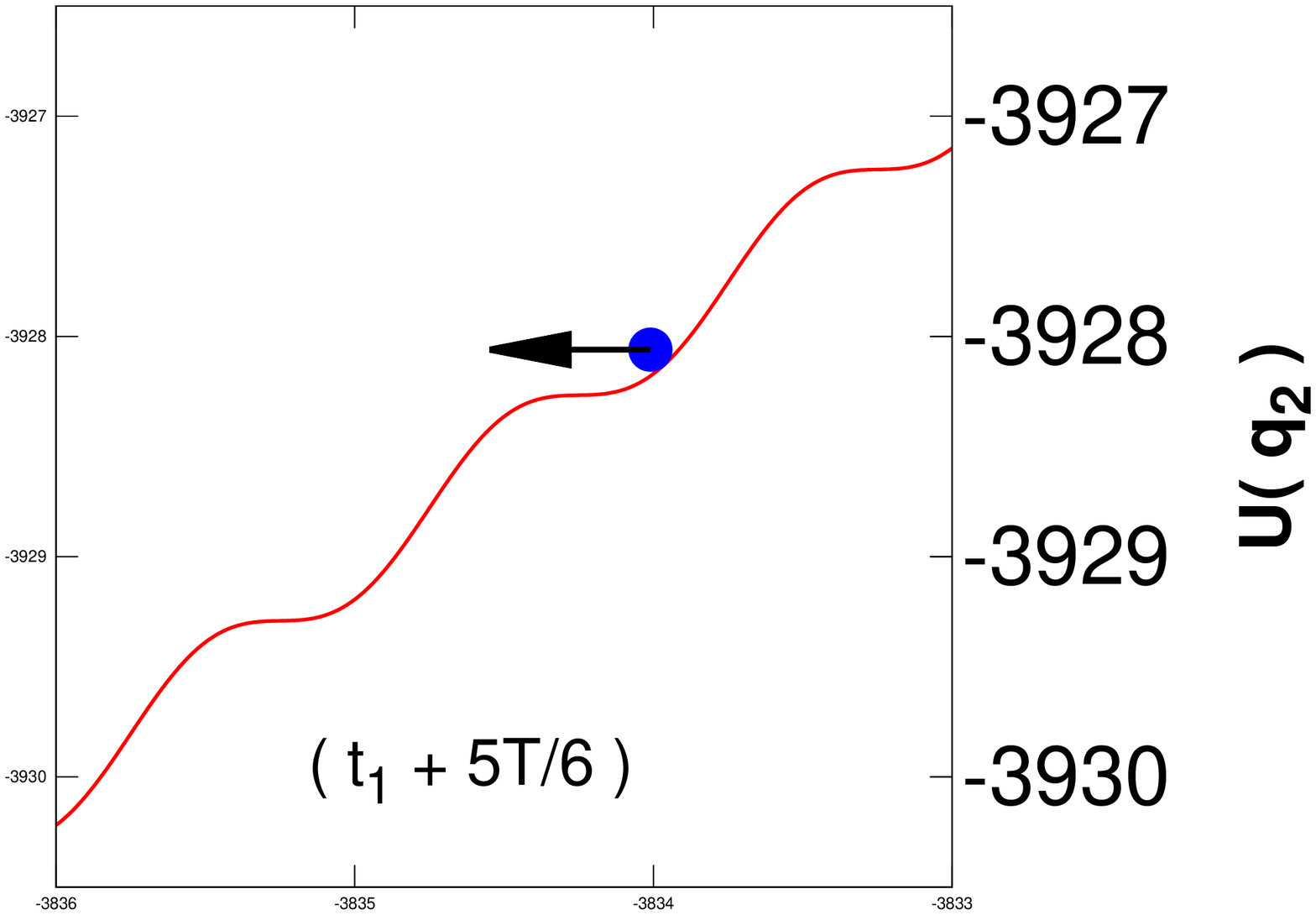}\\
\includegraphics[height=2.6cm, width=3.9cm]{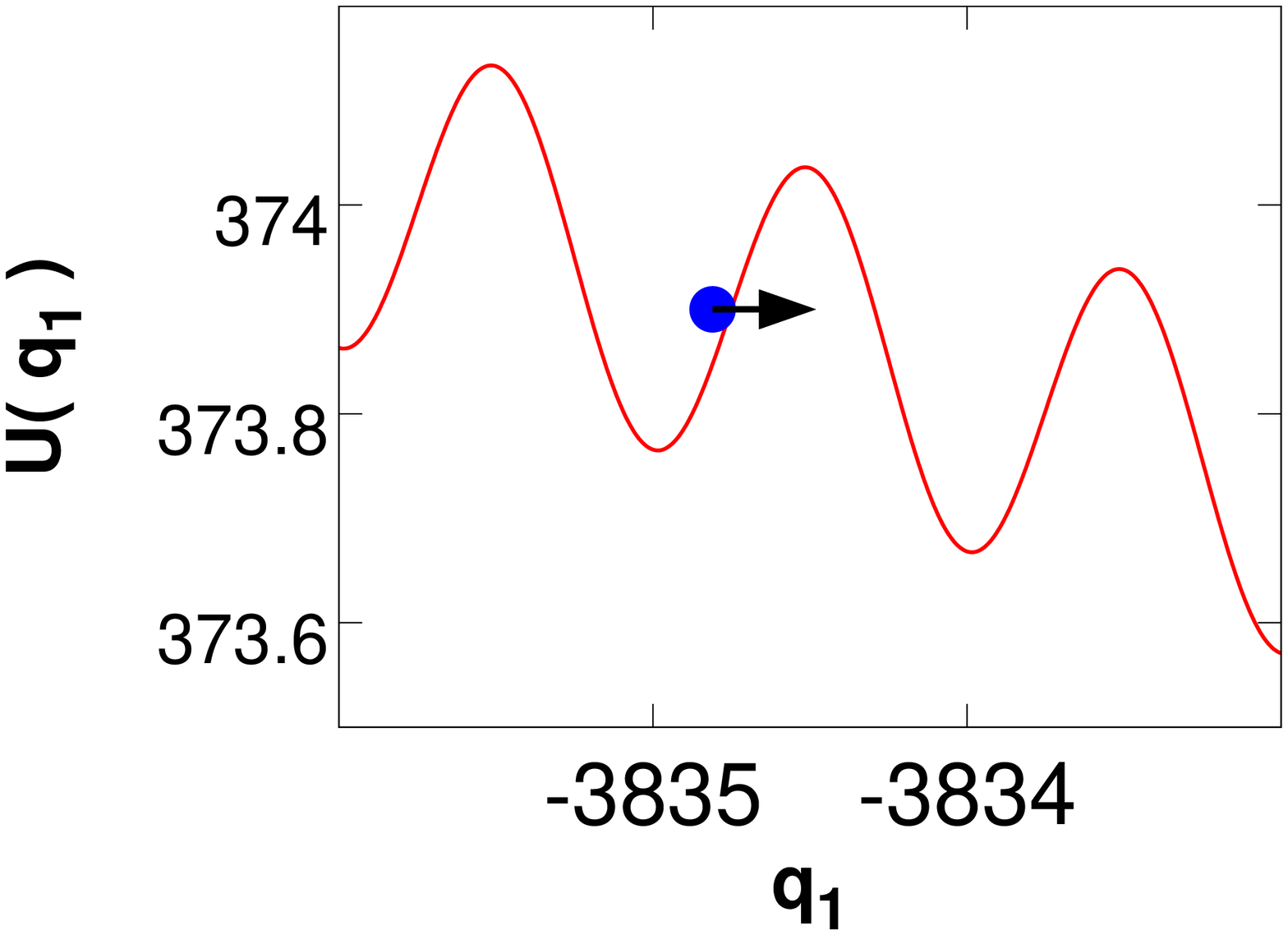}
\includegraphics[height=2.6cm, width=3.9cm]{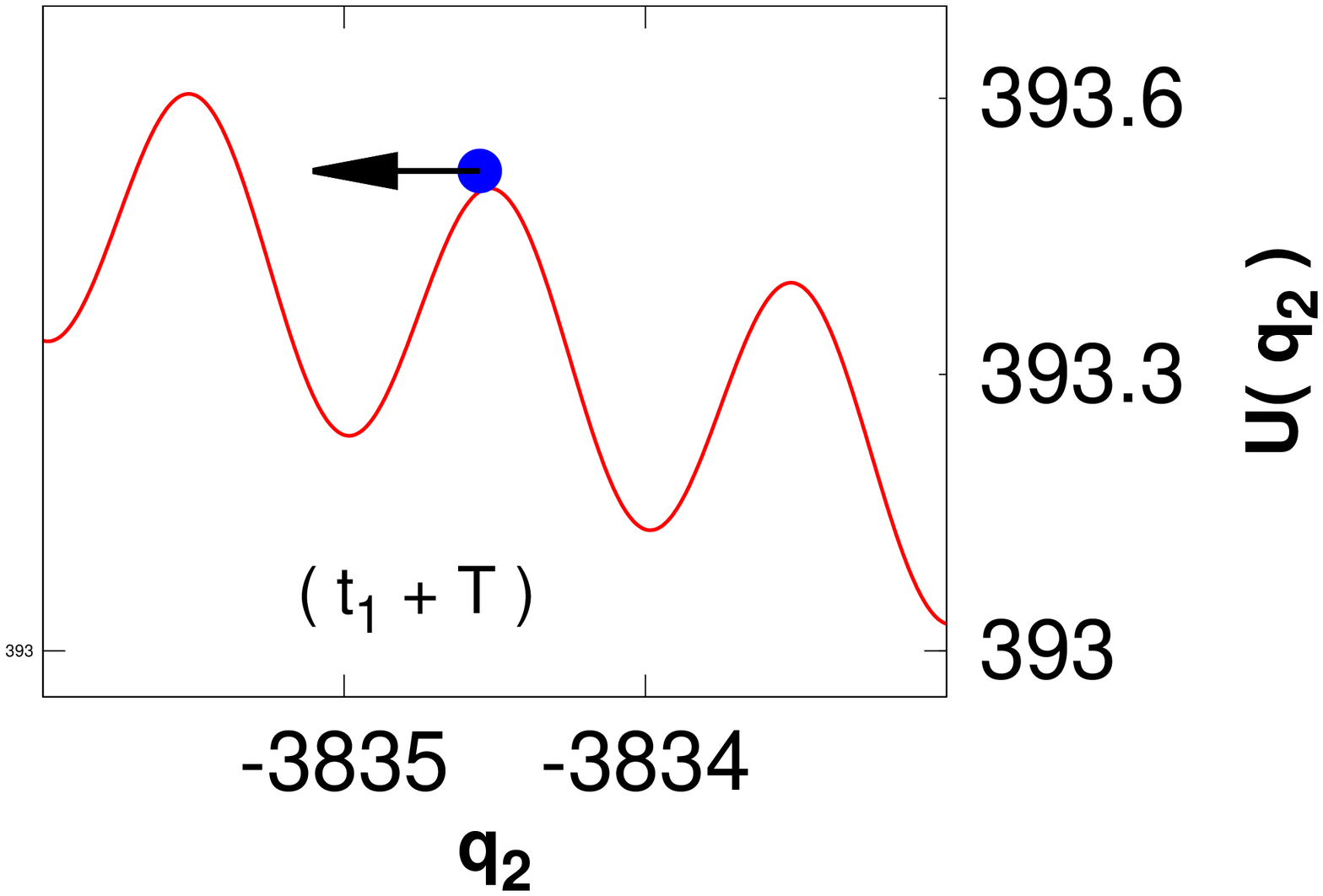}

\caption{(Colour online) Snapshots of driven and damped dimer motion going against the bias force, taken over a period $T$ of the
 external time-periodic modulation. The corresponding equations of motion are Eqs.~(\ref{eq:e.o.m1a})-(\ref{eq:e.o.m1b}). Reading from
top to bottom, each row shows the evolution advanced by a time $T/6$ starting from an initial time $t=t_1$. The bottom
axis is $q$ and the left/right axes $U(q)$. Arrows indicate direction and magnitude of a units momentum.
Left (right) column: unit $1$ ($2$). The parameters are set as $F_0 = 0.1$, $\gamma = 0.11$, $F = 1.3$, $\Omega = 2.22$,
$\theta_0 = 0$, and $\kappa = 0.372$. Source: Figure adapted from Ref.~\cite{Mulhern.2011.PRE}.}\hfill\label{fig:snaps}
\end{figure}

\noindent 
As an explanation of this phenomena, we describe the mechanism that makes negative mobility possible.
 Fig.~\ref{fig:snaps}
 shows snapshots of a two-unit compound moving in the opposite direction of the bias force
where the units move in the respective potential landscape given by
\begin{eqnarray}
U(q_{1},t)&=&\frac{1}{2\pi}\big(1-\cos(2\pi q_{1})\big)+F\sin({\Omega \,t+\theta_0}) q_{1}-F_{0}q_{1}\,,\\
U(q_{2},t)&=&\frac{1}{2\pi}\big(1-\cos(2\pi q_{2})\big)-F\sin({\Omega \,t+\theta_0}) q_{2}-F_{0}q_{2}\,.
\end{eqnarray}
(Note that the potential energy as given above
relates to the on-site potential not containing the unit interaction part).
 These seven snapshots, taken over one period of the driving, show the relative position of each unit
 (henceforth unit $1$ - left panels and unit $2$ - right panels) versus its position in the potential landscape.
In addition, arrows where present indicate the direction and magnitude of momentum for the respective units, with no
 visible arrow indicating a vanishing momentum. It can be seen that negative mobility is the product of coupling between
 the units and the effect of the time modulated potential. For example, at the beginning of the period unit $1$
has a positive momentum in the direction of the bias force. However, this is countered by the height of the potential
 barrier, and by the coupling to unit $2$ which has an even stronger negative momentum. Thus motion in the
direction of bias is hindered.
Regarding negative mobility we underline that the opposite time-periodic forces make it only
possible that for one unit the current inclination of the
washboard potential is of such form that the unit is temporarily
locked in a potential well (thus hampering its dragging influence on the other unit in the unwanted direction
 of the tilt) while the other
 unit experiences a washboard potential whose current inclination favours motion against the static
tilt force $F_0$. These
phases of temporary locking and running against the tilt alternate between
the units.
This cooperative effect between the units, together with the finely tuned modulations of the potential combine,
 for the duration of the period, aiding motion against the bias force. Consequently, in one period of the driving,
the dimer moves one spatial period against the bias force.

We remark here that in the work \cite{Mulhern.2011.PRE} we have also demonstrated the robustness of the  motion against the bias of the tilt
force with respect to additionally present  thermal fluctuations.

\FloatBarrier
\subsubsection[Adiabatic driving and directed transport in many unit systems]{Adiabatic driving and directed transport in many unit
systems}
\label{subsubsection:adiabatic}

\noindent Here we report on the Hamiltonian dynamics of a one-dimensional
chain of linearly coupled units in a spatially
periodic potential which is subjected to a slowly varying time-periodic, mono-frequency external force.
The average over time and space of the
related force vanishes and hence, the
system is effectively without bias which excludes any ratchet effect.
We pay special attention to
the escape of the entire chain when initially all of its units are
distributed in a potential well. Moreover for an escaping chain we explore the
possibility of the successive generation of a directed flow based on
large accelerations.

\subsubsection{The driven chain system of interacting units} We study a one-dimensional chain system consisting of
linearly coupled units with Hamiltonian of the following form \cite{Hennig.2008.PLA}
\begin{eqnarray}
H&=&\sum_{n=1}^{N}\left[\frac{p_n^{2}}{2}+U_0(q_n)+U_1(q_n,t)\right]\nonumber\\
&+&\frac{\kappa}{2}\sum_{n=1}^{N-1}\left(q_{n+1}-q_{n}\right)^2
\,,\label{equation:Hamiltonian}
\end{eqnarray}
wherein $p_n$ and $q_n$ denote the canonically conjugate momenta and
positions of the units evolving in the periodic, spatially-symmetric (washboard)
potential of unit period, i.e.,
\begin{equation}
U_0(q)=U_0(q+1)=-\cos(2\pi q)/(2\pi)\,.
\end{equation}
The external, time-dependent forcing field
\begin{equation}
U_1(q,t)=-F\sin({\Omega \,t+\Theta_0})q
\end{equation}
causes time-periodic modulations of the slope of the potential. It
has to be stressed that there is no additional bias force involved in the sense that
the following average over time and space vanishes, i.e.
\begin{equation}
\int\limits_{0}^{1}dq\int\limits_0^{T=2\pi/\Omega}dt
\frac{\partial U(q,t)}{\partial q}=0\,,
\end{equation}
with $U(q,t)=U_0(q)+U_1(q,t)$. The units interact linearly
with coupling strength $\kappa$.
Remarkably, as pointed out in  prior literature \cite{Yevtushenko.2000.PRE}
in the limiting case of uncoupled units, i.e. $\kappa=0$, there results an (unexpected)
asymmetry of the flux of units, emanating from one potential
well, and flowing to the left and right potential wells which
indicates the existence of directed transport without breaking the
reflection symmetry in space and time in this system. One reason for
the occurrence of phase-dependent directed transport is the lowering
of the symmetry of the flow in phase space by the ac-field where
this asymmetry vanishes only for specific values of the initial
phase $\Theta_0$ \cite{Yevtushenko.2000.PRE}. In Ref.~\cite{Soskin.2005.PRL}
the authors report further on this exceptional situation and show
that directed transport is sustained on fairly long time scales
despite the presence of chaos. In particular it has been
demonstrated that for sufficiently small forcing frequencies,
$\Omega \ll 1$, the width of the arising chaotic layer diverges
leading to a strong enhancement of the chaotic transport~\cite{Soskin.2005.PRL}.

The equations of motion derived from the Hamiltonian in Eq.~(\ref{equation:Hamiltonian}) read as
\begin{equation}
\frac{d^2q_n}{dt^2}\,+\,\sin(2\pi q_n)\,=\,F\sin(\Omega\, t +\Theta_0)+\kappa[q_{n+1}+q_{n-1}-2q_n]\,\,.\label{eq:qdot}
\end{equation}
The analysis here deals with zero initial phase of the external
force term, viz. $\Theta_0=0$.
The influence of the phase $\Theta_0$ on the generation of directed motion in the
periodic potential $U_0(q)$ has been studied under adiabatically slow driving in detail in Ref. \cite{Hennig.2008.EPJB}.
We stress that averaging over the initial phase $\Theta_0$ yields vanishing net
current in the periodically driven system.

\subsubsection{Extremely long transients of directed transport} In our dynamical studies the initial conditions are chosen such
that the units are initially contained in a potential well.
Furthermore, in the uncoupled case, i.e $\kappa=0$, escape from
the potential well induced by the driving force, as discussed in
\cite{Hennig.2008.EPJB}, is excluded. To ensure trapping in the driven but
uncoupled dynamics the orbits have to lie fairly deep inside the separatrix, yielding there
 a large island of stability with dynamics that is still regular.

Likewise, without the {\it weak} external ac-field and when the units are coupled the chain
is supposed to remain
trapped in the potential well despite the arising chaotic motion.

In terms of phase space structure we recall that in the case
of individual units, i.e. $\kappa=0$, in the corresponding adiabatically
driven system of one-and-a-half degree of freedom  there
arises a broad stochastic layer from the connection of various zones of instability
due to resonance overlap.
Non-contractible KAM tori confine the stochastic layer from below and above and form
impenetrable barriers
for motion in phase space. For the weakly nonintegrable system the chaotic sea contains still
islands of
regular motion. Provided these islands possess
non-zero winding numbers orbits with initial condition inside such an island facilitate
transport. Moreover, the motion
around these islands is characterised by the stickiness to them that
can lead to trapping of the trajectory for a long time \cite{Zaslavsky.1998.ImpColPress,MACKAY.1984.PhysicaD,Meiss.1986.PhysicaD,Denisov.2006.epl}.
This is due the intricate structure of
the stochastic layer where close to resonances at the boundary between
regular and chaotic regions there exists a hierarchy of smaller and
smaller islands and surrounding cantori.  The latter can severely restrict the transport
in phase space and thus effectively partition the chaotic layer \cite{MACKAY.1984.PhysicaD,radons.1989.ACP}.

It seems that the cantori are the less leaky the smaller the modulation
frequency $\Omega$. Hence they
form almost impenetrable barriers that confine trajectories for a very long but
{\it transient} period. One should
remark that eventually this transient period of directed motion terminates because
the trajectory escapes through one of the holes in the cantori and
accesses other regions of the chaotic layer.  Therefore the motion does not necessarily
proceed unidirectionally and
unless the trajectory becomes captured by ballistic channels \cite{Denisov.2001.PRE,Denisov.2002.PhysicaD,Denisov.2002.PREb}
it itinerates within the chaotic layer going along with changes of the direction of motion.

For nonlinear Hamiltonian systems with $N\ge 2$ degrees of freedom
only a few numerical results addressing the existence of an enhanced trapping
regime are known \cite{KANTZ.1987.PLA},\cite{Altmann.2007.EPL}.
It is supposed that the role played by cantori in driven systems with $N=1$ is played by
families of $N-$dimensional
tori, constituting partial barriers in the $2N-$dimensional phase space, where the chaotic
trajectory can stick to \cite{KANTZ.1987.PLA},\cite{Altmann.2007.EPL}.
On the other hand Arnold diffusion is possible and hence in principle a chaotic trajectory,
wandering along the entire stochastic layer,
can explore the whole phase space \cite{Arnold.1964.SovMathDok},\cite{Chirikov.1979.PR}.
However, due to stickiness to higher dimensional invariant
tori Arnold diffusion can be suppressed so that
certain stochastic regions are distinguished in which the trajectories
become trapped for longer times \cite{KANTZ.1987.PLA}.

We  demonstrate that it is possible to
generate directed motion with adiabatic periodic modulations of
the slope of the spatially periodic potential.

Concerning the initial conditions we proceed in the following way: Initially an amount
of energy $E_{n}=0.5 p_{n}^2+U_0(q_{n})\langle\Delta E$ is applied per unit such that the
whole chain is elongated homogeneously along a fixed
position  $\tilde{q}_0$ near the bottom of the well. Then, the
position and/or momenta of all units are
randomised.  The random position values are
chosen from a bounded interval $|q_n(0)-\tilde{q}_0| \leq \Delta q$ and,
likewise, the random initial momenta, $|p_n(0)-\tilde{p}_0|\leq \Delta p$.
The whole chain is thus initialised close to an almost
homogeneous state, but yet sufficiently displaced  ($\Delta q \ne
0$) in order to generate non-vanishing interactions, entailing the
exchange of energy among the coupled units.

First we note that in the case without external
modulation of the slope there occurs the formation of a pattern of localised
states due to modulational instability
(not depicted here). Due the irregular dynamics it happens that
occasionally a unit overcomes the potential barrier but no
coordinated motion of the chain results.

\begin{figure}
\centering
\includegraphics[scale=0.35]{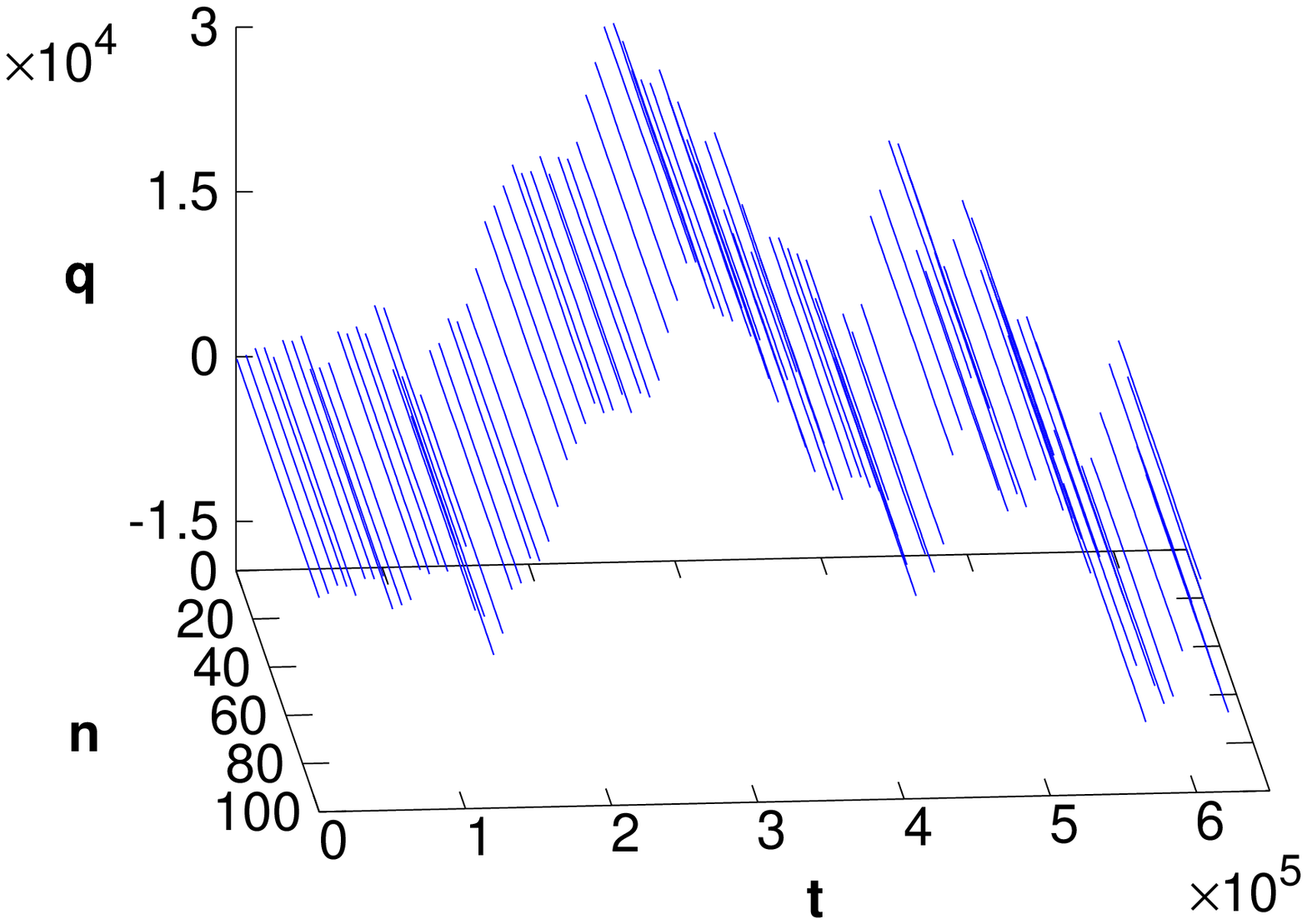}
\includegraphics[scale=0.35]{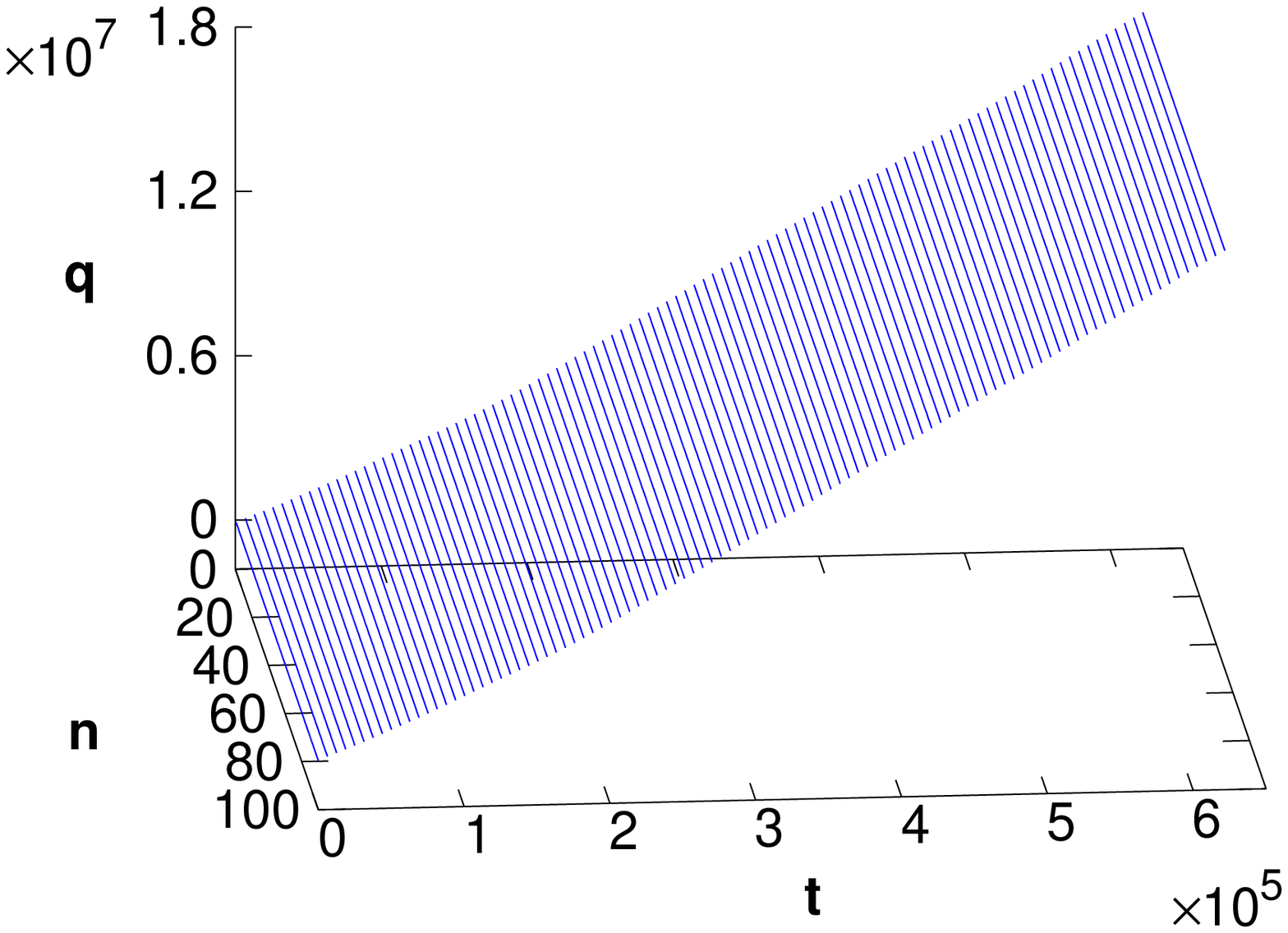}
\caption{(Colour online) Typical spatio-temporal evolution of the coordinates $q_n(t)$ for a
chain of $N=100$ coupled driven units, each evolving in a periodic potential - see Eq.~(\ref{eq:qdot}). Initially the coordinates are
uniformly distributed within the interval
$|q_n(0)-\tilde{q}_0| < \Delta q$ with an average $\tilde{q}_0=-0.35$ and  width $\Delta q=0.01$
and zero momenta, i.e. $p_n(0)=\tilde{p}_0=\Delta p=0$. The coupling strength is $\kappa=0.3$ and
the periodic external driving strength and phase are $F=0.05$ and $\Theta_0=0$, respectively.
In the left (right) panel, the driving frequency is $\Omega=10^{-1}$ ($\Omega=10^{-3}$). Source: Figure adapted from
Ref.~\cite{Hennig.2008.PLA}.}
\label{fig:Fig5}
\end{figure}

\begin{figure}
\centering
\includegraphics[scale=0.35]{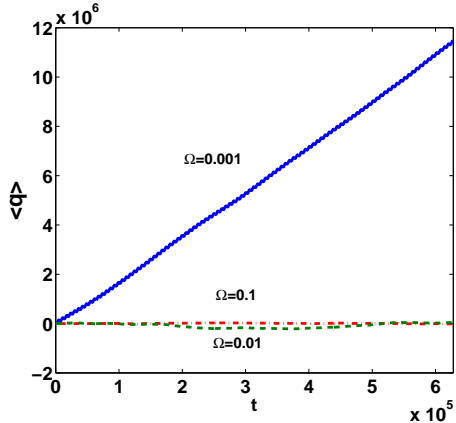}
\caption{(Colour Online) Temporal evolution of the mean value $\langle{q}\rangle=\langle\frac{1}{N}\sum_{n=1}^N\,q_n\rangle$ for chains
of
$N=100$ coupled units, each evolving in a periodic potential, driven with different driving
frequencies as indicated in the plot. The corresponding equations of motion are described by Eq.~(\ref{eq:qdot}). Ensemble averages,
denoted by $\langle\cdot \cdot\rangle$ were performed over
$100$ realisations of the initial conditions: Initially the coordinates of each unit are
uniformly distributed within the interval
$|q_n(0)-\tilde{q}_0| < \Delta q$ with an average $\tilde{q}_0=-0.35$ and  width $\Delta q=0.01$
and zero momenta, i.e. $p_n(0)=\tilde{p}_0=\Delta p=0$. The coupling strength is $\kappa=0.3$ and
the periodic external driving strength and phase are $F=0.05$ and $\Theta_0=0$, respectively. Source: Figure adapted from
Ref.~\cite{Hennig.2008.PLA}.}
\label{fig:Fig6}
\end{figure}

Remarkably, applying
the adiabatic modulation the entire chain not only escapes from the potential
well but manages also to travel freely and unidirectionally over a giant distance as seen from the spatio-temporal
evolution
of $q_n(t)$ in Fig.~\ref{fig:Fig5}.  In comparison for faster modulations no directed motion
of the chain is obtained. The chain consists of $100$ coupled
units and open boundary conditions are imposed.
The profile of the chain continuously undergoes changes with ensuing deviations from a flat state.
Nevertheless the intriguing feature of transients of extremely long-range transport of the chain
is provided by collective motion
which is also reflected in the temporal behaviour of the mean value of the coordinate
$\langle{q}\rangle=\langle\frac{1}{N}\sum_{n=1}^N\,q_n\rangle$ shown in Fig.~\ref{fig:Fig6}.
Ensemble averages, denoted by $\langle \cdot \cdot \rangle$ were performed over $100$ realisations of trajectories
whose initial condition lie in the
range given in the caption of Fig.~\ref{fig:Fig5}.
Note that while for driving with
$\Omega \lesssim 10^{-2}$ on average no directed motion results for sufficiently slow modulations
(here illustrated for
$\Omega=10^{-3}$) a large directed current is observed.
Conclusively, concerning transport, the many unit system exhibits features equivalent to that
observed in the system of individual units \cite{Hennig.2008.EPJB}.

In fact, our findings imply that in the driven $N$-unit
system the motion takes place in ballistic channels
\cite{Denisov.2001.PRE} associated with stickiness to $N$-dimensional invariant tori.
Notably, in the $N-$unit system the mean value $\langle{q}\rangle$ evolves in the same manner as
the single coordinate in the individual unit counterpart, viz. it exhibits effective growth.
Thus, the collective directed motion of the numerous microscopic degrees of freedom
is manifested at a
collective level in the evolution of the mean value of ${q}$.

\section{Summary and outlook}\label{sect:summary_final}
This review has considered systems of coupled oscillatory units, both in the absence and presence of
time-dependent forcing and/or environmental noise sources. The main focus was on the objective of dynamics driven escape events and in the case of non-confining
potential landscapes on the possibility of undergoing an escape driven  directed transport.
By use of both analytical and numerical means,  the main goal has been to develop a thorough understanding of
the coupled nonlinear dynamics in such systems. Of particular interest were the conditions under
which collective escape and/or directed transport may emerge.
Sometimes this is a trivial artefact of the initial conditions, including the choice of suitable parameter regimes.
Given the case, for example,
that there is insufficient energy available for the particles to overcome the potential barriers, would then imply a vanishing transport and consequently
no emerging directed current. It may also be the case  that there is sufficient energy available, but the coupling between the particles is rather weak
so that the initial
direction of motion may be maintained, yielding a  non-zero current. However, quite often it is the
case that escape and subsequent directed transport
can only emerge within a synergy where the particles work together. That is, cooperative effects become then
prominent in yielding possible escape and/or finite transport.

A key point that has been evident throughout this review is that the constructive cooperation between
the particles allows for novel escape mechanisms and transport scenarios that are absent in the case of a single particle dynamics.
Consider three examples; a first one elaborated in   Sec.~\ref{subsubsection:1d-chain_model}, a second one
from Sec.~\ref{subsection:autonomous}, and yet another from
Sec.~\ref{subsection:non-autonomous}. In Sec.~\ref{subsubsection:1d-chain_model}, we elucidated the case of a
noise-free, i.e. a purely deterministic scenario, as well as a thermally activated, noise-assisted
escape process of a chain of coupled units out of a metastable potential well.  Strikingly,
in the noise-free situation
the chain is not only able to generate  a critical
elongation in a very efficient self-organised manner, thereby
successfully surpassing the potentials local maximum, but generally can also outperform its thermally activated counterpart
by achieving (especially for smaller energies)  a considerably enhanced
escape. In the deterministic (noise-free) case, a substantial cooperation between the units expedites
the formation of the critical configuration, viz. the transition
state, whereas thermal noise may impinge on such cooperativity.
In Sec.~\ref{subsection:autonomous} it was
observed that a resting particle is able to escape from its initial confining well and undergo
directed transport,
after a transient period of chaotic energy exchange with a second particle. If
the particles were uncoupled,
this initially dwelling particle would remain trapped indefinitely as this system
obtains no assistance from external
perturbations. Therefore, it is exclusively cooperation between the particles that allows the
initially resting
particle to become transporting. Conversely, the process of a permanent bond formation
(dimerisation) may result,
as a consequence of the coupling, yielding no net transport. In view of a potential for beneficial   applications,
this process may be
seen as a filter, capturing/blocking a particle by
impeding its movement. As a second example, let us turn to
Sec.~\ref{subsubsection:running}. For
the particular model considered there, if the subsystems are uncoupled, then
the dynamics is chaotic with
the individual particles exploring  strange attractor dynamics. However, when the
particles are coupled, it
is possible that the motion can become regular and periodic, cf. Fig.~\ref{fig:running}.
Therefore, working synergetically
together the particles can suppress the effects of chaos. This effect of suppression
of chaos
has been  observed in \cite{Mulhern.2013.EPJB}, where the systems only positive Lyapunov exponent
(in the uncoupled limit)
decreased almost to zero as a result of coupling between the particles.

This review also examined two different coupling regimes. In
Sec.~\ref{subsection:autonomous}
the coupling
was short-range in nature, allowing for the effective decoupling of the particles
(through increasing
distance between the particles). In contrast, the coupling used in Sec.~\ref{subsection:autonomous}
was such that the
particles remain coupled even when the distance between them is large. It is worth
(briefly) discussing
the differences between these regimes.

The conservative Hamiltonian systems of Sec.~\ref{subsection:autonomous}
were endowed with an initial
small energy. Using
a long range coupling regime in this case would preclude the unbounded directed motion
of a single particle,
as the energy would  quickly become usurped by the interaction potential. Thus, many of the
transport scenarios
discussed in Sec.~\ref{subsection:autonomous} would not occur. Therefore, it is preferable to work with a short
range coupling in this case.

For the time-dependent driven and damped systems of Sec.~\ref{subsection:non-autonomous} the external driving force can
cause the rapid
separation of the particles. For short-range couplings this results in
effectively decoupling the particle dynamics
after short periods of time, even for relatively strong coupling strengths. The dynamics
in this case are
not of potential use. This, however, no longer holds true for long-range coupling where the dynamics have
proven to be beneficial.

While the second part of this review looked mainly at systems
of two coupled oscillators, it is certainly worthwhile
to extend
these results to similar systems with an arbitrary number of degrees-of-freedom. One
of the challenges
is to develop (reduction) tools, alternative to the method of Poincar\'{e} maps,
with which the underlying geometrical phase space
structures such as invariant hypertori in higher-dimensional phase-spaces  can be
 visualized appropriately.
However, this task becomes hopeless rather quickly with increasing particle number $N$ (or, more generally increasing  number
of degrees-of-freedom). This feature is a
consequence of the Froschl\'{e} conjecture \citep{Lichtenberg.1983.sprverny}. Therefore, it
is appropriate to have
a gradual approach to dimensionsional increase (as suggested by \cite{jung.2010.njop} for the
problem of chaotic scattering
in higher dimensional systems). Work in this direction has already been carried
out; see, for example, in Ref.
\cite{katsanikas.2011.ijobc}. Those authors used a technique known as \emph{colour and rotation}, where
$3$D projections of a
$4$D space are produced and the fourth dimension is represented by colour. While
this approach does provide useful information, it is not applicable to large
dimensional systems.

Furthermore, the transition from regular to chaotic motion in higher-dimensional
phase-spaces,  assisted
by unstable periodic orbits needs to be investigated. That is, in order to gain insight
into the transition
from stability to complex instability the structure of invariant manifolds associated
with unstable periodic
orbits needs to be explored. In particular with regard to the emergence of  transport
scenarios in higher-dimensional phase-spaces, the question of whether the phenomenon of
stickiness of chaotic orbits to the vicinity of periodic orbits is exhibited by
higher-dimensional dynamical systems is of  interest.
Moreover, diffusive features in higher dimensional systems entailed by motion along
unstable invariant manifolds, corresponding to unstable periodic orbits with distinct
magnitudes of eigenvalues of the linearised systems merit to be addressed. A question
of interest is whether multiple channels supporting diffusive behaviour coexist, or whether there emerges
a single dominant channel that prevails.

Further, it is worth considering the stochastic counterpart of the systems considered in the
second part of this review.
The introduction of noise to a system affects the dynamics in numerous ways, depending on the
type of noise and its strength. With regard to solutions of a system, particularly periodic
solutions, noise can be added to test the solutions stability. However, of interest is the
study of how trajectories that are close to a separatrix, which separates bounded from
unbounded motion, behave. For example, can those trajectories that are in bounded regions
of phase-space (or at a localisation/de-localisation transition in parameter space) be
subsequently kicked, under the influence of noise, into unbounded regions of phase-space?
Conversely, for those trajectories in unbounded regions of phase-space, is noise enhanced
trapping \cite{Altmann.2010.prl} a feature of the stochastic system?

Last but not least, our review here has been restricted to an interacting classical nonlinear dynamics.
The inclusion of quantum effects are much more demanding, both analytically and numerically. On the other hand, the phenomenon of
quantum tunneling and chaos assisted tunneling does reveal new routes for escape that are not classically available. This may well 
enrich the characteristics for escape and  directed  and driven quantum transport \cite{kohler.2005.pr}. In this spirit there 
remains plentiful space and time for exciting future developments.

\section{Acknowledgements}\label{sect:acknowledgements}
The authors thank Sergej Flach and Anna Deluca Silberberg for constructive and insightful discussions. Further, we would like 
to give acknowledgement to former coauthors including S. Fugmann, S. Martens, T. Gross, A.D. Brubanks, and A.H. Osbaldestin. 
This research was supported by the 
Volkswagen-Foundation projects I/80425 (L.S.-G.)  and I/80424 (P.H.) the  DFG Sachbeihilfen HA1517/31-2, and HA1517/35 (P.H.). This 
work was partially supported by the European Union, Seventh FrameworkProgramme  (FP7-REGPOT-2012-2013-1) under grant agreement  316165 
(G.T.). L.S-G would like to acknowledge DFG-IRT1740 for financial support.

\FloatBarrier
\section{References}
\bibliographystyle{elsarticle-num2}
\bibliography{PR-bibliography}

\end{document}